\documentclass[a4paper,11pt]{article}
\pdfoutput=1 

\usepackage{jheppub} 

\usepackage[T1]{fontenc} 
\usepackage[usenames,dvipsnames]{xcolor}
\usepackage[force]{feynmp-auto}

\newcommand{\half}{\frac{1}{2}}
\newcommand{\la}[1] {\left\langle #1 \right\rvert}
\newcommand{\ls}[1] {\left\lbrack #1 \bf \right\rvert}
\newcommand{\ra}[1] {\left\lvert #1 \right\rangle}
\newcommand{\rs}[1] {\left\lvert #1 \bf \right\rbrack}
\newcommand{\da}[1] {\left\langle #1 \right\rangle}
\newcommand{\ds}[1] {\left\lbrack #1 \bf \right\rbrack}

\usepackage{verbatim}
\usepackage{amssymb}
\usepackage{amstext} 
\usepackage{array}   
\newcolumntype{L}{>{$}l<{$}} 

\usepackage{comment}

 \includecomment{feynfig} 

\allowdisplaybreaks[1]

\title{Consistent Scattering Amplitudes, Yang-Mills, the Higgs Mechanism and the EFTs Beyond}

\author[a,1]{Timothy Trott%
\note{Currently at private address.}}
\affiliation[a]{Department of Physics, University of California, \\
Santa Barbara, CA 93106, U.S.A.}

\emailAdd{drtimothytrott@gmail.com}

\abstract{I study constraints on fundamental physics emerging from consistency of a unitary, local and perturbative $S$-matrix in $4d$. For massless particles, some new constraints arising from consistent complex factorisation of $2\rightarrow 2$ amplitudes are derived, leading, in particular, to the complete structure of gluonic three-particle amplitudes. 
For massive particles, a hierarchy of constraints may be derived instead by imposing a maximum rate of unitarity-violating growth in the high energy limit. All $2\rightarrow 2$ tree-level amplitudes of massive particles with spin $s\leq 1$ are calculated in generality using on-shell methods and presented with manifest high energy dependence. The anatomy of these amplitudes' helicity sectors is dissected in order to identify conditions under which their energy growth is limited or eliminated. Using these results, it is shown that the scattering of massive vector bosons has suppressed, but not fully unitarised, high energy dependence if
the parity-conserving parts of their self-couplings are Lie algebra structure constants, possibly non-semisimple or non-compact, and the parity-violating parts are ``generalised Chern-Simons terms''. Full unitarisation then requires the standard Yang-Mills Lie algebra properties and, for a gapped spectrum, the Higgs mechanism. These amplitudes are assembled, embedded and unified into elegant superamplitudes in theories with extended supersymmetry when the particles are BPS. In doing so, I elucidate cancellations that otherwise remain hidden. More generally, a broader landscape of EFTs is charted through various combinations of constraints and coupling hierarchies. 
}

\begin{document} 
\maketitle
\flushbottom

\section{Introduction}\label{sec:intro}

Quantum field theory (QFT) is widely believed to emerge as the reconciliation between quantum mechanics and special relativity \cite{Coleman:2011xi,weinberg2005quantum}. Effective field theories (EFTs) are QFT approximations to low energy phenomena and are identified by a sequence of Wilson coefficients defining a hierarchy of interaction strengths. Many open problems in fundamental physics, such as the meaning of the couplings and spectra of the Standard Model (SM) of particle physics, have a multitude of possible QFT and EFT solutions \cite{Weinberg:1979sa,Brivio:2017vri,DiLuzio:2020wdo,Cheung:2007st}. Others, such as the quantum evolution of black holes and spacetime, are much less clear, although low energy gravity should admit an EFT description on a flat enough background \cite{Burgess:2003jk}. These problems plausibly require a revision of some of the assumptions which led to the QFT picture in the first place and it would be interesting if this had implications or signatures in the space of low energy gravitational EFTs. It is, for these reasons, desirable to understand the breadth of possible phenomena that can be accommodated by QFTs and EFTs in as great a degree of generality as possible and how they emerge from requirements of fundamental principles. 

The $S$-matrix provides a framework for distilling the inevitability of QFT from fundamental assumptions of relativity and quantum mechanics. Before the acceptance of Yang-Mills (YM) theory as the QFT underpinning nuclear physics, $S$-matrix principles were studied as a strategy for describing strongly coupled particles. When targeted instead at specifically perturbative (i.e. weakly coupled) theories of particles, it was argued that massless particles with helicities $\geq 1$ could have self-consistent interactions for only a highly restricted class of renowned theories. These conclusions were drawn by taking soft limits of scattering amplitudes in which massless spinning particles are absorbed or emitted. The limits would be dysfunctional (yet should be theoretically observable) without special structure imposed upon the coupling constants. In particular, a single soft massless vector boson had to be a photon of the electromagnetic field with coupling strength determined by a conserved electric charge \cite{weinberg1964photons}. Only a single massless helicity-$2$ particle was permissible and it had to be the graviton, universally coupling to matter in obedience to the equivalence principle \cite{weinberg1964photons}. Massless helicity-$3/2$ particles had to be gravitinos in a theory of supergravity \cite{GRISARU1977323}. Gravitating massless higher spin particles were altogether inconsistent \cite{Weinberg:1980kq}. These arguments were, however, posed through off-shell gauge theoretic formulations. This both obscured the physical root of the tension and the generality to which these arguments could be most fully extended. 

Much later, it was discovered that the perturbative $S$-matrix was generated by analytic continuations of trivalent on-shell scattering amplitudes that were otherwise invisible because of their triviality with Minkowskian kinematics \cite{Witten:2003nn}. Powerful methods were developed for performing computations of scattering amplitudes by recursing down to elementary on-shell amplitudes that could be determined entirely from the quantum properties of the scattered particles in conjunction with Lorentz invariance. These elementary amplitudes could be fused together using the rules of quantum unitarity and locality to determine more complicated transition amplitudes. These methods bypassed use of the Feynman rules and off-shell actions which, in order to preserve manifest Lorentz invariance of the generating functionals, were hampered by the introduction and removal of unphysical degrees of freedom known as ``gauge invariance''. These unphysical degrees of freedom are absent from the transition amplitudes between physical states of particles characterised by their little group representation. On-shell methods therefore avoid the complications of gauge redundancy by only making use of building-blocks involving physical particle states. Significant progress was made in the computation of scattering amplitudes in gauge theories (including gravity) as a consequence of this \cite{elvang2015scattering,Dixon:2013uaa}. 

In light of the new on-shell methods, interest in the old bootstrap arguments were resuscitated when it was noticed that some of its conclusions could be drawn from requiring on-shell recursion to work consistently \cite{Benincasa:2007xk}. The kinematic parts of hypothetical elementary on-shell three-particle amplitudes could be constructed from Lorentz invariance and covariance under the little groups of the external particles. However, only for coupling constants obeying certain conditions could the amplitudes be sensibly combined into higher leg amplitudes obeying the fundamental rules expected of the $S$-matrix. These rules are: Lorentz invariance (and little group covariance), locality, in the form that the amplitude is an analytic function of the external particles' helicity spinors with simple Mandlestam poles as its only singularities, and finally unitarity, in the form of factorisation of the amplitude into a product of transition amplitudes multiplied by a wave propagator when the external kinematics allow for the production and propagation of an intermediate on-shell particle. This will be reviewed in Section \ref{AllChannel} below. Of particular note is the expected property of crossing, which is presumably part of the required analytic structure of the amplitudes \cite{GellMann:1954ttj,Mizera:2021fap,Caron-Huot:2023ikn}. 

In addition to the consistency conditions from soft limits listed above, further identifying constraints were shown to follow from these $S$-matrix rules. Multiple species of massless vector bosons could have self-interactions only if they were identified with gluons of non-Abelian YM theory. Furthermore, other particles had to interact with the gluons through couplings identified as generators of some Lie algebra. It was subsequently realised \cite{Schuster:2008nh} that these consistency constraints stemmed from tension in the capacity of a $4$-particle amplitude to be able to factorise correctly in each possible Mandelstam channel. This is because the residues of the Mandelstam poles for each channel can themselves contain poles, and therefore contain information about the other channels. A systematic study of consistency with on-shell complex factorisation was conducted in \cite{McGady:2013sga}, from which much of perturbative particle physics was able to be reconstructed. Contact between complex factorisation and the historical soft-limit arguments was made by the formulation presented in \cite{Elvang:2016qvq}. This provided a systematic method for generating these consistency constraints and extending them to higher leg processes. Nevertheless, some components of the complete argument remained elusive. 

Extension of some of these results to theories including massive particles has been made in \cite{Arkani-Hamed:2017jhn,Chung:2018kqs}. On-shell amplitudes of massive particles are more complicated because they admit many more possible independent Lorentz structures (in particular, the on-shell $3$-particle amplitudes are not generally kinematically unique as they (almost) are for massless particles). Their construction, even in cases with only three or four external particles, is closely tied to the problem of operator basis construction in effective field theories. This is also a topic that has also received significant recent attention, for both massless and massive theories, because of its applications to classifying deviations from the SM of particle physics produced by possible extensions \cite{Brivio:2017vri,DeAngelis:2022qco,Dong:2022mcv}, or possibly less structured departures to the properties of the Higgs and weak bosons \cite{Pich:2018ltt,Alonso:2023upf,Liu:2023jbq,Durieux:2019eor}, understanding the gravitational scattering of macroscopic bodies \cite{Ema:2025qgd,Chung:2018kqs,Chiodaroli:2021eug} and the determination of other constraints, stemming from causality and unitarity, imposed from the $S$-matrix bootstrap beyond three-legs \cite{Adams:2006sv,Tolley:2020gtv,Caron-Huot:2020cmc,Bellazzini:2020cot}. The full $S$-matrix bootstrap, beyond the simplifications offered by leading order perturbation theory, has itself received significant renewed interest in recent years, see e.g. \cite{Kruczenski:2022lot} for overview and references. This has been stimulated by advances in applications to CFTs \cite{Simmons-Duffin:2016gjk,Poland:2018epd} and motivated by seeking a better understanding of the landscape of consistent EFTs \cite{Bellazzini:2021oaj,Bellazzini:2025shd,Bertucci:2024qzt,Haring:2022sdp} and the possible of implications for string theory and quantum gravity \cite{Caron-Huot:2022ugt,Caron-Huot:2024lbf,Calisto:2025tjo,Hillman:2024ouy} (a small sample of many references), among other reasons.

Most of the progress in efficient construction of observables from on-shell methods has been made for massless theories. Massive amplitudes are much less well understood. Fewer techniques for systematically computing on-shell amplitudes are known \cite{Wu:2021nmq} (certainly compared to massless theories) and little successful use has been made of those that are (although \cite{Ema:2024rss,Ema:2024vww} present some interesting new advances). A better understanding of the energy dependence and structure of massive amplitudes could help with this. 

The singularity structure of tree-level amplitudes of massive particles is much less constraining than for massless particles. For massless particles, $4$-leg the residues consisting of factorised $3$-particle amplitudes frequently contain cross-channel poles. This does not happen in purely massive theories, so consistent factorisation is much less constraining (although interactions mixed with massless particles can still impose some demands \cite{Arkani-Hamed:2017jhn,Chung:2018kqs}). Together with the expanded sets of permissible Lorentz structures, massive $3$- and $4$-particle amplitudes have much more parametric freedom consistent with locality. The majority of this new freedom describes interactions that induce power law growth in energy that would lead to violations of perturbative unitarity if extrapolated too far beyond the masses of the participating particles. It was shown in \cite{Cornwall:1974km} that demanding such divergences in $4$-particle amplitudes be eliminated places restrictions on the underlying $3$-particle interactions resembling those obtained from requiring consistently factorising massless theories. This led, in particular, to the structure of spontaneously broken Yang-Mills and the Higgs mechanism. However, it is nevertheless understood that the enlarged space of massive $3$-particle interactions can be consistent within an effective field theory that approximates particle physics below some maximum energy scale, at least once supplemented by some power counting scheme to provide a sense of order. The high energy analysis specifically identifies theories that can be perturbatively extended to higher energies to varying degrees. However, as for the massless constraints, past studies have always involved some degree of question begging which led to the overlooking of some new results that will be derived here.

This work has three main objectives. The first is to complete the derivation of the rules of perturbative particle physics from the fundamental factorisation properties of the massless $S$-matrix. This is performed in Section \ref{AllChannel} using a thus far neglected category of factorisation channels. This allows for the reconstruction of the complete derivation of perturbative Yang-Mills from $3$-particle gluon amplitudes, including the required constraints on the properties of the Lie algebra, parity ($P$) and time-reversal ($T$) symmetries. The properties of the matter couplings are also fully determined. While this work was in preparation, a similar argument was independently presented by \cite{Fonseca:2025mzj}. While there is significant overlap in the results and some of the core elements of the argument, I retain the exposition presented here because there are some differences/complementarity in our approaches and for its thematic continuity with the new results in remainder of the study. Furthermore, I include some additional details (such as couplings to massive particles). 

The second goal is to explicitly construct, again through direct recourse to unitarity and locality, all $2\rightarrow 2$ exchange-mediated scattering amplitudes involving only massive particles of spin $\leq 1$ at tree-level. In preparation for the calculations involving external massive vector bosons, I begin in Section \ref{Sec:LowSpinAmp} by warming-up with the amplitudes involving only external particles of spin $\leq 1/2$ and examine their basic properties. These have been calculated previously by \cite{Christensen:2022nja}. However, along the way, I demonstrate how they assemble into the especially elegant expression predicted for $\mathcal{N}=2$ super-electrodynamics (SQED). This Section also contains a review of superamplitudes and the special kinematic properties of massive $3$-particle amplitudes of BPS(-like) particles. 

Sections \ref{Sec:MassiveVectors}, \ref{VectorHiggs}, \ref{VecSca} and \ref{VecFer} are dedicated to the calculations of the amplitudes involving external massive vector bosons. In Section \ref{Sec:MassiveVectors}, I introduce the general structure of the massive on-shell three vector amplitude. Most of this treatment is new. In particular, I give a general explanation of the coupling structure and the nature of its reinterpretation in massless theory at high energies. 

In Section \ref{VectorHiggs}, the $3$-particle amplitudes are used to compute the four vector amplitude induced by massive vector exchanges. Various $2\rightarrow 2$ exchange amplitudes involving massive vector bosons have been calculated directly from unitarity methods and using chiral spinor variables in numerous recent studies \cite{Christensen:2022nja,Christensen:2024bdt,Christensen:2024xzs,Bachu:2019ehv,Liu:2022alx,Durieux:2019eor,Ema:2024rss,Ni:2026wiz,Ni:2026mia}, mostly for the specific context of the SM, but in some cases with particle spectra in greater generality. Unlike these previous studies, I perform the calculations without prior assumptions about the absence of Lorentz structures representing effective operators. Besides being of intrinsic interest, these general calculations facilitate the third goal of this study: to identify the conditions on the couplings under which the high energy dependence is suppressed. It is found that requiring complete cancellation of divergences in the high energy limit (HEL) leads to the expected conditions satisfied by spontaneously broken Yang-Mills and the Higgs mechanism (including the Lie algebra structure of the possible matter couplings). This is, of course, very well-known \cite{Cornwall:1974km}. However, intermediate regimes are also identified in which the high energy divergence is present but weakened. This partial suppression still requires that the parity-symmetric parts of the three vector self-couplings be Lie algebra structure constants, but without the additional conditions (compactness, semi-simplicity, homogeneous generator normalisations) necessary for standard YM. Parity-violating couplings are also allowed and these are identified as generalised Chern-Simons (GCS) terms. These requirements are more broadly satisfied by gauged non-linear sigma models (NL$\Sigma$Ms) in field theory. This demonstration of emergence from the requirement of a partially unitarised regime is, to my knowledge, new. When some of the vector bosons are massless gluons, it is shown that a subset of the Lie algebra structure and GCS relations follow instead from consistent factorisation.

The mixed amplitudes with external scalar and fermion legs are studied in the subsequent Sections \ref{VecSca} and \ref{VecFer}. The emergent Lie algebra structure of the scalar and fermion couplings to vectors is shown to automatically accommodate this enlarged space of partially unitarised vector self-interactions. Likewise for the covariance of the Yukawa couplings under these representations, although the covariance of other possible effective interactions is only required in the limit that they are regarded as small corrections. Generally, divergences in separate helicity sectors may be individually suppressed in assorted combinations to produce different classes of effective theories (presumably underpinned by different hierarchies or power counting schemes). 

Returning to the second goal, it remains to elucidate how the cancellations actually transpire in the amplitude. Derived from factorisation, the massive amplitudes are constructed in a manifestly local form (in the sense that each separate term contains a single pole that can be directly identified with a particular factorisation channel). However, if the appropriate conditions on the couplings are enforced, the corresponding high energy cancellations are hidden. To expose them, I explicitly convert the amplitude into a form in which the cancellations are manifest (while preserving little group covariance). This usually involves combining together terms from different factorisation channels. Once in this form, the massive amplitudes can be reconciled with the corresponding massless amplitudes onto which they are expected to match in the HEL (since these can often only be expressed covariantly through terms containing pairs of Mandelstam poles). The $4$-particle superamplitude of spontaneously broken $\mathcal{N}=4$ super-Yang-Mills (SYM) provides an especially elegant demonstration of this. High energy unitarity is automatic and manifest when it is presented in a form with manifest supersymmetry (SUSY) (which can be effortlessly inferred from first principles). The component amplitudes are compared to the general expressions that I construct as a cross-check. The $\mathcal{N}=4$ SYM components provide simple, idealised examples that can be used to help guide the general cases toward a form that trades manifest locality for manifest high-energy dependence. The general analysis of the explicit reconciliation between the different forms of the amplitudes (one with manifest locality and the other with manifest high energy unitary) is largely new (although \cite{Christensen:2024bdt} has previously discussed some limited examples in the SM involving gluons and photons). The calculations also help to elucidate a lot of technical structure underpinning the kinematics of massive amplitudes, which may have applications to further studies (it is put to significant use in \cite{Trott:2026ozo} toward the application of the $4$-particle test to massive superamplitudes).

Appendix \ref{sec:HPoB} is a compilation of useful identities involving Lorentz structures appearing in the massive $4$-particle amplitudes. Appendix \ref{sec:S3PMK} provides some example analytic continuations of the complex factorisation channels from the Minkowskian limit. Appendix \ref{sec:ExoticGT} gives a simple example of the relationship between the geometry of the gauge algebra and the multipole structure of the on-shell three vector amplitude. Finally, Appendix \ref{sec:HDOInsertions} compiles some self-contained subsectors of the general $2\rightarrow 2$ amplitudes that are of tangential relevance to the main narrative of the study.

This paper is the first of two parts dedicated to the three goals described above. The second part \cite{Trott:2026ozo} specifically focuses on the inclusion of supersymmetry, gravity and supergravity (SUGRA), still mostly focusing on theories with spin $\leq 1$, but with some additional comments beyond this on SUSY breaking and the super-Higgs mechanism. In particular, it develops the $4$-particle test for massive superamplitudes in theories of extended SUSY/SUGRA, enabling the derivation of new consistency constraints on the structure of theories with BPS particles.

\section{Consistency from Complex Factorisation and the All-channel Pole}\label{AllChannel}

\subsection{Complex factorisation}\label{sec:ComplexFac}

The $S$-matrix is a Lorentz invariant function of helicity spinors and must transform under the little group in the same way as the external particle states. See \cite{weinberg2005quantum,Arkani-Hamed:2017jhn} for an overview. Observationally, the $S$-matrix entries in perturbative theories have, at leading order in small coupling constants, simple poles from particle propagators as their only singularities. Potentially more complicated singularity structures, as would arise from loop integrals, are perturbatively suppressed and may be approximately neglected. I will assume this to be the case throughout this entire paper. Locality and unitarity are therefore posited as the following rules for the $S$-matrix:
\begin{itemize}
    \item The only permitted singularities are simple poles (in Mandelstam variables) with residues of the form: $s_{1\dots m}\rightarrow m_P^2$ 
    \begin{align}\label{GenFact}
    \Rightarrow A_n(1,\dots n)\sim\frac{-1}{s_{1\dots m}-m_P^2}\sum_PA_{m+1}(P, 1\dots m)A_{n-m+1}(P\rightarrow m+1\dots n),\end{align} 
    where $s_{1\dots m}=-(p_1+\dots p_m)^2$ and $m<n-1$. Here $P$ indexes the intermediate particle species of mass $m_P$. I assume here that external particles are all outgoing (unless specified otherwise) and that the amplitudes related to each other by crossed particles can be analytically related to each other (see more details below). 
    \item Any factorisation channel consistent with the spectrum of particles and their interactions necessarily occurs.
\end{itemize}
This is supported by non-perturbative polology \cite{weinberg2005quantum}, although this lies within the LSZ framework and does not necessarily apply to scattering amplitudes in massless theories. The analytic properties are linked to microcausality in the field-theoretic picture \cite{GellMann:1954ttj}.

Accepting these conditions on the amplitudes, then, with the special exception of $3$-particle amplitudes involving massless particles, the leading order amplitude can always be written in terms of manifestly local Lorentz (or spinor) structures multiplied by some number of factors (possibly zero) of Mandelstam poles. Here, ``manifestly local'' means constructed from products of spinor bilinears, never quotients. The simple poles are the only permitted singularities at finite momentum and they must respect the locality and unitarity rules posited above. The little group or spin polarisation information of the amplitude is encoded in the spinor dependence of the numerators. This includes possible ``contact terms'' - terms that do not contain any Mandlestam poles. For $4$-particle amplitudes in particular, the focus of this work, this factorisation information is enough to bootstrap candidate expressions for the amplitude consistent with unitarity. Further assumptions are then made to justify a particular choice or truncation of the remaining possible contact terms (which are typified by polynomials in Mandelstam invariants multiplying spinor structures). Numerous illustrative examples will be provided in the Sections that follow, in particular Sections \ref{Sec:LowSpinAmp} and \ref{VectorHiggs} where the construction procedure will be explicitly demonstrated for both massive and single pole massless exchange amplitudes.

As is now well-known, the factorisation rules above even include amplitudes between four massless particles. In this case, the Mandelstam pole (\ref{GenFact}) can only be non-trivially approached by momenta analytically continued away from Minkowskian kinematics, which is also required for the non-vanishing of the $3$-particle amplitudes into which the $4$-particle amplitudes factorise. For a given analytic continuation, massless $3$-leg amplitudes are given by a single possible Lorentz structure determined by the helicities of the particles (see numerous examples in Section \ref{sec:AmpList} below). For example, taking an amplitude $A_4(1,2,3,4)$, the limit $s=\ds{12}\da{12}=\ds{34}\da{34}\rightarrow 0$ could be implemented as $\ds{12},\da{34}\rightarrow 0$ or $\da{12},\ds{34}\rightarrow 0$. In the first case, the factorisation residue is given by $\sum_P A_3(\ra{1},\ra{2},\ra{P})A_3(\rs{P}\rightarrow \rs{3},\rs{4})$, where $A_3(\ra{1},\ra{2},\ra{P})$ is entirely constructed out of right-handed spinors (which, for complex kinematics, are free and independent of the left-handed spinors, the latter of which have all aligned: $\rs{1}\propto\rs{2}\propto\rs{P}$) and $A_3(\rs{P}\rightarrow \rs{3},\rs{4})$ out of left-handed spinors. Here $\{P\}$ is the set of possible intermediate particles that can appear in the non-zero $3$-particle amplitudes \cite{elvang2015scattering}. 

Massless $3$-particle amplitudes are entirely fixed by Lorentz and little group symmetries for a given analytic continuation. In many cases, they unavoidably contain reciprocal dependence on spinor bilinears. As a result, the residues that they induce in 
(\ref{GenFact}) can themselves contain Mandelstam poles. This introduces internal tension within the unitarity conditions: the full $4$-particle amplitude is prohibited from have higher order poles, while the simple poles describing factorisation in each channel must be consistent with those induced in other channels. This offered the opportunity in \cite{McGady:2013sga} to bootstrap constraints on candidate $3$-particle amplitudes by testing their ability to unitarily combine into consistent $4$-particle expressions. 

A general massless $4$-particle amplitude, once the spinor structures are factored out, is necessarily decomposable into terms with zero, one, two or three Mandelstam poles, schematically represented as
\begin{align}\label{4masslessSchem}
A_4(1,2,3,4)=f(\ds{ij},\da{ij})\left(\frac{\#}{stu}+\frac{\#}{st}+\frac{\#}{tu}+\frac{\#}{us}+\frac{\#}{s}+\frac{\#}{t}+\frac{\#}{u}+p(s,t)\right),
\end{align}
where $f(\ds{ij},\da{ij})$ is the factor that accounts for the polarisation information and $p(s,t)$ is some polynomial that is largely irrelevant for the subsequent analysis. I assume here (and throughout) that the massless particles are helicity eigenstates. The factoring out of the spinor structure $f(\ds{ij},\da{ij})$ is possible here because it happens to be unique for massless $4$-leg amplitudes. The $\#$s just stand-in for some unspecified (and to be determined) coupling constants (many of which are usually just zero). Simply using dimensional analysis and little group scaling, \cite{McGady:2013sga} was able to rule-out a multitude of hypothetical Lorentz covariant $3$-particle interactions that immediately failed to combined across a factorisation channel to produce only terms of the form (\ref{4masslessSchem}).

As emphasised in \cite{Arkani-Hamed:2017jhn}, the (consistent) $4$-particle amplitudes (\ref{4masslessSchem}) themselves may be directly bootstrapped by matching them with combinations of the $3$-particle amplitudes across each on-shell factorisation channel. This circumvents the requirement for an on-shell recursive procedure, at least in the simple case of $4$-particle amplitudes. Consistently matching (\ref{4masslessSchem}) onto each channel enforces relations between the possible coupling constants appearing in the three-particle amplitudes. This was first identified in \cite{Benincasa:2007xk} and built on by \cite{Schuster:2008nh,McGady:2013sga,Arkani-Hamed:2017jhn} (\cite{McGady:2013sga} being the most comprehensive of these studies). 
This led to the classic results that massless vector boson couplings must have the form of Lie algebra generators, the universality of the graviton's couplings, the identification of massless Rarita-Schwinger particles as gravitinos and the inconsistency of gravitating massless higher-spin particles. Nevertheless, there were a few outstanding gaps in these demonstrations. It is the aim of Section \ref{AllChannel} to complete the argument and, later on, make some extensions to massive theories. The companion work \cite{Trott:2026ozo} is dedicated specifically to results involving supersymmetry. 

To this end, I will begin by drawing attention to the fact that it is also possible to take the complexified limit $s=\ds{12}\da{12}=\ds{34}\da{34}\rightarrow 0$ with spinor bilinears of the same chirality sent to zero on both sides of the channel e.g. $\da{12},\da{34}\rightarrow 0$ (it is obviously also possible to separately take both left-handed bilinears to zero as well but I will make this choice without loss of generality). In this case, because an intermediate particle's spinor $\ra{P_s}$ must obey both $\da{P_s1}=\da{P_s2}=0$ and $\da{P_s3}=\da{P_s4}=0$ on each side of the channel, all of the right-handed spinors necessarily align and thus their bilinears all vanish: $\da{ij}=0$ for any $i,j$. All of the Mandelstam variables are therefore sent to zero simultaneously. 

Such a limit may nevertheless describe a sensible approach to a single simple pole of the form 
\begin{align}\label{AllChanPole}
    \frac{\ds{12}\ds{34}}{s}=\frac{\ds{14}\ds{23}}{u}=\frac{\ds{13}\ds{42}}{t}.
\end{align}
On this pole, all factorisation channels ($s$, $t$ and $u$) activate simultaneously. The residue is given by a coherent sum over all possible channels, in addition to all possible intermediate particles. I will refer to this as the ``all-channel pole''. This is special to massless amplitudes. The locality and unitarity assumptions above still apply to this pole: the amplitude must correctly factorise whenever a possible factorisation channel exists. Singularities of this form have been recently studied for the purposes of celestrial amplitudes in \cite{Ren:2022sws,Ball:2023sdz}. See Appendix \ref{sec:S3PMK} for an explicit realisation of a complex kinematic configuration describing this limit.

\subsection{Simple examples}\label{sec:simple}

As for amplitudes with single-channel poles, $4$-particle amplitudes exhibiting the all-channel pole may be entirely constructed from combining together the factorised $3$-particle amplitudes on the residue. This is because the residue depends only upon left-handed spinor bilinears, which are completely independent of the right-handed spinors that are constrained by the limit. Factors of right-handed bilinears in the residue would necessarily cancel the pole, so, by construction, cannot be present. As a result, it should be possible to directly lift the residue off-shell. 

A near-trivial example of an all-channel pole is that of same-helicity gluon scattering $A\left(g^+_A,g^+_B,g^+_C,g^+_D\right)$ mediated by scalar $\phi$ exchange with couplings of the form
\begin{align}
A\left(g^+_A,g^+_B,\phi\right)=\delta_{AB}\ds{12}^2
\end{align}
(I neglect bothering to write an overall coupling constant and I will henceforth drop the subscript on the amplitude denoting the number of legs, since it will only be $3$ or $4$ and obvious from context). Constructing the $4$-leg amplitude out of the all-channel pole and its residue by adding all three channels together gives:
\begin{align}
A\left(g^+_A,g^+_B,g^+_C,g^+_D\right)&=-\delta_{AB}\delta_{CD}\frac{\ds{12}^2\ds{34}^2}{s}-\delta_{AD}\delta_{BC}\frac{\ds{23}^2\ds{14}^2}{u}-\delta_{AC}\delta_{BD}\frac{\ds{13}^2\ds{24}^2}{t}\nonumber\\
&=-\frac{\ds{12}\ds{34}}{s}\left(\delta_{AB}\delta_{CD}\ds{12}\ds{34}+\delta_{AD}\delta_{BC}\ds{14}\ds{23}+\delta_{AC}\delta_{BD}\ds{13}\ds{42}\right).
\end{align}
The correspondence between each term and channel is obvious. If there is only one flavour of gluon (i.e. a photon), then this amplitude is zero by the Schouten identity. In this case, the sum of the channels in the residue cancels.

A less trivial example example can be selected from several possibilities of photon-graviton scattering with insertions of operators like $RF^2$, $F^3$ or $R^3$. For example, the $RF^2$ contribution to $A(\gamma^+,\gamma^+,h^+,h^+)$ is constructed out of the three-particle amplitudes
\begin{align}
A\left(h^+,h^+,h^-\right)&=\frac{1}{M_{Pl}}\frac{\ds{12}^6}{\ds{23}^2\ds{31}^2}\nonumber\\
A\left(\gamma^+,h^+,\gamma^-\right)&=\frac{1}{M_{Pl}}\frac{\ds{12}^4}{\ds{31}^2}\nonumber\\
A\left(\gamma^+,h^+,\gamma^+\right)&=\frac{1}{M_{Pl}\Lambda^2}\ds{12}^2\ds{23}^2,
\end{align}
where $1/\Lambda^2$ represents some, potentially complex, dimensionful coupling constant. Here $\gamma$ is the photon and $h$ is the graviton. The $4$-particle amplitude may then be constructed from the on-shell diagrams in Fig \ref{PhotGravAmp} to give (neglecting the overall coupling constants):
\begin{feynfig}
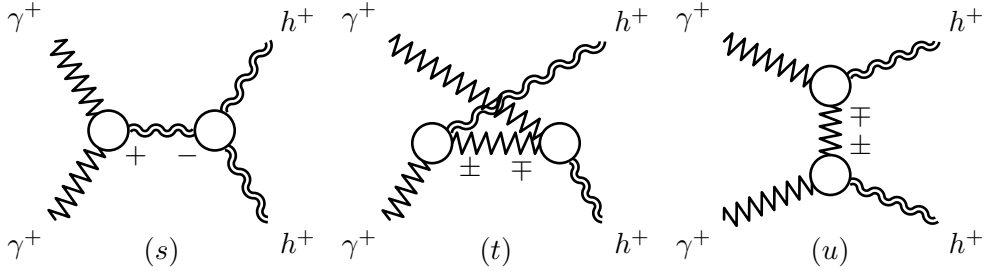
\begin{figure}[h]
\begin{fmffile}{gghh++++}

\begin{center}
\begin{tabular}{ c c c c c }
& & & & \\
   \begin{fmfgraph*}(100,67)
   \fmfleft{i1,i2}
   \fmfright{o1,o2}
   \fmf{zigzag}{i1,v1}
   \fmf{zigzag}{i2,v1}
   \fmf{dbl_wiggly}{v2,o1}
   \fmf{dbl_wiggly}{v2,o2}
   \fmf{dbl_wiggly,label=$+\quad-$}{v1,v2}
   \fmfv{decor.shape=circle,decor.filled=empty,decor.size=0.15w}{v1,v2}
   \fmflabel{$\gamma^+$}{i1}
   \fmflabel{$\gamma^+$}{i2}
   \fmflabel{$h^+$}{o2}
   \fmflabel{$h^+$}{o1}
 \end{fmfgraph*} 
 & \, & \begin{fmfgraph*}(100,67)
   \fmfleft{i1,i2}
   \fmfright{o1,o2}
   \fmf{zigzag}{i1,v1}
   \fmf{dbl_wiggly}{v2,o1}
   \fmf{phantom}{v1,i2}
   \fmf{phantom}{v2,o2}
   \fmf{zigzag,tension=-0.25}{v2,i2}
   \fmf{dbl_wiggly,tension=-0.25}{v1,o2}
   \fmf{zigzag,label=$\pm\quad\mp$}{v1,v2}
   \fmfv{decor.shape=circle,decor.filled=empty,decor.size=0.15w}{v1,v2}
   \fmflabel{$\gamma^+$}{i1}
   \fmflabel{$\gamma^+$}{i2}
   \fmflabel{$h^+$}{o2}
   \fmflabel{$h^+$}{o1}
 \end{fmfgraph*} & \, & 
 \begin{fmfgraph*}(100,67)
   \fmfleft{i1,i2}
   \fmfright{o1,o2}
   \fmf{zigzag}{i1,v1,v2,i2}
   \fmf{dbl_wiggly}{o1,v1}
   \fmf{dbl_wiggly}{o2,v2} \fmfv{decor.shape=circle,decor.filled=empty,decor.size=0.15w}{v1,v2}
   \fmflabel{$\gamma^+$}{i1}
   \fmflabel{$\gamma^+$}{i2}
   \fmflabel{$h^+$}{o2}
   \fmflabel{$h^+$}{o1}
   \fmfv{label=$\mp$,label.angle=-45,label.dist=0.1w}{v2}
   \fmfv{label=$\pm$,label.angle=45,label.dist=0.1w}{v1}
 \end{fmfgraph*} \\
 $(s)$ & \, &  $(t)$ & \, &  $(u)$
\end{tabular}
\end{center}
\end{fmffile}
\caption{On-shell diagrams for photon-graviton scattering mediated by $RF^2$ vertex.}\label{PhotGravAmp}
\end{figure}
\end{feynfig}
\begin{align}\label{PhotonGravHDO}
    A(\gamma^+,\gamma^+,h^+,h^+)&=\frac{-1}{s}\ds{1P_s}^2\ds{2P_s}^2\frac{\ds{34}^6}{\ds{3P_s}^2\ds{4P_s}^2}\nonumber\\
    &\qquad\qquad-\frac{1}{t}\left(\ds{13}^2\ds{3P_t}^2\frac{\ds{42}^4}{\ds{2P_t}^2}+\ds{24}^2\ds{4P_t}^2\frac{\ds{13}^4}{\ds{1P_t}^2}\right)\nonumber\\
    &\qquad\qquad-
    \frac{1}{u}\left(\ds{14}^2\ds{4P_u}^2\frac{\ds{32}^4}{\ds{2P_u}^2}+\ds{23}^2\ds{3P_u}^2\frac{\ds{14}^4}{\ds{1P_u}^2}\right),
\end{align}    
where $P_s$, $P_t$ and $P_u$ and are the internal momenta for the $s$, $t$ and $u$-channels respectively (more precisely, I define them as $P_s=p_3+p_4$, $P_t=p_2+p_4$ and $P_u=p_1+p_4$). A reference spinor $\ra{q}$ satisfying $\da{qi}\neq 0$ for each $i$ on the residue can be introduced. Introducing factors of $\da{P_{s}q}$, $\da{P_{t}q}$ and $\da{P_{u}q}$ into each respective term and applying conservation of momentum, the expression can be manipulated into
\begin{align}
A(\gamma^+,\gamma^+,h^+,h^+)&=-\frac{\ds{12}\ds{34}}{s\da{3q}^2\da{4q}^2\da{1q}^2}\Big(\ds{12}^3\ds{34}\da{2q}^2\da{1q}^4-\ds{14}^3\ds{32}\da{1q}^4\da{4q}^2\nonumber\\
    &\qquad\qquad\qquad\qquad\quad\,\,\,-\ds{32}^3\ds{14}\da{2q}^2\da{3q}^2\da{1q}^2+\ds{13}^3\ds{42}\da{1q}^4\da{3q}^2\nonumber\\
    &\qquad\qquad\qquad\qquad\qquad\qquad\qquad\qquad\qquad\quad+\ds{42}^3\ds{13}\da{2q}^2\da{4q}^2\da{1q}^2\Big)\nonumber\\
    &=\frac{\ds{12}^2\ds{34}^4}{s}.
\end{align}
The final line can be obtained by e.g. eliminating appearances of $\da{2q}$ with momentum conservation and then further eliminating terms with the Schouten identity. The universal coupling of gravitons to other gravitons and photons in the first two amplitudes in (\ref{PhotonGravHDO}) was essential for this final expression to be independent of the intermediary spinor $\ra{q}$, which otherwise would have obstructed the residue from being analytically continued off-shell into a Lorentz invariant, little group covariant expression. This reflects the sensitivity of consistent factorisation on these types of poles to the underlying coupling structure of the theory. There is no contribution to this amplitude purely from the regular minimal photon-graviton couplings $A\left(\gamma^+,h^\pm,\gamma^-\right)$ because there are no factorisation channels consistent with only this helicity configuration, hence $A(\gamma^+,\gamma^+,h^+,h^+)$ would be zero without the higher dimensional insertions. 

Same-sign helicity electromagnetic Compton scattering off massless matter provides a final illustrative example. For simplicity, choosing photon scattering off a (charged) scalar ($\varphi$), then the amplitude is given by a sum over $s$ and $t$ channels (see Fig \ref{scalarOSCS}):
\begin{feynfig}
\begin{figure}[h]
\begin{fmffile}{sQEDCompton}
\begin{center}
\begin{tabular}{ c c c }
& & \\
   \begin{fmfgraph*}(120,80)
   \fmfleft{i1,i2}
   \fmfright{o1,o2}
   \fmf{scalar}{i1,v1,v2,o1}
   \fmf{zigzag}{i2,v1}
   \fmf{zigzag}{o2,v2}
   \fmfv{decor.shape=circle,decor.filled=empty,decor.size=0.15w}{v1,v2}
   \fmflabel{$\gamma^+$}{i2}
   \fmflabel{$\gamma^+$}{o2}
   \fmflabel{$\varphi$}{i1}
   \fmflabel{$\overline{\varphi}$}{o1}
 \end{fmfgraph*} & \, & 
\begin{fmfgraph*}(120,80)
   \fmfleft{i1,i2}
   \fmfright{o1,o2}
   \fmf{scalar}{i1,v1,v2,o1}
   \fmf{phantom}{v1,i2}
   \fmf{phantom}{v2,o2}
   \fmf{zigzag,tension=-0.25}{v2,i2}
   \fmf{zigzag,tension=-0.25}{v1,o2}
   \fmfv{decor.shape=circle,decor.filled=empty,decor.size=0.15w}{v1,v2}
   \fmflabel{$\gamma^+$}{i2}
   \fmflabel{$\gamma^+$}{o2}
   \fmflabel{$\varphi$}{i1}
   \fmflabel{$\overline{\varphi}$}{o1}
 \end{fmfgraph*} \\ 
 $(s)$ & \, & $(t)$   
\end{tabular}
\end{center}
\end{fmffile}
\caption{On-shell diagrams for Compton scattering.}\label{scalarOSCS}
\end{figure}
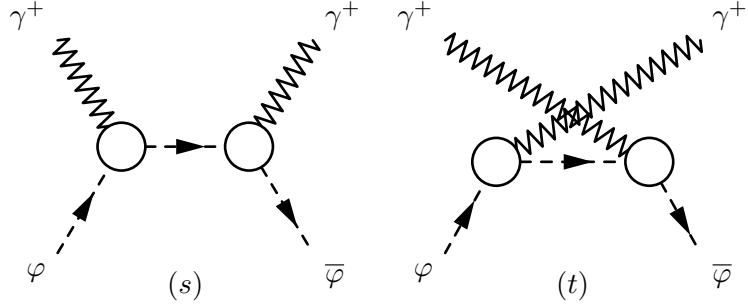
\end{feynfig}
\begin{align}
A(\varphi,\gamma^+,\gamma^+,\bar{\varphi})=\frac{\ds{21}\ds{2P_s}}{\ds{1P_s}}\frac{1}{s}\frac{\ds{34}\ds{3P_s}}{\ds{4P_s}}+\frac{\ds{31}\ds{3P_t}}{\ds{1P_t}}\frac{1}{t}\frac{\ds{24}\ds{2P_t}}{\ds{4P_t}}.
\end{align}
See (\ref{QEDmassless3}) below for the scalar QED on-shell $3$-particle amplitude used to construct this. I am neglecting to write the overall coupling. Performing a similar computation to the example above, 
\begin{align}\label{ComptonRes}
A(\varphi,\gamma^+,\gamma^+,\bar{\varphi})=\frac{\da{1q}\da{4q}}{\da{3q}\da{2q}}\left(\frac{\ds{12}\ds{34}}{s}-\frac{\ds{13}\ds{24}}{t}\right)=0.
\end{align}
By (\ref{AllChanPole}), the $s$ and the $t$ channel terms exactly cancel to ensure that the amplitude is zero. This computation demonstrates how, just like the opposite-sign helicity configuration, this amplitude is still directly generated from the elementary $3$-particle amplitudes (and how the amplitude is consistent with factorisation despite being equal to zero). The non-Abelian case will be presented further below, as well as more examples in Section \ref{AllChannelAmps}. However, I will first make use of the all-channel pole to give a complete derivation of the perturbative structure of Yang-Mills theory from consistent factorisation.


\subsection{3-particle amplitudes, unitarity and crossing}\label{sec:AmpList}
Here is a compilation of $3$-leg amplitudes and their relations under unitarity and crossing. This Section is intended to serve as a summary and reference for elementary results of the low spin $S$-matrix used throughout the rest of this paper. I generally follow the conventions of \cite{Srednicki:2007qs}. See \cite{Arkani-Hamed:2017jhn,Bachu:2019ehv,Durieux:2019eor,Balkin:2021dko,Liu:2022alx} for previous work subsumed in the results summarised here.

For $3$-leg amplitudes (and higher), unitarity of the $S$-matrix is posited as the rule
\begin{align}\label{unitarity}
&A(X_a(\{\rs{p_{\text{in}}},\ra{p_{\text{in}}}\})\rightarrow Y^b(\{\rs{p_{\text{out}}},\ra{p_{\text{out}}}\}))\nonumber\\
&\qquad\qquad\qquad\qquad\qquad\qquad\qquad=A(Y_b(\{\rs{p_{\text{out}}}^*,\ra{p_{\text{out}}}^*\})\rightarrow X^a(\{\rs{p_{\text{in}}}^*,\ra{p_{\text{in}}}^*\}))^*,
\end{align}
where $X$ and $Y$ denote the collective incoming and outgoing states. The indices $a$ and $b$ denote generic tensor components of some representation of observables and their heights reverse as a consequence of the conjugation. This includes massive $SU(2)$ little group spin indices. The amplitude depends upon the momentum and spin information of each particle through its spinors, which it is best regarded as a function of. It is emphasised in (\ref{unitarity}) that the spinors of the particles, denoted schematically as $\{\rs{p_{\text{in}}},\ra{p_{\text{in}}}\}$ and $\{\rs{p_{\text{out}}},\ra{p_{\text{out}}}\}$, are separately conjugated on either side of the expression (and hence so are their corresponding momenta when complexified). Of course, spinor chirality switches under complex conjugation. When the kinematics are real, then the conjugated spinors simply switch bracket shapes e.g. $\rs{p_{\text{out}}}^*=\la{p_{\text{out}}}$ (I suppress a possible massive spin index). However, with complexified kinematics, as is usually necessary with three particles, $\rs{p_{\text{out}}}^*$ is usually implicitly identified with the $\ra{p_{\text{out}}}$ of some distinct, complex conjugated momentum configuration. 

All states are chosen to be outgoing in the convention adopted here. The implicit massive $SU(2)$ spin indices introduced in \cite{Arkani-Hamed:2017jhn} are assumed to be raised for outgoing particles and are to be lowered when inferring the corresponding expressions with crossed (incoming) particles. By ``crossing'' I mean writing down the analogous expression for the amplitude as a function of the external particle momenta and spins but with some subset of particles pronounced as incoming instead of outgoing. The overall phases that must be included for this reinterpretation (i.e. phases produced by crossing individual legs) are inferred from comparison with the Feynman rules. These are denoted in the expressions below by factors of $(-1)^{n_i}$, which indicates a sign flip arising from crossing leg $i$ in the amplitude (in other words, $n_i=0$ if $i$ is outgoing, while $n_i=1$ if $i$ is incoming). I am unaware of a first principles derivation of these phases and am following a similar approach to the one proposed in \cite{Hebbar:2020ukp}. People frequently claim that this can all be accounted for by some simple analytic continuation of the spinors to negative energies \cite{elvang2015scattering}, but I am not aware of a derivation of this and need to keep careful track of negative signs. There is presumably some way of accounting for this from the composition of a spinor structure, but such a prescription will have to wait for another time. 

Denoting self-conjugate scalar states as $\varphi_i$ (labeled by some set of observable eigenvalues implicit in some index $i$), the amplitudes for interactions with vectors are
\begin{align*}
A(\varphi_i,\varphi_j,g^+_A)&=-(t_A)_{ij}\frac{\ds{13}\ds{32}}{\ds{12}}&
A(\varphi_i,\varphi_j,g^-_A)&=-((t_A)_{ji})^*\frac{\da{13}\da{32}}{\da{12}}\\
A(\varphi_i,\varphi_j,g^+_A)&=-(-1)^{n_1+n_3}(t_A)_{ij}\frac{m}{x}&
A(\varphi_i,\varphi_j,g^-_A)&=-(-1)^{n_1+n_3}((t_A)_{ji})^*mx
\end{align*}
\begin{align}\label{ScalarMat3leg}
A(\varphi_i,\varphi_j,W_A)=(-1)^{n_1+n_3}(t_A)_{ij}\frac{1}{m_3}\la{\bf{3}}p_1\rs{\bf{3}}.
\end{align}
Here $g_A$ are massless gluons and $W_A$ are massive vector bosons. I am using the massive spinor representations proposed in \cite{Arkani-Hamed:2017jhn} and conventions stated in Appendix A of \cite{Herderschee:2019ofc} (which were formulated for consistency with the conventions of \cite{Srednicki:2007qs}). The top line corresponds to massless scalars, the next is for massive scalars (of mass $m$) and the last is consistent with either. The scalar masses in the second line must be equal. Compared to the Feynman rules derived in \cite{Srednicki:2007qs}, I am omitting an overall factor of $\sqrt{2}$ which can be absorbed into or out of the coupling constants at will. The $x$ factor was introduced in \cite{Arkani-Hamed:2017jhn} and, for its purpose in this study, is defined as a constant of proportionality, 
\begin{align}\label{Prelimx}
x\rs{3}=\frac{p_1}{m}\ra{3}\qquad\Leftrightarrow\qquad x=\frac{\ls{q}p_1\ra{3}}{m\ds{q3}},
\end{align}
which is independent of the reference spinor $\rs{q}$ chosen to obey $\ds{q3}\neq 0$. The origin of this object is connected to the massive adaptation of the special complex $3$-particle kinematics underpinning the existence of massless $3$-leg amplitudes. This is elaborated upon further in \cite{Trott:2026ozo}, the results of which are summarised in Section \ref{Sec:LowSpinAmp} below, where a redundancy-free expression, independent of the reference spinor, is presented. However, of pertinence here to establishing the action of crossing is that it is defined specifically in (\ref{Prelimx})
to contain the momentum $p_1$. 

In the Feynman rules, crossing the vector introduces an overall $(-)$ (from the polarisation). Crossing either of the scalars in the massless cases does not introduce a sign, although the explicit factor of the momentum appearing in the massive cases (including through $x$) incurs a $(-)$ if it corresponds to that of a crossed particle (this will be conventionally chosen here to be particle $1$). In the fully massless amplitudes stated above, this sign is canceled by an application of momentum conservation. Bosonic exchange symmetry implies that $(t_A)_{ij}=-(t_A)_{ji}$ (if the masses are distinct, then this is implicit in the definition of the amplitude). Finally, the assumption of self-conjugacy in the case where all of the particles are massive implies directly that $((t_A)_{ji})^*=(t_A)_{ij}$ through unitarity.

The massive vector amplitude in (\ref{ScalarMat3leg}) is ideal for illustrating the matching of a longitudinally polarised massive vector boson onto a massless scalar particle in the massless limit. The longitudinal polarisation, in the helicity spin quantisation basis, is given by the $(+,-)$ spin index configuration in the little group tensor amplitude. The massless limit of the spinors can then be taken directly. See Appendix A of \cite{Herderschee:2019ofc} for the limits of the massive spinors in the conventions adopted here. An extra normalisation factor of $\sqrt{2}$ is also required (in matching the fundamental $SU(2)$ tensor indices onto a normalised spin and helicity eigenstate), leading to a trivalent scalar amplitude
\begin{align}\label{3leglonglimit}
A(\varphi_i,\varphi_j,W_A^L)\rightarrow A(\varphi_i,\varphi_j,\phi_A)=-(-1)^{n_3}\sqrt{2}(t_A)_{ij}\frac{m_1-m_2}{m_3}\frac{m_1+m_2}{2}
\end{align}
as $m_3\rightarrow 0$. This is clearly non-divergent (and hence consistent with perturbative unitarity) only if $m_1-m_2= \mathcal{O}(m_3)$ as $m_3\rightarrow 0$, which is also required in order for the transverse amplitudes to consistently match onto the massless vector amplitudes in (\ref{ScalarMat3leg}) (the result of the limit is otherwise not Lorentz invariant). More remarks will be made on this below with the analogous fermion amplitudes. For the purposes here of establishing the crossing rules, of interest is the factor of $(-1)^{n_3}$, which arose from the Feynman rules for crossing the vector polarisation, but which is unexpected in an ordinary three-scalar interaction (where the Feynman rules for incoming and outgoing scalars are trivially the same). This sign can be removed, however, by absorbing a factor of $i$ into the quantum state describing the longitudinal mode (so that a factor of $(-1)$, arising from complex conjugation when this amplitude is subjected to the the unitarity constraint (\ref{unitarity}), is removed). As a result, the factor of $-(-1)^{n_3}$ in (\ref{3leglonglimit}) should be replaced with a factor of $i$. So in summary, a factor of $-i$ should be introduced into the amplitude when taking the massless (or high energy) limit of a longitudinal outgoing vector leg. This is both consistent and necessary further below in Section \ref{Sec:MassiveVectors} for correctly matching the three-massive vector amplitude onto a theory of massless vectors interacting with scalar matter. 

For fermions $\psi_i$, the possible amplitudes are
\begin{align*}
A(\psi_i^-,\psi_j^+,g^+_A)&=(T_A)_{i}^{\,\,j}\frac{\ds{32}^2}{\ds{12}}&
A(\psi_i^-,\psi_j^+,g^-_A)&=-(-1)^{n_1+n_2}((T_A)_{j}^{\,\,i})^*\frac{\da{31}^2}{\da{12}}\\
A(\psi_i^+,\psi_j^+,g^+_A)&=-(-1)^{n_1}\frac{2((m_A)_{ji})^*}{\Lambda}\ds{13}\ds{23}&
A(\psi_i^-,\psi_j^-,g^-_A)&=-(-1)^{n_1}\frac{2(m_A)_{ij}}{\Lambda}\da{13}\da{23}
\end{align*}
\begin{align}\label{FermionMat3leg}
A(\psi_i,\psi_j,g^+_A)&=-(-1)^{n_1+n_2+n_3}(t_A)_{ij}\frac{1}{x}\da{\bf{12}}-(-1)^{n_1}\frac{2((m_A)_{ji})^*}{\Lambda}\ds{3\bf{1}}\ds{3\bf{2}}\nonumber\\
A(\psi_i,\psi_j,g^-_A)&=-(-1)^{n_3}((t_A)_{ji})^*x\ds{\bf{12}}-(-1)^{n_2}\frac{2(m_A)_{ij}}{\Lambda}\da{3\bf{1}}\da{3\bf{2}}\nonumber\\
A(\psi_i^-,\psi_j^+,W_A)&=(-1)^{n_1+n_3}(T_A)_{i}^{\,\,j}\frac{1}{m_3}\da{1\bf{3}}\ds{2\bf{3}}\nonumber\\
A(\psi_i^+,\psi_j^+,W_A)&=-(-1)^{n_1}\frac{2((m_A)_{ji})^*}{\Lambda}\ds{1\bf{3}}\ds{2\bf{3}}\nonumber\\
A(\psi_i^-,\psi_j^-,W_A)&=-(-1)^{n_1}\frac{2(m_A)_{ij}}{\Lambda}\da{1\bf{3}}\da{2\bf{3}}\nonumber\\
A(\psi_i,\psi_j,W_A)&=-(-1)^{n_3}(t_A)_{ij}\frac{1}{m_3}\da{\bf{13}}\ds{\bf{23}}+(-1)^{n_1+n_2+n_3}((t_{A})_{ij})^*\frac{1}{m_3}\da{\bf{23}}\ds{\bf{13}}\nonumber\\
&\qquad-(-1)^{n_1}\frac{2((m_A)_{ji})^*}{\Lambda}\ds{\bf{13}}\ds{\bf{23}}-(-1)^{n_2}\frac{2(m_A)_{ij}}{\Lambda}\da{\bf{13}}\da{\bf{23}}.
\end{align}
I am choosing a self-conjugate basis for the massive fermions. Identical fermion exchange antisymmetry implies that both $(m_A)_{ij}=-(m_A)_{ji}$ and $(t_A)_{ij}=-(t_A)_{ji}$, except for the last case, where instead $(t_A)_{ij}=\left((t_A)_{ji}\right)^*$. If the fermion masses are distinct and the vector is massless, then only the anomalous dipole terms are permitted (these are the terms with ``holomorphic'' spinor chiralities). I define the fermion ordering in the states so that the first listed particles are the last to be created from the vacuum. The crossing rules stated account for this. 


The scalar-fermion Yukawa amplitude has the form
\begin{align}\label{Yukawa}
A\left(\psi_i,\psi_j,\varphi_m\right)=(-1)^{n_1}(y^m_{ij})^*\ds{\bf{12}}+(-1)^{n_2}y^m_{ij}\da{\bf{12}}.
\end{align}
Fermionic exchange antisymmetry implies that $y^m_{ij}=y^m_{ji}$. The components $\Re\,y^m_{ij}$ represent ``scalar'' type couplings and $\Im\,y^m_{ij}$ are ``pseudoscalar'' couplings. The massless limits can be easily taken directly.

With the Yukawa amplitudes established, it is now convenient to return to the general massive fermion-vector coupling in (\ref{FermionMat3leg}), which will be reproduced here for convenience (omitting the anomalous dipole terms and crossing signs):
\begin{align}\label{MassiveFermVec3leg}
A(\psi_i,\psi_j,W_A)&=-(t_A)_{ij}\frac{1}{m_3}\da{\bf{13}}\ds{\bf{23}}+((t_{A})_{ij})^*\frac{1}{m_3}\da{\bf{23}}\ds{\bf{13}}.
\end{align}
Self-adjointness of the couplings implies that $\Re (t_A)_{ij}=\Re (t_A)_{ji}$ and $\Im (t_A)_{ij}=-\Im (t_A)_{ji}$. The real parts of the coupling constants represent the axial-vector coupling and the imaginary parts are the vector coupling. Only the latter have the required exchange symmetry to be consistent couplings in the limit that the vector boson is massless (as is clear in (\ref{FermionMat3leg})).

Taking the limit $m_3\rightarrow 0$ while keeping the fermions massive, the amplitude for transverse vector polarisation becomes
\begin{align}
A(\psi_i,\psi_j,W^+_A)&\rightarrow \frac{1}{\da{q3}}\left(\Re (t_A)_{ij}\left(\ds{3\bf{2}}\da{q\bf{1}}-\ds{3\bf{1}}\da{q\bf{2}}\right)+i\Im (t_A)_{ij}\left(\ds{3\bf{2}}\da{q\bf{1}}+\ds{3\bf{1}}\da{q\bf{2}}\right)\right),
\end{align}
where $\ra{q}$ is some reference spinor obeying $\da{q3}\neq0$ (it varies with the precise path in which the limit is taken). The term 
\begin{align}
\frac{1}{\da{q3}}\left(\ds{3\bf{2}}\da{q\bf{1}}+\ds{3\bf{1}}\da{q\bf{2}}\right)\rightarrow \frac{-1}{x}\da{\bf{12}}
\end{align}
is only independent of the residual reference spinor $\ra{q}$ (and hence is Lorentz or ``gauge'' invariant) if $m_1\rightarrow m_2$ simultaneously. In contrast, the kinematic factor accompanying the coupling $\Re (t_A)_{ij}$ is always dependent upon $\ra{q}$. The amplitude with the longitudinal mode becomes 
\begin{align}\label{LongModeLim}
A\left(\psi_i,\psi_j,W^L_A\right)\rightarrow \frac{-i}{\sqrt{2}}\left(\frac{m_1+m_2}{m_3}\Re (t_A)_{ij}\left(\da{\bf{12}}-\ds{\bf{12}}\right)+i\frac{m_2-m_1}{m_3}\Im (t_A)_{ij}\left(\da{\bf{12}}+\ds{\bf{12}}\right)\right).
\end{align}
This does not diverge as long as the conditions for the consistency of the transverse polarisation limit above are met: $\Re (t_A)_{ij}= \mathcal{O}(m_3)\rightarrow 0$ and either the masses $m_1$ and $m_2$ are equal or $\Im (t_A)_{ij}= \mathcal{O}(m_3)\rightarrow 0$. In the case that the fermion masses converge, the transversely polarised amplitudes then match onto the massless gluon amplitudes in (\ref{FermionMat3leg}) with the identification $(t_A)_{ij}=i\Im (t_A)_{ij}$. If these conditions are not met, the vector boson cannot be consistently taken massless while keeping the fermions massive without the coupling to the longitudinally polarised state growing non-perturbatively large. The growth of the longitudinal amplitude therefore provides the obstruction to the spurious violation of Lorentz invariance that would otherwise appear in the transverse amplitudes. The only non-trivial alternative is that $\Im (t_A)_{ij}\sim \mathcal{O}(m_3)$ so that the transverse vector boson altogether decouples and only non-zero Yukawa couplings to the longitudinal mode remain. 


In contrast, the amplitude (\ref{MassiveFermVec3leg}) in general is perfectly well-defined in the high energy limit. In this case, the masses $m_1$ and $m_2$ effectively scale like $m_3$ so that $(m_2-m_1)/m_3$ is finite (and possibly zero), even if the couplings are chiral. The longitudinal vector amplitude converges to a Yukawa scalar amplitude with couplings that can be directly read-off (\ref{LongModeLim}). The scalar couplings share the parity of their vector counterparts from which they emerge. 

The $3$-particle amplitudes with two or three vector bosons are extensively studied below in Section \ref{Sec:Massless4vec} for the massless case and in Section \ref{Sec:MassiveVectors} for the massive case. The amplitudes involving two vectors and a single scalar, which includes the Higgs couplings, are also presented in Section \ref{Sec:MassiveVectors}

\subsection{Complete 4-particle test for vector bosons}\label{Sec:Massless4vec}

Massless $4$-particle amplitudes may contain simple Mandelstam poles as their only singularities. However, when analytically continued to complex momenta, they can instead contain up to $8$ distinct simple spinor bilinear poles in totality, corresponding to $8$ possible complex factorisation channels. These are given either by sending pairs of opposite chirality spinor bilinears of different pairs of particles to zero (e.g. $\ds{12},\da{34}\rightarrow 0$), of which there are $6$ possibilities, or by sending all spinor bilinears of a particular chirality to zero, corresponding to the $2$ all-channel poles. An amplitude consistent with this formulation of unitarity and locality must consistently factorise on all $8$ channels. For a typical amplitude, only a small subset of these factorisation channels are non-trivial. However, full use of them across a set of several different helicity configurations is nevertheless required for deriving the complete structure a consistent theory, as will be demonstrated next. 

To reiterate comments made in the introduction, many of the results presented in this subsection were derived recently by \cite{Fonseca:2025mzj} while this work was in preparation. The authors of \cite{Fonseca:2025mzj} were able to draw the conclusions of the analysis entirely from studying the consistency of MHV amplitudes. While this also plays a role in the argument presented here, my derivation differs by also making use of consistency of the all-channel pole, which aids in the clarity of the argument. In addition to being thematically relevant to the work on massive particles presented in the Sections below, my exposition differs from \cite{Fonseca:2025mzj} by the inclusion of massive particles and otherwise provides an independent complementary perspective on the argument.

\subsubsection{Gluon scattering}

I here give the complete $S$-matrix derivation of the emergence of non-Abelian Yang-Mills from Lorentz-invariance, unitarity and locality in the form of consistent complex factorisation. The general three-leg amplitude between $N$ multiple ``flavours'' of massless vector bosons is 
\begin{align}\label{3gluons}
A\left(g^{+}_A,g^{+}_B,g^{-}_{C}\right)&=if_{AB}^{\,\,\,\,\,\,\,\,\,C}\frac{\ds{12}^3}{\ds{23}\ds{31}}\nonumber\\
A\left(g^{-}_A,g^{-}_B,g^{+}_C\right)&=i\left(f_{AB}^{\,\,\,\,\,\,\,\,\,C}\right)^*\frac{\da{12}^3}{\da{23}\da{31}}
\end{align}
I take all particles outgoing, as usual. There are no sign flips under crossing. Unitarity relates the couplings of each helicity configuration as complex conjugates. The coupling constants satisfy
\begin{align}
f_{AB}^{\,\,\,\,\,\,\,\,\,C}=-f_{BA}^{\,\,\,\,\,\,\,\,\,C}
\end{align}
by identical boson exchange symmetry (because the kinematic factor is clearly antisymmetric under exchange $1\leftrightarrow 2$). At this point, the index heights merely distinguish helicities, which, for a given flavour index, are particle/anti-particle pairs. For this reason, they represent fundamental and anti-fundamental tensor indices for $U(N)$ basis rotations on the external quantum scattering states. 

The next step is to demand that these $3$-leg amplitudes combine by unitarity into consistently factorising $4$-leg amplitudes. The ultra helicity violating (UHV) amplitude $A(+,+,+,-)$ (I just label particles by helicities here for brevity) can be fully constructed from $A(+,+,-)$ across an all-channel pole as in Fig \ref{UHVgluon}:
\begin{feynfig}
\begin{figure}[h]
\begin{fmffile}{GluonScattering+++-}

\begin{center}
\begin{tabular}{ c c c c c }
& & & & \\
   \begin{fmfgraph*}(100,67)
   \fmfleft{i1,i2}
   \fmfright{o1,o2}
   \fmf{gluon}{i1,v1}
   \fmf{gluon}{i2,v1}
   \fmf{gluon}{v2,o1}
   \fmf{gluon}{v2,o2}
   \fmf{gluon,label=$-\quad+$}{v1,v2}
   \fmfv{decor.shape=circle,decor.filled=empty,decor.size=0.15w}{v1,v2}
   \fmflabel{$+A$}{i1}
   \fmflabel{$+B$}{i2}
   \fmflabel{$+C$}{o2}
   \fmflabel{$-D$}{o1}
 \end{fmfgraph*} 
 &\,\,\,\,& \begin{fmfgraph*}(100,67)
   \fmfleft{i1,i2}
   \fmfright{o1,o2}
   \fmf{gluon}{i1,v1}
   \fmf{gluon}{v2,o1}
   \fmf{phantom}{v1,i2}
   \fmf{phantom}{v2,o2}
   \fmf{gluon,tension=-0.25}{v2,i2}
   \fmf{gluon,tension=-0.25}{v1,o2}
   \fmf{gluon,label=$-\quad+$}{v1,v2}
   \fmfv{decor.shape=circle,decor.filled=empty,decor.size=0.15w}{v1,v2}
   \fmflabel{$+A$}{i1}
   \fmflabel{$+B$}{i2}
   \fmflabel{$+C$}{o2}
   \fmflabel{$-D$}{o1}
 \end{fmfgraph*} &\,\,\,\,&
 \begin{fmfgraph*}(100,67)
   \fmfleft{i1,i2}
   \fmfright{o1,o2}
   \fmf{gluon}{i1,v1,o1}
   \fmf{gluon}{i2,v2,o2}
   \fmf{gluon}{v1,v2}
   \fmfv{decor.shape=circle,decor.filled=empty,decor.size=0.15w}{v1,v2}
   \fmflabel{$+A$}{i1}
   \fmflabel{$+B$}{i2}
   \fmflabel{$+C$}{o2}
   \fmflabel{$-D$}{o1}
   \fmfv{label=$-$,label.angle=-45,label.dist=0.1w}{v2}
   \fmfv{label=$+$,label.angle=45,label.dist=0.1w}{v1}
 \end{fmfgraph*}\\
 $(s)$ &\,\,\,\,& $(t)$ &\,\,\,\,& $(u)$
 \end{tabular}
\end{center}
\end{fmffile}
\caption{On-shell diagrams for UHV gluon scattering.}\label{UHVgluon}
\end{figure}
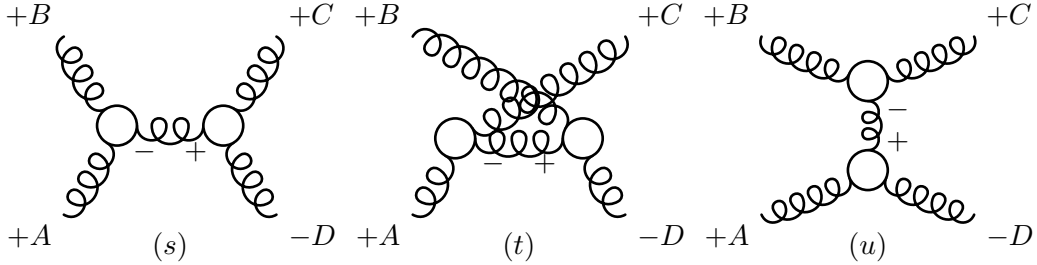
\end{feynfig}
\begin{align}\label{+++-glu}
A(g^+_A,g^+_B,g^+_C,g^-_D)&=\frac{1}{s}f_{AB}^{\,\,\,\,\,\,\,\,\,E} \frac{\ds{12}^3}{\ds{2P_s}\ds{P_s1}}f_{EC}^{\,\,\,\,\,\,\,\,\,D}\frac{\ds{P_s3}^3}{\ds{34}\ds{4P_s}}+\frac{1}{t}f_{AC}^{\,\,\,\,\,\,\,\,\,E}\frac{\ds{13}^3}{\ds{3P_t}\ds{P_t1}}f_{EB}^{\,\,\,\,\,\,\,\,\,D}\frac{\ds{P_t2}^3}{\ds{4P_t}\ds{24}}\nonumber\\
&\qquad\qquad+\frac{1}{u}f_{BC}^{\,\,\,\,\,\,\,\,\,E}\frac{\ds{23}^3}{\ds{P_u2}\ds{3P_u}}f_{AE}^{\,\,\,\,\,\,\,\,\,D}\frac{\ds{1P_u}^3}{\ds{P_u4}\ds{41}}\nonumber\\
&=\frac{\da{4q}^3}{\da{1q}\da{2q}\da{3q}}\Big(f_{AB}^{\,\,\,\,\,\,\,\,\,E}f_{EC}^{\,\,\,\,\,\,\,\,\,D}\frac{\ds{12}\ds{34}}{s}+f_{CA}^{\,\,\,\,\,\,\,\,\,E}f_{EB}^{\,\,\,\,\,\,\,\,\,D}\frac{\ds{13}\ds{42}}{t}\nonumber\\
&\qquad\qquad\qquad\qquad\qquad+f_{BC}^{\,\,\,\,\,\,\,\,\,E}f_{EA}^{\,\,\,\,\,\,\,\,\,D}\frac{\ds{14}\ds{23}}{u}\Big)\nonumber\\
&=\frac{\da{4q}^3}{\da{1q}\da{2q}\da{3q}}\frac{\ds{12}\ds{34}}{s}\left(f_{AB}^{\,\,\,\,\,\,\,\,\,E}f_{EC}^{\,\,\,\,\,\,\,\,\,D}+f_{CA}^{\,\,\,\,\,\,\,\,\,E}f_{EB}^{\,\,\,\,\,\,\,\,\,D}+f_{BC}^{\,\,\,\,\,\,\,\,\,E}f_{EA}^{\,\,\,\,\,\,\,\,\,D}\right)
\end{align}
using (\ref{AllChanPole}). The only way in which this residue can be lifted off-shell in a way consistent with Lorentz invariance is if the prefactor of coupling constants cancels to zero, giving:
\begin{align}
f_{AB}^{\,\,\,\,\,\,\,\,\,E}f_{EC}^{\,\,\,\,\,\,\,\,\,D}+f_{BC}^{\,\,\,\,\,\,\,\,\,E}f_{EA}^{\,\,\,\,\,\,\,\,\,D}+f_{CA}^{\,\,\,\,\,\,\,\,\,E}f_{EB}^{\,\,\,\,\,\,\,\,\,D}=0,
\end{align}
which is the Jacobi identity. The coupling constants $f_{AB}^{\,\,\,\,\,\,\,\,\,C}$ are therefore structure constants of a Lie algebra. The amplitude $A(+,+,+,-)$ and its emergence from trivalent interactions is therefore directly interpretable as the Jacobi identity. A related observation has been made previously by \cite{Herrmann:2016qea} in the context of on-shell diagrams.

This does not yet exhaust all of the available information, just as Yang-Mills theory requires further specifications on the Lie algebra. Consistent factorisation of the MHV amplitude implies two sets of constraints on the couplings. Generally, as described in \cite{Arkani-Hamed:2017jhn}, the amplitude consists of terms that each contain pairs of poles:
\begin{align}
A(g^+_A,g^+_B,g^-_C,g^-_D)=\ds{12}^2\da{34}^2\left(\frac{X_{AB}^{\,\,\,\,\,\,\,\,\,CD}}{st}+\frac{Y_{AB}^{\,\,\,\,\,\,\,\,\,CD}}{su}+\frac{Z_{AB}^{\,\,\,\,\,\,\,\,\,CD}}{tu}\right),
\end{align}
where $X,Y$ and $Z$ are coupling constants that are determined by demanding consistent (complex) factorisation:
\begin{align}
X_{AB}^{\,\,\,\,\,\,\,\,\,CD}-Y_{AB}^{\,\,\,\,\,\,\,\,\,CD}&=-f_{AB}^{\,\,\,\,\,\,\,\,\,E}\left(f_{CD}^{\,\,\,\,\,\,\,\,\,E}\right)^*\\
Y_{AB}^{\,\,\,\,\,\,\,\,\,CD}-Z_{AB}^{\,\,\,\,\,\,\,\,\,CD}&=f_{EA}^{\,\,\,\,\,\,\,\,\,D}\left(f_{EC}^{\,\,\,\,\,\,\,\,\,B}\right)^*=\left(f_{ED}^{\,\,\,\,\,\,\,\,\,A}\right)^*f_{EB}^{\,\,\,\,\,\,\,\,\,C}\\
Z_{AB}^{\,\,\,\,\,\,\,\,\,CD}-X_{AB}^{\,\,\,\,\,\,\,\,\,CD}&=-f_{EA}^{\,\,\,\,\,\,\,\,\,C}\left(f_{ED}^{\,\,\,\,\,\,\,\,\,B}\right)^*=-\left(f_{EC}^{\,\,\,\,\,\,\,\,\,A}\right)^*f_{EB}^{\,\,\,\,\,\,\,\,\,D}.\label{uResEq}
\end{align}
There are two possible non-trivial $t$ and $u$ factorisation channels given by sending opposite chirality bilinears to zero on each side, which gives the two equations in the second and third lines above. See Fig \ref{MHV2Limits}.
\begin{feynfig}
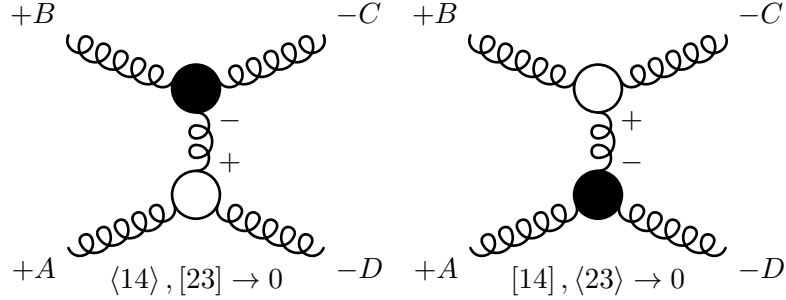
\begin{figure}[h]
\begin{fmffile}{GluonScattering++--}

\begin{center}
\begin{tabular}{ c c c }
& & \\
   \begin{fmfgraph*}(120,80)
   \fmfleft{i1,i2}
   \fmfright{o1,o2}
   \fmf{gluon}{i1,v1,o1}
   \fmf{gluon}{i2,v2,o2}
   \fmf{gluon}{v1,v2}
   \fmfv{decor.shape=circle,decor.filled=empty,decor.size=0.15w}{v1}
   \fmfv{decor.shape=circle,decor.filled=full,decor.size=0.15w}{v2}
   \fmflabel{$+A$}{i1}
   \fmflabel{$+B$}{i2}
   \fmflabel{$-C$}{o2}
   \fmflabel{$-D$}{o1}
   \fmfv{label=$-$,label.angle=-45,label.dist=0.1w}{v2}
   \fmfv{label=$+$,label.angle=45,label.dist=0.1w}{v1}
 \end{fmfgraph*}
 &\,\,\,\,& \begin{fmfgraph*}(120,80)
   \fmfleft{i1,i2}
   \fmfright{o1,o2}
   \fmf{gluon}{i1,v1,o1}
   \fmf{gluon}{i2,v2,o2}
   \fmf{gluon}{v1,v2}
   \fmfv{decor.shape=circle,decor.filled=empty,decor.size=0.15w}{v2}
   \fmfv{decor.shape=circle,decor.filled=full,decor.size=0.15w}{v1}
   \fmflabel{$+A$}{i1}
   \fmflabel{$+B$}{i2}
   \fmflabel{$-C$}{o2}
   \fmflabel{$-D$}{o1}
   \fmfv{label=$+$,label.angle=-45,label.dist=0.1w}{v2}
   \fmfv{label=$-$,label.angle=45,label.dist=0.1w}{v1}
 \end{fmfgraph*}\nonumber\\
 $\da{14},\ds{23}\rightarrow 0$ &\,\,\,\,& $\ds{14},\da{23}\rightarrow 0$
 \end{tabular}
\end{center}
\end{fmffile}
\caption{On-shell diagrams for MHV gluon scattering on both $u$-channels.}\label{MHV2Limits}
\end{figure}
\end{feynfig}
Consistency of these relations additionally demands that the sum of the
equations vanishes:
\begin{align}\label{ConjJac}
f_{AB}^{\,\,\,\,\,\,\,\,\,E}\left(f_{CD}^{\,\,\,\,\,\,\,\,\,E}\right)^*-f_{EA}^{\,\,\,\,\,\,\,\,\,D}\left(f_{EC}^{\,\,\,\,\,\,\,\,\,B}\right)^*+f_{EA}^{\,\,\,\,\,\,\,\,\,C}\left(f_{ED}^{\,\,\,\,\,\,\,\,\,B}\right)^*=0.
\end{align}

Multiplying both sides of (\ref{uResEq}) by $f_{FC}^{\,\,\,\,\,\,\,\,\,A}\left(f_{GB}^{\,\,\,\,\,\,\,\,\,D}\right)^*$ gives 
\begin{align}\label{KFSA}
\kappa_{AE}\left(\kappa_{EB}\right)^*=\lambda_A^{\,\,\,\,E}\lambda_E^{\,\,\,\,B},
\end{align}
where $\kappa_{FE}=-f_{FC}^{\,\,\,\,\,\,\,\,\,A}f_{EA}^{\,\,\,\,\,\,\,\,\,C}$ is the standard Killing form and $\lambda_F^{\,\,\,\,E}=f_{FC}^{\,\,\,\,\,\,\,\,\,A}\left(f_{EC}^{\,\,\,\,\,\,\,\,\,A}\right)^*$. The $\lambda$ matrices are self-adjoint, so can be diagonalised by an appropriate $U(N)$ rotation of the states. In this basis, $\lambda=\text{diag}(d_1,d_2,\ldots)$, where each $d_i$ is non-negative, because $\lambda$ is clearly positive semi-definite from its definition.

Because of (\ref{KFSA}), the eigenvalues $d_i=0$ correspond to degenerate directions in the Killing form - deleting the $i$th row and column of $\kappa$ for each $d_i=0$ leaves behind an invertible residual. Then by definition of $\lambda$, $d_i=0$ implies that $f_{iA}^{\,\,\,\,\,\,\,\,\,C}=0$ for all $A,C$. Multiplying the equation (\ref{uResEq}) by $f_{FC}^{\,\,\,\,\,\,\,\,\,A}$ implies that
\begin{align}\label{StrucConj}
\kappa_{FE}\left(f_{ED}^{\,\,\,\,\,\,\,\,\,B}\right)^*=-\lambda_F^{\,\,\,\,E}f_{EB}^{\,\,\,\,\,\,\,\,\,D},
\end{align}
This then implies that $\lambda_F^{\,\,\,\,E}f_{EB}^{\,\,\,\,\,\,\,\,\,i}=0$. Restricted to the non-zero eigenspaces, which are the only directions in which the lower indices of $f_{EB}^{\,\,\,\,\,\,\,\,\,C}$ can correspond to non-zero entries, the $\lambda$ matrix can be inverted to imply that $f_{EA}^{\,\,\,\,\,\,\,\,\,i}=0$ as well, for all $E,A$. So degenerate directions in the Killing form are only possible if the structure constants are vanishing, or, in other words, the Lie algebra describing interacting (or non-Abelian) vector bosons must be semi-simple. This establishes that the Killing form (and $\lambda$) is invertible on every simple non-Abelian subalgebra. I will henceforth restrict to the simple non-Abelian case. 

Multiplying (\ref{ConjJac}) by $f_{FD}^{\,\,\,\,\,\,\,\,\,A}$ (and then switching the $D$ and $C$ index labels), in conjunction with using the Jacobi identity, implies that 
\begin{align}
\kappa_{FE}\left(f_{ED}^{\,\,\,\,\,\,\,\,\,B}\right)^*=-f_{FB}^{\,\,\,\,\,\,\,\,\,A}\lambda_A^{\,\,\,\,D}.
\end{align}
Combining this with (\ref{StrucConj}) gives
\begin{align}
f_{iB}^{\,\,\,\,\,\,\,\,\,j}&=\left(\lambda^{-1}\right)_i^{\,\,\,\,F}f_{FB}^{\,\,\,\,\,\,\,\,\,A}\lambda_A^{\,\,\,\,j}\nonumber\\
&=\frac{1}{d_i}f_{iB}^{\,\,\,\,\,\,\,\,\,j} d_j
\end{align}
for all $i,j$ (which are not summed over in the last line). This is only possible if $d_i=d_j=d$. Thus $\lambda_A^{\,\,\,\,B}=d\delta_A^{\,\,\,\,B}$ for some positive real constant $d$. This is $U(N)$ invariant, so the freedom to rotate the external scattering states remains. The $U(N)$ freedom of basis choice can therefore be used to Takagi diagonalise the Killing form so that it has real, positive diagonal entries. Then (\ref{KFSA}) necessitates that all of its diagonal entries are equal: $\kappa_{AB}=d\delta_{AB}$. This demonstrates both that the Lie algebra is compact and that the generators are homogeneously normalised. The constant $d$ determines the overall normalisation of all of the Lie algebra generators and can be identified as the square of the gauge coupling.

Finally, given the determination of $\kappa$ and $\lambda$, (\ref{StrucConj}) implies that 
\begin{align}
\left(f_{ED}^{\,\,\,\,\,\,\,\,\,B}\right)^*=-f_{EB}^{\,\,\,\,\,\,\,\,\,D}=-f_{EBD}=f_{EDB}=f_{ED}^{\,\,\,\,\,\,\,\,\,B}.
\end{align}
The usual definition of $f_{EBD}=\frac{1}{d}\kappa_{DA}f_{EB}^{\,\,\,\,\,\,\,\,\,A}$ has been made here, where $f_{EBD}$ is fully antisymmetric for a semi-simple Lie algebra. So the structure constants are purely real. This represents the parity and time-reversal symmetry of the Yang-Mills couplings. This concludes the derivation of pure perturbative Yang-Mills theory from $S$-matrix self-consistency. 

\subsubsection{Compton scattering and matter couplings}

More is required to subsequently show that matter couplings, higher-leg and higher dimensional interactions must also couple to massless vectors through invariant tensors of the Lie algebra. As will be commented upon further below, soft-limits provide a systematic method for deriving these constraints on higher-leg and higher dimensional interactions \cite{Elvang:2016qvq}. I will here merely note that, while it is well-known that consistency of Compton scattering of gluons off matter demands that the matter-gluon $3$-leg coupling be a Lie algebra generator in some representation, the fact that the matter representations have to be unitary also follows simultaneously, as will now be explained. Later, in Section \ref{sec:PartialU} I will make some further comments about implications of soft gluon limits for massive particle couplings. 

For massless scalars, same-sign gluon Compton scattering can be constructed from the residue of the all-channel pole (see Fig \ref{scalarCSGen}): 
\begin{feynfig}
\begin{figure}[h]
\begin{fmffile}{SSGluonCompton}

\begin{center}
\begin{tabular}{ c c c c c }
& & & & \\
   \begin{fmfgraph*}(100,67)
   \fmfleft{i1,i2}
   \fmfright{o1,o2}
   \fmf{gluon}{i2,v1}
   \fmf{gluon}{v2,o2}
   \fmf{dashes}{i1,v1,v2,o1}
   \fmfv{decor.shape=circle,decor.filled=empty,decor.size=0.15w}{v1,v2}
   \fmflabel{$i$}{i1}
   \fmflabel{$+A$}{i2}
   \fmflabel{$+B$}{o2}
   \fmflabel{$l$}{o1}
 \end{fmfgraph*} 
 &\,\,\,\, & \begin{fmfgraph*}(100,67)
   \fmfleft{i1,i2}
   \fmfright{o1,o2}
   \fmf{dashes}{i1,v1,v2,o1}
   \fmf{phantom}{v1,i2}
   \fmf{phantom}{v2,o2}
   \fmf{gluon,tension=-0.25}{v2,i2}
   \fmf{gluon,tension=-0.25}{v1,o2}
   \fmfv{decor.shape=circle,decor.filled=empty,decor.size=0.15w}{v1,v2}
   \fmflabel{$i$}{i1}
   \fmflabel{$+A$}{i2}
   \fmflabel{$+B$}{o2}
   \fmflabel{$l$}{o1}
 \end{fmfgraph*} &\,\,\,\, & 
 \begin{fmfgraph*}(100,67)
   \fmfleft{i1,i2}
   \fmfright{o1,o2}
   \fmf{dashes}{i1,v1,o1}
   \fmf{gluon}{i2,v2,o2}
   \fmf{gluon}{v1,v2}
   \fmfv{decor.shape=circle,decor.filled=empty,decor.size=0.15w}{v1,v2}
   \fmflabel{$i$}{i1}
   \fmflabel{$+A$}{i2}
   \fmflabel{$+B$}{o2}
   \fmflabel{$l$}{o1}
   \fmfv{label=$-$,label.angle=-45,label.dist=0.1w}{v2}
   \fmfv{label=$+$,label.angle=45,label.dist=0.1w}{v1}
 \end{fmfgraph*}\\
 $(s)$ &\,\,\,\, & $(t)$ &\,\,\,\, & $(u)$
 \end{tabular}
\end{center}
\end{fmffile}
\caption{On-shell diagrams for gluon-scalar Compton scattering.}\label{scalarCSGen}
\end{figure}
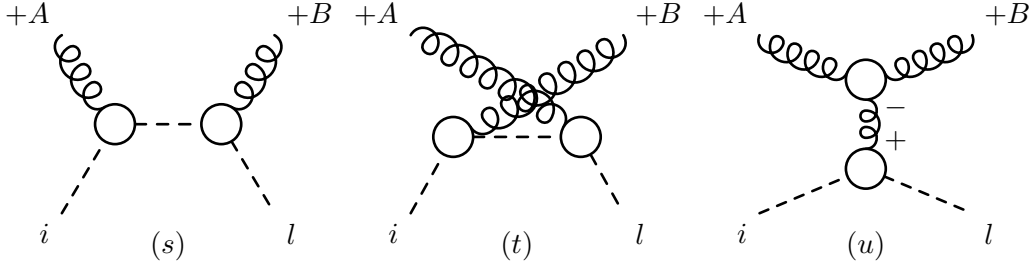
\end{feynfig}
\begin{align}
A(\varphi_i,g^+_A,g^+_B,\varphi_l)&=\frac{-1}{s}(t_A)_{im}\frac{\ds{12}\ds{2P_s}}{\ds{1P_s}}(t_B)_{ml}\frac{\ds{P_s3}\ds{34}}{\ds{P_s4}}-\frac{1}{t}(t_B)_{im}\frac{\ds{13}\ds{3P_t}}{\ds{1P_t}}(t_A)_{ml}\frac{\ds{P_t2}\ds{24}}{\ds{P_t4}}\nonumber\\
&\qquad-\frac{1}{u}if_{AB}^{\,\,\,\,\,\,\,\,\,C}\frac{\ds{23}^3}{\ds{3P_u}\ds{P_u2}}(-(t_C)_{il})\frac{\ds{1P_u}\ds{P_u4}}{\ds{14}}\nonumber\\
&=-\frac{\da{1q}\da{4q}}{\da{2q}\da{3q}}\frac{\ds{12}\ds{34}}{s}\left((t_A)_{im}(t_B)_{ml}-(t_B)_{im}(t_A)_{ml}-if_{AB}^{\,\,\,\,\,\,\,\,\,C}(t_C)_{il}\right).
\end{align}
Like the pure gluon example earlier, the only way in which this can be Lorentz invariant is if the overall combination of coupling constants is zero, implying that 
\begin{align}
(t_A)_{im}(t_B)_{ml}-(t_B)_{im}(t_A)_{ml}=if_{AB}^{\,\,\,\,\,\,\,\,\,C}(t_C)_{il},
\end{align}
so the scalar-gluon couplings are generators of the Lie algebra in some representation. 

Opposite sign Compton scattering can be computed to derive the analogous result, as has been described extensively previously \cite{Benincasa:2007xk}. Notably however, this amplitude contains the additional condition that residues of the $u$-channel pole in the limits $\ds{23}\rightarrow 0$ and $\da{23}\rightarrow 0$ must agree. These are shown in Fig \ref{OppositeCS}.
\begin{feynfig}
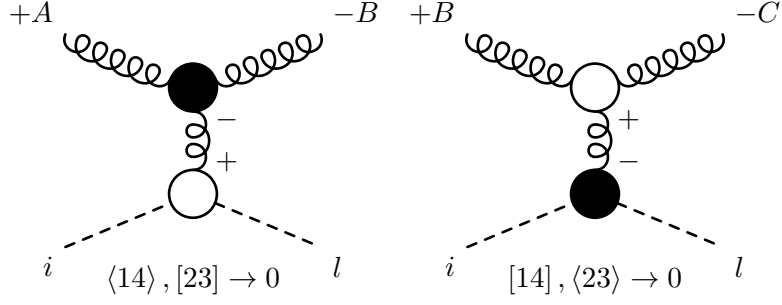
\begin{figure}[h]
\begin{fmffile}{OSGluonCompton}
\begin{center}
\begin{tabular}{ c c c }
& & \\
   \begin{fmfgraph*}(120,80)
   \fmfleft{i1,i2}
   \fmfright{o1,o2}
   \fmf{dashes}{i1,v1,o1}
   \fmf{gluon}{i2,v2,o2}
   \fmf{gluon}{v1,v2}
   \fmfv{decor.shape=circle,decor.filled=empty,decor.size=0.15w}{v1}
   \fmfv{decor.shape=circle,decor.filled=full,decor.size=0.15w}{v2}
   \fmflabel{$i$}{i1}
   \fmflabel{$+A$}{i2}
   \fmflabel{$-B$}{o2}
   \fmflabel{$l$}{o1}
   \fmfv{label=$-$,label.angle=-45,label.dist=0.1w}{v2}
   \fmfv{label=$+$,label.angle=45,label.dist=0.1w}{v1}
 \end{fmfgraph*}
 &\,\,\,\, &  \begin{fmfgraph*}(120,80)
   \fmfleft{i1,i2}
   \fmfright{o1,o2}
   \fmf{dashes}{i1,v1,o1}
   \fmf{gluon}{i2,v2,o2}
   \fmf{gluon}{v1,v2}
   \fmfv{decor.shape=circle,decor.filled=empty,decor.size=0.15w}{v2}
   \fmfv{decor.shape=circle,decor.filled=full,decor.size=0.15w}{v1}
   \fmflabel{$i$}{i1}
   \fmflabel{$+B$}{i2}
   \fmflabel{$-C$}{o2}
   \fmflabel{$l$}{o1}
   \fmfv{label=$+$,label.angle=-45,label.dist=0.1w}{v2}
   \fmfv{label=$-$,label.angle=45,label.dist=0.1w}{v1}
 \end{fmfgraph*}\nonumber\\
 $\da{14},\ds{23}\rightarrow 0$ &\,\,\,\, & $\ds{14},\da{23}\rightarrow 0$
 \end{tabular}
\end{center}
\end{fmffile}
\caption{On-shell diagrams for opposite sign Compton scattering on both $u$-channels.}\label{OppositeCS}
\end{figure}
\end{feynfig}
Reconstructing the singular part of the amplitude from these limits gives
\begin{align}
A(\varphi_i,g^+_A,g^-_B,\varphi_l)&\sim\frac{-1}{u}if_{MA}^{\,\,\,\,\,\,\,\,\,B}\frac{\ds{P_u2}^3}{\ds{23}\ds{3P_u}}(-(t_M)_{li})^*\frac{\da{1P_u}\da{P_u4}}{\da{14}}\nonumber\\
&=\frac{-1}{su}if_{MA}^{\,\,\,\,\,\,\,\,\,B}((t_M)_{li})^*\ds{12}^2\da{31}^2
\end{align}
for $\da{23},\ds{14}\rightarrow 0$ and likewise
\begin{align}
A(\varphi_i,g^+_A,g^-_B,\varphi_l)&\sim\frac{-1}{su}i(f_{BM}^{\,\,\,\,\,\,\,\,\,A})^*(t_M)_{il}\ds{12}^2\da{31}^2
\end{align}
for $\ds{23},\da{14}\rightarrow 0$. Demanding that they agree up to contact terms implies that 
\begin{align}
f_{MA}^{\,\,\,\,\,\,\,\,\,B}((t_M)_{li})^*=(f_{BM}^{\,\,\,\,\,\,\,\,\,A})^*(t_M)_{il}.
\end{align}
Using the results established above for the gluon couplings, the factors of the structure constants may be removed using the invertibility of the Killing form to give
\begin{align}\label{GenSA}
((t_M)_{li})^*=(t_M)_{il},
\end{align}
demonstrating that the generators must be self-adjoint, or equivalently, that the Lie algebra representation must be unitary. The case of massless fermion matter is entirely analogous. The Peter-Weyl theorem then implies that, since the Lie algebra is compact, these matter representations must be finite dimensional (or, more precisely, decomposable into a sum of finite-dimensional irreps). A finite number of species of elementary particles coupled to YM theory is therefore possible. 

If the matter particles are massive, and minimally coupled, then consistent factorisation again leads to the couplings being Lie algebra generators. I will just concentrate on the self-adjointness of the generators. Beginning with massive scalars, the $u$-channel residues in each complex limit are:
\begin{align}
A(\phi_i,g^+_A,g^-_B,\phi_l)&\sim\frac{-1}{u}if_{MA}^{\,\,\,\,\,\,\,\,\,B}\frac{\ds{P_u2}^3}{\ds{23}\ds{3P_u}}(-(t_M)_{li})^*mx\nonumber\\
&\sim\frac{-1}{u(s-m^2)}if_{MA}^{\,\,\,\,\,\,\,\,\,B}((t_M)_{li})^*\ls{2}p_1\ra{3}^2
\end{align}
for $\da{23}\rightarrow0$, while 
\begin{align}
A(\phi_i,g^+_A,g^-_B,\phi_l)&\sim\frac{-1}{u(s-m^2)}i(f_{BM}^{\,\,\,\,\,\,\,\,\,A})^*(t_M)_{il}\ls{2}p_1\ra{3}^2
\end{align}
for $\ds{23}\rightarrow 0$. Requiring agreement of the two expressions clearly gives the same result as the massless case (\ref{GenSA}). Here however, unitarity of the generators was already required for reasons explained below (\ref{ScalarMat3leg}).

If the massive matter has spin, the minimal coupling to gluons is given by
\begin{align}\label{MinGluon}
A(S_i,S_j,g^+_A)&=-(t_A)_{ij}\frac{1}{x}\frac{\da{\bf{12}}^{2s}}{m^{2s-1}}\nonumber\\
A(S_i,S_j,g^-_A)&=-((t_A)_{ji})^*x\frac{\ds{\bf{12}}^{2s}}{m^{2s-1}}.
\end{align}
I assume that the sign incurred from crossing these amplitudes for bosons and fermions is the same as that for spin-0 and spin-1/2 cases respectively. The $u$-channel residues of the Compton amplitude in either complex limit now differ by the chirality of the massive spinor bilinears in (\ref{MinGluon}). However, these factors are equivalent up to terms that vanish on the opposite limit. Using 
\begin{align}\label{SimpleId}
\ds{\bf{14}}=\da{\bf{14}}-\frac{\da{P_u\bf{4}}\ds{P_u\bf{1}}}{m}
\end{align}
to eliminate $\ds{\bf{14}}^{2s}$ in the residue of the $\da{23}\rightarrow 0$ pole, then all terms besides $\da{\bf{14}}^{2s}$ are proportional to
\begin{align}
\da{P_u\bf{4}}\ds{P_u\bf{1}}x\frac{\ds{P_u2}^3}{\ds{23}\ds{3P_u}}\sim\ds{23}\frac{\da{3\bf{4}}\da{3\bf{1}}\la{3}p_1\rs{2}}{s-m^2},
\end{align}
which clearly vanishes in the limit $\ds{23}\rightarrow 0$. The derivation of (\ref{GenSA}) is therefore unaffected by the presence of spin and is analogous to the scalar calculation above. 



If a subset of $n$ vectors are photons (vanishing structure constants), then consistency of same and opposite sign Compton scattering implies that $[t_A,t_B]=[t_A,t_B^\dagger]=0$ for each of the photons $A,B$. This implies that the real and imaginary parts of each generator commute. A real, antisymmetric matrix can be rotated into normal form by conjugation by orthogonal matrices. If a set of such matrices commute, then this can be performed simultaneously for all of them. Each Abelian generator therefore has the form 
\begin{align}\label{AbelianGen}
t_A=\begin{pmatrix}
0 && -i({q_A}_1+i{p_A}_1) && && &&\\
i({q_A}_1+i{p_A}_1) && 0 && && &&\\
&& && 0 && -i({q_A}_2+i{p_A}_2) &&\\
&& && i({q_A}_2+i{p_A}_2) && 0 &&\\
&& && && && \ddots
\end{pmatrix}
\end{align}
in some basis. I define $q_A,p_A\in\mathbb{R}$.

The pairings of matter fields in (\ref{AbelianGen}) can be rotated by the matrix \begin{align}\label{ConjBas}
\mathbb{U}=\frac{1}{\sqrt{2}}\begin{bmatrix}
 1 &  i  \\ 
 1 &  -i  \\ 
\end{bmatrix}:\qquad\qquad\qquad\begin{bmatrix}
 S_{\frac{i+1}{2}} \\ 
 \overline{S}_{\frac{i+1}{2}} \\ 
\end{bmatrix}=\mathbb{U} \begin{bmatrix}
 s_i \\ 
 s_{i+1} \\ 
\end{bmatrix},
\end{align}
where $s_i$ labels the self-conjugate matter particles' identities and $S_\frac{i+1}{2}$ and $\overline{S}_\frac{i+1}{2}$ are the charge eigenstates. In this basis, the generators have the same form as (\ref{AbelianGen}), but with the overall factor of $-i$ dropped. The charged eigenstates transform into each other under crossing. For convenience of reference, the amplitudes for minimal coupling of a single charged scalar or fermion in unbroken electrodynamics (calling $e$ the electric charge) are:
\begin{align}\label{QEDmassless3}
\begin{split}
A(\varphi,\overline{\varphi},\gamma^+)&=-e\frac{\ds{13}\ds{32}}{\ds{12}}\\
A(\varphi,\overline{\varphi},\gamma^+)&=-e\frac{m}{x}\\
A(\psi^-,\overline{\psi}^+,\gamma^+)&=e\frac{\ds{32}^2}{\ds{12}}\\
A(\psi,\overline{\psi},\gamma^+)&=-e\frac{1}{x}\da{\bf{12}}\\
&\qquad-2(\Delta\mu+id)\ds{3\bf{2}}\ds{3\bf{1}}
\end{split}
\begin{split}
A(\varphi,\overline{\varphi},\gamma^-)&=-e\frac{\da{13}\da{32}}{\da{12}}\\
A(\varphi,\overline{\varphi},\gamma^-)&=-emx\\
A(\psi^-,\overline{\psi}^+,\gamma^-)&=-e\frac{\da{31}^2}{\da{12}}\\
A(\psi,\overline{\psi},\gamma^-)&=-ex\ds{\bf{12}}\\
&\qquad-2(\Delta\mu-id)\da{3\bf{2}}\da{3\bf{1}}.
\end{split}
\end{align}
These have the same crossing rules as the their counterparts in the self-conjugate basis, except that they must now also be charge conjugated. The required sign flips are the same as in (\ref{ScalarMat3leg}) and (\ref{FermionMat3leg}) and are omitted for simplicity. For the massive fermions, $\Delta\mu$ and $d$ are the anomalous magnetic and electric dipole moments.

The generators are not self-adjoint if ${p_A}_i\neq 0$, modulo the freedom to absorb phases by the choice of basis for the photon states (which is an overall $U(n)$ ``electric-magnetic duality''). These parameters are magnetic charges, discussed in \cite{Caron-Huot:2018ape}. However, they are not consistent with perturbative photon exchanges. For example, when scattering two massless scalars off each other, the residue of the photon exchange channel must agree when reconstructed from both complex factorisation limits (see Fig \ref{OppositeMoller}): 
\begin{feynfig}
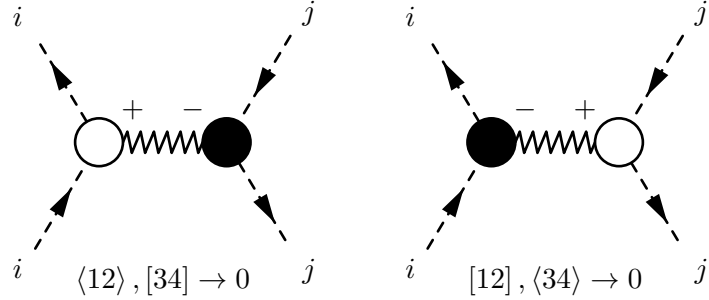
\begin{figure}[h]
\begin{fmffile}{sQEDMoller}
\begin{center}
\begin{tabular}{ c c c }
& & \\
   \begin{fmfgraph*}(120,80)
   \fmfleft{i1,i2}
   \fmfright{o1,o2}
   \fmf{scalar}{i1,v1,i2}
   \fmf{scalar}{o2,v2,o1}
   \fmf{zigzag}{v1,v2}
   \fmfv{decor.shape=circle,decor.filled=empty,decor.size=0.15w}{v1}
   \fmfv{decor.shape=circle,decor.filled=full,decor.size=0.15w}{v2}
   \fmflabel{$i$}{i2}
   \fmflabel{$j$}{o2}
   \fmflabel{$i$}{i1}
   \fmflabel{$j$}{o1}
   \fmfv{label=$+$,label.angle=45,label.dist=0.1w}{v1}
   \fmfv{label=$-$,label.angle=135,label.dist=0.1w}{v2}
 \end{fmfgraph*} 
 &\,\, & \begin{fmfgraph*}(120,80)
   \fmfleft{i1,i2}
   \fmfright{o1,o2}
   \fmf{scalar}{i1,v1,i2}
   \fmf{scalar}{o2,v2,o1}
   \fmf{zigzag}{v1,v2}
   \fmfv{decor.shape=circle,decor.filled=empty,decor.size=0.15w}{v2}
   \fmfv{decor.shape=circle,decor.filled=full,decor.size=0.15w}{v1}
   \fmflabel{$i$}{i2}
   \fmflabel{$j$}{o2}
   \fmflabel{$i$}{i1}
   \fmflabel{$j$}{o1}
   \fmfv{label=$-$,label.angle=45,label.dist=0.1w}{v1}
   \fmfv{label=$+$,label.angle=135,label.dist=0.1w}{v2}
 \end{fmfgraph*}\\
 $\da{12},\ds{34}\rightarrow 0$ &\,\, & $\ds{12},\da{34}\rightarrow 0$
 \end{tabular}
\end{center}
\end{fmffile}
\caption{On-shell diagrams for scalar Moller scattering on both $s$-channels.}\label{OppositeMoller}
\end{figure}
\end{feynfig}
\begin{align}
A(\varphi_i,\overline{\varphi}_i,\varphi_j,\overline{\varphi}_j)&\sim\frac{-1}{s}\sum_A({q_A}_i+i{p_A}_i)({q_A}_j-i{p_A}_j)\frac{\ds{1P_s}\ds{P_s2}}{\ds{12}}\frac{\da{3P_s}\da{P_s4}}{\da{34}}\nonumber\\
&=\frac{-t}{s}\mathfrak{q}_j\cdot\mathfrak{q}_i
\end{align}
for $\da{12},\ds{34}\rightarrow 0$, where I am calling the complex charge vector of particle $i$ $\mathfrak{q}_i=({q_1}_i+i{p_1}_i,{q_2}_i+i{p_2}_i\ldots,{q_n}_i+i{p_n}_i)$ and using the standard complex dot product, and 
\begin{align}
A(\varphi_i,\overline{\varphi}_i,\varphi_j,\overline{\varphi}_j)&\sim\frac{-1}{s}\sum_A({q_A}_i-i{p_A}_i)({q_A}_j+i{p_A}_j)\frac{\da{1P_s}\da{P_s2}}{\da{12}}\frac{\ds{3P_s}\ds{P_s4}}{\ds{34}}\nonumber\\
&=\frac{-t}{s}\mathfrak{q}_i\cdot\mathfrak{q}_j
\end{align}
for $\ds{12},\da{34}\rightarrow 0$. Consistency therefore requires that 
\begin{align}\label{ElectricConsist}
\mathfrak{q}_i\cdot\mathfrak{q}_j=\mathfrak{q}_j\cdot\mathfrak{q}_i 
\end{align}
for any matter particles $i$ and $j$. The fermion argument is similar. If massive, both helicity states are exchanged on the factorisation channel, but they cannot be combined in a way consistent with Lorentz invariance unless this reality condition holds, as explained in \cite{Caron-Huot:2018ape} (and using (\ref{SimpleId}) if the states are spinning). 

The $U(n)$ photon basis choice can be used to rotate the charge vector for a particular particle (call this $i$) to the form $\mathfrak{q}_i=(q,0\ldots,0)$ for some real electric charge $q$. Then (\ref{ElectricConsist}) implies that ${p_1}_j=0$ for all $j$, so there are no magnetic charges that couple to photon number $1$. This argument can then be subsequently repeated for each photon to show that ${p_A}_i=0$ for all $A$ and $i$. The Abelian generators are therefore now also self-adjoint. 

This entire discussion has shown that there are no consistent (perturbative) theories of chiral photons or gluons. The argument for gravitons is easier, accepting the conclusions from the soft theorems that the graviton is unique and couples with universal strength. The absence of a complex phase in the gravitational constant follows simply from the freedom to choose the phase of the external graviton helicity eigenstates in the standard three graviton amplitudes $A\left(h^+,h^+,h^-\right)$. Unitarity then fixes the opposite helicity amplitude. The required universality of the graviton coupling for both helicities (as is easily derived from soft limits \cite{Elvang:2016qvq}) then implies that all minimal graviton-matter couplings are also free of parity or time-reversal violating phases (this was assumed in the calculation of (\ref{PhotonGravHDO}) above in Section \ref{sec:simple}).

\subsection{Construction from the all-channel pole}\label{AllChannelAmps}

In the $S$-matrix derivation of Yang-Mills above, the four-gluon amplitudes with helicity configurations $A(++--)$ and $A(+++-)$ were constructed directly from unitarity by combining the elementary amplitudes $A(++-)$ and $A(+--)$, in the course of which constraints on the couplings from self-consistency were identified. The remaining configuration $A(++++)=0$ directly because it does have any consistent factorisation channels at all. These amplitudes then provide the base that ensures, through on-shell recursion such as BCFW, that all higher leg UHV amplitudes (those of the form $A(+++\ldots +-)$) also vanish. This well-known result is frequently attributed to accidental supersymmetry of Yang-Mills at tree-level \cite{elvang2015scattering}, but this symmetry is really an emergent feature of the way in which the $3$-particle amplitudes across the different channels combine and cancel, as shown here (in other words, Lorentz invariance and locality). Some recent discussion of some of these amplitudes and their connection to celestial holography has been given in \cite{Ren:2022sws,Ball:2023sdz}.

The vanishing amplitudes of same-sign Compton scattering are unified under supersymmetry. Analogous calculations to  (\ref{+++-glu}) with superamplitudes can be performed with similar results. While supersymmetric Ward identities (SWIs) necessitate that these amplitudes vanish (as well as their higher leg counterparts), supersymmetry seems to be incidental in the direct gluing calculation, merely organising the simultaneous factorisation of different spin Compton amplitudes into the same generating function. 

More examples of amplitudes constructed from the all-channel pole can be calculated. Allowing now for the three gluon amplitudes
\begin{align}\label{F3Amps}
A(g^+_A,g^+_B,g^+_C)&=i\frac{h_{ABC}}{\Lambda^2}\ds{12}\ds{23}\ds{31}\nonumber\\
A(g^-_A,g^-_B,g^-_C)&=i\frac{(h_{ABC})^*}{\Lambda^2}\da{12}\da{23}\da{31}
\end{align}
(or equivalently, off-shell $F^3$ operators), both $A(++++)$ and $A(+++-)$ receive non-zero contributions. Here $h_{ABC}$ is fully antisymmetric in particle indices in order for the amplitude to have bosonic exchange symmetry. The structure of the factorisation channels for $A(+++-)$ are analogous to the MHV amplitude in pure YM. The residues for each channel themselves contain single poles and consistent factorisation requires that the couplings obey the relation
\begin{align}
f_{ABE}\,h_{ECD}+f_{ACE}\,h_{EDB}+f_{ADE}\,h_{EBC}=0.
\end{align}
This implies that $h_{ABC}$ are invariant tensors under the action of the gauge algebra, and assuming that the gluons belong to a single simple non-Abelian algebra, then 
$h_{ABC}=cf_{ABC}$ for some complex, dimensionful constant $c$, as expected for ``gauge invariance''. 

Since the amplitude $A(+++-)$ is of the form discussed in \cite{McGady:2013sga}, I will illustrate the computation of $A(++++)$ from the all-channel pole. The $s$-channel terms are: 
\begin{align}\label{A4s++++}
A_{s}(g_A^+,g_B^+,g_C^+,g_D^+)&=f_{ABE}f_{ECD}\frac{1}{s}\left(\ds{12}\ds{2P_s}\ds{1P_s}\frac{\ds{34}^3}{\ds{3P_s}\ds{4P_s}}+\ds{34}\ds{3P_s}\ds{4P_s}\frac{\ds{12}^3}{\ds{1P_s}\ds{2P_s}}\right)\nonumber\\
&=f_{ABE}f_{ECD}\frac{\ds{12}\ds{34}}{s}\left(\ds{12}^2\frac{\da{1q}\da{2q}}{\da{3q}\da{4q}}+\ds{34}^2\frac{\da{3q}\da{4q}}{\da{1q}\da{2q}}\right)\nonumber\\
&=\frac{\ds{12}\ds{34}}{s}\frac{1}{\da{1q}\da{2q}\da{3q}\da{4q}}f_{ABE}f_{ECD}\nonumber\\
&\qquad\times\left(\left(\ds{12}\da{1q}\da{2q}+\ds{34}\da{3q}\da{4q}\right)^2-2\ds{12}\ds{34}\da{1q}\da{2q}\da{3q}\da{4q}\right).
\end{align}
The other channels are analogous and can be obtained from identical bosonic exchange symmetry (I am keeping the structure constants, but dropping the overall constant factors for brevity). Adding them all together, the term squared in the last line of (\ref{A4s++++}) is invariant under exchange of any two particles, so its sum over each channel cancels by the Jacobi identity. This leaves
\begin{align}\label{A4++++}
&A(g_A^+,g_B^+,g_C^+,g_D^+)\nonumber\\
&\quad=-2\frac{\ds{12}\ds{34}}{s}\left(f_{ABE}f_{ECD}\ds{12}\ds{34}+f_{BCE}f_{EAD}\ds{14}\ds{23}+f_{ACE}f_{EBD}\ds{13}\ds{24}\right).
\end{align}
Off the pole, the amplitude cannot depend upon the reference spinor $\ra{q}$, so all appearances of this in the combined residue must cancel. This has clearly occurred in (\ref{A4++++}). This amplitude is in agreement with the result presented in \cite{Dixon:1993xd}. The calculation presented here is equivalent to a Feynman graph calculation in which each external polarisation is defined in a gauge with the same reference spinor. It is also equivalent to a Risager shift \cite{Cachazo:2004kj,Risager:2005vk} calculation in which all right-handed spinors are shifted by the reference spinor that appearing above. The formulation here demonstrates how the result follows directly from unitarity and locality. The fact that it is non-zero is consistent with the fact that the constituent amplitudes (\ref{F3Amps}) violate supersymmetry. 

The covariance of the massless fermion dipole (\ref{FermionMat3leg}) and the $\varphi F^2$ scalar-vector type couplings (see (\ref{2vec1scalar}) below), both higher dimension operator couplings, can also be easily established from consistent factorisation with similar arguments, using amplitudes with either the all-channel pole or the single-channel complex factorisation poles (the Yukawa case was performed in \cite{Bachu:2023fjn}). When the amplitudes $A(g^+_A,g^+_B,g^+_C,\varphi_i)$ and $A(g^+_A,g^+_B,\psi^+_i,\psi_j^+)$ are computed from these interactions, covariance is required/ensures that they are Lorentz invariant and the only form that they can have that is consistent with this property is $0$ (which is expected on supersymmetric grounds for the same reasons as described above). For higher-dimension effective contact interactions involving more legs, the soft limit procedure proposed in \cite{Elvang:2016qvq} can be used with soft gluons to systematically establish gauge covariance. I will make some more comments about this regarding massive theories further below in Section \ref{sec:PartialU} (and will make more practical use of these methods in \cite{Trott:2026ozo} to establish the structure of supersymmetric theories from consistency of soft gravitinos).

Gravitational amplitudes can be calculated similarly to the gluonic ones. The calculation of the UHV four-graviton amplitude, analogous to (\ref{+++-glu}) for gluons, involves adding all three channels:
\begin{align}\label{+++-grav}
    A(h^+,h^+,h^+,h^-)&=\frac{-1}{M_{Pl}^2}\frac{\ds{12}^6}{\ds{1P_s}^2\ds{2P_s}^2}\frac{1}{s}\frac{\ds{P_s3}^6}{\ds{43}^2\ds{P_s4}^2}-\frac{1}{M_{Pl}^2}\frac{\ds{13}^6}{\ds{1P_t}^2\ds{3P_t}^2}\frac{1}{t}\frac{\ds{P_t2}^6}{\ds{42}^2\ds{P_t4}^2}\nonumber\\
    &\qquad-\frac{1}{M_{Pl}^2}\frac{\ds{23}^6}{\ds{3P_u}^2\ds{2P_u}^2}\frac{1}{u}\frac{\ds{P_u1}^6}{\ds{14}^2\ds{P_u4}^2}\nonumber\\
    &=\frac{-1}{M_{Pl}^2}\left(\frac{\da{q4}^3}{\da{q1}\da{q2}\da{q3}}\right)^2\left(\frac{\ds{12}^2\ds{34}^2}{s}+\frac{\ds{31}^2\ds{24}^2}{t}+\frac{\ds{23}^2\ds{14}^2}{u}\right)\nonumber\\
    &=\frac{-1}{M_{Pl}^2}\left(\frac{\da{q4}^3}{\da{q1}\da{q2}\da{q3}}\right)^2\frac{\ds{12}\ds{34}}{s}\left(\ds{12}\ds{34}+\ds{31}\ds{24}+\ds{23}\ds{14}\right)=0,
\end{align}
firstly using (\ref{AllChannel}) and secondly the Schouten identity. The calculation for same-helicity graviton Compton scattering off massless matter is very similar, where each channel combines to cancel again through the Schouten identity. 

These relations underpin the ``merger'' rule for combining on-shell diagrams in (super-)Yang-Mills \cite{Arkani-Hamed:2012zlh} and its generalisation to (super-)gravity identified in \cite{Herrmann:2016qea}. The latter reference derives the relationships between ``channels'' contributing to the all-channel pole as an equivalence of configurations of on-shell diagrams. If the particles have helicity one, it is required that the coupling constants obey the Jacobi identity, while if they have helicity two, the kinematic factors arrange into the Schouten identity, ensuring consistency provided that the gravitons are unique (and the couplings are universal constants in all channels). Again, similar observations have been made recently for multiparticle splitting functions in \cite{Ball:2023sdz}.

It has been observed very recently \cite{Guevara:2026qzd} that the general class of UHV gluon amplitudes are not actually completely zero, but rather have a non-analytic delta-function momentum dependence that restricts them to a measure-zero subspace of the external kinematics. This can be accounted for in all of the calculations presented in this Section by restoring the $+i\epsilon$ terms to the Mandelstam poles. For example, doing this for the same-helicity Compton amplitude derived in Section \ref{sec:simple}, then (\ref{ComptonRes}) is modified to
\begin{align}
A(\varphi,\gamma^+,\gamma^+,\bar{\varphi})&=\frac{\da{1q}\da{4q}}{\da{3q}\da{2q}}\left(\frac{\ds{12}\ds{34}}{s+i\epsilon}-\frac{\ds{13}\ds{24}}{t+i\epsilon}\right)\nonumber\\
&=\frac{\da{1q}\da{4q}}{\da{3q}\da{2q}}\left(\frac{1}{\frac{s}{\ds{12}\ds{34}}+i\epsilon sg_{12}sg_{34}}-\frac{1}{\frac{s}{\ds{12}\ds{34}}+i\epsilon sg_{13}sg_{24}}\right)\nonumber\\
&=\frac{\da{1q}\da{4q}}{\da{3q}\da{2q}}(-i\pi)\left(sg_{12}sg_{34}-sg_{13}sg_{24}\right)|\ds{12}|\delta\left(\da{34}\right),
\end{align}
where $sg_{ij}=\text{sign}\{\ds{ij}\}$ and 
$\frac{1}{z+i\epsilon}=P\frac{1}{z}-i\pi\delta(z)$ has been applied. The failure of the spinor prefactor to be compatible with Lorentz invariance away from the all-channel kinematic configuration is this time resolved by the presence of the delta-function. On the support of the delta-function appearing in the amplitude, the overall momentum-conserving delta-function that must implicitly accompany the amplitude becomes
\begin{align}
\delta\left(\sum_ip_i\right)=\frac{\da{4q}^2}{|\ds{12}|}\delta\left(\la{q}\sum_ip_i\right)\delta\left(\da{41}\right)\delta\left(\da{42}\right)
\end{align}
(I chose particle $4$ somewhat arbitrarily for this representation). Of course, this is all only possible for complexified momenta away from complete collinearity or, alternatively, $(-,-,+,+)$ signature spacetime on which the left- and right-handed spinors are independent real-valued functions. This is latter scenario has been assumed in the expressions above for ease of comparison with \cite{Guevara:2026qzd}, as has a convention in which only the right-handed spinors carry mass dimension and the left-handed spinor are dimensionless. I will restrict my use of these conventions to only this digression addressing \cite{Guevara:2026qzd}. The UHV gluon amplitude $A(+++-)$ can likewise be corrected for in the same way in order to identify the finite residual and verify agreement with the expression given in \cite{Guevara:2026qzd}, once all of the standard Lie algebra properties of the gluon self-couplings are applied. This demonstrates how these measure-zero amplitudes are still produced through on-shell factorisation. Higher leg amplitudes can be presumably computed on-shell as well. It would be interesting to explore this further.

\section{Low Spin Four Particle Matter Amplitudes}\label{Sec:LowSpinAmp}

As argued in \cite{Arkani-Hamed:2017jhn}, all $4$-leg, ``tree-level'' massive amplitudes can be fully constructed by adding 
\begin{enumerate}
    \item Terms that give correctly factorising expressions on each simple pole independently. For an exchange of a mass $m$ particle $P$ in the $s$-channel, unitarity implies that, as $s\rightarrow m^2$, the amplitude has the form $A(12\rightarrow34)\sim\frac{-A(12\rightarrow P)A(3^*4^*\rightarrow P^*)^*}{s-m^2}$. In this expression, it is being emphasised that the momenta (or, more precisely, the spinors as in (\ref{unitarity})) appearing in the second factor (involving the outgoing particles) are themselves complex conjugated.
    \item Contact terms determined by polynomials of Mandelstam invariants multiplying spinor structures (tensors of the little groups).
\end{enumerate}
This is in contrast to massless amplitudes, where the inverse kinematical dependence of the complex, on-shell 3-particle amplitudes can prevent 4-particle amplitudes from being decomposed into separate Lorentz invariant terms containing only a single simple pole.

The construction of the amplitude is thus decomposed into two separate problems. The first is the construction of a guess that correctly reproduces the singularity and factorisation structure of the amplitude, while the second is the determination of a basis of contact interactions. In general, there is no well-defined separation between the two classes, the former usually only being defined modulo terms that vanish on the pole, which are restored through the latter. However, power counting schemes, which are usually required to justify truncation of the infinite number of contact terms, can often be used to guide the specification of sensible factorisation terms on some physical grounds. 

One of the goals of this study is to construct all $2\rightarrow2$ scattering amplitudes between massive particles of spin $s\leq 1$ mediated by exchanges. Scattering of massive vector bosons will be the focus of most of the attention, since these are the most technically demanding and physically intricate. In this Section however, I will warm-up by constructing and analysing the $4$-particle amplitudes between scalars and fermions, predominantly in QED. These have been derived and presented previously in \cite{Christensen:2022nja} using massive chiral spinors and unitarity methods, although I have some additional comments to make beyond including them merely for completeness of the catalogue. Some of the results of this analysis, less interesting to the main narrative presented here, can be found separately in Appendix \ref{app:LowSpinGen}. In \cite{Trott:2026ozo}, I will extend these calculations to include gravity and supersymmetry. The low spin QED residues are subsequently useful as seeds from which amplitudes involving higher spin particles can be constructed (the double copy being a particularly efficient manifestation of this). 

Supersymmetric theories have especially elegant amplitudes, especially those involving BPS particles, where the restrictions placed by the enhanced symmetry outweigh complications introduced by the enlargement of the multiplets. The BPS condition and central charge conservation provide additional restrictions on the structure of the amplitudes and, in the examples of relevance here, this can be reinterpreted as higher dimensional Poincare invariance and supersymmetry \cite{Osborn:1979tq,Dennen:2009vk}. The reconciliation between the general massive vector boson scattering amplitude, derived from factorisation through the procedure described above, and the simple and compact expression that $\mathcal{N}=4$ SYM directly predicts is highly non-trivial and requires understanding precisely how the cancellation of high energy divergences proceeds. These special cases therefore provide idealised benchmarks that the general expressions can be compared to, which can help identify the structure of complicating features like the relaxation of the BPS constraints, as well as the specifically adjoint Higgs potential. For this reason, I will begin this Section by summarising the basic results derived in \cite{Trott:2026ozo} necessary to present and use the superamplitudes for $\mathcal{N}=2$ SQED Moller scattering and $\mathcal{N}=4$ SYM massive vector boson scattering relevant for Section \ref{VectorHiggs} further below.

\subsection{Review of special massive kinematics}

I begin by reviewing some of the special kinematic properties of massive $3$-particle amplitudes exhibiting BPS-like (complex) mass conservation. This can be viewed as a dimensional reduction of the kinematics of massless $3$-particle amplitudes in $6d$ \cite{Cheung:2008dn}. The little group covariant massive Dirac spinors for particle $i$ have the form
\begin{equation}\label{DiracSpinors}
\begin{split}
u_i^I=
\begin{pmatrix}
\rs{i^I}\\
e^{-i\varphi_i}\ra{i^I}
\end{pmatrix}
\end{split},\qquad
\begin{split}
v_i^I=
\begin{pmatrix}
e^{-i\varphi_i}\rs{i^I}\\
-\ra{i^I}
\end{pmatrix},
\end{split}
\end{equation}
These obey spin sums
\begin{align}\label{DiracSpinSums}
    u_I(p)\bar{u}^I(p)=-p^\mu\gamma_\mu + \mathbf{m}\qquad v^I(p)\bar{v}_I(p)=-p^\mu\gamma_\mu - \mathbf{m}^*,
\end{align}
where
\begin{align}
    \mathbf{m}=m\begin{pmatrix}
e^{i\varphi}I && 0\\
0 && e^{-i\varphi}I
\end{pmatrix}.
\end{align}
The phase $e^{i\varphi}$ is a possible complex phase given to the mass that interpolates between particle and antiparticle solutions to the Dirac equation. In the context of BPS particles with extended SUSY, this complex mass $me^{i\varphi}$ is more properly interpreted as a pair of central charges. 

For $3$-particle amplitudes in which the external particle masses obey complex mass conservation (this applies to the special case of the two equal mass, one massless configuration, where the two massive particles are assigned a relative $-1$ mass phase), then bilinears of the external legs' Dirac spinors degenerate and can be expressed as 
\begin{align}\label{SpecialSVD}
\overline{v}_i^Iu_j^J&=\mathbf{v}_{i}^I\mathbf{u}_{j}^J,
\end{align}
where the legs are chosen so that $i=j+1$ (mod $3$). This equation implicitly defines altogether new objects $\mathbf{v}_{i}^I$ and $\mathbf{u}_{j}^J$. They transform purely as spinors under the little groups of the particles that label them and have no other suppressed indices. These spinors project the spacetime spinors of each leg onto a common aligned direction:
\begin{align}\label{Trimmed3PSK}
\begin{split}
\rs{u}&=e^{i\varphi_i}\mathbf{u}_i^I\rs{i_I}=e^{i\varphi_j}\mathbf{u}_j^J\rs{j_J}\\
\ra{u}&=\mathbf{u}_i^I\ra{i_I}=\mathbf{u}_j^J\ra{j_J}
\end{split}
\begin{split}
\rs{v}&=e^{-i\varphi_i}\mathbf{v}_i^I\rs{i_I}=e^{-i\varphi_j}\mathbf{v}_j^J\rs{j_J}\\
\ra{v}&=\mathbf{v}_i^I\ra{i_I}=\mathbf{v}_j^J\ra{j_J},
\end{split}
\end{align}
providing the massive generalisation of the massless special $3$-particle complex kinematics underpinning the existence of on-shell massless $3$-particle amplitudes. 

There remains an unfixed ``tiny group'' (stabiliser of a pair of momenta \cite{Boels:2012ie}) redundancy in the definition of the $\mathbf{u}_{i}^I$ and $\mathbf{v}_{i}^I$ little group spinors, analogous to the little group redundancy for massless spacetime spinors. Under this scaling ambiguity, $\mathbf{u}_{i}^I\rightarrow \lambda\mathbf{u}_{i}^I$ and $\mathbf{v}_{i}^I\rightarrow \lambda^{-1}\mathbf{v}_{i}^I$, for any $\lambda\in\mathbb{C}\backslash\{0\}$. Since external scattering states do not form non-trivial tiny group representations, $\mathbf{u}_{i}^I$ and $\mathbf{v}_{i}^I$ can only appear in an amplitude in a composite form free of the rescaling ambiguity. 

In the special case in which one particle (say particle $3$) is massless, then the little group indices are naturally identified with invariant helicity indices. The little group spinors of the massless leg can then be combined into a special Lorentz scalar, little group charged and tiny group neutral object
\begin{align}\label{xExp}
x=-e^{-i\varphi}\mathbf{u}_{3}^-/\mathbf{u}_{3}^+=-e^{i\varphi}\mathbf{v}_{3}^-/\mathbf{v}_{3}^+=\frac{\ls{q}p_1\ra{3}}{m\rs{q3}}.
\end{align}
This provides a redundancy-free alternative definition to (\ref{Prelimx}). The relations
\begin{align}
\mathbf{v}_{3}^+\mathbf{u}_{3}^-&=me^{i\varphi}\nonumber\\
\mathbf{v}_{3}^-\mathbf{u}_{3}^+&=me^{-i\varphi}.
\end{align}
are also useful (where here $me^{i\varphi}$ is the complex mass of particle $1$, while particle $2$ has complex mass given by the negative of this).

These special little group spinors will have little use in this study, beyond their brief appearance in the $\mathcal{N}=4$ SYM $3$-particle superamplitude defined below, although their existence is intriguing. They become more relevant for the extended SUSY theories studied in \cite{Trott:2026ozo}.

\subsection{Review of superamplitudes}

Supermultiplets can be represented on-shell as fermionic coherent states or ``on-shell superfields''. These are Grassmann polynomials in which each coefficient of the independent Grassmann terms is identified with a distinct state in the multiplet. The supersymmetry algebra reduces to an algebra of fermionic harmonic oscillators when represented on these states. Each Grassmann variable is associated with a particular pair of ladder operators of an oscillator that acts non-trivally on the multiplet. See \cite{Ferrara:1980ra,Herderschee:2019dmc,Herderschee:2019ofc,Chen:2021hjl,Chen:2021huj,Caron-Huot:2018ape,KNBalasubramanian:2022sae} for further relevant background on superamplitudes (scattering amplitudes between fermionic coherent states) of massive particles and \cite{elvang2015scattering} for a general review of superamplitudes. 

When choosing a representation for the fermionic coherent states (tantamount to picking the Clifford vacuum), it is natural to preserve as much symmetry as possible. Since the subject of this study is little group covariant amplitudes, it seems obvious to represent the SUSY algebra on the massive particles in a little group covariant way. Unfortunately, for BPS particles in extended SUSY, of relevance to here and explained further below, this will involve breaking the $R$-symmetry, although parity will be manifest. For example, the $\frac{1}{2}$BPS $\mathcal{N}=4$ massive vector multiplet will be represented as
\begin{align}
\mathcal{W} =\phi+\eta^a_I\lambda^I_a+\half \eta^a_I \eta^b_J (\epsilon^{IJ} \phi_{(ab)} + \epsilon_{ab} W^{(IJ)})+\frac{1}{2^2}\eta_{bI}\eta_J^b\eta^J_a\widetilde{\lambda}^{aI} +\frac{1}{2^4}\eta_{aI}\eta_{J}^a\eta_{b}^I\eta^{bJ}\widetilde{\phi}.
\label{LongMultNtwo}
\end{align}
Here $a\in\{1,2\}$ is a (not necessarily covariant) $R$-index - pairs of the supercharges are being represented differently (see below). The spin eigenstate content consists of a massive vector $W$, gauginos $\lambda$ and $\widetilde{\lambda}$ and five scalars and $\phi$, $\widetilde{\phi}$ and $\phi_{(ab)}$. The four Grassmann variables are denoted by $\eta_{I}^a$. The individual particles in the multiplet can be extracted by acting upon it with some number of Grassmann derivatives before setting them to zero (in this sense, the on-shell superfields are generating functions of the multiplet). 

I will concentrate on the specific case of superampltidues of $\frac{1}{2}$BPS massive vector multiplets in $\mathcal{N}=4$ spontaneously broken SYM in this review, since this is the theory of relevance to this work, with some remarks about $\mathcal{N}=2$ SQED at the end. This is all discussed in greater detail in \cite{Trott:2026ozo}. For the purposes of this study, I will simply quote the relevant results and the information necessary to interpret them. 

For a single particle $i$, the BPS condition is 
\begin{align}\label{BPS}
p^{\dot{\alpha}\alpha}_iQ_{i,A\alpha}=\frac{1}{2}Z_{i,AB}Q^{\dagger B\dot{\alpha}}_i,
\end{align}
where $Z_{i,AB}$ is the matrix of central charge eigenvalues for the particle, while $p_i$, $Q_i$ and $Q_i^\dagger$ are its momentum and supercharges respectively. I will assume that the central charges have the form
\begin{align}\label{CentralChargesEleN=4}
Z_{i,AB}=2m_i\begin{bmatrix}
 0 &  0 & -e^{i\varphi_i} & 0 \\ 
 0 &  0 & 0 & -e^{-i\varphi_i} \\ 
 e^{i\varphi_i} &  0 & 0 & 0 \\ 
 0 & e^{-i\varphi_i} & 0 & 0 \\ 
\end{bmatrix},
\end{align}
that is, only two electric central charges are active. The supermultiplet represented above in (\ref{LongMultNtwo}) is a central charge eigenstate. Conservation of central charge then manifests itself as the constraint
\begin{align}
\sum_im_ie^{i\varphi_i}=\sum_im_ie^{-i\varphi_i}=0.
\end{align}
Under these assumptions, when the $\frac{1}{2}$BPS vector multiplets are represented by the form (\ref{LongMultNtwo}), the total supercharges arrange into Dirac spinors and are represented on the superamplitudes by the expressions
\begin{align}\label{N=4Charges}
\begin{split}
\frac{1}{\sqrt{2}}\mathcal{Q}^1&=\frac{1}{\sqrt{2}}\begin{pmatrix}
Q_3\\
Q^{\dagger 1}
\end{pmatrix}=\sum_iv^{-\varphi,I}_{i}\eta^{1}_{i,I}\\
\frac{1}{\sqrt{2}}\mathcal{Q}^2&=\frac{1}{\sqrt{2}}\begin{pmatrix}
Q_4\\
Q^{\dagger 2}
\end{pmatrix}=\sum_i v^{\varphi,I}_{i}\eta^{2}_{i,I}
\end{split}\qquad
\begin{split}
\frac{1}{\sqrt{2}}\mathcal{C}\overline{\mathcal{Q}}_{1}^T&=\frac{1}{\sqrt{2}}\begin{pmatrix}
Q_1\\
Q^{\dagger 3}
\end{pmatrix}=\sum_i u_{i,I}^\varphi\frac{\partial}{\partial\eta^1_{i,I}}\\
\frac{1}{\sqrt{2}}\mathcal{C}\overline{\mathcal{Q}}_{2}^T&=\frac{1}{\sqrt{2}}\begin{pmatrix}
Q_2\\
Q^{\dagger 4}
\end{pmatrix}=\sum_i u_{i,I}^{-\varphi}\frac{\partial}{\partial\eta^2_{i,I}}.
\end{split}
\end{align}
I specifically indicate the mass phase appearing in the spinors (\ref{DiracSpinors}). Here $Q_A$ and $Q^{\dagger A}$ are the total supercharges of the legs as they are usually defined in the form of chiral spinors, while $\mathcal{C}$ is the charge conjugation matrix \cite{Srednicki:2007qs}. These representations of the supersymmetry algebra can again be viewed as dimensional reductions of previous formulations of massless on-shell superspaces for half-maximal SUSY in $6d$ \cite{Dennen:2009vk}. 

Superamplitudes are scattering amplitudes with fermionic coherent states as external particle legs. They are effectively Grassmann generating functions in which coeffecients of the independent Grassmann terms are interpreted as scattering amplitudes between corresponding component particles in the multiplets. The component amplitudes are guaranteed to obey the supersymmetric Ward identities if the superamplitude is invariant under the action of the supercharges (e.g. as represented in (\ref{N=4Charges})). The Grassmann structure of the superamplitude must encode invariance under the action these supercharges. The lowest degree Grassmann invariant for superamplitudes with at least four legs is given by the product of each of the SUSY delta functions
$\delta^{(2)}\left(Q_{a+2}\right)=\frac{1}{2}\ds{Q_{a+2}Q_{a+2}}$ and $\delta^{(2)}\left(Q^{\dagger a}\right)=\frac{1}{2}\da{Q^{\dagger a}Q^{\dagger a}}$, where $Q_{a+2}$ and $Q^{\dagger a}$ are the multiplicatively represented supercharges. At four legs, this invariant is unique. The delta functions can be grouped into the Dirac form
\begin{align}
\delta^{(2)}\left(Q_{a+2}\right)\delta^{(2)}\left(Q^{\dagger a}\right)=-\epsilon\left(\mathcal{Q}^a,\mathcal{Q}^a,\mathcal{Q}^a,\mathcal{Q}^a\right),
\end{align}
where the Levi-Civita contraction of Dirac spinors is defined as 
\begin{align}
\epsilon(X,Y,Z,W)=\epsilon^{ABCD}X_AY_BZ_CW_D.
\end{align}
This latter form of the supercharges and the invariants parallel the expressions proposed in \cite{Dennen:2009vk} in $6d$ SYM.

For the special case of $3$-particle superampltiudes, two pairs of supercharges degenerate
\begin{align}\label{SuperDegen}
\da{v Q^{\dagger 1}}&=\ds{v Q_3}\nonumber\\ \da{u Q^{\dagger 2}}&=\ds{u Q_4}.
\end{align}
The SUSY delta functions in this special case are unique invariants that can be expressed in a form using Dirac spinors as
\begin{align}\label{3Delta}
\Delta_v&=\frac{1}{\ds{q_vv}\da{q_vv}}\epsilon\left(q_v,\mathcal{Q}^1,\mathcal{Q}^1,\mathcal{Q}^1\right)\nonumber\\
\Delta_u&=\frac{1}{\ds{q_uu}\da{q_uu}}\epsilon\left(q_u,\mathcal{Q}^2,\mathcal{Q}^2,\mathcal{Q}^2\right).
\end{align}
The prefactors are necessary for the expression to be independent of the reference spinors $\rs{q_v}$, $\ra{q_v}$, $\rs{q_u}$ and $\ra{q_u}$, which satisfy $\ds{q_vv},\da{q_vv},\ds{q_uu},\da{q_uu}\neq 0$. These can be assembled into Dirac spinors $q_v$ and $q_u$. The SUSY delta functions $\Delta_u$ and $\Delta_v$ each carry opposite tiny group charge. The three-particle superamplitude between $\frac{1}{2}$BPS vector multiplets is fully fixed by SUSY, little group and tiny group invariance:
\begin{align}\label{N=43vec}
\mathcal{A}(\mathcal{W}_A,\mathcal{W}_B,\mathcal{W}_C)=-if_{ABC}\Delta_u\Delta_v.
\end{align}
The couplings $f_{ABC}$ are fully antisymmetric in colour indices. Consistent superfactorisation of these superamplitudes, which is non-trivial because of the kinematic dependence in the denominators outside of the supercharges in (\ref{3Delta}), necessitates their interpretation as Yang-Mills Lie algebra structure constants \cite{Trott:2026ozo}. 

In $\mathcal{N}=4$ SYM, the colour-ordered partial superamplitude for $4$-leg $\frac{1}{2}$BPS vector multiplet scattering is 
\begin{align}\label{N=4SYM}
\mathcal{A}[\mathcal{W},\mathcal{W},\mathcal{W},\mathcal{W}]=\frac{-1}{(s-m_s^2)(u-m_u^2)}\prod_a\epsilon\left(\mathcal{Q}^a,\mathcal{Q}^a,\mathcal{Q}^a,\mathcal{Q}^a\right).
\end{align}
The internal masses are $m_s=|m_3e^{i\varphi_3}+m_4e^{i\varphi_4}|$ and $m_u=|m_1e^{i\varphi_1}+m_4e^{i\varphi_4}|$. This superamplitude is fully constrained by SUSY and may be produced directly by constructing it from superfactorisation into $3$-particle superamplitudes. The numerator is identical for all colour orderings. Of note here is that the superamplitude, and hence each of its components, has manifestly unitary high-energy dependence, being at most constant in the HEL. As will be described extensively below in Section \ref{VectorHiggs}, this is not at all automatic for the non-SUSY component amplitudes when constructed from the procedure described at the start of this Section and requires considerable computation to demonstrate.

With $\mathcal{N}=2$ SUSY, only one pair of supercharges in (\ref{N=4Charges}) exists, which I choose to be $Q_1$ and $Q_3$ (and conjugates). Representing the algebra in otherwise the same way as its embedding in the $\mathcal{N}=4$ algebra above, the $\frac{1}{2}$BPS massive ($1/2$-)hypermultiplet has structure 
\begin{align}
K=\phi + \eta_I \chi^I + \half \eta_I \eta^I \widetilde{\phi},
\end{align}
where $\chi$ is a fermion and $\phi$ and $\widetilde{\phi}$ are two scalars. This is paired with charge conjugate multiplet $\overline{K}$ with opposite central charge. Moller scattering of $\frac{1}{2}$BPS hypermultiplets in $\mathcal{N}=2$ super-QED is described by the superamplitude
\begin{align}\label{SQEDFact}
\mathcal{A}(K,\overline{K},K',\overline{K'})=\epsilon\left(\mathcal{Q},\mathcal{Q},\mathcal{Q},\mathcal{Q}\right)e^{-i\left(\varphi_1+\varphi_3\right)}\frac{-1}{s},
\end{align}
where $K$ and $K'$ are two possibly distinct species of hypermultiplet and $\varphi_1$ and $\varphi_3$ are their respective central charge phases. The appearance of these phases and the construction of this superamplitude is explained in \cite{Trott:2026ozo}. The expression (\ref{SQEDFact}) also holds for exchange of a massive $\frac{1}{2}$BPS photon, in which case the hypermultiplets are no longer conjugate pairs and the pole is modified to $s\rightarrow s-m_s^2$, where $m_s$ is the mass of the photon.

\subsection{Electrodynamics and unification in $\mathcal{N}=2$ SQED}\label{sec:QED}

For scattering in scalar electrodynamics, with two distinct scalar fields $\varphi$ and $\varphi'$ with masses $m$ and $M$ respectively, the Feynman rules give 
\begin{align}\label{sQED}
A(\overline{\varphi},\varphi\rightarrow\varphi',\overline{\varphi'})=-\left(t-m^2-M^2\right)\frac{-1}{s}
\end{align}
(I won't usually bother to write the overall factor of the electric charges in this context). The amplitude is ambiguous up to a contact term that has the same (constant) high energy scaling - this represents the free quartic coupling that is consistent with the theory being ``tree-unitary''. 

The forward limit corresponds here to $p_4\rightarrow p_1$ and $p_3\rightarrow p_2$ (specifically for $\varphi'=\varphi$ and $M=m$, although I will not assume this in the algebra that follows in order to maintain generality, in which case all remarks are to be interpreted only as the appropriate analytic continuations of Mandelstam invariants). The residue in (\ref{sQED}) appears to be negative, because $u=-(p_2-p_3)^2$ and $t=-(p_1-p_3)^2$, so that $u\rightarrow 0$ and $t\rightarrow 2(m^2+M^2)$. However, the residue of the $s\rightarrow 0$ pole should be a sum of squares. The reconciliation is that the pole exists deep in an unphysical kinematic region accessible only for complexified momentum. One simple complex momentum configuration that both activates the pole and leaves the intermediate photon's momentum $P=p_1+p_2$ null and real is given by $P=E(1,0,0,1)$, $p_1=P+(0,0,im,0)$ and $p_2=(0,0,-im,0)$ (and similarly for $p_3$ and $p_4$ on the opposite side of the factorisation channel). Here $E$ is some real, positive energy. Taking the $s\rightarrow 0$ limit in this way, the residue (with particles $1$ and $2$ incoming) is 
\begin{align}\label{sQEDRes}
\sum_{p=\pm} A(\overline{\varphi},\varphi\rightarrow\gamma^p)\left(A(\varphi',\overline{\varphi'}\rightarrow \gamma^{p})\right)^*=\pm mM\left(\frac{x_{34}}{x_{12}}+\frac{x_{12}}{x_{34}}\right)\sim (m^2+M^2-t)\sim \frac{(u-t)}{2}.
\end{align}
As mentioned above, the momenta appearing in the factor $A(\varphi',\overline{\varphi'}\rightarrow \gamma^{p})$ are conjugated, so the residue is not a sum of squares in this limit, so need not be non-negative. The residue must instead be non-negative for the alternative complexified forward limit $p_4\rightarrow p_1^*$ and $p_3\rightarrow p_2^*$. This complexified forward limit therefore instead corresponds to 
\begin{align}\label{ForwardLim}
u\rightarrow(m+M)^2\nonumber\\
t\rightarrow (m-M)^2,
\end{align}
for which the residue is clearly positive, as required for unitarity. 
Note that with all particles outgoing, momenta $p_1$ and $p_2$ are to be crossed in the expressions above. This introduces a (-) sign in the middle expression of (\ref{sQEDRes}) (represented by the lower choice in the $\pm$ symbol), but the others remain unchanged. Expressions in what follows will generally obey the convention in which all particles are outgoing and this will be left implicit unless stated otherwise. The final equivalence in (\ref{sQEDRes}) holds on the pole $s\sim 0$ and represents two possible forms of the residue that would differ in the amplitude by a modification to the contact term. 

It will be convenient to define abbreviations for some common spinor structures for the $4$-particle amplitudes:
\begin{align}\label{SpinStrucDef}
\mathbf{\Pi}_s&=\ds{\bf{12}}\da{\bf{34}}+\ds{\bf{34}}\da{\bf{12}}\nonumber\\
\mathbf{\Pi}_t&=\ds{\bf{13}}\da{\bf{24}}+\ds{\bf{24}}\da{\bf{13}}\nonumber\\
\mathbf{\Pi}_u&=\ds{\bf{14}}\da{\bf{23}}+\ds{\bf{23}}\da{\bf{14}}.
\end{align}
Employing these definitions, fermion scattering in quantum electrodynamics is \cite{Christensen:2022nja}
\begin{align}\label{FermionQED}
A(\psi,\overline{\psi},\psi',\overline{\psi'})=-\frac{\mathbf{\Pi}_t+\mathbf{\Pi}_u}{s}.
\end{align}
There are no contact terms with the same (constant) order of high energy scaling, so this expression is unambiguous up to the potential inclusion of higher order terms in an EFT expansion. On the pole, the residue can be identified as
\begin{align}\label{QEDRes}
-\left(\frac{x_{34}}{x_{12}}\da{\bf{12}}\ds{\bf{34}}+\frac{x_{12}}{x_{34}}\da{\bf{34}}\ds{\bf{12}}\right)\sim\left(\mathbf{\Pi}_t+\mathbf{\Pi}_u\right).
\end{align}
The spin of the exchanged photon is manifest in the fact that the spinor bilinears in the residue involve two cross-channel contractions (one for each half-integer unit of spin). In contrast, a purely $s$-channel scalar exchange amplitude is
\begin{align}\label{4Yukawa}
A(\psi,\overline{\psi},\psi',\overline{\psi'})=-\frac{\mathbf{\Pi}_s}{s},
\end{align}
provided by specific Yukawa couplings
\begin{align}\label{3Yukawa}
A(\psi,\overline{\psi},\phi)&=(-1)^{n_1}\ds{\bf{12}}\nonumber\\
A(\psi,\overline{\psi},\overline{\phi})&=(-1)^{n_2}\da{\bf{12}},
\end{align}
where $\phi$ is a complex scalar particle (and I am also indicating the crossing sign). The absence of angular momentum in this exchange means that the amplitude's factorised form on the pole can be trivially lifted off-shell. Under crossing fermions $1$ and $2$ to incoming, the residue of (\ref{4Yukawa}) flips sign, which easily follows from the crossing rules in (\ref{3Yukawa}). In the complexified forward limit described above, the residue converges to $-s A\left(\overline{\psi},\psi\rightarrow \psi',\overline{\psi'}\right)=-\left(|\ds{\bf{12}}|^2+|\da{\bf{12}}|^2\right)$ (for any implicitly specific polarisation configuration), which is negative. This is expected and accounted for here by the convention that I use for fermion ordering of the external states, where the incoming and outgoing states are defined respectively as 
$|\overline{\psi},\psi\rangle=d^\dagger b^\dagger|0\rangle$ and $\langle\psi',\overline{\psi'}|=\langle 0|d'b'$ (calling $b^{(\dagger)}$ and $d^{(\dagger)}$ respectively the fermion and antifermion annihilation/creation operators), which, for a single fermion species $\psi'=\psi$, differ from being mutual duals by a fermion ordering exchange. 

Adding the photon to the scalar exchange amplitudes and assuming that they have equal sized couplings gives the $\mathcal{N}=2$ super-QED Moller amplitude:
\begin{align}\label{N=2Moller}
A(\psi,\overline{\psi},\psi',\overline{\psi'})=\left(\mathbf{\Pi}_s+\mathbf{\Pi}_t+\mathbf{\Pi}_u\right)\frac{-1}{s}=\epsilon(1,2,3,4)\frac{-1}{s},
\end{align}
where 
\begin{align}
\epsilon(1,2,3,4)=\epsilon^{ABCD}v_{1A}^Iv_{2B}^Jv_{3B}^Kv_{4B}^L=\mathbf{\Pi}_s+\mathbf{\Pi}_t+\mathbf{\Pi}_u
\end{align}
is a contraction of the Dirac spinors associated to each of the external particles (where the electrons have mass phase $0$ and the positrons have mass phase $\pi$). In this form, the relationship with the superamplitude of hypermultiplets
\begin{align}\label{HyperQED4}
\mathcal{A}(K,\overline{K},K',\overline{K'})=\epsilon\left(\mathcal{Q},\mathcal{Q},\mathcal{Q},\mathcal{Q}\right)\frac{-1}{s},
\end{align}
entirely fixed by supersymmetry, is clear. 

The set of low spin QED amplitudes is completed by mixed fermion-scalar scattering, which is given by
\begin{align}\label{sfQED}
A(\psi,\overline{\psi},\varphi,\overline{\varphi})=\frac{1}{2}\left(\la{\bf{1}}p_4-p_3\rs{\bf{2}}+\la{\bf{2}}p_4-p_3\rs{\bf{1}}\right)\frac{-1}{s},
\end{align}
with $s$-channel residue
\begin{align}\label{sfSQED}
-\left(\frac{x_{34}}{x_{12}}\da{\bf{12}}+\frac{x_{12}}{x_{34}}\ds{\bf{12}}\right)\sim\frac{1}{M}\left(\la{\bf{1}}p_4\rs{\bf{2}}+\la{\bf{2}}p_4\rs{\bf{1}}\right)
\end{align}
where here $M$ is the mass of the scalars. Note that
\begin{align}
\la{\bf{1}}p_4\rs{\bf{2}}+\la{\bf{2}}p_4\rs{\bf{1}}=-\left(\la{\bf{1}}p_3\rs{\bf{2}}+\la{\bf{2}}p_3\rs{\bf{1}}\right)
\end{align}
manifests the inherent exchange (anti)symmetry of the kinematics of the interactions.

When the Yukawa interactions are restored for $\mathcal{N}=2$ super-QED, the amplitude, as contained in (\ref{HyperQED4}), is 
\begin{align}\label{sfHyperQED}
A(\psi,\overline{\psi},\varphi,\overline{\varphi})=-\left(M\left(\ds{\bf{12}}+\da{\bf{12}}\right)+\frac{1}{2}\left(\la{\bf{1}}p_4-p_3\rs{\bf{2}}+\la{\bf{2}}p_4-p_3\rs{\bf{1}}\right)\right)\frac{1}{s}.
\end{align}
If the two hypermultiplets are identical, then the amplitude (\ref{HyperQED4}) should be extended to include a $u$-channel pole with the same residue. For the mixed scalar-fermion component amplitude, this residue is instead fully generated by Yukawa emission of scalars directly off the fermion line. SUSY then equates this process to the $s$-channel photon and scalar exchange.

For completeness, (\ref{HyperQED4}) also contains pure contact amplitudes between non-conjugate scalars belonging to conjugate hypermultiplets. These are not produced by the exchange of any particle in the exchanged supermultiplet, but are necessarily present for consistency with SUSY.

\subsection{Fermion scattering by massive vector exchange}\label{sec:4fermion}

I will conclude this Section by calculating the four fermion amplitude mediated by the exchange of a massive vector boson with the general (non-dipole) couplings given in (\ref{FermionMat3leg}). This generalises the QED calculation to allow for axial-vector couplings as well. These amplitudes have been presented before \cite{Christensen:2024bdt} in the specific context of the SM. In Appendix \ref{app:LowSpinGen}, I give the additional contributions generated by the dipole coupling terms in (\ref{FermionMat3leg}), which I omit for brevity in this Section. 
\begin{feynfig}
\begin{figure}[h]
\begin{fmffile}{FermionMassiveQED}
\begin{center}
\begin{tabular}{ c c c }
& & \\
\begin{fmfgraph*}(120,80)
   \fmfleft{i1,i2}
   \fmfright{o1,o2}
   \fmf{plain}{i2,v1}
   \fmf{plain}{v2,o2}
   \fmf{plain}{i1,v1}
   \fmf{plain}{v2,o1}
   \fmf{boson}{v1,v2}
   \fmfv{decor.shape=circle,decor.filled=gray50,decor.size=0.15w}{v1,v2}
   \fmflabel{$i$}{i1}
   \fmflabel{$j$}{i2}
   \fmflabel{$k$}{o2}
   \fmflabel{$l$}{o1}
 \end{fmfgraph*} &\,&
 \begin{fmfgraph*}(120,80)
   \fmfleft{i1,i2}
   \fmfright{o1,o2}
   \fmf{plain}{i2,v1}
   \fmf{plain}{v2,o2}
   \fmf{plain}{i1,v1}
   \fmf{plain}{v2,o1}
   \fmf{dashes}{v1,v2}
   \fmfv{decor.shape=circle,decor.filled=gray50,decor.size=0.15w}{v1,v2}
   \fmflabel{$i$}{i1}
   \fmflabel{$j$}{i2}
   \fmflabel{$k$}{o2}
   \fmflabel{$l$}{o1}
 \end{fmfgraph*}
\end{tabular}
\end{center}
\end{fmffile}
\caption{On-shell contributions to massive fermion scattering.}\label{4fermionDi}
\end{figure}
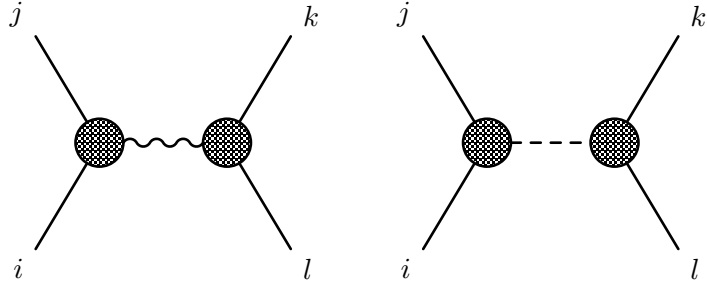
\end{feynfig}

The general four fermion amplitude mediated by $s$-channel massive vector exchange is
\begin{align}\label{GenMoller}
A\left(\psi_i,\psi_j,\psi_k,\psi_l\right)&=\frac{-1}{s-m_{s_M}^2}\Big((t_M)_{ij}(t_M)_{kl}\da{\bf{13}}\ds{\bf{24}}+(t_M)_{ji}(t_M)_{lk}\ds{\bf{13}}\da{\bf{24}}\nonumber\\
&\qquad\qquad\qquad-(t_M)_{ij}(t_M)_{lk}\da{\bf{14}}\ds{\bf{23}}-(t_M)_{ji}(t_M)_{kl}\ds{\bf{14}}\da{\bf{23}}\Big)\nonumber\\
&+\frac{1}{s-m_{s_M}^2}\frac{1}{2m_{s_M}^2}\Big((t_M)_{ij}(t_M)_{kl}\left(m_1\ds{\bf{12}}-m_2\da{\bf{12}}\right)\left(m_3\ds{\bf{34}}-m_4\da{\bf{34}}\right)\nonumber\\
&\qquad\qquad\qquad\,\,+(t_M)_{ji}(t_M)_{lk}\left(m_2\ds{\bf{12}}-m_1\da{\bf{12}}\right)\left(m_4\ds{\bf{34}}-m_3\da{\bf{34}}\right)\nonumber\\
&\qquad\qquad\qquad\,\,+(t_M)_{ij}(t_M)_{lk}\left(m_1\ds{\bf{12}}-m_2\da{\bf{12}}\right)\left(m_4\ds{\bf{34}}-m_3\da{\bf{34}}\right)\nonumber\\
&\qquad\qquad\qquad\,\,+(t_M)_{ji}(t_M)_{kl}\left(m_2\ds{\bf{12}}-m_1\da{\bf{12}}\right)\left(m_3\ds{\bf{34}}-m_4\da{\bf{34}}\right)\Big).
\end{align}
See Fig \ref{4fermionDi} for on-shell diagram. Here, $m_i$ is the mass of external particle $i$ and $m_{s_M}$ denotes the mass of an exchanged particle labeled by species index $M$ (of which there can be multiple). Repeated indices are implicitly summed over, even if they appear more than once. In the special case of a purely vectorial coupling (i.e. antisymmetric couplings), the amplitude simplifies to 
\begin{align}\label{MollerVec}
A\left(\psi_i,\psi_j,\psi_k,\psi_l\right)&=(t_M)_{ij}(t_M)_{kl}\nonumber\\
&\qquad\times\left(-\frac{\mathbf{\Pi}_t+\mathbf{\Pi}_u}{s-m_{s_M}^2}+\frac{(m_1-m_2)(m_3-m_4)(\ds{\bf{12}}+\da{\bf{12}})(\ds{\bf{34}}+\da{\bf{34}})}{2m_{s_M}^2(s-m_{s_M}^2)}\right).
\end{align}
In the limit that $m_1=m_2$ and $m_3=m_4$, this agrees with the standard QED amplitude (\ref{FermionQED}) modified merely by shifting the pole(s) to $m_{s_M}^2$. In comparison, with a purely axial-vector coupling (symmetric coupling constants),
\begin{align}\label{MollerAxVec}
A\left(\psi_i,\psi_j,\psi_k,\psi_l\right)&=(t_M)_{ij}(t_M)_{kl}\nonumber\\
&\qquad\times\left(-\frac{\mathbf{\Pi}_t-\mathbf{\Pi}_u}{s-m_{s_M}^2}+\frac{(m_1+m_2)(m_3+m_4)(\ds{\bf{12}}-\da{\bf{12}})(\ds{\bf{34}}-\da{\bf{34}})}{2m_{s_M}^2(s-m_{s_M}^2)}\right).
\end{align}

While these processes are clearly already unitary in the HEL, it is amusing to add a contribution from a Higgs boson exchange in order to reconcile them with the simple result contained within the general $\mathcal{N}=2$ superamplitude (\ref{HyperQED4}). A term proportional to $\mathbf{\Pi}_s$ can be obtained by adding a pseudoscalar exchange amplitude with appropriate masses and couplings to the last term in (\ref{MollerVec}):
\begin{align}\label{PseudoMoller}
A_P\left(\psi_i,\psi_j,\psi_k,\psi_l\right)=-(t_M)_{ij}(t_M)_{kl}\frac{(m_1-m_2)(m_3-m_4)(\ds{\bf{12}}-\da{\bf{12}})(\ds{\bf{34}}-\da{\bf{34}})}{2m_{s_M}^2(s-m_{s_M}^2)},
\end{align}
or similarly a scalar exchange to (\ref{MollerAxVec}). The required mass and couplings of the pseudoscalar(s) should be implicitly obvious in (\ref{PseudoMoller}) and will not be restated. The BPS condition on the masses then forces the overall coupling constant of the $\mathbf{\Pi}_s$ term into agreement with the $\mathbf{\Pi}_t$ and $\mathbf{\Pi}_u$ terms. A general combination of scalars of either parity can be added to the general case (\ref{GenMoller}) to reach a similar result. 

In the HEL, the matching of (\ref{GenMoller}) onto massless amplitudes is clear. The first set of terms each describe gluon exchange between different helicity configurations of fermions (in which the fermion pair on each side of the channel have opposite helicities). The remaining terms represent scalar exchanges that emerge as the longitudinal modes of the vectors. These have Yukawa couplings consistent with (\ref{LongModeLim}).

\section{General Amplitude for Three Vector Bosons}\label{Sec:MassiveVectors}

Before studying the three vector amplitude, it is first relevant and useful to establish the general amplitudes between two vector bosons and a scalar. These couplings have the general form and crossing properties:
\begin{align*}
A\left(g^+_A,g^+_B,\varphi_i\right)&=\frac{c_{AB}^i}{\Lambda}\ds{12}^2 & A\left(g^-_A,g^-_B,\varphi_i\right)&=\frac{(c^i_{AB})^*}{\Lambda}\da{12}^2\\
A\left(g^+_A,W_B,\varphi_i\right)&=\frac{c_{AB}^i}{\Lambda}\ds{1\bf{2}}^2 & A\left(g^-_A,W_B,\varphi_i\right)&=\frac{(c_{AB}^i)^*}{\Lambda}\da{1\bf{2}}^2\nonumber
\end{align*}
\begin{align}\label{2vec1scalar}
A\left(W_A,W_B,\varphi_i\right)=\frac{c_{AB}^i}{\Lambda}\ds{\bf{12}}^2+\frac{(c_{AB}^i)^*}{\Lambda}\da{\bf{12}}^2+(-1)^{n_1+n_2}\frac{1}{\sqrt{2}}\lambda^i_{AB}\ds{\bf{12}}\da{\bf{12}}.
\end{align}
The terms that are holomorphic in a particular spinor chirality correspond off-shell to dimension-$5$ effective operators of the schematic form $\varphi F^2$ (which I will mean to implicitly include the $\varphi F\widetilde{F}$ counterpart as well). Bosonic exchange symmetry implies that $c_{AB}^i=c_{BA}^i$. The remaining coupling, special to the case of two massive vectors, can describe the interaction with Higgs bosons \cite{Arkani-Hamed:2017jhn} provided that it is chosen to obey a particular constraint, as will be derived further below in Section \ref{sec:FullUni}. This coupling is generally flavour symmetric $\lambda^i_{AB}=\lambda^i_{BA}$ and real $(\lambda^i_{AB})^*=\lambda^i_{AB}$. I will generally refer to this as a Higgs coupling, and the associated scalar a Higgs boson, even if the further constraints demanded by the Higgs mechanism are not met.

The general $3$-particle amplitude for massive vector bosons decomposes into (mostly) independent Lorentz structures as
\begin{align}\label{3VecPre}
&A(W_A,W_B,W_C)=\frac{ih_{ABC}}{\Lambda^2}\ds{\bf{12}}\ds{\bf{23}}\ds{\bf{31}}+\frac{i\tilde{h}_{ABC}}{\Lambda^2}\da{\bf{12}}\da{\bf{23}}\da{\bf{31}}\nonumber\\
&\qquad\qquad\qquad\qquad+(-1)^{n_1+n_2}\frac{i{f}_{AB}^{\,\,\,\,\,\,\,\,\,C}}{2m_1m_2}\ds{\bf{23}}\ds{\bf{13}}\da{\bf{12}}+(-1)^{n_1+n_2}\frac{i{\tilde{f}}_{AB}^{\,\,\,\,\,\,\,\,\,C}}{2m_1m_2}\da{\bf{23}}\da{\bf{13}}\ds{\bf{12}}\nonumber\\
&\qquad\qquad\qquad\qquad+(-1)^{n_1+n_3}\frac{i{f}_{CA}^{\,\,\,\,\,\,\,\,\,B}}{2m_3m_1}\ds{\bf{12}}\ds{\bf{32}}\da{\bf{31}}+(-1)^{n_1+n_3}\frac{i{\tilde{f}}_{CA}^{\,\,\,\,\,\,\,\,\,B}}{2m_3m_1}\da{\bf{12}}\da{\bf{32}}\ds{\bf{31}}\nonumber\\
&\qquad\qquad\qquad\qquad+(-1)^{n_2+n_3}\frac{i{f}_{BC}^{\,\,\,\,\,\,\,\,\,A}}{2m_2m_3}\ds{\bf{31}}\ds{\bf{21}}\da{\bf{23}}+(-1)^{n_2+n_3}\frac{i{\tilde{f}}_{BC}^{\,\,\,\,\,\,\,\,\,A}}{2m_2m_3}\da{\bf{31}}\da{\bf{21}}\ds{\bf{23}}.
\end{align}
There is, however, one remaining redundancy among the terms with non-holomorphic spinor chiralities due to an identity derived by \cite{Durieux:2019eor}:
\begin{align}\label{3PMassRed}
&(-1)^{n_2+n_3}m_1\ds{\bf{12}}\da{\bf{23}}\ds{\bf{31}}+(-1)^{n_1+n_3}m_2\ds{\bf{23}}\da{\bf{31}}\ds{\bf{12}}+(-1)^{n_1+n_2}m_3\ds{\bf{31}}\da{\bf{12}}\ds{\bf{23}}\nonumber\\
&-(-1)^{n_2+n_3}m_1\da{\bf{12}}\ds{\bf{23}}\da{\bf{31}}-(-1)^{n_1+n_3}m_2\da{\bf{23}}\ds{\bf{31}}\da{\bf{12}}-(-1)^{n_1+n_2}m_3\da{\bf{31}}\ds{\bf{12}}\da{\bf{23}}\nonumber\\
&\qquad\qquad\qquad\qquad\qquad\qquad\qquad\qquad\qquad\qquad\qquad\qquad\qquad\qquad\qquad\qquad\qquad=0.
\end{align}
This redundancy can be accounted for in a way preserving the identical particle exchange symmetry by introducing a shift invariance into the definitions of the $f$ and $\tilde{f}$ couplings:
\begin{align}
&{f}_{AB}^{\,\,\,\,\,\,\,\,\,C}\mapsto {f}_{AB}^{\,\,\,\,\,\,\,\,\,C}+a_{ABC}\qquad {f}_{BC}^{\,\,\,\,\,\,\,\,\,A}\mapsto {f}_{BC}^{\,\,\,\,\,\,\,\,\,A}+a_{ABC}\qquad{f}_{CA}^{\,\,\,\,\,\,\,\,\,B}\mapsto {f}_{CA}^{\,\,\,\,\,\,\,\,\,B}+a_{ABC}\nonumber\\
&\tilde{f}_{AB}^{\,\,\,\,\,\,\,\,\,C}\mapsto \tilde{f}_{AB}^{\,\,\,\,\,\,\,\,\,C}-a_{ABC}\qquad \tilde{f}_{BC}^{\,\,\,\,\,\,\,\,\,A}\mapsto \tilde{f}_{BC}^{\,\,\,\,\,\,\,\,\,A}-a_{ABC}\qquad\tilde{f}_{CA}^{\,\,\,\,\,\,\,\,\,B}\mapsto \tilde{f}_{CA}^{\,\,\,\,\,\,\,\,\,B}-a_{ABC}
\end{align}
for some collection of constants $a_{ABC}\in\mathbb{C}$ that are fully antisymmetric under index exchanges. With self-conjugate states, unitarity relates the amplitude to its complex conjugate. Unitarity would therefore be manifest if the couplings were related so that $\tilde{h}_{ABC}=(h_{ABC})^*$ and $\tilde{{f}}_{AB}^{\,\,\,\,\,\,\,\,\,C}=({f}_{AB}^{\,\,\,\,\,\,\,\,\,C})^*$. However, this latter equality may be spoiled by the shift redundancy in the definitions of these couplings. In order to preserve the appearance of manifest unitarity of the amplitude, it is therefore convenient to partially fix the redundancy so that the shift $a$ is restricted to being purely imaginary. 

In summary, the self-conjugate three vector amplitude can be written as:
\begin{align}\label{3legVec}
&A(W_A,W_B,W_C)=\frac{ih_{ABC}}{\Lambda^2}\ds{\bf{12}}\ds{\bf{23}}\ds{\bf{31}}+\frac{i(h_{ABC})^*}{\Lambda^2}\da{\bf{12}}\da{\bf{23}}\da{\bf{31}}\nonumber\\
&\qquad\qquad\qquad\quad+(-1)^{n_1+n_2}\frac{i{f}_{AB}^{\,\,\,\,\,\,\,\,\,C}}{2m_1m_2}\ds{\bf{23}}\ds{\bf{13}}\da{\bf{12}}+(-1)^{n_1+n_2}\frac{i({f}_{AB}^{\,\,\,\,\,\,\,\,\,C})^*}{2m_1m_2}\da{\bf{23}}\da{\bf{13}}\ds{\bf{12}}\nonumber\\
&\qquad\qquad\qquad\quad+(-1)^{n_1+n_3}\frac{i{f}_{CA}^{\,\,\,\,\,\,\,\,\,B}}{2m_3m_1}\ds{\bf{12}}\ds{\bf{32}}\da{\bf{31}}+(-1)^{n_1+n_3}\frac{i({f}_{CA}^{\,\,\,\,\,\,\,\,\,B})^*}{2m_3m_1}\da{\bf{12}}\da{\bf{32}}\ds{\bf{31}}\nonumber\\
&\qquad\qquad\qquad\quad+(-1)^{n_2+n_3}\frac{i{f}_{BC}^{\,\,\,\,\,\,\,\,\,A}}{2m_2m_3}\ds{\bf{31}}\ds{\bf{21}}\da{\bf{23}}+(-1)^{n_2+n_3}\frac{i({f}_{BC}^{\,\,\,\,\,\,\,\,\,A})^*}{2m_2m_3}\da{\bf{31}}\da{\bf{21}}\ds{\bf{23}}.
\end{align}
In this form, bosonic exchange symmetry clearly implies that the coupling constants are antisymmetric among exchanges of lowered indices, which, at this point, is the only significance of the index heights:
\begin{align}\label{YMantisym}
{f}_{AB}^{\,\,\,\,\,\,\,\,\,C}=-{f}_{BA}^{\,\,\,\,\,\,\,\,\,C}.
\end{align}
Furthermore, the couplings have the shift invariance
\begin{align}\label{GCSShiftInvar}
\Im f_{AB}^{\,\,\,\,\,\,\,\,\,C}\mapsto \Im f_{AB}^{\,\,\,\,\,\,\,\,\,C}+a_{ABC}\qquad\Im f_{BC}^{\,\,\,\,\,\,\,\,\,A}\mapsto \Im f_{BC}^{\,\,\,\,\,\,\,\,\,A}+a_{ABC}\qquad\Im f_{CA}^{\,\,\,\,\,\,\,\,\,B}\mapsto \Im f_{CA}^{\,\,\,\,\,\,\,\,\,B}+a_{ABC}
\end{align}
for any fully antisymmetric collection of constants $a_{ABC}\in\mathbb{R}$.

If the $f$-type couplings in (\ref{3legVec}) are all real and equal, then they can describe the coupling of three massive vector bosons in ``spontaneously broken'' YM \cite{Arkani-Hamed:2017jhn}:
\begin{align}\label{NormalBrokenYM}
&A\left(W_A,W_B,W_C\right)\nonumber\\
&=-\frac{if_{ABC}}{2m_1m_2m_3}\big((-1)^{n_2+n_3}m_1\ds{\bf{12}}\da{\bf{23}}\ds{\bf{31}}+(-1)^{n_1+n_3}m_2\ds{\bf{23}}\da{\bf{31}}\ds{\bf{12}}\nonumber\\
&\qquad\qquad\qquad\qquad+(-1)^{n_1+n_2}m_3\ds{\bf{31}}\da{\bf{12}}\ds{\bf{23}}+(-1)^{n_2+n_3}m_1\da{\bf{12}}\ds{\bf{23}}\da{\bf{31}}\nonumber\\
&\qquad\qquad\qquad\qquad+(-1)^{n_1+n_3}m_2\da{\bf{23}}\ds{\bf{31}}\da{\bf{12}}+(-1)^{n_1+n_2}m_3\da{\bf{31}}\ds{\bf{12}}\da{\bf{23}}\big),
\end{align}
where $f_{ABC}={f}_{AB}^{\,\,\,\,\,\,\,\,\,C}={f}_{BC}^{\,\,\,\,\,\,\,\,\,A}={f}_{CA}^{\,\,\,\,\,\,\,\,\,B}$ is real and fully antisymmetric. This expression is the kinematically unique structure contained in the $\mathcal{N}=4$ SYM superamplitude (\ref{N=43vec}) (in \cite{Trott:2026ozo}, I explain more generally how the vector amplitude (\ref{3legVec}) embeds and is constrained in supersymmetric theories). This special case (\ref{NormalBrokenYM}) is distinguished from all of its general deformations presented in (\ref{3legVec}) by the property that it has the softest possible HEL, scaling at worst as $\sim E$ (which would qualify it as ``tree unitary'' if the standards of \cite{Cornwall:1974km} were extrapolated to $3$-particle amplitudes with complex momenta). The high-energy dependence of the general amplitude will be elaborated upon further below - I will momentarily concentrate here on reviewing the limits specifically for (\ref{NormalBrokenYM}).

In the HEL, the amplitude (\ref{NormalBrokenYM}) converges as expected to the massless gluon amplitudes (\ref{3gluons}) when the vectors are all transversely polarised. The other nontrivial helicity configuration in this limit occurs when two of the vectors are longitudinal. In this case, it is easy to verify that the amplitude converges kinematically to that of massless scalars emitting a gluon in (\ref{ScalarMat3leg}). Matching the coupling constants for the limit $A\left(W_A,W_B,W_C\right)\rightarrow A\left(\phi_A,\phi_B,g^+_C\right)$ gives
\begin{align}\label{BrokenGen}
(t_C)_{AB}=\frac{-if_{ABC}}{2m_1m_2}\left(m_1^2+m_2^2-m_3^2\right).
\end{align}
This is the expected form of the Lie algebra generators describing the coupling of the emergent longitudinal scalar modes to gluons in the basis of mass eigenstates of the initial massive theory, as will be shown below. In this context, the labels $A$ and $B$ are to be reinterpreted as indices describing some components of the representation space of scalar states while otherwise being directly identified with the adjoint indices of the broken generators associated with the vectors. 

Note that the explicit factors of particle masses in (\ref{NormalBrokenYM}) have been chosen to ensure HEL convergence to the massless gluon amplitudes (\ref{3gluons}). Other choices lead to helicity configurations with either $\sim E/m$ divergences (which would obstruct perturbative unitarity in the UV) or trivial three gluon amplitudes. This latter case amounts to ``ungauging'' the theory in the massless limit by having a mass-dependent gauge coupling. The mass dependence of the general expression (\ref{3legVec}) has been chosen to agree with (\ref{NormalBrokenYM}) if full permutation antisymmetry is restored to the couplings. However, if the couplings to the different Lorentz structures in (\ref{3legVec}) are not equal, then there are helicity configurations involving longitudinally polarised modes that produce terms that diverge as $\sim E/m$ and unitarily obstruct a consistent massless gluon amplitude from emerging. These divergent amplitudes have a UV form resembling those in (\ref{2vec1scalar}) describing effective dimension-$5$ interactions. This will be elaborated upon further below, but here it suffices to remark that these obstructions ensure that the only consistent massless gluon amplitudes that can emerge from (\ref{3legVec}) must have fully antisymmetric couplings, in agreement with the conclusions drawn in Section \ref{Sec:Massless4vec} about the inconsistency of other possibilities. Interestingly, in spite of this, the helicity configurations involving two longitudinal and one transverse polarisation have sensible limits for the general amplitude (\ref{3legVec}) regardless of the problems with the transverse configurations, assuming the explicit mass dependence as stated. In this general case, the gluon-scalar coupling that the amplitude matches onto is given by 
\begin{align}\label{BrokenGenGeneral}
(t_C)_{AB}=\frac{-i}{2m_1m_2}\left(m_1^2\Re f_{BC}^{\,\,\,\,\,\,\,\,\,A}+m_2^2\Re f_{CA}^{\,\,\,\,\,\,\,\,\,B}-m_3^2\Re f_{AB}^{\,\,\,\,\,\,\,\,\,C}\right).
\end{align}

It will also be important to establish the HEL for the Higgs-like amplitude in (\ref{2vec1scalar}). The coupling $\lambda$ implicitly contains some inverse dependence on some mass scale. The leading high energy divergences of the amplitude occur when one of the vectors is longitudinal and the other is transverse. Matching the coupling constants from the limit $A\left(W_A,W_B,\varphi_i\right)\rightarrow A\left(g^+_A,\phi_B,\varphi_i\right)$ gives
\begin{align}\label{HiggsGen}
(t_A)_{Bi}=\frac{im_A}{2}\lambda^i_{AB}.
\end{align}
In order to be consistent with the limit in which the helicity assignments are switched (like $A\left(W_A,W_B,\varphi_i\right)\rightarrow A\left(\phi_A,g^+_B,\varphi_i\right)$), 
\begin{align}
\frac{1}{m_A}(t_A)_{Bi}=\frac{1}{m_B}(t_B)_{Ai}.
\end{align}
Again, the expected relation to Lie algebra representations will be derived in the Sections below. 

I now return to the general amplitude between three massive vector bosons. In the limit that only one of the vectors is taken massless (choose here particle $3$), (\ref{3legVec}) becomes 
\begin{align}
A\left(W_A,W_B,g_C^+\right)&=i(-1)^{n_1+n_3}\,\Re f_{BC}^{\,\,\,\,\,\,\,\,\,A}\frac{1}{mx}\da{\bf{12}}^2\nonumber\\
&\qquad\qquad\qquad-i(-1)^{n_1+n_2}\left(f_{BC}^{\,\,\,\,\,\,\,\,\,A}-f_{AB}^{\,\,\,\,\,\,\,\,\,C}\right)\frac{1}{2m^2}\ds{3\bf{2}}\ds{3\bf{1}}\da{\bf{12}}\label{2Mass1MasslessVec}\\
A\left(W_A,W_B,g_C^-\right)&=i(-1)^{n_1+n_3}\,\Re f_{BC}^{\,\,\,\,\,\,\,\,\,A}\frac{x}{m}\ds{\bf{12}}^2\nonumber\\
&\qquad\qquad\qquad-i(-1)^{n_1+n_2}\left((f_{BC}^{\,\,\,\,\,\,\,\,\,A})^*-(f_{AB}^{\,\,\,\,\,\,\,\,\,C})^*\right)\frac{1}{2m^2}\da{3\bf{2}}\da{3\bf{1}}\ds{\bf{12}}\\
A\left(W_A,W_B,\phi_C\right)&=-\sqrt{2}(-1)^{n_1+n_2}\,\Re f_{BC}^{\,\,\,\,\,\,\,\,\,A}\frac{\delta m}{m_3}\frac{1}{m}\ds{\bf{12}}\da{\bf{12}}\label{2Mass1MasslessLong}
\end{align}
for each possible helicity configuration (I neglect the ``holomorphic'' terms because they are trivially affected by the limit and do not contain anything substantially new beyond the fully massive case). For this limit to work, the masses of the remaining massive vectors must converge to the same value at least as fast as $m_3\rightarrow 0$, so that $\delta m=(m_2-m_1)=\mathcal{O}(m_3)$ (and $m$ is the common mass that $m_1$ and $m_2$ converge to). The ratio $\delta m/m_3$ is then some dimensionless coupling constant (possibly zero). Additionally, the couplings must also satisfy
\begin{align}\label{PartAntiSym}
f_{CB}^{\,\,\,\,\,\,\,\,\,A}=-f_{CA}^{\,\,\,\,\,\,\,\,\,B}, 
\end{align}
which has been assumed in the expressions above. This is required for identical boson exchange symmetry, but does not completely enforce the full antisymmetry of the couplings that would be required for standard YM (\ref{NormalBrokenYM}).

By ``limit to work'', I mean that the transversely polarised amplitudes be Lorentz invariant (void of spurious spinors or ``gauge artifacts'' that may appear as path dependent choices of the limit), and the longitudinally polarised amplitude not diverge as $\sim 1/m_3$ (which would otherwise provide a physical obstruction to taking the limit in a way consistent with perturbative unitarity). If the couplings are structure constants of a standard YM Lie algebra (and are therefore real and equal under cyclic permutations of indices at their default heights) then (\ref{2Mass1MasslessVec}) reduces to minimal coupling. See \cite{Durieux:2019eor} for previous discussion of this issue and earlier analysis of these amplitudes. Deviations from this correspond to anomalous multipole moments of the form given by the second term in (\ref{2Mass1MasslessVec}). See Appendix \ref{sec:ExoticGT} for a simple example. 


Finally, it should be noted that the limit of the amplitude with the longitudinal mode (\ref{2Mass1MasslessLong}) has the form of the Higgs-like coupling in (\ref{2vec1scalar}). This suggests that the longitudinal mode of the massive vector bosons could possibly play the role of a Higgs boson in unitarising vector boson scattering. Further comments on this will be made below. 

I conclude this subsection by providing an alternative representation of the three massive vector amplitude in which high-energy dependence is also made manifest:
\begin{align}\label{3legVecHEL}
A(W_A,W_B,W_C)&=\frac{ih_{ABC}}{\Lambda^2}\ds{\bf{12}}\ds{\bf{23}}\ds{\bf{31}}+\frac{i(h_{ABC})^*}{\Lambda^2}\da{\bf{12}}\da{\bf{23}}\da{\bf{31}}\nonumber\\
&\qquad+\frac{i\sqrt{2}}{3}\frac{c^A_{BC}}{\Lambda}\left(\frac{1}{m_3}\ds{\bf{23}}\da{\bf{31}}\ds{\bf{12}}-\frac{1}{m_2}\ds{\bf{31}}\da{\bf{12}}\ds{\bf{23}}\right)\nonumber\\
&\qquad+\frac{i\sqrt{2}}{3}\frac{(c^A_{BC})^*}{\Lambda}\left(\frac{1}{m_3}\da{\bf{23}}\ds{\bf{31}}\da{\bf{12}}-\frac{1}{m_2}\da{\bf{31}}\ds{\bf{12}}\da{\bf{23}}\right)\nonumber\\
&\qquad+\frac{i\sqrt{2}}{3}\frac{c^B_{CA}}{\Lambda}\left(\frac{1}{m_1}\ds{\bf{31}}\da{\bf{12}}\ds{\bf{23}}-\frac{1}{m_3}\ds{\bf{12}}\da{\bf{23}}\ds{\bf{31}}\right)\nonumber\\
&\qquad+\frac{i\sqrt{2}}{3}\frac{(c^B_{CA})^*}{\Lambda}\left(\frac{1}{m_1}\da{\bf{31}}\ds{\bf{12}}\da{\bf{23}}-\frac{1}{m_3}\da{\bf{12}}\ds{\bf{23}}\da{\bf{31}}\right)\nonumber\\
&\qquad+\frac{i\sqrt{2}}{3}\frac{c^C_{AB}}{\Lambda}\left(\frac{1}{m_2}\ds{\bf{12}}\da{\bf{23}}\ds{\bf{31}}-\frac{1}{m_1}\ds{\bf{23}}\da{\bf{31}}\ds{\bf{12}}\right)\nonumber\\
&\qquad+\frac{i\sqrt{2}}{3}\frac{(c^C_{AB})^*}{\Lambda}\left(\frac{1}{m_2}\da{\bf{12}}\ds{\bf{23}}\da{\bf{31}}-\frac{1}{m_1}\da{\bf{23}}\ds{\bf{31}}\da{\bf{12}}\right)\nonumber\\
&\quad-\frac{if_{ABC}}{2m_1m_2m_3}\big(m_1\ds{\bf{12}}\da{\bf{23}}\ds{\bf{31}}+m_2\ds{\bf{23}}\da{\bf{31}}\ds{\bf{12}}+m_3\ds{\bf{31}}\da{\bf{12}}\ds{\bf{23}}\nonumber\\
&\qquad\qquad\quad+m_1\da{\bf{12}}\ds{\bf{23}}\da{\bf{31}}+m_2\da{\bf{23}}\ds{\bf{31}}\da{\bf{12}}+m_3\da{\bf{31}}\ds{\bf{12}}\da{\bf{23}}\big).
\end{align}
I am not bothering to restate the crossing signs for brevity. The couplings introduced in (\ref{3legVecHEL}) can be matched onto (\ref{3legVec}). Firstly, I define
\begin{align}
f_{ABC}&=\frac{1}{3}\left({f}_{BC}^{\,\,\,\,\,\,\,\,\,A}+{f}_{CA}^{\,\,\,\,\,\,\,\,\,B}+{f}_{AB}^{\,\,\,\,\,\,\,\,\,C}\right).
\end{align}
This is fully antisymmetric in its index exchanges. It is natural to fix the shift invariance (\ref{GCSShiftInvar}) by setting $f_{ABC}$ real and this has been done implicitly in (\ref{3legVecHEL}). This coupling now corresponds directly to the spontaneously broken YM structure in (\ref{NormalBrokenYM}) and isolates the terms with softened high energy dependence. However, further redundancy has been introduced elsewhere. The couplings of the form $c_{AB}^C$ are complex, symmetric in their lowered indices ($c_{AB}^C=c_{BA}^C$) and obey the constraint
\begin{align}\label{CSumCons}
m_1c_{BC}^A+m_2c_{CA}^B+m_3c_{AB}^C=0.
\end{align}
They can be matched onto (\ref{3legVec}) through 
\begin{align}\label{ctof}
\frac{c_{AB}^C}{\Lambda}&=\frac{1}{2\sqrt{2}m_3}\left({f}_{CA}^{\,\,\,\,\,\,\,\,\,B}-{f}_{BC}^{\,\,\,\,\,\,\,\,\,A}\right)\nonumber\\
\frac{c_{BC}^A}{\Lambda}&=\frac{1}{2\sqrt{2}m_1}\left({f}_{AB}^{\,\,\,\,\,\,\,\,\,C}-{f}_{CA}^{\,\,\,\,\,\,\,\,\,B}\right)\nonumber\\
\frac{c_{CA}^B}{\Lambda}&=\frac{1}{2\sqrt{2}m_2}\left({f}_{BC}^{\,\,\,\,\,\,\,\,\,A}-{f}_{AB}^{\,\,\,\,\,\,\,\,\,C}\right).
\end{align}
Alternatively, 
\begin{align}\label{ftoc}
{f}_{AB}^{\,\,\,\,\,\,\,\,\,C}=f_{ABC}+\frac{2\sqrt{2}}{3}\left(m_1\frac{c^A_{BC}}{\Lambda}-m_2\frac{c^B_{CA}}{\Lambda}\right)
\end{align}
and similarly for the other permutations of indices. These equalities arise from direct comparison of each term in (\ref{3legVec}) with (\ref{3legVecHEL}). Of course, these couplings are not linearly independent, so these equalities need only hold for a particular choice that are otherwise part of an equivalence class generated by (\ref{GCSShiftInvar}) and (\ref{CSumCons}). The matching presented above is nevertheless useful for making quick exchanges between the two descriptions (\ref{3legVec}) and (\ref{3legVecHEL}) of the same amplitude. 

The ``holomorphic'' terms in the first line of either (\ref{3legVec}) or (\ref{3legVecHEL}) correspond off-shell to $F^3$ and $F^2\widetilde{F}$ operators (giving the real and imaginary parts of $h_{ABC}$). Their behaviour has a clear high energy dependence, scaling as $\sim E^3$ in the HEL for transverse helicities all of the same sign and matching onto the massless amplitudes (\ref{F3Amps}). The scaling of the remaining terms in the HEL is not so obvious. 

Neglecting the $F^3$ terms, there are two classes of terms contained in the remainder of the amplitude: those that scale as $\sim E^2$  in the HEL and those that scale as the suppressed $\sim E$. Note that while this distinction is relevant to the present analysis, it does not automatically follow that the softer $3$-particle terms will generate softer higher leg amplitudes - this instead requires the further constraints on the couplings derived in the Sections below. The $\sim E$ terms are isolated to the $f_{ABC}$ term in (\ref{3legVecHEL}) and its limits have been discussed above. 

The class of terms proportional to the $c$-type constants in (\ref{3legVecHEL}) have leading high-energy scaling $\sim E^2$ for helicity configurations with two same-sign transverse vectors and one longitudinal vector. Assuming that the $c$-type constants do not contain inverse dependence on any of the particle masses, then, in the HEL, these terms match onto the corresponding massless amplitudes in (\ref{2vec1scalar}) (hence the notational choices), while they vanish for other helicity configurations. I will denote by $(h_1,h_2,\dots)$ a list of particle helicities describing a specific helicity configuration. To exemplify this matching, take the helicity configuration $(++,++,L)$ (The $L$ denotes longitudinal vector polarisation corresponding to helicity $0$ or, equivalently, $(+-)$-valued spin indices), then the amplitude $A(W_A,W_B,W_C)$ converges to $A(g^+_A,g^+_B,\phi_C)$ and the coupling $c_{AB}^C$ matches directly onto that in (\ref{2vec1scalar}) with $C$ reinterpreted as a ``matter'' index in a particular basis of scalar states. Through (\ref{ftoc}), this assumption about the (in)dependence of the $c$-type constants on the vector masses is tantamount to the assumption that the departure of the $f$-type couplings in (\ref{3legVec}) from full antisymmetry scales as $\mathcal{O}(\frac{m}{\Lambda})$, where $\Lambda$ is some higher mass scale characterising the break-down of perturbative unitarity. However, if the $c$-type couplings are less mass suppressed, as suggested by (\ref{ctof}), then they will obstruct the HEL of the subleading helicity configurations to which the YM terms would otherwise provide the unique, leading order contribution toward. This would altogether prevent a sensible HEL from existing and signify a theory with a low cut-off. Finally, I note that in the special case in which one of the vectors is massless (particle $3$, as above), then the constraint (\ref{PartAntiSym}) translates into $c^A_{BC}=-c^B_{AC}$ and the parameter $c^{C}_{AB}$ disappears. 

Off-shell, the real and imaginary parts of the non-holomorphic Lorentz structures encapsulated in each term of (\ref{3legVec}) correspond individually to operators (at leading order in the number of constituent fields) of the form $A_\mu A_\nu F^{\mu\nu}$ and $A_\mu A_\nu \widetilde{F}^{\mu\nu}$. By themselves, in the high-energy limit, they converge at leading order to $\varphi FF$ and $\varphi F\widetilde{F}$ operators (from the Stuckelberg trick). However, the special gauge invariant linear combination that can be assembled into the non-kinetic parts of $F^2$ corresponds to Yang-Mills theory and has the softened energy dependence. In this sense, the $c$-type couplings in (\ref{3legVecHEL}) are interpreted as analogues of $\varphi F^2$ dimension $5$ operators in which the scalar is identified as the longitudinal mode of a massive vector boson. This is corroborated by the further results of this Section below.

\section{Massive Vector Boson Scattering and the Higgs Mechanism}\label{VectorHiggs}

\subsection{Mass and the high energy limit}\label{sec:4vpreamble}

In contrast to massless $3$-particle amplitudes, massive amplitudes admit the possibility of multiple independent Lorentz structures, each of which may scale differently with high energies. The factorisable terms in $4$-leg amplitudes are likewise afflicted with this complication (in addition to the contact terms, for which this was obviously always an issue). The structures with the strongest energy dependence will eventually dominate the amplitude at high enough energies. At $4$-legs, as long as energy growing terms exist, perturbative unitarity will inevitably be violated at some point. In order to consistently extrapolate the theory to higher energies without obstruction from perturbative unitarity, these terms must be suppressed or eliminated in succession of their growth rates. This imposes a sequence of conditions on the underlying couplings of the theory required for the cancellation of the strongest to weakest growing terms. This sequence will be derived in the subsequent Sections. However, accepting some finite energy cut-off, the hierarchy in energy growth can generally be complicated by variations in the size of the associated coupling constants. Lorentz structures with stronger high-energy dependence can be suppressed by smaller couplings in order to make them comparable to terms with weaker energy dependence, provided that there is some energy ceiling. This allows for the possibility of a variety of low energy effective theories defined by different coupling hierarchies and which will be elaborated upon further below. 

The ``high energy limit'' of a scattering amplitude is precisely defined here as $E\gg m$, where $E$ stands generically for the energy of the external particles and $m$ the particle masses, be they external or internal. Implicitly, this assumes the existence of a mass gap for the particles involved in each individual amplitude to which this is applied (I consider only scattering of mass eigenstates). Some more comments about this will be made below in Section \ref{sec:2vec2sca}. If the theory violates perturbative unitarity in the HEL, I will use $\Lambda$ to describe the scale at which this occurs
, although it should be noted that this can depend upon the helicity sector under examination. If the theory admits an energy regime $\Lambda\gg E\gg m$, then it should match onto a sensible massless theory in the HEL, although this need not happen if $\Lambda$ is not too much higher than the masses. 

In the HEL, the factorisable terms in the amplitude fall into one of two possible categories: either they match onto massless amplitudes containing poles describing the exchange of a massless particle between two $3$-particle amplitudes, or they match onto $4$-particle contact terms. These contact terms always contain energy dependence of the form $(E/m)^n$ (and multiplied by some combinations of coupling constants) for some $n\geq 0$, where $m$ is the mass of an involved particle. If the contact terms involve scalars emerging from the longitudinal polarisations of the vector bosons, then it is possible that there is no corresponding contact term in the massive theory that would endow this emergent Lorentz structure with independent parametric freedom. In this case, the emergent $4$-particle contact coupling is entirely determined by the $3$-particle couplings of the massive amplitudes. 

The massive terms matching onto massless exchange amplitudes can only be suppressed by suppressing the accompanying couplings constants. However, the emergent $4$-particle contact terms typically receive contributions from multiple factorisation channels and these can interfere with each other. The contact terms can therefore be suppressed by tuning the $3$-particle couplings to obey particular relations that ensures their cancellation. These relations usually resemble analogous constraints on the properties of massless $3$-particle couplings discussed above from consistency. This provides an alternative to suppressing the contact interactions independently of weakening the constituent $3$-particle couplings.

Massive Lorentz structures generally match onto a particular massless Lorentz structure at leading order in the HEL for a particular set of helicity configurations. However, they will also contribute spurious path-dependent spinors (or ``gauge artifacts'') at subleading order in other helicity configurations. The existence of terms like this prevents the amplitude from being interpreted as matching onto a massless amplitude for these other helicity configurations. However, these limits are always obstructed by the leading divergences hitting the perturbative unitarity boundary. The unitarity bound and the appearance of Lorentz violating gauge artifacts are two separate manifestations of the same problem which is resolved by suppressing the couplings involved. This both extends the unitarity cut-off for the leading terms while relegating the spurious terms to ignorable subleading effects. Once these divergences have been suppressed or eliminated, the theory can be extrapolated to higher energies until it is obstructed again by the next strongest divergence in another helicity sector. This continues until all divergent terms have been eliminated and the theory is ``unitarised''. 

The goal of the present Section is to find the conditions under which, at tree-level, $2\rightarrow 2$ vector boson scattering has the tamest high energy scaling. Doing so should uncover a sequence of conditions on the theory required to ameliorate the energy growth in distinct helicity sectors. These conditions delineate different classes of EFTs. The general $4$-leg massive vector scattering amplitude at tree-level decomposes into terms generated by two insertions of the $3$-leg amplitudes (\ref{3legVec}) and (\ref{2vec1scalar}) and then, additionally, $4$-leg contact interactions. To identify the theory with the weakest HEL, the coefficients of the contact terms can be tuned to cancel divergences of the terms that correctly account for the factorisation properties. However, the lowest dimension contact interactions have mass dimension $4$ (corresponding to dim-$8$ operators) and, with one notable exception, cannot eliminate $\sim E^3$ (or weaker) high-energy divergences in the amplitude's factorisable sector. As will be shown below, this is inadequate for unitarising the amplitude, so further constraints on the $3$-particle amplitudes underpinning the factorising terms are required. This will lead to the emergence of Yang-Mills theory and the Higgs mechanism, but also a few other intriguing details. 

It will be useful to introduce some new spinor structure building blocks. A basis of maximally symmetric, dimension $3$ Lorentz structures consisting of a single spinor for each external particle is given by terms of the form
\begin{align}\label{L}
\mathbf{L}_1/m_1&=\la{\bf{1}}p_2-p_3\rs{\bf{4}}\ds{\bf{23}}+\la{\bf{1}}p_2-p_4\rs{\bf{3}}\ds{\bf{24}}+\la{\bf{1}}p_4-p_3\rs{\bf{2}}\ds{\bf{43}}\nonumber\\
&\qquad+\la{\bf{1}}p_4-p_2\rs{\bf{3}}\ds{\bf{42}}+\la{\bf{1}}p_3-p_2\rs{\bf{4}}\ds{\bf{32}}+\la{\bf{1}}p_3-p_4\rs{\bf{2}}\ds{\bf{34}}
\end{align}
and similarly for $\mathbf{L}_2=\mathbf{L}_1|_{1\leftrightarrow 2}$, $\mathbf{L}_3=\mathbf{L}_1|_{1\leftrightarrow 3}$ and $\mathbf{L}_4=\mathbf{L}_1|_{1\leftrightarrow 4}$. The structure $\mathbf{L}_i$ is explicitly symmetric under exchanges of any pair of particles besides $i$. Parity conjugate structures $\hat{\mathbf{L}}_i$ can be defined analogously with the bracket shapes switched. It will also be useful to define the following abbreviations: 
\begin{align}\label{SpinStrucDef2}
\mathbf{P}_s&=\da{\bf{12}}\ds{\bf{12}}\da{\bf{34}}\ds{\bf{34}}\nonumber\\
\mathbf{P}_t&=\da{\bf{13}}\ds{\bf{13}}\da{\bf{24}}\ds{\bf{24}}\nonumber\\
\mathbf{P}_u&=\da{\bf{14}}\ds{\bf{14}}\da{\bf{23}}\ds{\bf{23}}.
\end{align}

Cancellations between terms generated in different factorisation channels can be manifested by the application of identities known as ``syzygies''. These are relations in which linear combinations of spinor structures dressed by Mandelstam invariants are decomposed into linear combinations of lower dimensional Lorentz structures (with factors of masses accounting for the remaining dimensionality). Without the Mandelstam invariants, the spinor structures would otherwise be linearly independent and such a decomposition would not exist. These relations can be identified as massive deformations of simple massless identities. The simplest example deforms the trivial relation
\begin{align}\label{SimpSyzFact}
\frac{\da{34}\ds{43}}{s}-\frac{\da{31}\ds{13}}{t}&=0\nonumber\\
\hookrightarrow\frac{\la{\bf{3}}p_4\rs{\bf{3}}}{s-m_s^2}-\frac{\la{\bf{3}}p_1\rs{\bf{3}}}{t-m_t^2}&=\frac{m_3\left(\ls{\bf{3}}p_1p_4\rs{\bf{3}}+\la{\bf{3}}p_1p_4\ra{\bf{3}}\right)}{(s-m_s^2)(t-m_t^2)}\nonumber\\
&\qquad+\frac{(m_1^2+m_3^2-m_t^2)\la{\bf{3}}p_4\rs{\bf{3}}}{s-m_s^2}-\frac{(m_3^2+m_4^2-m_s^2)\la{\bf{3}}p_1\rs{\bf{3}}}{t-m_t^2}.
\end{align}
As is clear from this example, these identities demonstrate the tension between manifest locality and high energy unitarity: one of these properties is obscured at the expense of the other. The cancellation of high energy dependence amalgamates terms from different factorisation channels. 

Throughout the ensuing calculations, extensive use will be made of the syzygy representing the massive deformation of the all-channel pole:
\begin{align}\label{OldSyzFact}
\frac{\ds{12}\ds{34}}{s}+\frac{\ds{13}\ds{24}}{t}&=0\nonumber\\
\hookrightarrow\frac{\ds{\bf{12}}\ds{\bf{34}}}{s-m_s^2}+\frac{\ds{\bf{13}}\ds{\bf{24}}}{t-m_t^2}&=\frac{\sum_i\mathbf{L}_i}{12(s-m_s^2)(t-m_t^2)}\nonumber\\
&\qquad+\frac{(\frac{1}{3}\sum_im_i^2-m_t^2)\ds{\bf{12}}\ds{\bf{34}}}{s-m_s^2}-\frac{(\frac{1}{3}\sum_im_i^2-m_s^2)\ds{\bf{13}}\ds{\bf{24}}}{t-m_t^2}.
\end{align}
This has been identified previously in a much less symmetric form in \cite{Durieux:2020gip}. Related identities can be derived by contracting (\ref{OldSyzFact}) with various combinations bilinears, which have application to amplitudes involving particles with various spins. In Appendix \ref{sec:HPoB}, I present a list of relevant syzygies and other useful kinematic identities between Lorentz structures appearing in the $4$-particle amplitudes constructed in this study.

\subsection{Partial unitarisation}\label{sec:PartialU}

For simplicity, I will ignore the $F^3$ terms in the calculation presented in this Section, although their inclusion is straightforward and is done in Appendix \ref{app:4V}. Like their massless counterparts, double insertions of these produce terms that scale in the HEL as $\sim E^4$ but cannot be canceled by contact interactions (this occurs, in particular, for the MHV helicity configuration). Searching for the theory with weakest high-energy scaling therefore immediately demands $h_{ABC}=0$. However, there nevertheless remains some further information that can be potentially extracted. Single insertions of $F^3$ paired with the $f$-type terms in (\ref{3legVec}) produce $\sim E^3$ high-energy divergences that can be reduced to $\sim E^2$ (but no further) by demanding some sense of gauge covariance of the $h_{ABC}$  couplings, as was demonstrated in the massless case in Section \ref{AllChannelAmps}. The complete analysis of this is provided in Appendix \ref{app:4V}, but does not affect the conclusions derived here. 

\subsubsection{General amplitude from factorisation}

\begin{feynfig}
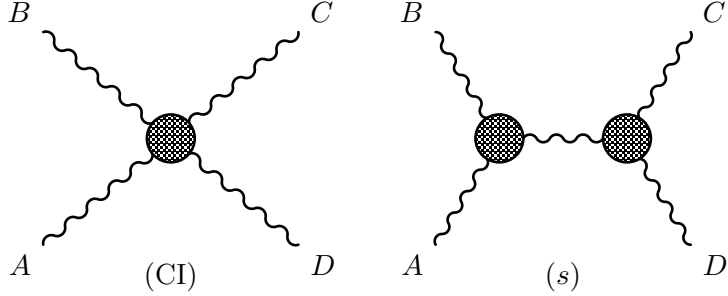
\begin{figure}[h]
\begin{fmffile}{4Vec}
 \begin{center}
 \begin{tabular}{ c c c }
 & & \\
   \begin{fmfgraph*}(120,80)
   \fmfleft{i1,i2}
   \fmfright{o1,o2}
   \fmf{boson}{i2,v1}
   \fmf{boson}{v1,o2}
   \fmf{boson}{i1,v1}
   \fmf{boson}{v1,o1}
   \fmfv{decor.shape=circle,decor.filled=gray50,decor.size=0.15w}{v1}
   \fmflabel{$A$}{i1}
   \fmflabel{$B$}{i2}
   \fmflabel{$C$}{o2}
   \fmflabel{$D$}{o1}
 \end{fmfgraph*} 
&\,\,&\begin{fmfgraph*}(120,80)
   \fmfleft{i1,i2}
   \fmfright{o1,o2}
   \fmf{boson}{i2,v1}
   \fmf{boson}{v2,o2}
   \fmf{boson}{i1,v1}
   \fmf{boson}{v2,o1}
   \fmf{boson}{v1,v2}
   \fmfv{decor.shape=circle,decor.filled=gray50,decor.size=0.15w}{v1,v2}
   \fmflabel{$A$}{i1}
   \fmflabel{$B$}{i2}
   \fmflabel{$C$}{o2}
   \fmflabel{$D$}{o1}
 \end{fmfgraph*} \\
  $(\text{CI})$ &\,\,& $(s)$
  \end{tabular}
\end{center}
\caption{On-shell contributions to vector boson scattering (omitting $t$ and $u$-channels).}\label{4vectorOSDi}
\end{fmffile}
\end{figure}
\end{feynfig}

Terms accounting for the factorisation structure of the amplitude can be constructed in the usual way by combining the $3$-leg amplitudes (\ref{3legVec}) across each channel on-shell. See Fig \label{4vectorOSDi}. Making use of particle permutation symmetries and parity helps to economise the calculation, while keeping the intermediate particle's momentum on-shell for as long as possible minimises the high-energy scaling. In the case at hand, the terms constructed from factorisation in this way have directly $\sim E^3$ manifest high-energy scaling. As mentioned above, the contact terms for this amplitude have mass dimension $4$ or greater, so (with one notable exception) scale as $\sim E^4$ in the HEL at their weakest. Demanding the elimination of $\mathcal{O}(E^4)$ divergences therefore amounts to neglecting the separate inclusion of contact terms into the amplitude when the factorisation terms are constructed in this way. However, converting the expression to the $\{\mathbf{L}_i\}$ basis (to manifest exchange symmetries) can reintroduce the need to tune the contact terms to cancel the divergences. Of utility in conversion to the basis of $\{\mathbf{L}_i\}$ Lorentz structures are the identities 
\begin{align}\label{E4toL}
&\ls{\bf{1}}P_s\ra{\bf{3}}\ls{\bf{2}}P_s\ra{\bf{4}}+\ls{\bf{1}}P_s\ra{\bf{4}}\ls{\bf{2}}P_s\ra{\bf{3}}\nonumber\\
&=\left(u-t\right)\ds{\bf{12}}\da{\bf{34}}+\frac{1}{6}\left(\frac{m_2}{m_1}\hat{\mathbf{L}}_1+\frac{m_1}{m_2}\hat{\mathbf{L}}_2+\frac{m_4}{m_3}\mathbf{L}_3+\frac{m_3}{m_4}\mathbf{L}_4\right)\nonumber\\
&\qquad+\frac{2}{3}m_1m_2\left(\da{\bf{13}}\da{\bf{24}}+\da{\bf{14}}\da{\bf{23}}\right)+\frac{2}{3}m_3m_4\left(\ds{\bf{13}}\ds{\bf{24}}+\ds{\bf{14}}\ds{\bf{23}}\right)
\end{align}
and
\begin{align}\label{E3toL}
&\da{\bf{13}}\la{\bf{2}}P_s\rs{\bf{4}}+\da{\bf{23}}\la{\bf{1}}P_s\rs{\bf{4}}\nonumber\\
&=-\frac{1}{6m_4}\hat{\mathbf{L}}_4-\frac{1}{3}m_4\left(\da{\bf{14}}\da{\bf{23}}+\da{\bf{13}}\da{\bf{24}}\right)-m_1\da{\bf{23}}\ds{\bf{14}}-m_2\da{\bf{13}}\ds{\bf{24}},
\end{align}
where $P_s=p_3+p_4$ is the intermediate $s$-channel on-shell momentum as defined in Section \ref{sec:simple} above.

Regardless of the precise order of operations, an amplitude free of $\sim E^4$ HEL divergences can be constructed that otherwise satisfies all factorisation requirements. The leading order $\sim E^3$ terms occur for helicity configurations with single transverse and three longitudinal polarisations. For the helicity configuration $(L,L,L,++)$, the terms producing these divergences are, for the $s$-channel,
\begin{align}\label{E34Vec}
&A_{E^3s}(W_A,W_B,W_C,W_D)\nonumber\\
&=\frac{-1}{24\prod_i m_i}\frac{1}{s-m_{s_M}^2}\ds{\bf{12}}\ds{\bf{34}}\hat{\mathbf{L}}_4\nonumber\\
&\quad\times\left(\Re f_{AB}^{\,\,\,\,\,\,\,\,\,M}\Re f_{CM}^{\,\,\,\,\,\,\,\,\,D}+i\left(\Re f_{AB}^{\,\,\,\,\,\,\,\,\,M}\Im f_{CM}^{\,\,\,\,\,\,\,\,\,D}+\frac{1}{2}\left(\Re f_{AB}^{\,\,\,\,\,\,\,\,\,M}\Im f_{CD}^{\,\,\,\,\,\,\,\,\,M}-\Im f_{AB}^{\,\,\,\,\,\,\,\,\,M}\Re f_{CD}^{\,\,\,\,\,\,\,\,\,M}\right)\right)\right),
\end{align}
while $t$ and $u$-channel terms are given by exchanging $2\leftrightarrow 3$ and $1\leftrightarrow 3$ respectively (including the corresponding internal indices). All other $\sim E^3$ terms may be inferred from exchange symmetries and parity (implemented by switching the bracket shapes and complex conjugating the coupling constants). 

Before presenting the remainder of the amplitude generated by factorisation, there remains some freedom among the contact interactions to be addressed. As alluded to above, there still exists a single independent contact interaction with $\sim E^3$ high energy scaling \cite{Durieux:2020gip}. Labeling
\begin{align}
\mathbf{C}=\ds{\bf{12}}\da{\bf{23}}\ds{\bf{34}}\da{\bf{41}}-\da{\bf{12}}\ds{\bf{23}}\da{\bf{34}}\ds{\bf{41}},
\end{align}
then the contact amplitude to be included is
\begin{align}\label{vectorE3CI}
A_{c}(W_A,W_B,W_C,W_D)=\frac{iC_{ABCD}}{\prod_i m_i}\mathbf{C},
\end{align}
which is parity violating. Here, $C_{ABCD}$ are some coupling constants which must be fully antisymmetric in their indices in order to reflect the full antisymmetry of the accompanying kinematic factor under particle exchange (which is true but not manifest in the expression presented above). The choice of explicit mass dependence of the coefficient in (\ref{vectorE3CI}) has been made purely for convenience. The leading divergences occur for helicity configurations with one transverse and three longitudinal polarisations, like for (\ref{E34Vec}) above. The weaker HEL dependence than its mass dimension occurs because it is the difference of two linearly independent massive Lorentz structures that converge to the same massless Lorentz structure for the helicity configuration in which each massive spinor is unsuppressed (which is the 4$L$ configuration). The contact amplitude (\ref{vectorE3CI}) should be added to the terms constructed from factorisation in order to complete the full amplitude. Note that the kinematic factor is simply $\epsilon_{\mu\nu\rho\sigma}\varepsilon_1^\mu\varepsilon_2^\nu\varepsilon_3^\rho\varepsilon_4^\sigma$, where $\varepsilon_i$ are just the polarisations of the external vector bosons (see Appendix \ref{sec:ExoticGT}). 

Returning to the amplitude generated by factorisation, the remaining terms can be collected together into independent Lorentz structures identified by the helicity configurations for which they have the strongest high energy scaling. As for (\ref{E34Vec}), I will present the $s$-channel expressions from which the complete $t$ and $u$-channel expressions can be inferred from particle exchanges. For fully transverse helicities, the relevant terms are given by
\begin{align}\label{4Tfact}
&A_{4TE^2s}(W_A,W_B,W_C,W_D)\nonumber\\
&=-\frac{1}{8m_{s_M}^2}\frac{1}{s-m_{s_M}^2}\left(\left(f_{MB}^{\,\,\,\,\,\,\,\,\,A}+f_{MA}^{\,\,\,\,\,\,\,\,\,B}\right)\ds{\bf{12}}^2+\left((f_{MB}^{\,\,\,\,\,\,\,\,\,A})^*+(f_{MA}^{\,\,\,\,\,\,\,\,\,B})^*\right)\da{\bf{12}}^2\right)\nonumber\\
&\qquad\qquad\qquad\qquad\qquad\times\left(\left(f_{MD}^{\,\,\,\,\,\,\,\,\,C}+f_{MC}^{\,\,\,\,\,\,\,\,\,D}\right)\ds{\bf{34}}^2+\left((f_{MD}^{\,\,\,\,\,\,\,\,\,C})^*+(f_{MC}^{\,\,\,\,\,\,\,\,\,D})^*\right)\da{\bf{34}}^2\right).
\end{align}
The terms corresponding to two longitudinal polarisations and two opposite transverse helicities are given by (for $(--,L,L,++)$)
\begin{align}\label{2T2LOppfact}
&A_{2L2T\text{opp}E^2s}(W_A,W_B,W_C,W_D)\nonumber\\
&=\frac{1}{4m_2m_3}\frac{1}{s-m_{s_M}^2}\left(f_{AB}^{\,\,\,\,\,\,\,\,\,M}+(f_{BM}^{\,\,\,\,\,\,\,\,\,A})^*\right)\left((f_{CD}^{\,\,\,\,\,\,\,\,\,M})^*+f_{MC}^{\,\,\,\,\,\,\,\,\,D}\right)\da{\bf{12}}\da{\bf{13}}\ds{\bf{24}}\ds{\bf{34}}
\end{align}
plus all distinct terms generated particle exchanges (there are none in which both longitudinal modes appear on the same side of the channel). When the transverse helicities have the same sign, the terms are 
\begin{align}\label{2T2LSamefact}
&A_{2L2T\text{same}E^2s}(W_A,W_B,W_C,W_D)\nonumber\\
&=\frac{1}{4m_3m_4m_{s_M}^2}\frac{1}{s-m_{s_M}^2}\left(f_{MB}^{\,\,\,\,\,\,\,\,\,A}+f_{MA}^{\,\,\,\,\,\,\,\,\,B}\right)\left(m_3^2\Re f_{MD}^{\,\,\,\,\,\,\,\,\,C}+m_4^2\Re f_{MC}^{\,\,\,\,\,\,\,\,\,D}\right)\ds{\bf{12}}^2\da{\bf{34}}\ds{\bf{34}}\nonumber\\
&\quad+\frac{1}{4m_2m_4}\frac{1}{s-m_{s_M}^2}\Big(f_{BM}^{\,\,\,\,\,\,\,\,\,A}f_{DM}^{\,\,\,\,\,\,\,\,\,C}+i\Im f_{AB}^{\,\,\,\,\,\,\,\,\,M}f_{MD}^{\,\,\,\,\,\,\,\,\,C}+i\Im f_{CD}^{\,\,\,\,\,\,\,\,\,M}f_{MB}^{\,\,\,\,\,\,\,\,\,A}\nonumber\\
&\qquad\qquad\qquad\qquad\qquad\qquad\qquad-\Re f_{AB}^{\,\,\,\,\,\,\,\,\,M}\Re f_{CD}^{\,\,\,\,\,\,\,\,\,M}-\Im f_{AB}^{\,\,\,\,\,\,\,\,\,M}\Im f_{CD}^{\,\,\,\,\,\,\,\,\,M}\Big)\ds{\bf{12}}\ds{\bf{13}}\da{\bf{24}}\ds{\bf{34}}
\end{align}
for $(++,++,L,L)$ and $(++,L,++,L)$, plus all distinct terms generated particle exchanges and parity conjugate terms. For fully longitudinal helicity configuration, the relevant terms are 
\begin{align}\label{4LEvenfact}
&A_{4LE^2s}(W_A,W_B,W_C,W_D)\nonumber\\
&=\frac{1}{2\prod_im_i}\frac{1}{s-m_{s_M}^2}\nonumber\\
&\qquad\times\bigg(\bigg(\frac{1}{3}\left(\Re f_{AB}^{\,\,\,\,\,\,\,\,\,M}\left(m_4^2\Re f_{MC}^{\,\,\,\,\,\,\,\,\,D}-m_3^2\Re f_{MD}^{\,\,\,\,\,\,\,\,\,C}\right)+\Re f_{CD}^{\,\,\,\,\,\,\,\,\,M}\left(m_2^2\Re f_{MA}^{\,\,\,\,\,\,\,\,\,B}-m_1^2\Re f_{MB}^{\,\,\,\,\,\,\,\,\,A}\right)\right)\nonumber\\
&\qquad\qquad\qquad\qquad\qquad\qquad\qquad\qquad\qquad\qquad\qquad+m_{s_M}^2\Re f_{AB}^{\,\,\,\,\,\,\,\,\,M}\Re f_{CD}^{\,\,\,\,\,\,\,\,\,M}\bigg)\left(\mathbf{P}_t-\mathbf{P}_u\right)\nonumber\\
&\qquad\qquad\qquad\qquad\qquad-\frac{1}{m_{s_M}^2}\left(m_1^2\Re f_{MB}^{\,\,\,\,\,\,\,\,\,A}+m_2^2\Re f_{MA}^{\,\,\,\,\,\,\,\,\,B}\right)\left(m_3^2\Re f_{MD}^{\,\,\,\,\,\,\,\,\,C}+m_4^2\Re f_{MC}^{\,\,\,\,\,\,\,\,\,D}\right)\mathbf{P}_s\bigg).
\end{align}
These are automatically $P$- and $T$-symmetric. The final class of terms is given by the expression
\begin{align}\label{4LOddfact}
&A_{3L1TEs}(W_A,W_B,W_C,W_D)\nonumber\\
&=\frac{i}{4\prod_im_i}\frac{1}{s-m_{s_M}^2}\nonumber\\
&\quad\times\bigg(\bigg(\frac{1}{3}\left(m_3^2+m_4^2-m_1^2-m_2^2\right)\left(\Re f_{AB}^{\,\,\,\,\,\,\,\,\,M}\Im f_{CD}^{\,\,\,\,\,\,\,\,\,M}-\Re f_{CD}^{\,\,\,\,\,\,\,\,\,M}\Im f_{AB}^{\,\,\,\,\,\,\,\,\,M}\right)\nonumber\\
&\qquad\qquad\qquad\qquad\qquad\qquad\qquad\quad+m_{s_M}^2\left(\Re f_{AB}^{\,\,\,\,\,\,\,\,\,M}\Im f_{CD}^{\,\,\,\,\,\,\,\,\,M}+\Re f_{CD}^{\,\,\,\,\,\,\,\,\,M}\Im f_{AB}^{\,\,\,\,\,\,\,\,\,M}\right)\bigg)\nonumber\\
&\qquad\quad-\frac{2}{3}\Big(\Re f_{AB}^{\,\,\,\,\,\,\,\,\,M}\left(m_4^2\Im f_{MC}^{\,\,\,\,\,\,\,\,\,D}-m_3^2\Im f_{MD}^{\,\,\,\,\,\,\,\,\,C}\right)+\Re f_{CD}^{\,\,\,\,\,\,\,\,\,M}\left(m_2^2\Re f_{MA}^{\,\,\,\,\,\,\,\,\,B}-m_1^2\Re f_{MB}^{\,\,\,\,\,\,\,\,\,A}\right)\Big)\bigg)\mathbf{C}.
\end{align}
Being proportional to $\mathbf{C}$, this has weaker $\sim E$ HEL scaling, which occurs for helicity configurations in which three of the polarisations are longitudinal and one is transverse. This class is $P$ and $T$-violating. 

At this point, it is simple to include the possible contribution to the amplitude from the tree-level exchange of scalars with the couplings given in (\ref{2vec1scalar}). The $s$-channel exchange amplitude (Fig \ref{4VecScalarEx}) is given by
\begin{feynfig}
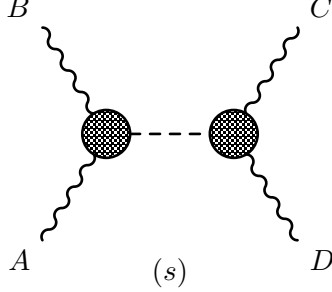
\begin{figure}[h]
\begin{fmffile}{4VecHiggs}
\begin{center}
   \begin{fmfgraph*}(120,80)
   \fmfleft{i1,i2}
   \fmfright{o1,o2}
   \fmf{boson}{i2,v1}
   \fmf{boson}{v2,o2}
   \fmf{boson}{i1,v1}
   \fmf{boson}{v2,o1}
   \fmf{dashes}{v1,v2}
   \fmfv{decor.shape=circle,decor.filled=gray50,decor.size=0.15w}{v1,v2}
   \fmflabel{$A$}{i1}
   \fmflabel{$B$}{i2}
   \fmflabel{$C$}{o2}
   \fmflabel{$D$}{o1}
 \end{fmfgraph*}\\
  $(s)$
\end{center}
\end{fmffile}
\caption{Scalar $s$-channel exchange contributions to vector boson scattering.}\label{4VecScalarEx}
\end{figure}
\end{feynfig}
\begin{align}\label{4vecScalar}
&A_{\varphi s}(W_A,W_B,W_C,W_D)\nonumber\\
&=\frac{-1}{s-m_{s_i}^2}\left(\frac{c_{AB}^i}{\Lambda}\ds{\bf{12}}^2+\frac{(c_{AB}^i)^*}{\Lambda}\da{\bf{12}}^2+\frac{1}{\sqrt{2}}\lambda^i_{AB}\ds{\bf{12}}\da{\bf{12}}\right)\nonumber\\
&\qquad\qquad\qquad\times\left(\frac{c_{CD}^i}{\Lambda}\ds{\bf{34}}^2+\frac{(c_{CD}^i)^*}{\Lambda}\da{\bf{34}}^2+\frac{1}{\sqrt{2}}\lambda^i_{CD}\ds{\bf{34}}\da{\bf{34}}\right).
\end{align}
These clearly have HEL divergences that scale as $\sim E^2$ which could potentially interfere with the high-energy behaviour of the vector exchange terms.

\subsubsection{Emergence of gauged non-linear sigma models}

Having now constructed a complete and correctly factorising amplitude, the next step is to identify the conditions under which it can be extrapolated to high energies with minimal obstruction from perturbative unitarity. Since the contact terms have been (mostly) fixed, demanding that the terms diverging as $\sim E^3$ cancel necessitates constraints on the three vector self-couplings. Choosing, for example, the helicity configuration $(L,L,L,++)$, then the cancellation of the real (or $P$-conserving) part of the amplitude's couplings (contained in (\ref{E34Vec})) across the three factorisation channels implies that
\begin{align}\label{MassiveLA}
\Re f_{AB}^{\,\,\,\,\,\,\,\,\,M}\Re f_{MC}^{\,\,\,\,\,\,\,\,\,D}+\Re f_{BC}^{\,\,\,\,\,\,\,\,\,M}\Re f_{MA}^{\,\,\,\,\,\,\,\,\,D}+\Re f_{CA}^{\,\,\,\,\,\,\,\,\,M}\Re f_{MB}^{\,\,\,\,\,\,\,\,\,D}=0.
\end{align}
This is the Jacobi identity. The real parts of the couplings must therefore be structure constants of a Lie algebra, as expected for Yang-Mills. However, unlike the massless case, there is no requirement that the Lie algebra be semi-simple or compact, even if the couplings are purely real. The coefficients of each of the Lorentz structures in (\ref{3legVec}) may differ. For the special case in which one of the vectors is a massless photon, departures from the standard properties materialise as large anomalous multipole moments of the charged massive vector bosons (see Appendix \ref{sec:ExoticGT}). While the remaining $\sim E^2$ HEL scaling of the amplitude is not unitary by the standards of renormalisable gauge theory, it is unitary by the standards of gravity, which explains the prevalence of non-semisimple or non-compact gauge algebras in theories of supergravity (see e.g. \cite{VanProeyen:2003zj,DAuria:1998emc,Trigiante:2016mnt} - amplitudes for some theories with these features were also recently studied in $5d$ \cite{Chiodaroli:2023tvo}).

Cancellation of the imaginary parts of the $\sim E^3$ divergent couplings requires tuning the quartic (\ref{vectorE3CI}) against the $P$-violating parts of (\ref{E34Vec}) to fix:
\begin{align}\label{GCSquarticfix}
C_{ABCD}&=-\frac{1}{4}\big(\Re f_{AB}^{\,\,\,\,\,\,\,\,\,M}\Im f_{CD}^{\,\,\,\,\,\,\,\,\,M}-\Re f_{CD}^{\,\,\,\,\,\,\,\,\,M}\Im f_{AB}^{\,\,\,\,\,\,\,\,\,M}+\Re f_{BC}^{\,\,\,\,\,\,\,\,\,M}\Im f_{AD}^{\,\,\,\,\,\,\,\,\,M}\nonumber\\
&\qquad\qquad-\Re f_{AD}^{\,\,\,\,\,\,\,\,\,M}\Im f_{BC}^{\,\,\,\,\,\,\,\,\,M}-\Re f_{AC}^{\,\,\,\,\,\,\,\,\,M}\Im f_{BD}^{\,\,\,\,\,\,\,\,\,M}+\Re f_{BD}^{\,\,\,\,\,\,\,\,\,M}\Im f_{AC}^{\,\,\,\,\,\,\,\,\,M}\nonumber\\
&\qquad\qquad+2\left(\Re f_{AB}^{\,\,\,\,\,\,\,\,\,M}\Im f_{CM}^{\,\,\,\,\,\,\,\,\,D}-\Re f_{AC}^{\,\,\,\,\,\,\,\,\,M}\Im f_{BM}^{\,\,\,\,\,\,\,\,\,D}+\Re f_{BC}^{\,\,\,\,\,\,\,\,\,M}\Im f_{AM}^{\,\,\,\,\,\,\,\,\,D}\right)\big).
\end{align}
This is only consistent with the full antisymmetry of $C_{ABCD}$ if 
\begin{align}\label{nonAbGCScons}
&\Re f_{BC}^{\,\,\,\,\,\,\,\,\,M}\left(\Im f_{AD}^{\,\,\,\,\,\,\,\,\,M}+\Im f_{AM}^{\,\,\,\,\,\,\,\,\,D}\right)+\Re f_{BD}^{\,\,\,\,\,\,\,\,\,M}\left(\Im f_{AC}^{\,\,\,\,\,\,\,\,\,M}+\Im f_{AM}^{\,\,\,\,\,\,\,\,\,C}\right)+\Re f_{CA}^{\,\,\,\,\,\,\,\,\,M}\left(\Im f_{BM}^{\,\,\,\,\,\,\,\,\,D}+\Im f_{BD}^{\,\,\,\,\,\,\,\,\,M}\right)\nonumber\\
&\qquad+\Re f_{DA}^{\,\,\,\,\,\,\,\,\,M}\left(\Im f_{BM}^{\,\,\,\,\,\,\,\,\,C}+\Im f_{BC}^{\,\,\,\,\,\,\,\,\,M}\right)+\Re f_{AB}^{\,\,\,\,\,\,\,\,\,M}\left(\Im f_{CM}^{\,\,\,\,\,\,\,\,\,D}+\Im f_{DM}^{\,\,\,\,\,\,\,\,\,C}\right)=0.
\end{align}
Equivalently, this relation is required for the contact term to be sufficient to cancel the $P$-violating divergences for all $\sim E^3$ helicity configurations. The constraint (\ref{nonAbGCScons}), in conjunction with (\ref{YMantisym}) and (\ref{GCSShiftInvar}), identifies the imaginary parts of the couplings off-shell as ``generalised Chern-Simons terms'' \cite{deWit:1984rvr,Anastasopoulos:2006cz}, which are known to arise, for instance, in anomalous EFTs of massive vectors that can be UV completed to non-anomalous Yang-Mills. Deviations to the special quartic coupling (\ref{vectorE3CI}) fixed by (\ref{GCSquarticfix}) may be regarded as ``anomalous non-Abelian Generalised Chern-Simons quartic couplings'' (aGCSQC), analogous to ``anomalous quartic gauge couplings'' described by different choices of $4$-leg contact interactions from spontaneously broken YM, except that these are distinguished as having slightly tamer high-energy scaling (``anomalous'' in the sense of ``irregular'', not in the sense of the chiral anomaly referred to in the previous sentence).

The Jacobi identity possesses a symmetrical structure in which each term clearly arises from the factorisation channel corresponding to the partitioning of the indices in the couplings. Each such term/channel is otherwise on identical footing as the others. Since the relation (\ref{nonAbGCScons}) has similar origins here, it is insightful to find a form with analogous symmetries. Defining $F_{AB;C}^{\,\,\,\,\,\,\,\,\,\,\,\,\,\,\,D}=\Re f_{AB}^{\,\,\,\,\,\,\,\,\,M}\Im f_{CM}^{\,\,\,\,\,\,\,\,\,D}+\frac{1}{2}\left(\Re f_{AB}^{\,\,\,\,\,\,\,\,\,M}\Im f_{CD}^{\,\,\,\,\,\,\,\,\,M}-\Re f_{CD}^{\,\,\,\,\,\,\,\,\,M}\Im f_{AB}^{\,\,\,\,\,\,\,\,\,M}\right)$, then the GCS constraint becomes
\begin{align}
&F_{BC;A}^{\,\,\,\,\,\,\,\,\,\,\,\,\,\,\,D}+F_{AB;C}^{\,\,\,\,\,\,\,\,\,\,\,\,\,\,\,D}+F_{CA;B}^{\,\,\,\,\,\,\,\,\,\,\,\,\,\,\,D}\nonumber\\
&\qquad+F_{DA;B}^{\,\,\,\,\,\,\,\,\,\,\,\,\,\,\,C}+F_{AB;D}^{\,\,\,\,\,\,\,\,\,\,\,\,\,\,\,C}+F_{BD;A}^{\,\,\,\,\,\,\,\,\,\,\,\,\,\,\,C}=0,
\end{align}
which clearly has a similar $stu$-like and cyclic structure. The quartic coupling is also then
\begin{align}
C_{ABCD}=\frac{1}{2}\left(F_{AB;C}^{\,\,\,\,\,\,\,\,\,\,\,\,\,\,\,D}+F_{CA;B}^{\,\,\,\,\,\,\,\,\,\,\,\,\,\,\,D}+F_{BC;A}^{\,\,\,\,\,\,\,\,\,\,\,\,\,\,\,D}\right).
\end{align}

\subsubsection{Leading high energy cancellations}

Having established (\ref{MassiveLA}), (\ref{GCSquarticfix}) and (\ref{nonAbGCScons}), the remaining amplitude now scales at worst as $\sim E^2$ in the HEL. There are three sources of terms arising in the calculation at this point: 
\begin{enumerate}
\item terms constructed immediately from factorisation
\item terms arising from cancellation between the GCS terms in (\ref{E34Vec}) and the contact term (\ref{vectorE3CI})
\item terms arising from cancellation between the pure Lie algebra terms in (\ref{E34Vec}) across different channels.
\end{enumerate}

The terms belonging to category $1$ were constructed and presented above. The remaining $\sim E^2$ terms require implementing the $\sim E^3$ cancellations. To begin with, it is convenient to homogenise the masses appearing in the propagators in each channel. Selecting a single specific particle mass for each channel and denoting these simply as $m_s$, $m_t$ and $m_u$ (which are effectively free parameters), then e.g. 
\begin{align}\label{HomoDen}
\frac{N_m}{s-m_{s_m}^2}=\frac{N_m}{s-m_{s}^2}+\frac{N_m\left(m_{s_m}^2-m_s^2\right)}{\left(s-m_{s}^2\right)\left(s-m_{s_m}^2\right)}.
\end{align}
Here $N_m$ simply denotes the numerator associated to the $s$-channel exchange of the particles of mass $m_{s_m}$. I will refer to the terms proportional to $m_{s_m}^2-m_s^2$ as the ``mass splitting'' terms. If $N_m\sim E^4$ (or weaker) in the HEL, then these terms are clearly manifestly unitary and can therefore be left alone as they are. This will be the case in the examples covered in the subsequent Sections \ref{VecSca} and \ref{VecFer} below, where these terms will be acknowledged but mostly ignored (although they will play a role in matching the unitarised expressions onto their massless counterparts). However, in the case at hand, the numerator in (\ref{E34Vec}) scales as $\sim E^5$, so they must be (temporarily) retained.

The mass eigenstate basis is the natural basis with which to describe the particle states involved in transition amplitudes. However, this may generally be disorganised and unnatural from the perspective of other internal quantum numbers. In this study, I do not intend to provide a detailed investigation into the question of representation reducibility and its physical manifestation as the phenomenon of mass mixing. This issue is secondary to establishing the algebraic structure of the couplings and has been the subject of \cite{Bachu:2023fjn}. However, the mass splitting terms invite the opportunity for some comments as they can describe finite mass mixing effects. 

The four-fermion amplitude (\ref{GenMoller}) in Section \ref{sec:4fermion} provides a simple example with which this relation can be illustrated. For example, in the UV, the fermions could couple to the vector bosons (and therefore group into a non-trivial representation of the YM Lie algebra), but not have Yukawa couplings to the scalars (which may possibly be forbidden by symmetries again). However, in the IR, the vector-fermion coupling necessarily draws in the scalar identified as the longitudinal mode of the vector. There are two possible resolutions to make this consistent with vanishing UV Yukawa couplings. The first is that the coupling is vector-like and the fermions on each side of the channel have equal mass (as explained below (\ref{LongModeLim})), so that the exchanged longitudinal mode directly decouples in the HEL. In the IR massive amplitude, this corresponds to the direct elimination of the $s$-wave term in (\ref{MollerVec}), in which case the UV massless vector exchange amplitudes directly match onto the limits of the first term in (\ref{MollerVec}). The second possibility is that additional scalar exchanges are introduced into the amplitude that cancel the UV effects of the longitudinal mode. These have Yukawa couplings determined from those required to cancel the $s$-wave terms in (\ref{GenMoller}). However, the cancellation may not be complete against (\ref{GenMoller}) in the IR because of the possible differences between masses of the scalars and the vectors in the propagators. The mass splitting terms, applied to this example, represent the residual mass mixing effects. 

Returning to the four vector amplitude having homogenised the propagator masses, the $\sim E^3$ cancellation between the terms with imaginary couplings occurs directly between the leading divergent terms in (\ref{E34Vec}) and the contact term (\ref{vectorE3CI}). This can be demonstrated by converting (\ref{vectorE3CI}) to a form with manifest high energy dependence at the expense of manifest locality. Using (\ref{OldSyzFact}), (\ref{vectorE3CI}) can be expressed as e.g.
\begin{align}\label{UnpackedGCSquart}
&\frac{iC_{ABCD}}{\prod_i m_i}\mathbf{C}\nonumber\\
&\,\,\,\,=\frac{-1}{s-m_s^2}\frac{iC_{ABCD}}{\prod_im_i}\Bigg(\frac{1}{12}\left(\ds{\bf{12}}\ds{\bf{34}}\sum_j\hat{\bf{L}}_j-\da{\bf{12}}\da{\bf{34}}\sum_j\mathbf{L}_j\right)+\left(m_s^2-\frac{1}{3}\sum_j m_j^2\right)\mathbf{C}\Bigg)
\end{align}
and there are analogous expressions with spurious $t$- and $u$-channel poles that may be easily arrived at by particle exchanges. The contact term (\ref{vectorE3CI}) is most symmetrically represented as an average over all three such Mandelstam channel representations and I will assume this in the following. Hence (\ref{UnpackedGCSquart}) should be multiplied by $1/3$ and added to the corresponding expressions with $1\leftrightarrow 3$ and $2\leftrightarrow 3$. The HEL is now manifest and it is clear how the terms cancel against the leading energy terms in (\ref{E34Vec}). The $s$, $t$ and $u$-channel coupling partitions in (\ref{GCSquarticfix}), when substituted into (\ref{UnpackedGCSquart}), each cancel against the respective $s$, $t$ and $u$-channel terms of the form (\ref{E34Vec}). For example, the cancellation between the $s$-channel $\hat{\mathbf{L}}_4$ term in (\ref{E34Vec}) with the corresponding $s$-channel couplings in the $t$-channel representation of (\ref{UnpackedGCSquart}) leaves 
\begin{align}\label{Cat2b}
&A_{E^2st}^{(GCSb)}(W_A,W_B,W_C,W_D)\nonumber\\
&=\frac{-i}{72\prod_im_i}\frac{1}{(s-m_s^2)(t-m_t^2)}\left(\Re f_{AB}^{\,\,\,\,\,\,\,\,\,M}\Im f_{CM}^{\,\,\,\,\,\,\,\,\,D}+\frac{1}{2}\left(\Re f_{AB}^{\,\,\,\,\,\,\,\,\,M}\Im f_{CD}^{\,\,\,\,\,\,\,\,\,M}-\Im f_{AB}^{\,\,\,\,\,\,\,\,\,M}\Re f_{CD}^{\,\,\,\,\,\,\,\,\,M}\right)\right)\nonumber\\
&\qquad\times\hat{\mathbf{L}}_4\left(\frac{1}{12}\sum_j\mathbf{L}_j-\left(m_s^2-\frac{1}{3}\sum_jm_j^2\right)\ds{\bf{13}}\ds{\bf{24}}-\left(m_t^2-\frac{1}{3}\sum_jm_j^2\right)\ds{\bf{12}}\ds{\bf{34}}\right)
\end{align}
after application of (\ref{OldSyzFact}). The corresponding cancellation with the $u$-channel representation of (\ref{UnpackedGCSquart}) is analogous and can be obtained from the above expression by exchanging $1\leftrightarrow 2$, while the cancellation with the $s$-channel representation is a direct elimination. Then the analogous argument can be applied to the cancellation of the $t$- and $u$-channel couplings. The left over terms $t$-coupling terms can be inferred from (\ref{Cat2b}) by applying the exchanges $2\leftrightarrow 3$ and also the sequence ($1\leftrightarrow 2$, $2\leftrightarrow 3$), while the $u$-coupling terms are obtained by adding both terms generated by the exchanges $1\leftrightarrow 3$ and ($1\leftrightarrow 2$, $1\leftrightarrow 3$). This then completes the collection of residual $P$-violating terms that descend from the $\sim E^3$ divergence in the $(L,L,L,++)$ helicity configuration. The residual terms produced in the cancellation for the other helicity configurations are obtained in the same way as the other $E^3$ terms generated from the action of exchanges and parity on (\ref{E34Vec}). Along with the subleading terms proportional to $\mathbf{C}$ on the right-hand side of (\ref{UnpackedGCSquart}), which I restate here for emphasis:
\begin{align}\label{Cat2a}
&A_{Est}^{(GCSa)}(W_A,W_B,W_C,W_D)\nonumber\\
&=\frac{-i}{3\prod_im_i}C_{ABCD}\mathbf{C}\Bigg(\frac{m_s^2-\frac{1}{3}\sum_jm_j^2}{s-m_s^2}+\frac{m_t^2-\frac{1}{3}\sum_jm_j^2}{t-m_t^2}+\frac{m_u^2-\frac{1}{3}\sum_jm_j^2}{u-m_u^2}\Bigg),
\end{align}
these constitute the category 2 terms descending from the reduction of the GCS divergences. 

Finally are the terms in category 3: the $\sim E^2$ terms that arise from reducing the $P$-conserving $\sim E^3$ terms in (\ref{E34Vec}). Generally, it is useful to write the amplitude in the form
\begin{align}\label{SymmRed}
&\frac{C_sN_s}{s-m_s^2}+\frac{C_tN_t}{t-m_t^2}+\frac{C_uN_u}{u-m_u^2}&\nonumber\\
&\qquad\qquad\quad=\frac{1}{3}\Bigg(\left(C_s-C_t\right)\left(\frac{N_s}{s-m_s^2}-\frac{N_t}{t-m_t^2}\right)+\left(C_t-C_u\right)\left(\frac{N_t}{t-m_t^2}-\frac{N_u}{u-m_u^2}\right)\nonumber\\
&\qquad\qquad\qquad\qquad\qquad\qquad\qquad\qquad\qquad\qquad\qquad\quad+\left(C_u-C_s\right)\left(\frac{N_u}{u-m_u^2}-\frac{N_s}{s-m_s^2}\right)\Bigg)
\end{align}
where $C_x$ are combinations of coupling constants for channel $x$ obeying $C_s+C_t+C_u=0$ and $N_x$ are the corresponding kinematical factors. If the pairings of kinematical terms in (\ref{SymmRed}) can be combined to produce expressions with manifestly reduced high energy scaling, then (\ref{SymmRed}) provides a symmetrical way of accounting for this. 

Applying (\ref{SymmRed}) to the $\hat{\mathbf{L}}_4$ terms generated by (\ref{E34Vec}) gives 
\begin{align}
&A^{(J)}_{E^2st}(W_A,W_B,W_C,W_D)\nonumber\\
&\qquad=\frac{1}{72\prod_i m_i}\left(\Re f_{AB}^{\,\,\,\,\,\,\,\,\,M}\Re f_{MC}^{\,\,\,\,\,\,\,\,\,D}-\Re f_{CA}^{\,\,\,\,\,\,\,\,\,M}\Re f_{MB}^{\,\,\,\,\,\,\,\,\,D}\right)\left(\frac{\ds{\bf{12}}\ds{\bf{34}}}{s-m_{s}^2}+\frac{\ds{\bf{13}}\ds{\bf{24}}}{t-m_{t}^2}\right)\hat{\bf{L}}_4
\end{align}
for the $s$ and $t$-channel pairings. The terms in (\ref{SymmRed}) grouping the $t$ and $u$ channels are given by the exchange $1\leftrightarrow 3$, while the $s$ and $u$ grouping is given by $1\leftrightarrow 2$. The kinematic factors can be reduced using (\ref{OldSyzFact}) to give 
\begin{align}\label{JacRed}
&A^{(J)}_{E^2st}(W_A,W_B,W_C,W_D)\nonumber\\
&=\frac{1}{72\prod_i m_i}\frac{1}{(s-m_{s}^2)(t-m_{t}^2)}\left(\Re f_{AB}^{\,\,\,\,\,\,\,\,\,M}\Re f_{MC}^{\,\,\,\,\,\,\,\,\,D}-\Re f_{CA}^{\,\,\,\,\,\,\,\,\,M}\Re f_{MB}^{\,\,\,\,\,\,\,\,\,D}\right)\nonumber\\
&\qquad\times\hat{\bf{L}}_4\left(\frac{1}{12}\sum_j\mathbf{L}_j-\left(m_s^2-\frac{1}{3}\sum_jm_j^2\right)\ds{\bf{13}}\ds{\bf{24}}-\left(m_t^2-\frac{1}{3}\sum_jm_j^2\right)\ds{\bf{12}}\ds{\bf{34}}\right),
\end{align}
which now has manifest $\sim E^2$ HEL scaling. As described with (\ref{E34Vec}), to this class of terms generated by $A^{(J)}_{E^2st}(W_A,W_B,W_C,W_D)$ and its symmetrisation of the form (\ref{SymmRed}) should be added the parity conjugate terms (obtained by swapping the bracket shapes) and the terms exchanging particle $4$ with the each of the others. This completes the remaining class of terms in the partially unitarised amplitude. 

The structures $\hat{\mathbf{L}}_4\mathbf{L}_j$ containing the leading energy dependence in (\ref{Cat2b}) and (\ref{JacRed}) can be decomposed into a basis of lower dimensional spinor structures multiplied by Mandelstam invariants and masses in (\ref{LLhatSame}) and (\ref{LLhatDiff}). The leading order $\sim E^2$ terms all have only a single pole and can be combined with the terms constructed from factorisation that are stated above. This will be pursued in subsection \ref{sec:FullUni} below in the course of identifying the conditions required for further unitarisation.

\subsubsection{Consistency with massless gluons}

In the special case in which some of the external vector bosons are massless, some of the conditions represented in (\ref{MassiveLA}) and (\ref{nonAbGCScons}) become mandatory consequences of consistent factorisation rather than features of
high-energy unitarity. This can be demonstrated directly and explicitly by using the systematic soft limit procedure formulated by \cite{Elvang:2016qvq}, which was subsequently used by \cite{Falkowski:2020aso} for massive amplitudes. I  will briefly summarise the key steps here and give a more extensive review in \cite{Trott:2026ozo}.

The holomorphic soft shift of \cite{Elvang:2016qvq} can be used to systematically implement the $4$-particle test. A massless particle is taken soft in a complexified way by shifting its momentum $p_s$ as
\begin{align}\label{MasslessSoftShift}
   \hat{p}_s&=-\ra{\hat{s}}\ls{s}=-\left(\epsilon\ra{s}-z\ra{X}\right)\ls{s}.
\end{align}
Here $\epsilon\rightarrow 0$ parameterises the soft limit. The shift parameter $z\ll\epsilon$ is introduced in order to separate each on-shell factorisation channel activated in the soft limit into separate soft poles located at
\begin{align}\label{softpole}
    \epsilon\rightarrow\epsilon^{(k)}_*=\frac{\la{X}p_k\rs{s}}{2p_s\cdot p_k}z
\end{align}
for each hard leg $k$ that the soft particle is emitted off. Without $z\neq 0$, these would all be indistinguishably blurred. Finally, the spurious spinor $\ra{X}$ is arbitrary provided that it obeys $\da{Xs}=0$. Some of the remaining hard momenta must also be shifted to absorb the recoil of the deformed soft momentum, but this can be practically ignored for the purposes here.

Taking this soft limit, the amplitude describing the soft particle's emission off a set of hard legs can be decomposed according to the master formula
\begin{align}\label{MasterSoft}
A_{n+1}(1,2,\dots)=\sum_k\sum_P A_3(\hat{P}_{sk}\rightarrow k,\hat{s}^+)\frac{1}{P_{sk}^2+m_{sk}^2} \frac{1}{\epsilon}\frac{1}{1-\frac{z}{\epsilon}\frac{\la{X}p_k\rs{s}}{2p_s\cdot p_k}}A_n^{(k)}(1,2,\ldots,\hat{P}_{sk},\ldots)
\end{align}
where $\hat{P}_{sk}=\left(p_k+\hat{p}_s\right)$ is an on-shell intermediate momentum and $P_{sk}=\left(p_k+p_s\right)$ is general. The sum over $P$ indicates the possible intermediate particle quantum numbers (internal or spin related) while $m_{P_{sk}}$ are their corresponding masses. The index $k$ represents the hard legs. The hard subamplitudes $A_n^{(k)}$ in the sum are evaluated on the poles (\ref{softpole}) and involve the hard leg $k$ being replaced with the intermediate particle. Finally, I assume without loss of generality that the soft particle is right-handed. 

The limit $z\rightarrow0$ applied to (\ref{MasterSoft}) should be finite in order for a sensible soft limit to exist. A singularity in $z$ reflects a failure of consistent factorisation. These can potentially arise through the kinematic structure of the on-shell $3$-particle amplitudes appearing in (\ref{MasterSoft}). Demanding that (\ref{MasterSoft}) be regular as $z\rightarrow0$ presents an unambiguous test of consistency of a hypothetical perturbative $S$-matrix with locality. Applied to a $4$-particle amplitude and taking one leg soft, this automatically reproduces the $4$-particle test. 

This procedure can be applied to a $4$-particle amplitude involving a massless gluon and three massive vector bosons. Taking the gluon soft, each term in (\ref{MasterSoft}) can be identified as a factorisation channel in which the gluon is radiated on-shell off one massive leg. The factorisation structure of this process is illustrated in Fig \ref{3vec1Glu}. 
\begin{feynfig}
\begin{figure}[h]
\begin{fmffile}{3Vec1Glu}
 \begin{center}
 \begin{tabular}{ c c c c c }
 & & \\
   \begin{fmfgraph*}(120,80)
   \fmfleft{i1,i2}
   \fmfright{o1,o2}
   \fmf{gluon}{i2,v1}
   \fmf{boson}{v2,o2}
   \fmf{boson}{i1,v1}
   \fmf{boson}{v2,o1}
   \fmf{boson}{v1,v2}
   \fmfv{decor.shape=circle,decor.filled=gray50,decor.size=0.15w}{v1,v2}
   \fmflabel{$A$}{i1}
   \fmflabel{$B+$}{i2}
   \fmflabel{$C$}{o2}
   \fmflabel{$D$}{o1}
 \end{fmfgraph*} 
&\,&\begin{fmfgraph*}(120,80)
   \fmfleft{i1,i2}
   \fmfright{o1,o2}
   \fmf{gluon}{i2,v1}
   \fmf{boson}{v2,o2}
   \fmf{boson}{i1,v1}
   \fmf{boson}{v2,o1}
   \fmf{boson}{v1,v2}
   \fmfv{decor.shape=circle,decor.filled=gray50,decor.size=0.15w}{v1,v2}
   \fmflabel{$D$}{i1}
   \fmflabel{$B+$}{i2}
   \fmflabel{$A$}{o2}
   \fmflabel{$C$}{o1}
 \end{fmfgraph*}
&\,&\begin{fmfgraph*}(120,80)
   \fmfleft{i1,i2}
   \fmfright{o1,o2}
   \fmf{gluon}{i2,v1}
   \fmf{boson}{v2,o2}
   \fmf{boson}{i1,v1}
   \fmf{boson}{v2,o1}
   \fmf{boson}{v1,v2}
   \fmfv{decor.shape=circle,decor.filled=gray50,decor.size=0.15w}{v1,v2}
   \fmflabel{$C$}{i1}
   \fmflabel{$B+$}{i2}
   \fmflabel{$A$}{o2}
   \fmflabel{$D$}{o1}
 \end{fmfgraph*} \\
  $(s)$ &\,\,& $(t)$ &\,\,& $(u)$
  \end{tabular}
\end{center}
\caption{Factorisation for soft gluon emission off massive vector hard scattering.}\label{3vec1Glu}
\end{fmffile}
\end{figure}
\end{feynfig}
The underlying hard amplitude describing the interaction of the three massive vectors is given by the general expression (\ref{3legVec}), while the soft gluon emission factors are described by the amplitude (\ref{2Mass1MasslessVec}). The Lorentz scalars appearing in these soft factors are evaluated as
\begin{align}\label{MassiveSoftBis}
    \da{k^K\hat{P}^M_{sk}}=m_k\epsilon^{KM}\qquad
    \frac{1}{\hat{x}}=\frac{\ls{s}p_k\ra{s}}{m_k\da{Xs}z}.
\end{align}
The $1/\hat{x}$ factor, found in the minimal coupling term in (\ref{2Mass1MasslessVec}), contains a potentially spurious $z$-pole. Substituting (\ref{2Mass1MasslessVec}) and (\ref{3legVec}) into the master formula, the terms containing a $z$ pole must cancel. This produces the requirement
\begin{align}\label{MasslessGluonConsFact}
&\Re f_{AB}^{\,\,\,\,\,\,\,\,\,M}\big((f_{MD}^{\,\,\,\,\,\,\,\,\,A})^*m_1\da{\bf{41}}\da{\bf{31}}\ds{\bf{34}}+(f_{AM}^{\,\,\,\,\,\,\,\,\,D})^*m_4\da{\bf{34}}\da{\bf{14}}\ds{\bf{13}}+(f_{DA}^{\,\,\,\,\,\,\,\,\,M})^*m_3\da{\bf{13}}\da{\bf{43}}\ds{\bf{41}}\nonumber\\
&\qquad\qquad+f_{MD}^{\,\,\,\,\,\,\,\,\,A}m_1\ds{\bf{41}}\ds{\bf{31}}\da{\bf{34}}+f_{AM}^{\,\,\,\,\,\,\,\,\,D}m_4\ds{\bf{34}}\ds{\bf{14}}\da{\bf{13}}+f_{DA}^{\,\,\,\,\,\,\,\,\,M}m_3\ds{\bf{13}}\ds{\bf{43}}\da{\bf{41}}\big)\nonumber\\
&\qquad\qquad\qquad\qquad\qquad\qquad\qquad+\left(1\leftrightarrow 3\right)+\left(3\leftrightarrow 4\right)=0.
\end{align}
Here, particle $2$ (colour $B$) has been chosen as the massless gluon and (\ref{PartAntiSym}) has been used. The non-minimal term in (\ref{2Mass1MasslessVec}) does not contribute to this.

The multiple independent Lorentz structures lead to multiple consistency conditions that can be read off once the redundancy (\ref{3PMassRed}) is fixed in some way. This reproduces the subset of Jacobi relations (\ref{MassiveLA}) and non-Abelian GCS constraints (\ref{nonAbGCScons}) in which the massless colour index is exclusively lowered. Because of (\ref{PartAntiSym}), the general non-Abelian GCS constraint equations (\ref{nonAbGCScons}) collapse from ten to six terms when colour $B$ is taken massless (one pair emerging from each of the three factorisation channels), which reconciles it with the constraints enforced here. 

The non-Abelian GCS constraints are a notable counterexample to the naive expectation that the couplings between massive particles should be covariant tensors of the Lie algebra. Even when some of the vectors are massless gluons, the consistency requirement from the soft gluon limit does not produce a covariance relation. This is possible because the Lorentz structures corresponding to each class of coupling constants do not remain linearly independent across factorisation channels. The resulting consistency conditions arising from single soft gluon limits can therefore mix couplings across different structures. This is in contrast to e.g. the Yukawa coupling amplitudes (\ref{Yukawa}), which do not kinematically change when a soft gluon is emitted off each leg.

\subsection{Full unitarisation}\label{sec:FullUni}

\subsubsection{Leading transverse divergences and Yang-Mills}

I will next advance toward establishing the conditions for unitarisation of the high energy amplitude beyond $E^2$. To begin with, the terms in the full amplitude diverging as $\sim E^2$ for purely transversely polarised vectors are given in (\ref{4Tfact}) and (\ref{4vecScalar}). It is clear from comparison of (\ref{4Tfact}) with (\ref{4vecScalar}) that the divergent terms generated by the vector exchange have the same structure as those produced from scalar exchange mediated by double insertions of the dim-$5$ $\varphi F^2$ type interactions. Through conversion of the vector self-couplings to the HEL unitarity basis with (\ref{ctof}), (\ref{4Tfact}) can be reinterpreted as (\ref{4vecScalar}) with the scalar index $i$ extended to include the longitudinal vector modes (ignoring the $\lambda$-type terms). For the $(++,++,--,--)$ helicity configuration, the leading divergence (which arises entirely in the $s$-channel) therefore only cancels if
\begin{align}
&\frac{1}{8m_{s_M}^2}\left(f_{MA}^{\,\,\,\,\,\,\,\,\,B}+f_{MB}^{\,\,\,\,\,\,\,\,\,A}\right)\left(f_{MD}^{\,\,\,\,\,\,\,\,\,C}+f_{MC}^{\,\,\,\,\,\,\,\,\,D}\right)^*+\frac{1}{\Lambda^2}c_{AB}^i(c_{CD}^i)^*\nonumber\\
&\qquad\qquad\qquad=\frac{1}{\Lambda^2}\left(c_{AB}^M(c_{CD}^M)^*+c_{AB}^i(c_{CD}^i)^*\right)=0,
\end{align}
The combination of couplings is positive semi-definite, so can only vanish if $c^{\{M,i\}}_{AB}=0$. The dimension-$5$ scalar couplings are therefore unsurprisingly ruled-out from a theory with unitary high-energy scaling. This includes those implicit in the general three vector amplitude, which is tantamount to requiring that $f_{AB}^{\,\,\,\,\,\,\,\,\,C}$ is fully antisymmetric upon any index exchange:
\begin{align}\label{StandardLA}
f_{AB}^{\,\,\,\,\,\,\,\,\,M}=f_{BM}^{\,\,\,\,\,\,\,\,\,A}=f_{MA}^{\,\,\,\,\,\,\,\,\,B}=f_{ABM}.
\end{align}
This also directly implies that the imaginary parts of the couplings vanish and that the Lie algebra must have the standard properties of ``unbroken'' Yang-Mills: semi-simplicity, compactness and homogeneously normalised generators. 

While inconsistent with absolute high energy unitarity, I will nevertheless retain these dim-$5$ couplings when determining the unitarity conditions required for the other helicity sectors. This is of interest for two reasons. The first reason is simply to establish how unitarisation of the other sectors depends on these couplings (if at all). Secondly, suppressing these couplings (as would be expected by taking $\Lambda\gg m_i$ for all participating particle masses $m_i$) while retaining them should allow for the possibility of a limited high energy regime in which $m_i\ll E\ll \Lambda$. In this case, it should be meaningful to treat the dim-$5$ couplings as ``irrelevant deformations'' of the otherwise unitarised theory and the inclusion of these couplings should enhance the generality of the analysis without interfering with the conclusions that would otherwise be reached in their absence. 

UV completions of GCS are typically associated with loop processes, so while it is not expected that they can be unitarised by a tree-level exchange of a particle, it may be possible with loops \cite{Preskill:1990fr,Anastasopoulos:2006cz,Bonnefoy:2020gyh}. I will leave a more complete study of this for further work, restricting the scope of the present analysis to tree-level processes. 

Next, collecting together each of the $\sim E^2$ divergent terms for the two longitudinal and opposite sign transverse helicities from (\ref{2T2LOppfact}), (\ref{Cat2b}) and (\ref{JacRed}) gives, for $(--,L,L,++)$ in the $s$-channel,
\begin{align}\label{2T2LOpp}
&\frac{1}{4m_2m_3}\frac{1}{s-m_s^2}\left((f_{AB}^{\,\,\,\,\,\,\,\,\,M})^*+(f_{MB}^{\,\,\,\,\,\,\,\,\,A})^*\right)\left(f_{CD}^{\,\,\,\,\,\,\,\,\,M}-f_{MC}^{\,\,\,\,\,\,\,\,\,D}\right)\da{\bf{12}}\da{\bf{13}}\ds{\bf{24}}\ds{\bf{34}}.
\end{align}
In the high energy limit, these terms match onto the expected massless amplitudes generated by double insertions of emergent dim-$5$ operators, since
\begin{align}
\frac{1}{4m_2m_3}\left((f_{AB}^{\,\,\,\,\,\,\,\,\,M})^*+(f_{MB}^{\,\,\,\,\,\,\,\,\,A})^*\right)\left(f_{CD}^{\,\,\,\,\,\,\,\,\,M}-f_{MC}^{\,\,\,\,\,\,\,\,\,D}\right)=-\frac{2(c^B_{AM})^*c^C_{MD}}{\Lambda^2}.
\end{align}
The $t$-channel is also non-zero, but linearly independent and cannot cancel the $s$-channel. Again, these divergences can only be tempered by suppressing these couplings. 

It should be noted that, in extracting the contributing single pole terms to (\ref{2T2LOpp}) from (\ref{Cat2b}) and (\ref{JacRed}) using (\ref{LLhatDiff}) (as mentioned above), some residual $\sim E^0$ terms are generated through the use of 
\begin{align}\label{mu2}
(s-m_s^2)+(t-m_t^2)+(u-m_u^2)=\mu^2=\sum_im_i^2-\left(m_s^2+m_t^2+m_u^2\right).
\end{align}
The combination of masses on the right-hand side defines the parameter $\mu^2$. These residual terms have the same leading energy helicity configuration (here $(--,L,L,++)$). They are included in the expression for the fully unitarised amplitude given further below. I make a note of them here because I do not intend to explicitly present the general partially unitarised amplitude in the Section above in the basis in which the Mandelstam invariant dependence of the products $\mathbf{L}_i\hat{\mathbf{L}}_j$ has been squeezed out (as given in (\ref{LLhatDiff}) and (\ref{LLhatSame})), although I do intend to make use of it throughout this Section in collecting together the leading divergences and presenting the final unitarised amplitude. The point is just to note that these terms proportional to $\mu^2$ are generated here when the amplitude is converted to a form in which it manifestly matches onto the appropriate massless amplitudes in the HEL.

\subsubsection{Longitudinal sector and the Higgs mechanism}

I next turn to the $\sim E^2$ divergent terms for purely longitudinal polarisations. Combining together (\ref{4LEvenfact}) and (\ref{JacRed}) (there are no contributions from the parity-violating GCS terms), these are 
\begin{align}\label{Amp4LE2}
&\frac{1}{2\prod_i m_i}\frac{-1}{s-m_{s}^2}\nonumber\\
&\,\,\times\Bigg(\Bigg(\frac{1}{m_{s_M}^2}\left(m_1^2\Re f_{BM}^{\,\,\,\,\,\,\,\,\,A}+m_2^2\Re f_{AM}^{\,\,\,\,\,\,\,\,\,B}\right)\left(m_3^2\Re f_{DM}^{\,\,\,\,\,\,\,\,\,C}+m_4^2\Re f_{CM}^{\,\,\,\,\,\,\,\,\,D}\right)+\left(\prod_jm_j\right)\lambda_{AB}^m\lambda_{CD}^m\Bigg)\mathbf{P}_s\nonumber\\
&\qquad+\bigg(\frac{1}{3}\left(\Re f_{AB}^{\,\,\,\,\,\,\,\,\,M}\left(m_4^2\Re f_{MC}^{\,\,\,\,\,\,\,\,\,D}-m_3^2\Re f_{MD}^{\,\,\,\,\,\,\,\,\,C}\right)+\Re f_{CD}^{\,\,\,\,\,\,\,\,\,M}\left(m_2^2\Re f_{MA}^{\,\,\,\,\,\,\,\,\,B}-m_1^2\Re f_{MB}^{\,\,\,\,\,\,\,\,\,A}\right)\right)\nonumber\\
&\qquad\qquad\qquad\qquad\qquad\qquad\qquad\qquad\qquad\qquad\qquad-m_{s_M}^2\Re f_{AB}^{\,\,\,\,\,\,\,\,\,M}\Re f_{CD}^{\,\,\,\,\,\,\,\,\,M}\bigg)(\mathbf{P}_t-\mathbf{P}_u)\Bigg),
\end{align}
with $t$ and $u$-channel terms inferable from particle exchange symmetries. 

Demanding that the $\sim E^2$ divergent terms cancel in the HEL imposes the following condition on the masses and coupling constants:
\begin{align}\label{EmLA}
&-\frac{1}{4}\frac{1}{m_{t_M}m_3}\left(m_{t_M}^2\Re f_{AC}^{\,\,\,\,\,\,\,\,\,M}+m_3^2\Re f_{MA}^{\,\,\,\,\,\,\,\,\,C}-m_1^2\Re f_{CM}^{\,\,\,\,\,\,\,\,\,A}\right)\nonumber\\
&\qquad\qquad\qquad\qquad\qquad\times\frac{1}{m_{t_M}m_4}\left(m_{t_M}^2\Re f_{DB}^{\,\,\,\,\,\,\,\,\,M}+m_4^2\Re f_{BM}^{\,\,\,\,\,\,\,\,\,D}-m_2^2\Re f_{MD}^{\,\,\,\,\,\,\,\,\,B}\right)\nonumber\\
&+\frac{1}{4}\frac{1}{m_{u_M}m_3}\left(m_{u_M}^2\Re f_{BC}^{\,\,\,\,\,\,\,\,\,M}+m_3^2\Re f_{MB}^{\,\,\,\,\,\,\,\,\,C}-m_2^2\Re f_{CM}^{\,\,\,\,\,\,\,\,\,B}\right)\nonumber\\
&\qquad\qquad\qquad\qquad\qquad\times\frac{1}{m_{u_M}m_4}\left(m_{u_M}^2\Re f_{DA}^{\,\,\,\,\,\,\,\,\,M}+m_4^2\Re f_{AM}^{\,\,\,\,\,\,\,\,\,D}-m_1^2\Re f_{MD}^{\,\,\,\,\,\,\,\,\,A}\right)\nonumber\\
&\qquad+\frac{1}{4}m_1m_2\left(\lambda_{AC}^m\lambda_{BD}^m-\lambda_{BC}^m\lambda_{AD}^m\right)\nonumber\\
&\qquad\qquad\qquad=\frac{1}{2m_3m_4}\Re f_{AB}^{\,\,\,\,\,\,\,\,\,M}\left(m_3^2\Re f_{DM}^{\,\,\,\,\,\,\,\,\,C}+m_4^2\Re f_{MC}^{\,\,\,\,\,\,\,\,\,D}-m_{s_M}^2\Re f_{CD}^{\,\,\,\,\,\,\,\,\,M}\right),
\end{align}
as well as analogous results with the vectors' colour indices permuted. Recognising the presence of (\ref{BrokenGen}) and (\ref{HiggsGen}), this can be reinterpreted as 
\begin{align}\label{LAbrokengen}
(t_A)_{CM}(t_B)_{MD}+(t_A)_{Cm}(t_B)_{mD}-(t_B)_{CM} (t_A)_{MD}-(t_B)_{Cm} (t_A)_{mD}=i\Re f_{AB}^{\,\,\,\,\,\,\,\,\,M}(t_M)_{CD},
\end{align}
which is the commutator of Lie algebra generators. So unitarisation of the longitudinal sector requires that the emergent scalar-vector $3$-particle couplings in the HEL (\ref{BrokenGen}) must be identified as Lie algebra generators (\ref{LAbrokengen}) in some representation of the (general) Lie algebra. This condition effectively constrains the possible mass spectrum of the vectors given the Lie algebra. The massive vector colours are also identified as the default coordinates on the representation space of these emergent scalar degrees of freedom. Massive scalar particles participating in the (tree-level) amplitude are also permitted provided that their couplings are also interpreted as generators (\ref{HiggsGen}). These scalars provide the coordinates that extend and complete the internal index sum in the commutator (\ref{LAbrokengen}). This identifies them as Higgs bosons. It remains to extend the commutation relations to external indices identified with scalars (either with Higgs couplings  (\ref{2vec1scalar}) or standard minimal coupling (\ref{ScalarMat3leg})) in order to show that they form a complete representation (possibly in combination with the longitudinal vector modes). This will be done below in Section \ref{VecSca} by unitarising the amplitudes with mixed scalar and vector legs. 

Are Higgs bosons always needed for (\ref{LAbrokengen}) or, equivalently, can the generators defined in (\ref{BrokenGen}) constitute a complete representation by themselves? For a general $4$-leg amplitude between vector bosons, even assuming the standard properties of spontaneously broken YM in (\ref{StandardLA}), the Higgs terms in (\ref{LAbrokengen}) are not always necessary. An obvious example is $\mathcal{N}=4$ SYM with aligned central charges (or simply adjoint Higgs theories, to the extent that it matters here), where the only Higgs boson in the theory is massless and part of the supermultiplet with the vev (the particle couplings in this theory will be extracted in the Sections below). This Higgs has no involvement in an amplitude involving four vector bosons of different masses. 

In a general theory obeying the standard YM Lie algebra properties (\ref{StandardLA}), setting $D=A$ and $C=B$ reduces the commutation relation (\ref{LAbrokengen}) without Higgs exchange to 
\begin{align}\label{HiggslessBrokenLA}
(f_{ABM})^2\left(m_{s_M}^4-\frac{2}{3}m_{s_M}^2\left(m_A^2+m_B^2\right)-\frac{1}{3}\left(m_A^2-m_B^2\right)^2\right)=0
\end{align}
($M$ is still summed over but the other indices are not). If vectors $A$ and $B$ are the most massive pairs with non-zero structure constants $f_{ABM}$ for any $M$, then the left-hand side of this equation is necessarily negative, which is a contradiction. This is true more generally as long as the masses are bounded above. So while every amplitude need not have a Higgs exchange in order to unitarise, every theory with an upper bound on the vector masses (i.e. a gap) contains amplitudes that necessarily do. The representation of the emergent generators in (\ref{BrokenGen}) and (\ref{HiggsGen}) then necessarily combines scalars with longitudinal vectors. 

If there are an infinite number of vectors with unbounded masses, then (\ref{LAbrokengen}) can possibly be satisfied without Higgses. This is a known feature of Kaluza-Klein (KK) reductions of higher-dimensional Yang-Mills \cite{Chivukula:2001esy,Csaki:2003dt}, where scattering amplitudes of vector resonances are unitarised by exchanges of higher level resonances in the towers. The sum rules proposed in \cite{Csaki:2003dt}, relating the masses and KK couplings, which themselves encode the geometry and boundary conditions of the extra dimension(s), are contained as special instances of (\ref{HiggslessBrokenLA}). I will not investigate here the interpretation of (\ref{LAbrokengen}) for towers of particles, but simply note this as a possible instance of this relation \cite{Bonifacio:2019ioc}. See \cite{Bonifacio:2019ioc} and references therein for further discussion on the constraints arising from this analysis and their implications for the spectra of KK theories. As is also well-known, while KK reduced YM can possess amplitudes that do not increase with energy, the growth in the number of possible production channels provides a perturbative unitarity cut-off, reflecting the fact that this is nevertheless an effective theory. 

Of course, implicit in the ``high energy limit'' as defined here is the assumption of a mass gap, so the fact that a theory with an infinite tower of particles obeys conditions like (\ref{LAbrokengen}) is not obviously to be expected. The explanation is that the tests of high energy unitarity conducted here assume this only for the specific amplitudes under consideration. It is not immediately required that the entire underlying theory be consistent with a gap, rather only the subset of particles participating in the tree-level amplitudes. This accords with the bootstrap philosophy of necessity instead of sufficiency. The requirement (\ref{LAbrokengen}) (and all other similar conditions derived in this massive amplitude analysis) must hold for a theory in which each of the amplitudes between mass eigenstates contain an upper bound on the exchanged particle masses. This need not necessarily apply to the overarching theory in which these amplitudes exist (although this assumption would be sufficient for these conclusions to hold). This is the reason that KK YM obeys the relations derived here in spite of not being gapped, as well as the reason that it is not perturbatively unitary in the HEL yet still consistent with these results. Similarly, as will be shown in Section \ref{sec:2vec2sca} below, the partial unitarisation of mixed scalar-vector amplitudes is predicated upon the assumption of an upper bound on the participating particle masses, yet the conclusions seem to accommodate theories that require an infinite number of states. Such theories may nevertheless be consistent with the analysis provided that the mass eigenstate amplitudes are gapped. The possible impacts of a tower within the amplitudes will remain, as far as this analysis is concerned, an open question. 

Notably, the interpretation of (\ref{LAbrokengen}) as a Lie algebra commutator does not actually depend upon any further conditions such as compactness or semi-simplicity of the Lie algebra. As long as (\ref{LAbrokengen}) is satisfied, then the purely longitudinal sector of the amplitude unitarises (although the cancellation will leave $\sim E$ divergences in other helicity configurations with a single transverse vector - see below). Presumably small departures from generator normalisation homogeneity do not affect the argument presented above for the necessity of a Higgs boson, but this is less clear if the departures are large. If the Lie algebra is altogether non-compact, then the representations must be infinite dimensional and towers become a necessity. In this case, the transverse sector will inevitably provide a unitarity cut-off anyway. 

Now accepting the condition (\ref{LAbrokengen}) needed for cancellation of the $\sim E^2$ divergence, it remains to combine the longitudinal terms in (\ref{Amp4LE2}) for each channel to make this cancellation explicit. In the HEL, these terms contain kinematic factors corresponding to the Mandelstam variables $s,t$ and $u$, of which only two of the three are linearly independent. The cancellation proceeds by choosing two such Mandelstam variables, converting the Lorentz structures to a form in which they directly converge in the HEL to either of the two Mandelstams selected, grouping the common terms together in the form of (\ref{SymmRed}) and finally repeated application of (\ref{OldSyzFact}). I additionally choose to average over all particle exchanges in order to preserve exchange symmetries. Altogether, this produces e.g. for the $t$-residue terms from the basis of Mandelstams given by $\{t,u\}$,
\begin{align}\label{4L(tu)tres}
&\frac{-1}{36\prod_im_i}\da{\bf{13}}\da{\bf{24}}
\Bigg((d_s-d_t)\left(\frac{\ds{\bf{12}}\ds{\bf{34}}}{s-m_s^2}+\frac{\ds{\bf{13}}\ds{\bf{24}}}{t-m_t^2}\right)-(d_t-d_u)\left(\frac{\ds{\bf{13}}\ds{\bf{24}}}{t-m_t^2}+\frac{\ds{\bf{14}}\ds{\bf{23}}}{u-m_u^2}\right)\nonumber\\
&\qquad\qquad\qquad\qquad\qquad\qquad\qquad\qquad\qquad\qquad\qquad\quad+(d_u-d_s)\left(\frac{\ds{\bf{14}}\ds{\bf{23}}}{u-m_u^2}-\frac{\ds{\bf{12}}\ds{\bf{34}}}{s-m_s^2}\right)\Bigg)\nonumber\\
&\qquad\qquad\qquad\qquad+P-\text{conjugate}\nonumber\\
&=\frac{-1}{36\prod_im_i}\da{\bf{13}}\da{\bf{24}}\nonumber\\
&\qquad\times\Bigg(\frac{1}{12}\sum_i\mathbf{L}_i\left(\frac{(d_s-d_t)}{(s-m_s^2)(t-m_t^2)}+\frac{(d_t-d_u)}{(t-m_t^2)(u-m_u^2)}+\frac{(d_u-d_s)}{(u-m_u^2)(s-m_s^2)}\right)\nonumber\\
&\qquad\quad+\frac{(d_s-d_t)}{(s-m_s^2)(t-m_t^2)}\left(\left(\frac{1}{3}\sum_im_i^2-m_t^2\right)\ds{\bf{12}}\ds{\bf{34}}+\left(\frac{1}{3}\sum_im_i^2-m_s^2\right)\ds{\bf{13}}\ds{\bf{24}}\right)\nonumber\\
&\qquad\quad-\frac{(d_t-d_u)}{(t-m_t^2)(u-m_u^2)}\left(\left(\frac{1}{3}\sum_im_i^2-m_u^2\right)\ds{\bf{13}}\ds{\bf{24}}+\left(\frac{1}{3}\sum_im_i^2-m_t^2\right)\ds{\bf{14}}\ds{\bf{23}}\right)\nonumber\\
&\qquad\quad+\frac{(d_s-d_t)}{(u-m_u^2)(s-m_s^2)}\left(\left(\frac{1}{3}\sum_im_i^2-m_s^2\right)\ds{\bf{14}}\ds{\bf{23}}-\left(\frac{1}{3}\sum_im_i^2-m_u^2\right)\ds{\bf{12}}\ds{\bf{34}}\right)\Bigg)\nonumber\\
&\qquad\qquad\qquad\qquad+P-\text{conjugate},
\end{align}
where
\begin{align}
d_s=A_s+B_s\qquad d_t=-A_t-B_t \qquad d_u=2A_u
\end{align}
with
\begin{align}\label{defAB}
A_s&=\bigg(\frac{1}{3}\left(\left(m_1^2\Re f_{BM}^{\,\,\,\,\,\,\,\,\,A}+m_2^2\Re f_{MA}^{\,\,\,\,\,\,\,\,\,B}\right)\Re f_{CD}^{\,\,\,\,\,\,\,\,\,M}+\Re f_{AB}^{\,\,\,\,\,\,\,\,\,M}\left(m_3^2\Re f_{DM}^{\,\,\,\,\,\,\,\,\,C}+m_4^2\Re f_{MC}^{\,\,\,\,\,\,\,\,\,D}\right)\right)\nonumber\\
&\qquad\qquad\qquad\qquad\qquad\qquad\qquad\qquad\qquad\qquad\qquad\qquad\qquad-m_{s_M}^2\Re f_{AB}^{\,\,\,\,\,\,\,\,\,M}\Re f_{CD}^{\,\,\,\,\,\,\,\,\,M}\bigg)\nonumber\\
B_s&=\frac{1}{m_{s_M}^2}\left(m_1^2\Re f_{BM}^{\,\,\,\,\,\,\,\,\,A}-m_2^2\Re f_{MA}^{\,\,\,\,\,\,\,\,\,B}\right)\left(m_3^2\Re f_{DM}^{\,\,\,\,\,\,\,\,\,C}-m_4^2\Re f_{MC}^{\,\,\,\,\,\,\,\,\,D}\right)+\left(\prod_i m_i\right)\lambda_{AB}^m\lambda_{CD}^m\nonumber\\
&\qquad\qquad A_t=A_s|_{2\leftrightarrow 3}\qquad A_u=A_s|_{1\leftrightarrow 3}\qquad B_t=B_s|_{2\leftrightarrow 3}\qquad B_u=B_s|_{1\leftrightarrow 3}.
\end{align}
Added to the expression (\ref{4L(tu)tres}) should be all terms generated by the five permutations: $3\leftrightarrow 4$, $1\leftrightarrow 3$, $3\leftrightarrow 4$ followed by $1\leftrightarrow 3$, $2\leftrightarrow 3$ and finally 
$3\leftrightarrow 4$ followed by $2\leftrightarrow 3$. This is to complete the full amplitude in a manifestly exchange symmetric form. Again I have used (\ref{OldSyzFact}) to implement the cross-channel cancellations.

\subsubsection{Mixed longitudinal and transverse divergences and covariance}

The final $\sim E^2$ divergent helicity configuration of the amplitude corresponds to two longitudinal and two transverse vectors with the same helicities. This is more interesting than the opposite helicity case identified above. In the same-sign configuration, the amplitude converges to contact terms in the HEL and receives contributions from all three channels. The relevant terms are entirely contained in (\ref{2T2LSamefact}) and the appropriate parts of (\ref{4vecScalar}). Choosing the $(++,++,L,L)$ configuration for example, cancellation of the emerging contact term requires
\begin{align}\label{CovSFEm}
&2\sqrt{2}m_3m_4\lambda^i_{CD}\frac{c^i_{AB}}{\Lambda}-\frac{1}{m_{s_M}^2}\left(f_{BM}^{\,\,\,\,\,\,\,\,\,A}+f_{AM}^{\,\,\,\,\,\,\,\,\,B}\right)\left(m_3^2\Re f_{DM}^{\,\,\,\,\,\,\,\,\,C}+m_4^2\Re f_{CM}^{\,\,\,\,\,\,\,\,\,D}\right)\nonumber\\
&+f_{CM}^{\,\,\,\,\,\,\,\,\,A}f_{DM}^{\,\,\,\,\,\,\,\,\,B}-\Re f_{AC}^{\,\,\,\,\,\,\,\,\,M}\Re f_{BD}^{\,\,\,\,\,\,\,\,\,M}-\Im f_{AC}^{\,\,\,\,\,\,\,\,\,M}\Im f_{BD}^{\,\,\,\,\,\,\,\,\,M}+i\left(\Im f_{AC}^{\,\,\,\,\,\,\,\,\,M}f_{MD}^{\,\,\,\,\,\,\,\,\,B}+\Im f_{BD}^{\,\,\,\,\,\,\,\,\,M}f_{MC}^{\,\,\,\,\,\,\,\,\,A}\right)\nonumber\\
&+f_{CM}^{\,\,\,\,\,\,\,\,\,B}f_{DM}^{\,\,\,\,\,\,\,\,\,A}-\Re f_{BC}^{\,\,\,\,\,\,\,\,\,M}\Re f_{AD}^{\,\,\,\,\,\,\,\,\,M}-\Im f_{BC}^{\,\,\,\,\,\,\,\,\,M}\Im f_{AD}^{\,\,\,\,\,\,\,\,\,M}+i\left(\Im f_{BC}^{\,\,\,\,\,\,\,\,\,M}f_{MD}^{\,\,\,\,\,\,\,\,\,A}+\Im f_{AD}^{\,\,\,\,\,\,\,\,\,M}f_{MC}^{\,\,\,\,\,\,\,\,\,B}\right)\nonumber\\
&=-8i\sqrt{2}\frac{m_3}{\Lambda}\left((t_D)_{Ci}c^i_{AB}+(t_D)_{CM}c^M_{AB}-i\Re f_{DA}^{\,\,\,\,\,\,\,\,\,M}c^C_{MB}-i\Re f_{DB}^{\,\,\,\,\,\,\,\,\,M}c^C_{AM}\right)\nonumber\\
&\qquad+2\left(f_{AD}^{\,\,\,\,\,\,\,\,\,M}-f_{DM}^{\,\,\,\,\,\,\,\,\,A}\right)\left(f_{BC}^{\,\,\,\,\,\,\,\,\,M}-f_{CM}^{\,\,\,\,\,\,\,\,\,B}\right)+2\left(f_{BD}^{\,\,\,\,\,\,\,\,\,M}-f_{DM}^{\,\,\,\,\,\,\,\,\,B}\right)\left(f_{AC}^{\,\,\,\,\,\,\,\,\,M}-f_{CM}^{\,\,\,\,\,\,\,\,\,A}\right)=0.
\end{align}
The identification of the broken generators (\ref{BrokenGenGeneral}) and (\ref{HiggsGen}) have been made here. If the left-hand side of the last line above can be ignored, then (\ref{CovSFEm}) would be the expected equation representing covariance of the $\{c_{AB}^i\}$ couplings under the emergent Lie algebra in the specific case in which the external scalar index $i$ is identified with the longitudinal modes of the vectors. Since the scalars emerging from the longitudinal vector polarisations are expected to combine with the Higgs bosons into complete representation spaces, the internal sum over the scalar coordinates correctly includes terms from both scalar and vector exchange. However, covariance remains to be established when the external scalar index represents a Higgs boson. This will be derived in Section \ref{3Vec1Scalar} below under analogous circumstances in the three vector, one scalar amplitude. 

This covariance appears to be spoiled, however, by the terms in the last line. In the limit that $m_i\ll\Lambda$, these terms become small corrections to the covariance condition of the order $m_i/\Lambda$. Their presence is nevertheless consistent with the order of truncation in an energy expansion implicit in the admission of the $c$-type couplings in (\ref{3legVecHEL}) and (\ref{2vec1scalar}) into the computation of the $4$-particle amplitude. In the HEL, the left-hand side of (\ref{CovSFEm}) is the coefficient of an emergent effective quartic contact amplitude of the form $\frac{i}{\sqrt{2}m_3m_4}\ds{12}^2$. It is therefore consistent to allow this coefficient to be $\mathcal{O}(\frac{m^2}{\Lambda^2})$ instead of exactly zero so that the emergent quartic amplitude becomes $\sim \frac{1}{\Lambda^2}\ds{12}^2$. This would then be of the same order as the contribution of the $c$-type couplings to the fully and partially transverse classes of terms described earlier around (\ref{4Tfact}), (\ref{4vecScalar}) and (\ref{2T2LOpp}). Since the contaminant terms in (\ref{CovSFEm}) are of the same order, their appearance in this emergent quartic coupling is consistent with the power counting and they can be effectively absorbed into its definition. This is in contrast to the emergent Lie algebra commutator (\ref{LAbrokengen}), which is notably free from subleading corrections that would otherwise spoil its form. This is arguably because, as an expansion in $m/\Lambda$, the couplings contributing to the relation (\ref{CovSFEm}) are of order $\sim m/\Lambda$ and the corrections are $\mathcal{O}\left(m^2/\Lambda^2\right)$, whereas the leading order part of (\ref{LAbrokengen}) is $\sim 1$ and corrections arising from emergent $c$-type couplings begin at order $\sim m/\Lambda$. These $m/\Lambda$ corrections are still dominant over the allowable $\mathcal{O}(m^2/\Lambda^2)$ truncation error, so must also be part of the cancellation. However, since each term in these constraints is a quadratic and the subleading terms arise as corrections to couplings in each factor, the form of the $m^2/\Lambda^2$ terms in (\ref{LAbrokengen}) is fixed and necessarily accounted for in the definitions of the emergent generators in the same way as the $m/\Lambda$ terms.

Converting the terms in the amplitude contributing to this limit to the form (\ref{SymmRed}), the divergence can be explicitly canceled by another application of (\ref{OldSyzFact}):
\begin{align}\label{++LLred}
&\frac{-1}{24m_3m_4}\ds{\bf{12}}\da{\bf{34}}\Bigg(\left(C_s-C_t\right)\left(\frac{\ds{\bf{12}}\ds{\bf{34}}}{s-m_s^2}+\frac{\ds{\bf{13}}\ds{\bf{24}}}{t-m_t^2}\right)-\left(C_t-C_u\right)\left(\frac{\ds{\bf{13}}\ds{\bf{24}}}{t-m_t^2}+\frac{\ds{\bf{14}}\ds{\bf{23}}}{u-m_u^2}\right)\nonumber\\
&\qquad\qquad\qquad\qquad\qquad\qquad\qquad\qquad\qquad\qquad\qquad+\left(C_u-C_s\right)\left(\frac{\ds{\bf{14}}\ds{\bf{23}}}{u-m_u^2}-\frac{\ds{\bf{12}}\ds{\bf{34}}}{s-m_s^2}\right)\Bigg)
\end{align}
where here $C_s$, $C_t$ and $C_u$ are given respectively by the first, second and third lines of (\ref{CovSFEm}) (assuming perfect cancellation and dropping the possible $\mathcal{O}(\frac{m}{\Lambda})$ corrections from a suppressed emergent quartic coupling). The cancellation proceeds as usual by applying (\ref{OldSyzFact}). However, the remaining dimension $5$ Lorentz structures are ideally decomposed into a uniform basis across the amplitude. The relation (\ref{LtoLhat}) demonstrates that there is some redundancy among these Lorentz structures, but it can also be used to convert them into a common basis in which the spinor prefactors accompanying the $\mathbf{L}_i$ structures are as uniform in chirality as possible (this happened to be automatic for the terms computed from (\ref{Cat2b}) and  (\ref{JacRed}) using (\ref{LLhatDiff}) and (\ref{LLhatSame}), as well as in reducing the purely longitudinal terms (\ref{4L(tu)tres})). This issue of basis choice also arose in the treatment of the $F^3$ contributions in Appendix \ref{app:4V} where it was similarly addressed. So applying (\ref{OldSyzFact}) and (\ref{LtoLhat}) to the kinematic structures here gives, for the $s$ and $t$ pairing term in (\ref{++LLred}):
\begin{align}\label{++LLredAlgebra}
&\ds{\bf{12}}\da{\bf{34}}\left(\frac{\ds{\bf{12}}\ds{\bf{34}}}{s-m_s^2}+\frac{\ds{\bf{13}}\ds{\bf{24}}}{t-m_t^2}\right)\nonumber\\
&=\frac{1}{(s-m_s^2)(t-m_t^2)}\nonumber\\
&\qquad\times\Bigg(\frac{1}{12}\ds{\bf{12}}\da{\bf{34}}\left(\mathbf{L}_3+\mathbf{L}_4\right)+\frac{1}{12}\ds{\bf{12}}\ds{\bf{34}}\left(\frac{m_2}{m_1}\hat{\mathbf{L}}_1+\frac{m_1}{m_2}\hat{\mathbf{L}}_2\right)\nonumber\\
&\qquad\qquad+\frac{1}{3}m_1m_2\ds{\bf{12}}\ds{\bf{34}}\left(\da{\bf{13}}\da{\bf{24}}+\da{\bf{14}}\da{\bf{23}}\right)\nonumber\\
&\qquad\qquad-\frac{1}{6}\left(m_1^2+m_2^2\right)\ds{\bf{12}}\da{\bf{34}}\left(\ds{\bf{13}}\ds{\bf{24}}+\ds{\bf{14}}\ds{\bf{23}}\right)\nonumber\\
&\qquad\qquad+\frac{1}{2}\ds{\bf{12}}\left(m_3\da{\bf{34}}-m_4\ds{\bf{34}}\right)\left(m_1\da{\bf{13}}\ds{\bf{24}}+m_2\da{\bf{23}}\ds{\bf{14}}\right)\nonumber\\
&\qquad\qquad+\frac{1}{2}\ds{\bf{12}}\left(m_4\da{\bf{34}}-m_3\ds{\bf{34}}\right)\left(m_1\da{\bf{14}}\ds{\bf{23}}+m_2\da{\bf{24}}\ds{\bf{13}}\right)\nonumber\\
&\qquad\qquad\qquad-\left(m_s^2-\frac{1}{3}\sum_jm_j^2\right)\ds{\bf{13}}\ds{\bf{24}}-\left(m_t^2-\frac{1}{3}\sum_jm_j^2\right)\ds{\bf{12}}\ds{\bf{34}}\Bigg).
\end{align}
Besides the last line, the numerator factors, including the remaining HEL divergent terms, are common to all of the channel pairings in (\ref{++LLred}).

This cancellation leaves residual $\sim E$ divergences for two classes of helicity configurations. The first class consists of $(++,++,L,--)$ and $(++,++,--,L)$ and the second class of $(++,L,L,L)$ and $(L,++,L,L)$. The first class of terms, consisting of two same-sign transverse, one opposite sign and one longitudinal polarisation, receive no contributions from any other sources in the amplitude. Choosing e.g. $(++,++,L,--)$, they are given by the terms above proportional to the kinematic structure $\ds{\bf{12}}\da{\bf{34}}\mathbf{L}_4$. In the HEL, these directly match onto the expected massless amplitudes involving single insertions of the $c$-type couplings. The $s$-channel is generated by a scalar exchange while the $t$ and $u$-channels arise from gluon exchanges. Each term in the corresponding massless amplitude contains two Mandelstam poles, which is responsible for the covariance constraint emerging from application of the $4$-particle test to the corresponding massless theory. Again, these divergent terms can only be eliminated from the massive amplitude by eliminating the $c$-type couplings.  

The remaining $\sim E$ divergences occur for helicity configurations with three longitudinal helicities and one transverse. Contributions to these are produced in both (\ref{4LOddfact}) and (\ref{Cat2a}) with (\ref{UnpackedGCSquart}), both (\ref{JacRed}) and (\ref{Cat2b}) with both (\ref{LLhatDiff}) and (\ref{LLhatSame}), (\ref{4L(tu)tres}) and finally (\ref{++LLred}) with (\ref{++LLredAlgebra}). This is also the first juncture at which the mass splitting
terms in (\ref{HomoDen}) must be included. In the HEL, these all combine to converge to massless amplitudes with three external scalars and one gluon that are generated by single insertions of dim-$5$ couplings. These massless amplitudes consist of terms with only single Mandelstam poles representing internal gluon exchanges. Taking, for example, the $(L,L,L,++)$ helicity configuration, the massive kinematic structure that converges to the $s$-channel massless amplitude is
\begin{align}
\frac{1}{s-m_s^2}\hat{\mathbf{L}}_4\frac{1}{2}\left(\frac{\ds{\bf{13}}\ds{\bf{24}}}{t-m_t^2}-\frac{\ds{\bf{23}}\ds{\bf{14}}}{u-m_u^2}\right)\rightarrow \frac{12\ds{24}\ds{34}\da{23}}{s}.
\end{align}
The other channels and helicity configurations can be obtained by particle exchanges, as usual. 

While I will mostly ignore the mass splitting terms in (\ref{HomoDen}) for this study, they nevertheless participate in the cancellations, typically serving to replace the dummy internal mass parameters with their physical values where required in the unitarised expressions. However, unlike the amplitudes computed in the subsequent Sections, the mass splitting terms here include divergent $\sim E$ terms, arising from (\ref{E34Vec}). For this reason, I explicitly include them in the cancellation among the $\sim E$ terms. This cancellation involves the terms descending from the reduction of the $\sim E^3$ divergences arranged by the Jacobi and GCS quartic constraints, which are presented in (\ref{JacRed}), (\ref{Cat2a}) and (\ref{Cat2b}) and using (\ref{LLhatDiff}) and (\ref{LLhatSame}). Combining the relevant terms with those describing the mass splitting produces
\begin{align}
&\frac{m_s^2}{48\prod_i m_i}\Re f_{AB}^{\,\,\,\,\,\,\,\,\,M}\left(\Re f_{CM}^{\,\,\,\,\,\,\,\,\,D}+iF_{AB;C}^{\,\,\,\,\,\,\,\,\,\,\,\,\,\,\,D}\right)\frac{\hat{\mathbf{L}}_4}{s-m_s^2}\left(\frac{\ds{\bf{13}}\ds{\bf{24}}}{t-m_t^2}-\frac{\ds{\bf{23}}\ds{\bf{14}}}{u-m_u^2}\right)\nonumber\\
&-\frac{(m_{s_M}^2-m_s^2)}{24\prod_i m_i}\Re f_{AB}^{\,\,\,\,\,\,\,\,\,M}\left(\Re f_{CM}^{\,\,\,\,\,\,\,\,\,D}+iF_{AB;C}^{\,\,\,\,\,\,\,\,\,\,\,\,\,\,\,D}\right)\frac{\hat{\mathbf{L}}_4\ds{\bf{12}}\ds{\bf{34}}}{(s-m_s^2)(s-m_{s_M}^2)}\nonumber\\
&=\frac{m_{s_M}^2}{48\prod_i m_i}\Re f_{AB}^{\,\,\,\,\,\,\,\,\,M}\left(\Re f_{CM}^{\,\,\,\,\,\,\,\,\,D}+iF_{AB;C}^{\,\,\,\,\,\,\,\,\,\,\,\,\,\,\,D}\right)\frac{\hat{\mathbf{L}}_4}{s-m_s^2}\left(\frac{\ds{\bf{13}}\ds{\bf{24}}}{t-m_t^2}-\frac{\ds{\bf{23}}\ds{\bf{14}}}{u-m_u^2}\right)\nonumber\\
&\qquad+\frac{1}{48\prod_i m_i}\frac{m_s^2-m_{s_M}^2}{s-m_s^2}\Re f_{AB}^{\,\,\,\,\,\,\,\,\,M}\left(\Re f_{CM}^{\,\,\,\,\,\,\,\,\,D}+iF_{AB;C}^{\,\,\,\,\,\,\,\,\,\,\,\,\,\,\,D}\right)\nonumber\\
&\qquad\times\Bigg(\frac{1}{12}\sum _j\mathbf{L}_j\frac{1}{s-m_{s_M}^2}\left(\frac{1}{t-m_t^2}-\frac{1}{u-m_u^2}\right)+\frac{\ds{\bf{12}}\ds{\bf{34}}\left(\frac{2}{3}\sum_jm_j^2-m_t^2-m_u^2\right)}{s-m_{s_M}^2}\nonumber\\
&\qquad\qquad\qquad\qquad\qquad\qquad\qquad\qquad+\left(m_{s_M}^2-\frac{1}{3}\sum_jm_j^2\right)\left(\frac{\ds{\bf{13}}\ds{\bf{24}}}{t-m_t^2}-\frac{\ds{\bf{23}}\ds{\bf{14}}}{u-m_u^2}\right)\Bigg),
\end{align}
again using (\ref{OldSyzFact}). As expected, the mass splitting terms replace the dummy internal masses with their physical values in the leading order diverging terms (as is clear from the first lines on the left and right-hand sides of the equation) while leaving behind a more mass-suppressed residual. 

Further mass splitting terms are expected from directly applying (\ref{HomoDen}) to the remaining terms in the original amplitude presented in (\ref{4Tfact}), (\ref{2T2LOppfact}), (\ref{2T2LSamefact}), (\ref{4LEvenfact}) and (\ref{4LOddfact}). These are expected to cancel against the $\sim E^0$ terms in the amplitude in a similar way to that just described for the $\sim E$ terms, in which they serve to replace the dummy internal masses with their physical values and leave mass-suppressed residuals. The restoration of the physical masses to the numerators ensures that the amplitude matches correctly onto the expected massless amplitudes in the HEL (see further below) with the correct dependence on the emergent broken generators. I will not bother to demonstrate the further mass splitting related cancellations involving the non-divergent terms for this amplitude and those that follow. However, it can be cross-checked that these terms correctly modify the expected residues in the HEL. 

Now, collecting together the complete set of contributions to the $(L,L,L,++)$ divergent terms produces
\begin{align}
&\frac{1}{48\prod_im_i}\frac{\hat{\mathbf{L}}_4}{s-m_s^2}\left(\frac{\ds{\bf{13}}\ds{\bf{24}}}{t-m_t^2}-\frac{\ds{\bf{23}}\ds{\bf{14}}}{u-m_u^2}\right)\nonumber\\
&\bigg(\left(m_{s_M}^2\Re f_{AB}^{\,\,\,\,\,\,\,\,\,M}-\frac{1}{2}m_1^2\left(\Re f_{AB}^{\,\,\,\,\,\,\,\,\,M}+\Re f_{BM}^{\,\,\,\,\,\,\,\,\,A}\right)-\frac{1}{2}m_2^2\left(\Re f_{AB}^{\,\,\,\,\,\,\,\,\,M}+\Re f_{MA}^{\,\,\,\,\,\,\,\,\,B}\right)\right)\left(f_{CD}^{\,\,\,\,\,\,\,\,\,M}-f_{MC}^{\,\,\,\,\,\,\,\,\,D}\right)\nonumber\\
&\qquad\qquad-\frac{1}{2}\left(m_1^2\left(\Im f_{AB}^{\,\,\,\,\,\,\,\,\,M}-\Im f_{BM}^{\,\,\,\,\,\,\,\,\,A}\right)+m_2^2\left(\Im f_{AB}^{\,\,\,\,\,\,\,\,\,M}-\Im f_{MA}^{\,\,\,\,\,\,\,\,\,B}\right)\right)\left(\Im f_{CD}^{\,\,\,\,\,\,\,\,\,M}-\Im f_{MC}^{\,\,\,\,\,\,\,\,\,D}\right)\bigg).
\end{align}
The coupling constants in the last two lines can be arranged as
\begin{align}
&4m_1m_2m_3\Bigg(\left(-i\sqrt{2}(t_M)_{AB}+m_1\frac{c^B_{AM}}{\Lambda}-m_2\frac{c^A_{BM}}{\Lambda}\right)\frac{c^C_{MD}}{\Lambda}\nonumber\\
&\qquad\qquad\qquad\qquad\qquad\qquad\qquad\qquad\qquad+\left(m_1\frac{\Im c^B_{AM}}{\Lambda}-m_2\frac{\Im c^A_{BM}}{\Lambda}\right)\frac{\Im c^C_{MD}}{\Lambda}\Bigg),
\end{align}
which agrees with the expected HEL once the $\mathcal{O}(m/\Lambda)$ corrections are neglected (all of the terms above with two factors of $c$-type couplings). It is now clear that full unitarisation of the amplitude is only possible if the $c$-type couplings are eliminated or, equivalent, full antisymmetry (and reality) is restored to the vector self-couplings (\ref{StandardLA}).

In summary, the standard antisymmetry of the vector self-couplings (\ref{StandardLA}) (and the elimination of the $c^i_{AB}$ vector-scalar couplings) is responsible for fully unitarising the transverse sectors of the amplitude. Violation of these conditions produces $\sim E^2$ divergences for some helicity configurations and $\sim E$ divergences for others (if the offending couplings are approximately covariant (\ref{CovSFEm})). In contrast, (\ref{LAbrokengen}) unitarises the fully longitudinal sector and is specifically responsible for eliminating only the purely longitudinal $\sim E^2$ divergences. So assuming (\ref{StandardLA}), then the further cancellation of the pure longitudinal $\sim E^2$ divergences is automatically sufficient for full unitarisation. This is the traditional Higgs mechanism and its success or failure in the real world has been the subject of extensive study in the context of electroweak symmetry breaking and departures from the Standard Model \cite{Pich:2018ltt,Brivio:2017vri,Contino:2010rs}. Without (\ref{LAbrokengen}), this theory converges to a NL$\Sigma$M model of Goldstone bosons coupled to standard YM. Restoration of (\ref{LAbrokengen}), along with the assumption of a mass gap, allows for full unitarisation at high energies and the UV emergence of a linear sigma model coupled to regular YM. 

Alternatively, it is possible to instead have (potentially small) violations of (\ref{StandardLA}) accompanying the longitudinal cancellations enforced by (\ref{LAbrokengen}). If the energy can be extrapolated hierarchically beyond the particle masses unobstructed (corresponding to $\Lambda\gg m$), then this theory should either converge to spontaneously broken YM supplemented with ``irrelevant'' deformations or, if the undeformed theory is (approximately) Abelian, a NL$\Sigma$M with scalar manifold isometries potentially free of the geometric restrictions required for the YM Lie algebra (as frequently arise in models of extended SUGRA). These scenarios have been left open throughout the analysis of this Section so far. I have explicitly presented the divergent Lorentz structures that match onto the expected massless amplitudes generated by ``higher dimensional operators'' of the form $\varphi F^2$. However, I will not bother compiling the remaining subleading terms of the amplitude under these conditions, but note that all of the necessary components are implicit in the expressions stated in the pages above. The interested reader can organise and collect them together if they please. I will instead proceed under the assumption of full unitarisation enforced by both 
(\ref{LAbrokengen}) and (\ref{StandardLA}) and present the final expression for the complete, manifestly unitarised amplitude.

\subsubsection{The fully unitarised amplitude and final cancellations}

The fully unitarised amplitude can be grouped into terms that match onto specific classes of helicity amplitudes in the HEL, with respect to which they can be compared and cross-checked. There are three groups of non-zero terms. It is convenient to abbreviate the following prefactor containing colour and singularity structure:
\begin{align}
C&=\frac{-1}{3}\Bigg(\frac{f_{ABM}f_{MCD}-f_{CAM}f_{MBD}}{(s-m_s^2)(t-m_t^2)}+\frac{f_{CAM}f_{MBD}-f_{BCM}f_{MAD}}{(t-m_t^2)(u-m_u^2)}\nonumber\\
&\qquad\qquad\qquad\qquad\qquad\qquad\qquad\qquad\qquad\qquad+\frac{f_{BCM}f_{MAD}-f_{ABM}f_{MCD}}{(u-m_u^2)(s-m_s^2)}\Bigg)
\end{align}
The first type of terms involves two pairs of opposite sign transverse polarisations e.g. $(--,--,++,++)$:
\begin{align}
A_{4T}(W_A,W_B,W_C,W_D)=C \da{\bf{12}}^2\ds{\bf{34}}^2.
\end{align}
These terms originate entirely from (\ref{JacRed}). They clearly correspond to different MHV configurations of Yang-Mills gluons in the HEL. Terms matching onto other fully transverse helicity configurations are simply obtained by particle permutations. 

The second type involves two opposite sign transverse polarisations and two longitudinal polarisations. It is useful to retain the definitions (\ref{defAB}) of some of the combinations of couplings and masses descending from the reduced $\sim E^2$ fully longitudinal terms:
\begin{align}
B_s&=f_{ABM}f_{MCD}\frac{1}{m_{s_M}^2}(m_1^2-m_2^2)(m_3^2-m_4^2)+\left(\prod_im_i\right)\lambda^m_{AB}\lambda^m_{CD}\nonumber\\
B_t&=B_s|_{2\leftrightarrow 3},\qquad B_u =B_s|_{1\leftrightarrow 3}\nonumber\\
B&=f_{ABM}f_{MCD}(m_u^2-m_t^2)-B_s\nonumber\\
&\qquad\qquad\qquad\qquad\qquad+(1\leftrightarrow 3)+(2\leftrightarrow 3).
\end{align}
For the special cases of SYM broken by Coulomb branch vevs, such as $\mathcal{N}=4$ SYM, $B=0$ by both the BPS condition and central charge conservation. The Compton-like Lorentz structures in the amplitude are e.g. for $(--,++,L,L)$, 
\begin{align}
&A_{2T2L}(W_A,W_B,W_C,W_D)\nonumber\\
&=\left(-C\left(m_s^2-m_3^2-m_4^2\right)-\frac{1}{3}\left(\mu^2\left(f_{BCM}f_{MAD}-f_{CAM}f_{MBD}\right)+\frac{1}{2}B\right)\frac{1}{(t-m_t^2)(u-m_u^2)}\right)\nonumber\\
&\qquad\qquad\qquad\qquad\qquad\qquad\qquad\qquad\qquad\qquad\qquad\qquad\times\frac{1}{m_3m_4}
\da{\bf{13}}\da{\bf{14}}\ds{\bf{23}}\ds{\bf{24}}.
\end{align}
Terms matching onto other spin configurations may be obtained by permutations. Again, for BPS kinematics, $\mu^2=0$ and only the term proportional to $C$ survives. In general however, using the emergent Lie algebra representations (\ref{EmLA}), it can be verified that these residues and numerators have the expected form of the couplings for Compton scattering of gluons off scalar matter in the HEL (with generators given by (\ref{BrokenGen})). 

The remaining $\sim E^0$ terms are those in which the leading energy dependence occurs for fully longitudinal polarisations. The remaining residuals from decomposing (\ref{JacRed}) combine with those from (\ref{4L(tu)tres}) to produce these terms:
\begin{align}\label{Amp4LE0}
&A_{4L}(W_A,W_B,W_C,W_D)\nonumber\\
&=\frac{1}{6\prod_j m_j}\frac{1}{(s-m_s^2)(u-m_u^2)}\mathbf{P}_t
\nonumber\\
&\quad\times\Bigg(\frac{1}{9}\left(f_{BCM}f_{MAD}-f_{ABM}f_{MCD}\right)
\Bigg(\sum_im_i^4+3\sum_im_i^2\left(m_s^2+m_u^2-m_t^2\right)\nonumber\\
&\qquad\qquad-2\left(8\left((m_1m_3)^2+(m_2m_4)^2\right)-(m_1m_2)^2-(m_3m_4)^2-(m_2m_3)^2-(m_1m_4)^2\right)\Bigg)\nonumber\\
&\qquad\qquad+\frac{1}{24}B\sum_im_i^2\nonumber\\
&\qquad\qquad+\frac{1}{2}\Bigg(\left(m_s^2-\frac{1}{3}\sum_im_i^2\right)\Bigg(f_{ABM}f_{CDM}\left(m_{s_M}^2-\frac{1}{3}\sum_im_i^2\right)\nonumber\\
&\qquad\qquad\qquad\qquad\qquad\qquad\qquad\qquad\qquad+2f_{BCM}f_{ADM}\left(m_{u_M}^2-\frac{1}{3}\sum_im_i^2\right)-B_s\Bigg)\nonumber\\
&\qquad\qquad\qquad-\left(m_u^2-\frac{1}{3}\sum_im_i^2\right)\Bigg(2f_{ABM}f_{CDM}\left(m_{s_M}^2-\frac{1}{3}\sum_im_i^2\right)\nonumber\\
&\qquad\qquad\qquad\qquad\qquad\qquad\qquad\qquad\qquad+f_{BCM}f_{ADM}\left(m_{u_M}^2-\frac{1}{3}\sum_im_i^2\right)+B_u\Bigg)\Bigg)\Bigg)\nonumber\\
&\quad+\frac{1}{6\prod_jm_j}\frac{1}{(s-m_s^2)(t-m_t^2)}\mathbf{P}_t\nonumber\\
&\quad\times\Bigg(\frac{1}{9}\left(f_{ABM}f_{MCD}-f_{CAM}f_{MBD}\right)\Bigg(\sum_im_i^4+3\sum_im_i^2\left(m_s^2+m_u^2-m_t^2\right)\nonumber\\
&\qquad\qquad-2\left(8\left((m_1m_3)^2+(m_2m_4)^2\right)-(m_1m_2)^2-(m_3m_4)^2-(m_2m_3)^2-(m_1m_4)^2\right)\Bigg)\nonumber\\
&\qquad\qquad-\frac{1}{24}B\sum_im_i^2\nonumber\\
&\qquad\qquad-\frac{1}{2}\Bigg(\left(m_t^2-\frac{1}{3}\sum_im_i^2\right)\Bigg(2f_{ABM}f_{CDM}\left(m_{s_M}^2-\frac{1}{3}\sum_im_i^2\right)\nonumber\\
&\qquad\qquad\qquad\qquad\qquad\qquad\qquad\qquad\qquad\quad+f_{CAM}f_{BDM}\left(m_{t_M}^2-\frac{1}{3}\sum_im_i^2\right)-B_t\Bigg)\nonumber\\
&\qquad\qquad\qquad+\left(m_s^2-\frac{1}{3}\sum_im_i^2\right)\Bigg(3f_{ABM}f_{CDM}\left(m_{s_M}^2-\frac{1}{3}\sum_im_i^2\right)-B_s-2B_t\Bigg)\Bigg)\Bigg)
\end{align}
plus all distinct terms obtained by particle exchanges. 

As explained in the derivation and discussion around the photon exchange amplitude for scalar QED (\ref{sQEDRes}), the terms associated with factorisation from photon exchange and the contact terms both have the same high-energy dependence and any distinction between the two is simply basis dependent. However, a basis in which particle exchange symmetries are manifest is nevertheless a natural choice. In this basis, it can be verified in the HEL that (\ref{Amp4LE0}) matches onto the required scalar QCD amplitudes with generators given by (\ref{BrokenGen}), while the emergent contact term is 
\begin{align}
&A_c(W_A^L,W^L_B,W^L_C,W^L_D)\nonumber\\
&\qquad\rightarrow\frac{1}{8\prod_im_i}\left(f_{ABM}f_{MCD}\big((m_1m_3)^2+(m_2m_4)^2-(m_1m_4)^2-(m_2m_3)^2\right)\nonumber\\
&\qquad\qquad\qquad\qquad+f_{CAM}f_{MBD}\left((m_1m_4)^2+(m_2m_3)^2-(m_1m_2)^2-(m_3m_4)^2\right)\nonumber\\
&\qquad\qquad\qquad\qquad+f_{BCM}f_{MAD}\left((m_1m_2)^2+(m_3m_4)^2-(m_1m_3)^2-(m_2m_4)^2\right)\nonumber\\
&\qquad\qquad\qquad\qquad\qquad\qquad\qquad\qquad\qquad\qquad\qquad\qquad-\left(m_s^2B_s+m_t^2B_t+m_u^2B_u\right)\big).
\end{align}

The final set of terms constituting the amplitude are those that vanish in the HEL: 
\begin{align}
&A_{\mathcal{O}(1/E)}(W_A,W_B,W_C,W_D)\nonumber\\
&=\frac{1}{54\prod_i m_i}\frac{1}{(s-m_s^2)(t-m_t^2)(u-m_u^2)}
\nonumber\\
&\quad\times\Bigg(\frac{-1}{2}B\Bigg(-18\prod_jm_j\da{\bf{12}}^2\ds{\bf{34}}^2+\text{ex.}\nonumber\\
&\quad\qquad\qquad\quad+18m_1m_2\left(m_s^2-m_3^2-m_4^2\right)\ds{\bf{13}}\ds{\bf{14}}\da{\bf{23}}\da{\bf{24}}+\text{ex.}\nonumber\\
&\quad\qquad\qquad\quad-2(m_1m_2)^2\left(8\mathbf{P}_s-\mathbf{P}_t-\mathbf{P}_u\right)+\text{ex.}\nonumber\\
&\quad\qquad\qquad\quad+3\left(\sum_jm_j^2\right)m_s^2\left(\mathbf{P}_t+\mathbf{P}_u-\mathbf{P}_s\right)+\text{ex.}\nonumber\\
&\quad\qquad\qquad\qquad\qquad\qquad\qquad\quad+\sum_jm_j^4\left(\mathbf{P}_s+\mathbf{P}_t+\mathbf{P}_u\right)\Bigg)\nonumber\\
&\quad\qquad+\frac{1}{8}B\Bigg(\sum_j\hat{\mathbf{L}}_j\big((m_u^2-m_t^2)\ds{\bf{12}}\ds{\bf{34}}-(m_s^2-m_u^2)\ds{\bf{13}}\ds{\bf{24}}\nonumber\\
&\quad\qquad\qquad\qquad\qquad\qquad\qquad\qquad\qquad\qquad+(m_t^2-m_s^2)\ds{\bf{23}}\ds{\bf{14}}\big)+\text{Parity conj.}\Bigg)\nonumber\\
&\quad\qquad+\mu^2\Bigg(\sum_jm_j^2((f_{ABM}f_{MCD}-f_{CAM}f_{MBD})\ds{\bf{12}}\ds{\bf{34}}\nonumber\\
&\qquad\qquad\qquad\qquad\qquad\qquad\qquad+(f_{BCM}f_{MAD}-f_{CAM}f_{MBD})\ds{\bf{14}}\ds{\bf{23}})\nonumber\\
&\quad\qquad\qquad\quad\times\Bigg(\frac{1}{12}\sum_k\hat{\mathbf{L}}_k+\left(m_u^2-\frac{1}{3}\sum_km_k^2\right)\da{\bf{12}}\da{\bf{34}}\nonumber\\
&\qquad\qquad\qquad\qquad\qquad\qquad\qquad\qquad\qquad\qquad\quad-\left(m_s^2-\frac{1}{3}\sum_km_k^2\Bigg)\da{\bf{23}}\da{\bf{14}}\right)\nonumber\\
&\quad\qquad\qquad\quad+\frac{1}{16}\Bigg(f_{CAM}f_{MBD}\left(m_{t_M}^2-\frac{1}{3}\sum_km_k^2-m_t^2+m_u^2\right)\nonumber\\
&\quad\qquad\qquad\qquad\quad\quad+f_{BCM}f_{MAD}\left(m_{u_M}^2-\frac{1}{3}\sum_km_k^2-m_u^2+m_t^2\right)\nonumber\\
&\quad\qquad\qquad\qquad\quad\quad-f_{ABM}f_{MCD}\left(2m_{s_M}^2+3m_t^2+3m_u^2-\frac{8}{3}\sum_km_k^2\right)\Bigg)\sum_j\hat{\mathbf{L}}_j\ds{\bf{12}}\ds{\bf{34}}\nonumber\\
&\qquad\qquad\qquad\qquad\qquad\qquad\qquad\qquad\qquad\qquad\qquad\qquad\qquad\qquad+\text{ex.}+\text{Parity conj.}\Bigg)\Bigg).
\end{align}
As in (\ref{OldSyz2}) (which is where the first set of terms originates), ``$+$ex.'' means add all distinct terms given by particle exchanges. ``Parity conj.'' means add terms obtained by switching bracket shapes and complex conjugating coefficients. These terms all vanish for $\mathcal{N}=4$ SYM. 

The four vector amplitude can be compared to the especially elegant (colour-stripped) expression for $\mathcal{N}=4$ SYM contained in (\ref{N=4SYM}), 
\begin{align}\label{N=44VecComp}
A[W,W,W,W]=\frac{-1}{(t-m_t^2)(u-m_u^2)}\epsilon_\varphi(1,2,3,4)\,\epsilon_{-\varphi}(1,2,3,4),
\end{align}
where 
\begin{align}
\epsilon_\varphi(1,2,3,4)=&\da{\bf{12}}\ds{\bf{34}}e^{i(\varphi_3+\varphi_4)}+\da{\bf{34}}\ds{\bf{12}}e^{i(\varphi_1+\varphi_2)}+\da{\bf{14}}\ds{\bf{23}}e^{i(\varphi_2+\varphi_3)}\nonumber\\
&+\da{\bf{23}}\ds{\bf{14}}e^{i(\varphi_1+\varphi_4)}+\da{\bf{31}}\ds{\bf{24}}e^{i(\varphi_2+\varphi_4)}+\da{\bf{24}}\ds{\bf{31}}e^{i(\varphi_1+\varphi_3)}
\end{align}
and $\epsilon_{-\varphi}(1,2,3,4)$ is analogous but with conjugated mass phases. As for the full superamplitude, the numerator of this component amplitude factorises into a product of analogous $\mathcal{N}=2$ expressions for vector-mediated hypermultiplet scattering (as contained in (\ref{SQEDFact})), each factor with conjugate central charge phases. The general expression for the massive vector amplitude above can be cross-checked against (\ref{N=44VecComp}) for verification. Only agreement with the fully longitudinal terms is not obvious (given the remarks already made above). This can be confirmed by using
\begin{align}
\cos\left(\varphi_1+\varphi_2-\varphi_3-\varphi_4\right)=\frac{-1}{2\prod_i m_i}(m_t^2m_u^2-(m_1m_2)^2-(m_3m_4)^2)
\end{align}
and completing the algebra.

\subsubsection{Coupling hierarchies and the space of theories}

\begin{table}[h!]
\begin{center}
\begin{tabular}{ c || c c c c c}
Growth & Helicity Config & Couplings & Structure & Removal \\
\hline\hline
$E^4$ & $++,++,--,--$ & $\frac{h^2}{\Lambda^4}$ & fact & suppress\\
\hline
 & $++,++,--,L$ & $\frac{h}{\Lambda^2}\frac{c}{\Lambda}$ & fact & suppress\\
\cline{2-5}
$E^3$ & $++,++,++,L$ & $\frac{h}{\Lambda^2}\{\frac{1}{m}f,\frac{c}{\Lambda}\}$ & contact & $F^3$ cov\\
\cline{2-5}
 & $++,L,L,L$ & $\frac{1}{m}f\{\frac{1}{m^2}f^2,\frac{c}{\Lambda}\frac{1}{m}f,\frac{c^2}{\Lambda^2}\}$ & contact & Jacobi and GCS\\
\hline
$\begin{array}{l}
 \\
 \\
E^2
\end{array}$ & $\begin{array}{l}
++,++,++,++ \\
++,++,--,-- \\
++,--,L,L
\end{array}$ & $\frac{c^2}{\Lambda^4}$ & fact & suppress\\
\cline{2-5}
 & $++,++,L,L$ & $\frac{c}{\Lambda}\{\frac{1}{m}t,\frac{c}{\Lambda}\}$ & contact & $\varphi F^2$ cov\\
 \cline{2-5}
 & $L,L,L,L$ & $\frac{1}{m^2}t\{f,t\}$ & contact & Lie algebra reps\\
 \hline
$E$ & $\begin{array}{l}
++,++,--,L \\
++,L,L,L
\end{array}$
 & $\frac{c}{\Lambda} \{f,t\}$ & fact & suppress
\end{tabular}
\end{center}
\caption{Summary of results for the four vector amplitude.}\label{tab:4V}
\end{table}

Table \ref{tab:4V} summarises the power counting and qualitative results of the analysis of the massive four vector amplitude. The listed helicity configuration labels implicitly include those related by both parity and particle exchanges. Typically, $f,t\sim g$, where $g$ is the strength of the gauge coupling. The $c$-type couplings indicated in these tables include those implicit in vector boson self-couplings through (\ref{3legVecHEL}) and (\ref{ctof}). Likewise, I denote by $f$ the standard soft coupling as defined in (\ref{3legVecHEL}). I make this choice in the Table to directly link the parameters to energy growth. The $t$-type couplings are those implicit in both (\ref{BrokenGenGeneral}) and (\ref{HiggsGen}).

More precisely, $\Lambda$ here is the unitarity cut-off scale characteristic of helicity sectors in which the amplitude converges to double insertions of $h$ or $c$-type couplings. Other sectors may potentially impose a lower cut-off, depending upon their associated couplings. ``Suppress'' means make $\Lambda$ hierarchically larger than the particle masses $m$ or possibly other cut-offs like $\sim m/g$ if the theory is not otherwise unitarised (this would be appropriate, for example, in comparing $F^3$ corrections to a gauged NL$\Sigma$M lacking the Higgs mechanism). This ensures that, over some finite energy range, the double insertion terms are suppressed compared to other terms with weaker high energy growth. Alternatives to suppressing $3$-particle couplings are stated if available. Unsurprisingly, they always occur for emergent contact terms where the divergence scales as $\mathcal{O}(E/m)$ (and never occur for ``fact'' terms that converge to massless particle exchanges). Typically, the less suppressed the couplings, the more interesting the theory. It is, however, always possible to have a theory in which the $\mathcal{O}(E/m)$ divergences are also comparably suppressed compared to the $\mathcal{O}(E/\Lambda)$ factorisation terms because of small coupling constants, rather than cancellations. This arises frequently in extended SUGRA. 

For extended SUGRA coupled to a pair of massive BPS vector multiplets, the mass of the vector is proportional to the gauge coupling. In the massless limit, the theory matches onto SUGRA with an otherwise free pair of massless vector multiplets. The vectors (both ``matter'' and those paired with the graviton by supersymmetry) are effectively Abelian, although supersymmetrisation of the minimal gravitational coupling gives them dimension $5$ couplings to the scalar and fermion partners (see e.g. \cite{Trott:2026ozo} for recapitulation of these couplings and further details of this theory). I say ``effectively'' Abelian, as how exactly the gauge coupling is accounted for in this limit is irrelevant, since it becomes a subleading ($g\sim m/M_{Pl}\ll E/M_{Pl}$) correction. The UV cut-off of the emergent NL$\Sigma$M obstructs an exact massless limit with a non-zero non-Abelian gauge coupling, for which the emergent $3$-particle amplitude would contain gauge artifacts. Since the gauge coupling is subleading in this case to the $E/M_{Pl}$ expansion parameter, the vector bosons effectively Abelianise and the theory is approximately the same as the the ungauged theory, which has a well-understood massless structure. 

Table \ref{tab:4V} (in conjunction with the results established below for massive vector scattering off scalars and fermions) offers a general picture of the space of theories: 
\begin{itemize}
\item There is a special class of unitary theories that are valid to arbitrarily high energies (spontaneously broken YM). These are characterised by:
\begin{enumerate}
\item $3$-particle amplitudes that scale as $\sim E$ at high energies.
\item $3$-vector couplings that are Lie algebra structure constants and have full antisymmetry. 
\item The assembly of scalars and fermions (see below) into unitary representations of the Lie algebra. The longitudinal components of the vectors also assemble into a complete representation with some supplementary scalars (Higgs bosons). 
\item Couplings constants that are covariant tensors under these representations.
\end{enumerate}

\item Small deformations to these theories by ``irrelevant operators'':

These induce $4$-particle amplitudes that grow with energy until perturbative unitarity invalidates them. Nevertheless, the energy scale $\Lambda$ at which this happens is much higher than the particle masses, allowing for the theory to describe a large range of energies $m\ll E\ll \Lambda$. The cut-off is also parametrically free and can be consistently taken arbitrarily high (by weakening their effects at low energies). The coupling constants must still transform as covariant tensors of the particle representations. 

\item Large deformations to these theories:
    
These describe the more general class of gauged NL$\Sigma$Ms. Covariance of the interactions with divergent HEL scaling is unnecessary, as is the Higgs mechanism. However, the conditions are otherwise satisfied to ensure that the amplitudes have $\sim E^2$ HEL scaling up to the cut-off. A significant hierarchy between the cut-off and the particle masses can still occur if the couplings are all generally small (which is required for the theory to match onto a sensible massless theory in the HEL). This can occur in extended SUGRA, where the underlying gauge sector is Abelian. However, this need not be the case for general NL$\Sigma$Ms, which can have large quartic couplings between emergent Goldstone bosons or large departures from the standard properties of the Lie algebra in the transverse sector. 

There are four independent types of couplings identified in Table \ref{tab:4V} that are potentially involved in these theories. The first is an underlying non-Abelian gauge coupling and the second consists of the deformations to the standard Lie algebra properties and GCS terms (these can be potentially large compared to the gauge coupling to the point that the algebra can be identified as non-semisimple or non-compact). Then there is also the four particle couplings between Goldstone bosons determined by the degree of failure of the Higgs mechanism and, finally, four particle coupling between two Goldstone bosons and two vectors that is roughly determined by the failure of covariance of the $c$-type couplings. The extended SUGRA examples are cases in which the regular YM vector self-coupling is subleading (and hence approximately Abelian) while the other three are comparable for energies $m\ll E\ll M_{Pl}$. This is generally true if the couplings obey $\Lambda\sim m/g$ and $g\ll 1$. 

With larger vector self-couplings, the non-Abelian YM structure cannot be ignored and the emergent $4$-particle couplings can only be controlled by cancellations (otherwise, they violate unitarity at energies close to the masses). These cancellations require the Higgs mechanism (which is also effectively necessary for the $\varphi F^2$ covariance). In this case, there are effectively two sectors: the transverse and the longitudinal. The transverse sector will force a low cut-off unless the standard YM Lie algebra properties are approximately obeyed, while the longitudinal sector will force a low cut-off if the couplings of the emergent scalars do not arrange into generator representations (which implicitly necessitates the Higgs mechanism, at least if there is a mass gap). The satisfaction of both of these conditions defines the theories described in the first dot point, while the theories for which either fails are presumably a broader class of gauged NL$\Sigma$Ms with a low cut-off. I won't actually demonstrate that these theories are NL$\Sigma$Ms in this study (which would require establishing the $S$-matrix characteristics of Goldstone bosons and spontaneously broken symmetries, see e.g. \cite{Derda:2024jvo,Bertuzzo:2023slg,Brauner:2024juy}), but simply identify agreement with the expected behaviour in the emergent $3$- and $4$-particle amplitudes of longitudinal modes. 

\item Structureless couplings - no Lie algebra, no representations. If the vector boson self-interactions are large, then the cut-off must be low and no consistent UV picture of approximately massless particles exists. If the couplings are suppressed, then it may be possible to match onto a massless theory in the HEL. Neglecting $h$-type couplings, a theory failing the Jacobi identity and non-Abelian GCS constraint would match onto dim-$7$ $4$-leg contact amplitudes in the HEL. These would still possess the $\sim E^3$ energy growth and dominate over massless exchange amplitudes. 
\end{itemize}

Various combinations are possible by selecting only certain subsets of couplings, such as an Abelian theory with multiple photon flavours that have only $h$-type couplings (which would technically fall under the third dot point). Of course, $4$-leg (and higher) massive contact interctions can also be explicitly introduced although, as emphasised above, they have distinctly stronger energy growth than the scenarios listed above and in Table \ref{tab:4V} (barring double $F^3$).

\section{Vector-Scalar Scattering}\label{VecSca}

In the remaining Sections, I will repeat the HEL analysis for mixed amplitudes of massive vector bosons with massive low spin matter ($s\leq\frac{1}{2}$). Both the emergence of the Lie algebra structure of possible matter couplings and the Higgs mechanism are shown to arise from demanding unitarisation in the HEL, just as they did in the old analysis of \cite{Cornwall:1974km,Liu:2022alx} involving only the kinematic structures of renormalisable interactions (this was recently revisited in \cite{Jeong:2026dzt} using the ALT shift of \cite{Ema:2024vww}). However, in extending the analysis to the general set of possible Lorentz structures described in Section \ref{Sec:MassiveVectors}, it is revealed that a partially unitarised regime also exists. This regime remains defined by the exact emergent Lie algebra commutator, but accommodates the presence of the deformations to the Lie algebra properties or GCS terms (that is, $\varphi F^2$-type operator insertions that may induce divergences in the transverse helicity sectors). In contrast, cancellations expected from demanding covariance of higher dimensional $\varphi F^2$-type interactions are shown to be inexact. This establishes the precise hierarchy of constraints on candidate EFTs, in particular leading to the picture of an emerging massless NL$\Sigma$M that a privileged class of theories match onto in the UV. Additionally, I also explicitly perform the cancellations of the divergences to reconcile the constructed $4$-particle amplitudes with their massless counterparts.

\subsection{Two vectors and two scalars}\label{sec:2vec2sca}

To begin with, in this Section, I calculate the mixed scattering of a vector boson off a scalar, first finding the conditions required for the HEL to be softened and unitarised and then reducing the amplitude to a form that explicitly manifests this. The results are cross-checked against the $\mathcal{N}=4$ SYM case contained in (\ref{N=4SYM}) when subject to the simplifying BPS relations between particle masses.

\subsubsection{Scalar exchange amplitudes and Lie algebra structure}

I will initially assume that only the amplitude in (\ref{ScalarMat3leg}) mediates interactions between the vector and scalar particles. The two vector and one scalar amplitudes in (\ref{2vec1scalar}) will be introduced subsequently below and mostly covered in Appendix \ref{app:2V2S}, along with contributions from the $F^3$-type interactions in (\ref{3legVec}). However, with the exception of the pure Higgs-like coupling in (\ref{2vec1scalar}), these interactions induce high energy divergences that are directly ascribable to their presence and can only be softened by their elimination. For this reason I mention them only in the Appendix and neglect their existence in the present analysis. Their impact on the $4$-particle amplitude is self-contained and unremarkably matches expectations. I will likewise do this in the subsequent Sections below for most cases in which the amplitudes are induced by $F^3$, $\varphi F^2$ (excluding the case where $\varphi$ represents a longitudinal vector mode) and fermionic dipole moments and where they otherwise have little impact on the analysis in the main text. An exception will be made in Section \ref{3Vec1Scalar} below, where constraints related to the covariance of the $\varphi F^2$ interaction will be derived and which are connected to the structure of the four vector amplitude analysed in the Section above. 

Accepting these restrictions, all three factorisation channels must still exist and contribute to the amplitude. These are shown in Fig \ref{2V2SOSDi}.
\begin{feynfig}
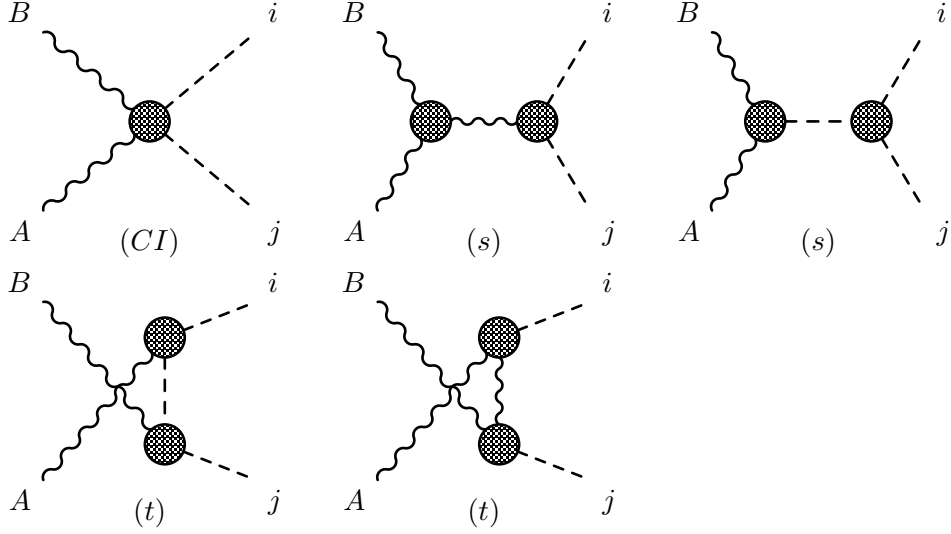
\begin{figure}[h]
\begin{fmffile}{VecScalar}

 \begin{center}
 \begin{tabular}{c c c c c}
 & & & & \\
   \begin{fmfgraph*}(100,67)
   \fmfleft{i1,i2}
   \fmfright{o1,o2}
   \fmf{boson}{i2,v1}
   \fmf{dashes}{v1,o2}
   \fmf{boson}{i1,v1}
   \fmf{dashes}{v1,o1}
   \fmfv{decor.shape=circle,decor.filled=gray50,decor.size=0.15w}{v1}
   \fmflabel{$A$}{i1}
   \fmflabel{$B$}{i2}
   \fmflabel{$i$}{o2}
   \fmflabel{$j$}{o1}
 \end{fmfgraph*} 
 &\,&\begin{fmfgraph*}(100,67)
   \fmfleft{i1,i2}
   \fmfright{o1,o2}
   \fmf{boson}{i2,v1}
   \fmf{dashes}{v2,o2}
   \fmf{boson}{i1,v1}
   \fmf{dashes}{v2,o1}
   \fmf{boson}{v1,v2}
   \fmfv{decor.shape=circle,decor.filled=gray50,decor.size=0.15w}{v1,v2}
   \fmflabel{$A$}{i1}
   \fmflabel{$B$}{i2}
   \fmflabel{$i$}{o2}
   \fmflabel{$j$}{o1}
 \end{fmfgraph*} 
 &\,& \begin{fmfgraph*}(100,67)
   \fmfleft{i1,i2}
   \fmfright{o1,o2}
   \fmf{boson}{i2,v1}
   \fmf{dashes}{v2,o2}
   \fmf{boson}{i1,v1}
   \fmf{dashes}{v2,o1}
   \fmf{dashes}{v1,v2}
   \fmfv{decor.shape=circle,decor.filled=gray50,decor.size=0.15w}{v1,v2}
   \fmflabel{$A$}{i1}
   \fmflabel{$B$}{i2}
   \fmflabel{$i$}{o2}
   \fmflabel{$j$}{o1}
 \end{fmfgraph*} \nonumber\\
$(CI)$ &\,& $(s)$ &\,& $(s)$\nonumber\\
& & & & \nonumber\\
   \begin{fmfgraph*}(100,67)
   \fmfleft{i1,i2}
   \fmfright{o1,o2}
   \fmf{phantom}{i2,v1}
   \fmf{dashes}{v1,o2}
   \fmf{phantom}{i1,v2}
   \fmf{dashes}{v2,o1}
   \fmf{dashes}{v1,v2}
   \fmf{boson,tension=-0.25}{v1,i1}
   \fmf{boson,tension=-0.25}{v2,i2}
   \fmfv{decor.shape=circle,decor.filled=gray50,decor.size=0.15w}{v1,v2}
   \fmflabel{$A$}{i1}
   \fmflabel{$B$}{i2}
   \fmflabel{$i$}{o2}
   \fmflabel{$j$}{o1}
 \end{fmfgraph*} 
 &\,& \begin{fmfgraph*}(100,67)
   \fmfleft{i1,i2}
   \fmfright{o1,o2}
   \fmf{phantom}{i2,v1}
   \fmf{dashes}{v1,o2}
   \fmf{phantom}{i1,v2}
   \fmf{dashes}{v2,o1}
   \fmf{boson}{v1,v2}
   \fmf{boson,tension=-0.25}{v1,i1}
   \fmf{boson,tension=-0.25}{v2,i2}
   \fmfv{decor.shape=circle,decor.filled=gray50,decor.size=0.15w}{v1,v2}
   \fmflabel{$A$}{i1}
   \fmflabel{$B$}{i2}
   \fmflabel{$i$}{o2}
   \fmflabel{$j$}{o1}
 \end{fmfgraph*} &\,& \\
 $(t)$ &\,& $(t)$ &\,&
 \end{tabular}
  \end{center}
\end{fmffile}
\caption{Vector-scalar scattering (omitting the $u$-channel).}\label{2V2SOSDi}
\end{figure}
\end{feynfig}
Unitarity implies that an amplitude with the correct $t$-channel factorisation structure has the form 
\begin{align}\label{2V2SScat}
A_t(W_A,W_B,\varphi_i,\varphi_j)&=(t_A)_{im}(t_B)_{mj}\frac{1}{t-m_{t_m}^2}\frac{1}{m_1m_2}\la{\bf{1}}p_3\rs{\bf{1}}\la{\bf{2}}p_4\rs{\bf{2}},
\end{align}
where $m_{t_m}$ is the mass of exchanged particle $m$ (this label is implicitly summed over). The $u$-channel amplitude is the same, but with the vector boson labels exchanged and obviously $t$ and $m_{t_m}$ replaced by $u$ and $m_{u_m}$. The $s$-channel amplitude is 
\begin{align}\label{ScaVecAmps}
&A_s(W_A,W_B,\varphi_i,\varphi_j)\nonumber\\
&\qquad=
\frac{-i}{4}\frac{1}{s-m_{s_M}^2}\frac{1}{m_{s_M}^2m_1m_2}(t_M)_{ij}\nonumber\\
&\qquad\qquad\times\big(m_{s_M}^2{f}_{AB}^{\,\,\,\,\,\,\,\,\,M}\da{\bf{12}}\left(m_1\la{\bf{1}}p_3-p_4\rs{\bf{2}}+m_2\la{\bf{2}}p_3-p_4\rs{\bf{1}}+\ds{\bf{12}}\left(u-t\right)\right)\nonumber\\
&\qquad\qquad\qquad+m_{s_M}^2({f}_{AB}^{\,\,\,\,\,\,\,\,\,M})^*\ds{\bf{12}}\left(m_1\ls{\bf{1}}p_3-p_4\ra{\bf{2}}+m_2\ls{\bf{2}}p_3-p_4\ra{\bf{1}}+\da{\bf{12}}\left(u-t\right)\right)\nonumber\\
&\qquad\qquad\qquad+\left(m_1{f}_{BM}^{\,\,\,\,\,\,\,\,\,A}\ds{\bf{12}}+m_2({f}_{MA}^{\,\,\,\,\,\,\,\,\,B})^*\da{\bf{12}}\right)\nonumber\\
&\qquad\qquad\qquad\qquad\times\left((m_3^2-m_4^2)\left(m_1\da{\bf{12}}-m_2\ds{\bf{12}}\right)+m_{s_M}^2\ls{\bf{1}}p_3-p_4\ra{\bf{2}}\right)\nonumber\\
&\qquad\qquad\qquad+\left(m_1({f}_{BM}^{\,\,\,\,\,\,\,\,\,A})^*\da{\bf{12}}+m_2{f}_{MA}^{\,\,\,\,\,\,\,\,\,B}\ds{\bf{12}}\right)\nonumber\\
&\qquad\qquad\qquad\qquad\times\left((m_3^2-m_4^2)\left(m_1\ds{\bf{12}}-m_2\da{\bf{12}}\right)+m_{s_M}^2\la{\bf{1}}p_3-p_4\rs{\bf{2}}\right)\big).
\end{align}
The masses $m_{s_M}$ are those of the exchanged vector bosons. The form presented here has manifest exchange symmetries.

These terms describing the factorisation channels all have high-energy scaling $\sim E^2$ when the vectors both have longitudinal helicity, but weaker $\sim E$ scaling for other helicity configurations. Contact terms for this amplitude have weakest high energy scaling $\sim E^2$. This is obvious from the fact that there is a unique, lowest dimension term with this scaling for each of three different helicity configurations. For longitudinal helicity, this term is 
\begin{align}
A_c(W_A,W_B,\varphi_i,\varphi_j)=\frac{c_{AB;ij}}{m_1m_2}\ds{\bf{12}}\da{\bf{12}}
\end{align}
and it contributes to the amplitude at the same order in the HEL as the terms constructed above, so should be included. Here $c_{AB;ij}$ is some coupling constant that is generally free, but also participates in any constraint arising from cancellation in the HEL. 

Taking the high energy limit of the combined set of terms in the amplitude and demanding that they cancel implies that:
\begin{align}
(t_A)_{im}(t_B)_{mj}t+(t_B)_{im}(t_A)_{mj}u+\frac{i}{2}\Re{f}_{AB}^{\,\,\,\,\,\,\,\,\,M}(t_M)_{ij}(u-t)-c_{AB;ij}s=0.
\end{align}
Equating to zero the coeffecients of the two independent Mandlestam variables fixes the contact term:
\begin{align}
c_{AB;ij}=-\frac{1}{2}\left((t_A)_{im}(t_B)_{mj}+(t_B)_{im}(t_A)_{mj}\right)
\end{align}
and subsequently necessitates the constraint:
\begin{align}\label{LAscalar}
(t_A)_{im}(t_B)_{mj}-(t_B)_{im}(t_A)_{mj}=i\Re{f}_{AB}^{\,\,\,\,\,\,\,\,\,M}(t_M)_{ij}.
\end{align}
The scalar-vector couplings must therefore be identified with Lie algebra generators in some representation, accepting that the real parts of the vector self-couplings are structure constants of the Lie algebra. This interpretation is independent of the further conditions needed to unitarise four vector scattering below $\sim E^2$ scaling, and is, in particular, not reliant upon assumptions about the imaginary parts of the vector self-couplings or the signature of the Lie algebra. However, since the generators are self-adjoint, any representation satisfying (\ref{LAscalar}) must be infinite dimensional if the Lie algebra is non-Abelian and not compact.

\subsubsection{High energy cancellations}

With the contact term fixed, the next task is to find a representation of the amplitude in which the diverging HEL terms are explicitly removed. The terms scaling as $\sim E^2$ in the complete amplitude can be grouped into two structures:
\begin{align}\label{AWWPPE2}
A_{E^2}(W_A,W_B,\varphi_i,\varphi_j)&=\frac{1}{2m_1m_2}\left((t_A)_{im}(t_B)_{mj}+(t_B)_{im}(t_A)_{mj}\right)\nonumber\\
&\qquad\times\left(\frac{\la{\bf{1}}p_3\rs{\bf{1}}\la{\bf{2}}p_4\rs{\bf{2}}}{t-m_t^2}+\frac{\la{\bf{1}}p_4\rs{\bf{1}}\la{\bf{2}}p_3\rs{\bf{2}}}{u-m_u^2}-\ds{\bf{12}}\da{\bf{12}}\right)\nonumber\\
&\qquad+\frac{1}{2m_1m_2}i\Re{f}_{AB}^{\,\,\,\,\,\,\,\,\,M}(t_M)_{ij}\nonumber\\
&\qquad\times\left(\frac{\la{\bf{1}}p_3\rs{\bf{1}}\la{\bf{2}}p_4\rs{\bf{2}}}{t-m_t^2}-\frac{\la{\bf{1}}p_4\rs{\bf{1}}\la{\bf{2}}p_3\rs{\bf{2}}}{u-m_u^2}+\frac{t-u}{s-m_s^2}\ds{\bf{12}}\da{\bf{12}}\right).
\end{align}
I have homogenised the masses in the propagators as in (\ref{HomoDen}) and omitted the terms proportional to the mass splittings in each channel, although I will use these briefly further below. The kinematic structure accompanying the anticommutator of generators (the ``Abelian'' part of the amplitude) can be algebraically reduced in order to manifest a unitary HEL. Using identity (\ref{BigSyz}), these terms can be combined as 
\begin{align}\label{MassiveComptAbelian}
&\left(u-m_u^2\right)\left(t-m_t^2\right)\ds{\bf{12}}\da{\bf{12}}-\left(u-m_u^2\right)\la{\bf{1}}p_3\rs{\bf{1}}\la{\bf{2}}p_4\rs{\bf{2}}-\left(t-m_t^2\right)\la{\bf{1}}p_4\rs{\bf{1}}\la{\bf{2}}p_3\rs{\bf{2}}\nonumber\\
&=\left(m_3^2+m_4^2-m_t^2-m_u^2\right)\la{\bf{1}}p_4\rs{\bf{2}}\la{\bf{2}}p_4\rs{\bf{1}}-m_1m_2\left(\la{\bf{1}}p_4\rs{\bf{2}}^2+\la{\bf{2}}p_4\rs{\bf{1}}^2\right)\nonumber\\
&\qquad+m_2\left(m_u^2-m_1^2-m_4^2\right)\left(\ds{\bf{12}}\la{\bf{1}}p_4\rs{\bf{2}}+\da{\bf{12}}\la{\bf{2}}p_4\rs{\bf{1}}\right)\nonumber\\
&\qquad-m_1\left(m_t^2-m_2^2-m_4^2\right)\left(\ds{\bf{12}}\la{\bf{2}}p_4\rs{\bf{1}}+\da{\bf{12}}\la{\bf{1}}p_4\rs{\bf{2}}\right)\nonumber\\
&\qquad+\left((m_1^2+m_2^2-m_3^2)m_4^2+m_t^2m_u^2-m_1^2m_t^2-m_2^2m_u^2+m_1^2m_2^2\right)\ds{\bf{12}}\da{\bf{12}}\nonumber\\
&\qquad-m_1m_2m_4^2\left(\ds{\bf{12}}^2+\da{\bf{12}}^2\right)
\end{align}
over a common denominator of $\frac{-1}{(t-m_t^2)(u-m_u^2)}$. This is now manifestly unitarised. 

The remaining terms in (\ref{AWWPPE2}) accompanying the factor of the structure constant can be likewise reduced. The terms involving explicit appearances of $t$ can be combined as
\begin{align}\label{AmpstComb}
& t\left(t-m_t^2\right)\ds{\bf{12}}\da{\bf{12}}+\left(s-m_s^2\right)\la{\bf{1}}p_3\rs{\bf{1}}\la{\bf{2}}p_4\rs{\bf{2}}\nonumber\\
&=-\left(t-m_t^2\right)\left(m_1\left(\ds{\bf{12}}\la{\bf{2}}p_4\rs{\bf{1}}+\da{\bf{12}}\la{\bf{1}}p_4\rs{\bf{2}}\right)+m_2\left(\ds{\bf{12}}\la{\bf{1}}p_4\rs{\bf{2}}+\da{\bf{12}}\la{\bf{2}}p_4\rs{\bf{1}}\right)\right)\nonumber\\
&\qquad+(2m_2^2+m_3^2+m_4^2-m_t^2-m_s^2)\left(t-m_t^2\right)\ds{\bf{12}}\da{\bf{12}}\nonumber\\
&\qquad+\left(m_s^2-m_1^2-m_2^2\right)\la{\bf{1}}p_4\rs{\bf{2}}\la{\bf{2}}p_4\rs{\bf{1}}-m_1m_2\left(\la{\bf{1}}p_4\rs{\bf{2}}^2+\la{\bf{2}}p_4\rs{\bf{1}}^2\right)\nonumber\\
&\qquad-m_2\left(m_s^2+m_t^2-m_2^2-m_3^2\right)\left(\ds{\bf{12}}\la{\bf{1}}p_4\rs{\bf{2}}+\da{\bf{12}}\la{\bf{2}}p_4\rs{\bf{1}}\right)\nonumber\\
&\qquad-m_1\left(m_t^2-m_2^2-m_4^2\right)\left(\ds{\bf{12}}\la{\bf{2}}p_4\rs{\bf{1}}+\da{\bf{12}}\la{\bf{1}}p_4\rs{\bf{2}}\right)\nonumber\\
&\qquad+\left(m_t^2(2m_2^2+m_3^2+m_4^2-m_t^2-m_s^2)+m_2^2\left(m_s^2-m_2^2-m_3^2\right)+m_4^2\left(m_1^2-m_3^2\right)\right)\ds{\bf{12}}\da{\bf{12}}\nonumber\\
&\qquad-m_4^2m_1m_2\left(\ds{\bf{12}}^2+\da{\bf{12}}^2\right)
\end{align}
multiplied by $\frac{1}{(s-m_s^2)(t-m_t^2)}$. The other terms given by combining $s$ and $u$ pole terms are obtained by exchanging particles $1$ and $2$. 

This leaves divergent terms that scale as $\sim E$ for a single transverse and longitudinally polarised vector. Contact terms with HEL scaling $\sim E$ don't exist, so unitarisation can only involve further constraints on the $3$-particle couplings. The remaining divergent terms in (\ref{AmpstComb}) directly cancel those in (\ref{ScaVecAmps}) provided that the couplings obey the relations:
\begin{align}\label{EqCoup}
\Re f_{AB}^{\,\,\,\,\,\,\,\,\,M}(t_M)_{ij}=f_{AB}^{\,\,\,\,\,\,\,\,\,M}(t_M)_{ij}=f_{BM}^{\,\,\,\,\,\,\,\,\,A}(t_M)_{ij}=f_{MA}^{\,\,\,\,\,\,\,\,\,B}(t_M)_{ij}.
\end{align}
Each adjoint index $A$ denotes a linearly independent vector particle. Momentarily assuming that no other particle exchanges contribute to the amplitude, then the fact that the couplings $\{(t_A)_{ij}\}$ constitute a representation of the Lie algebra implies that they are linearly independent. This subsequently implies (\ref{StandardLA}), which again is just a statement of the standard properties of the YM Lie algebra (i.e. that the structure constants are fully antisymmetric). Of course, this was necessary in the previous Section to unitarise the $4$-vector amplitude below $\sim E^2$ HEL scaling in the first place, so could be reasonably regarded as an assumption here rather than a conclusion. This becomes the necessary point of view when the scalars are Higgs bosons in the calculations further below in Section \ref{3Vec1Scalar}. The reason that these conditions only become necessary at $\sim E$ order here is that, as is clear from the three vector amplitude in the unitarity basis (\ref{3legVecHEL}), the emergent $c$-type couplings can only appear in this amplitude as single insertions. In the HEL, these $\sim E$ terms match onto massless amplitudes involving one external gluon and three external scalars in which a gluon is exchanged between a dim-$5$ coupling and the regular scalar-gluon coupling in (\ref{ScalarMat3leg}). 

The remaining terms in the amplitude after the final cancellation produced by (\ref{EqCoup}) are fully unitarised. The left-over single $s$-pole terms combine to give 
\begin{align}\label{ScaVecSPole}
\frac{1}{2m_1m_2}if_{ABM}(t_M)_{ij}\frac{1}{s-m_s^2}\ds{\bf{12}}\da{\bf{12}}\left((m_u^2-m_t^2)+\frac{1}{m_{s_M}^2}(m_3^2-m_4^2)(m_2^2-m_1^2)\right).
\end{align}
An $s$-channel exchange of a Higgs boson is consistent with unitarity in itself and does not modify the argument above. It can contribute a term 
\begin{align}\label{ScaVecHiggs}
A_h(W_A,W_B,\varphi_i,\varphi_j)=\frac{C_{ijm}\lambda_{AB}^m}{s-m_{h_m}^2}\frac{1}{\sqrt{2}}\ds{\bf{12}}\da{\bf{12}},
\end{align}
where $C_{ijm}$ is a possible scalar trilinear coupling and $m_{h_m}$ is the Higgs mass. In the high energy limit, this combination of coupling constants is reinterpreted as a quartic coupling between the scalars and the longitudinal modes of the vectors. 

There still remains some algebraic rearrangement necessary to reconcile this amplitude with the corresponding massless amplitudes that are expected to emerge in the HEL. For both vectors transversely polarised, this is already manifest. The terms proportional to $\la{\bf{1}}p_4\rs{\bf{2}}^2$ and $\la{\bf{2}}p_4\rs{\bf{1}}^2$ obtained by combining (\ref{AWWPPE2}) with (\ref{MassiveComptAbelian}) and (\ref{AmpstComb}) already clearly match onto the expected massless expressions for Compton scattering of gluons off a scalar. However, the case where the vectors are both longitudinally polarised requires more verification. 

The longitudinal case involves two sets of terms in the HEL: those describing internal particle exchange and the contact terms (see Section \ref{sec:QED} for discussion of this in the QED case). Firstly are the terms that retain a pole in the HEL. These are given by the first term in the third line of (\ref{AmpstComb}) (when substituted into (\ref{AWWPPE2})). However, there is also a further contribution from the mass splitting terms. These are given by the contributions from the leading-order terms in (\ref{ScaVecAmps}) to the mass splitting terms in (\ref{HomoDen}), which clearly share the same coupling constant prefactors assuming (\ref{StandardLA}). Combining together and reducing these two sets of terms requires another application of the spinor identity underpinning (\ref{AmpstComb}). I will not bother to present this here, but simply make note that the leading order $\sim E^0$ terms are simply modified by having $m_s^2$ replaced with $m_{s_M}^2$ in (\ref{AmpstComb}). These then match onto the expected expression for gluon exchange between scalars in scalar QCD (as can be seen from taking the massless limit of (\ref{sQCD}) with equal mass scalars in Appendix \ref{app:LowSpinGen}). Here, the emergent generator (\ref{BrokenGen}) appears describing the coupling of the internal vector to the two emergent longitudinal scalars. 

This leaves the terms that produce a quartic contact interaction between scalars in the HEL. The scalar quartic amplitude so obtained is 
\begin{align}
&A_c(W_A^L,W_B^L,\varphi_i,\varphi_j)\nonumber\\
&\rightarrow\frac{\sqrt{2}}{4}C_{ijm}\lambda_{AB}^m+\frac{i}{4m_1m_2}f_{ABM}(t_M)_{ij}\frac{1}{m_{s_M}^2}(m_3^2-m_4^2)(m_2^2-m_1^2)\nonumber\\
&\qquad+\frac{1}{4m_1m_2}\big((m_3^2+m_4^2)\left((t_A)_{im}(t_B)_{mj}+(t_B)_{im}(t_A)_{mj}\right)\nonumber\\
&\qquad\qquad\qquad\qquad\qquad\qquad\qquad-2m_{t_m}^2(t_A)_{im}(t_B)_{mj}-2m_{u_m}^2(t_B)_{im}(t_A)_{mj}\big).
\end{align}
The terms in the last line are supplied by the mass splitting terms canceling the internal mass terms appearing in the first line of (\ref{MassiveComptAbelian}) and the single-pole terms (\ref{ScaVecSPole}). This cancellation can be made manifest in the massive amplitude by applying (\ref{BigSyz}) to the latter pair of terms, although I will not bother to present the reduced result here.

\subsubsection{Inclusion of vector exchanges and more of the Higgs mechanism}

In order to complete the two massive vector, two scalar amplitude, possible contributions from $t$ and $u$-channel vector boson exchange remain to be included. Assuming that the scalar-vector coupling is restricted to the Higgs-like coupling in (\ref{2vec1scalar}), then the $t$-channel vector exchange can be calculated as
\begin{align}\label{2V2SVect}
A_t(W_A,W_B,\varphi_i,\varphi_j)&=\frac{1}{t-m_{t_M}^2}\frac{1}{4}\lambda^i_{AM}\lambda^j_{MB}\left(\la{\bf{1}}p_3\rs{\bf{1}}\la{\bf{2}}p_4\rs{\bf{2}}-2m_{t_M}^2\ds{\bf{12}}\da{\bf{12}}\right).
\end{align}
The $u$-channel term is analogous. The second term in (\ref{2V2SVect}) is manifestly unitary. The first has an identical form to (\ref{2V2SScat}), with the identification of $\lambda^i_{AM}=-2i(t_A)_{Mi}/m_1$ and $\lambda^j_{MB}=-2i(t_B)_{Mj}/m_2$, in agreement with the requirement from the study of the Higgs couplings in vector boson scattering above. Including this vector exchange term therefore modifies the unitarity constraints (\ref{LAscalar}) by simply extending the sum in the matrix product of generators to include the intermediate components indexed by vector colour $M$, while the kinematics otherwise algebraically reduces in exactly the same way. The intermediate states that provide the representation space for the generators can therefore be a mix of independent matter scalars (Higgs bosons) and longitudinal modes from the massive vectors. As long as they arrange into a consistent representation, they can collectively unitarise the amplitude (more precisely, it has at this point been established that the commutator holds when both external coordinates are identified either with external scalars or longitudinal vectors - the case in which they are one of each type remains to be derived in the next Section below). The new couplings also modify the unitarised amplitudes in (\ref{AWWPPE2}) by similarly extending the sum over internal generator indices appearing in the expression. However, the kinematically new term proportional to the internal mass in (\ref{2V2SVect}) matches onto a new $t$-channel gluon exchange between massless scalars in the HEL. The generators supplied by the Higgs couplings match onto the generators describing the couplings of the exchanged gluon to Higgs bosons and the emergent longitudinal scalars. Finally, contributions from the ``holomorphic'' terms in (\ref{2vec1scalar}) are calculated in Appendix \ref{app:2V2S}. They clearly give contributions to the amplitude that have $\sim E^2$ HEL divergences (for opposite transverse helicities) and can only be eliminated by switching them off. This justifies ignoring them here. 

Note that at least two separate channels are required for this amplitude to unitarise. In particular, the $s$-channel exchange of a vector boson must be accompanied by a cross-channel exchange of scalar, possibly the longitudinal mode of another massive vector, that Higgses the amplitude. This remains the case if the $s$-channel vector is massless, as in the case of electromagnetic $W$-boson scattering. This parallels the inescapable generation of cross-channel Mandelstam poles in the residues of massless gluon amplitudes. 

The results of the calculations can be checked against the elegant predictions from $\mathcal{N}=4$ SYM. The mixed vector-scalar $3$-particle component amplitudes can be extracted from the $3$-particle superamplitude (\ref{N=43vec}). For most of the scalars, these couplings are simply minimal. However, the minimal coupling of $\phi_{(12)}$ in (\ref{LongMultNtwo}) is accompanied by a mixing angle $A[\phi_{(12)},\phi_{(12)},W_A]\propto\cos\left(\varphi_1-\varphi_2\right)$, while this scalar also has a further Higgs interaction with coupling $A[\phi_{(12)},W_A,W_B]\propto\frac{1}{m_1}\sin\left(\varphi_2-\varphi_3\right)$. Note that the superamplitude with one massless and two massive vector multiplets always contains a Higgs coupling involving the massless $\phi_{[12]}$ particle. Higgs bosons are always present in massless multiplets in the theory, but need not exist in massive multiplets unless central charges are misaligned. 

With the dependence of the coupling constants on the central charges established, the $4$-particle amplitudes constructed out of $\mathcal{N}=4$ SYM $3$-particle component amplitudes can be compared to the expected result directly extracted from the $4$-particle superamplitude. The four vector superamplitude for a particular colour ordering with $s$ and $u$ poles was exhibited in (\ref{N=4SYM}), from which the mixed scalar-vector components can be extracted. Taking the scalars to be those in (\ref{LongMultNtwo}) corresponding to minimum and maximum Grassmann degree respectively, the amplitude is 
\begin{align}\label{N=4ScaVec}
&A[W,W,\phi,\widetilde{\phi}]\nonumber\\
&=\left(m_4\left(\da{\bf{12}}e^{2i\varphi_4}-\ds{\bf{12}}e^{i(\varphi_1+\varphi_2)}\right)+e^{i\varphi_4}\left(e^{i\varphi_2}\la{\bf{1}}p_4\rs{\bf{2}}-e^{i\varphi_1}\la{\bf{2}}p_4\rs{\bf{1}}\right)\right)\nonumber\\
&\qquad\times \left(m_4\left(\da{\bf{12}}e^{-2i\varphi_4}-\ds{\bf{12}}e^{-i(\varphi_1+\varphi_2)}\right)+e^{-i\varphi_4}\left(e^{-i\varphi_2}\la{\bf{1}}p_4\rs{\bf{2}}-e^{-i\varphi_1}\la{\bf{2}}p_4\rs{\bf{1}}\right)\right)\nonumber\\
&\qquad\times\frac{-1}{(s-m_s^2)(u-m_u^2)}.
\end{align}
The numerator is identical for the other channel pairings. The choice of external particles preserves the factorised form of the numerator as a product of numerators from $\mathcal{N}=2$ amplitudes: $N_{\mathcal{N}=2}[\chi,\chi,\phi,\widetilde{\phi}]\times N_{\mathcal{N}=2}[\chi,\chi,\phi,\widetilde{\phi}]$, where $N_{\mathcal{N}=2}$ denotes the numerator of the appropriate component amplitude contained (\ref{SQEDFact}). When the external scalars are both identified as $\phi_{(12)}$, then (\ref{N=4ScaVec}) has a further overall factor of $\cos(\varphi_3-\varphi_4)$. Comparing this to (\ref{MassiveComptAbelian}) and (\ref{AmpstComb}) after applying the BPS conditions to the masses gives agreement. With aligned central charges, the single pole term (\ref{ScaVecSPole}) vanishes, leaving only terms with two separate Mandelstam poles, as expected from (\ref{N=4ScaVec}). When the central charges are misaligned, the single $s$-pole term (\ref{ScaVecSPole}) is non-zero but is canceled by an $s$-channel Higgs exchange instead. 

\begin{table}[h!]
\begin{center}
\begin{tabular}{ c || c c c c c}
Growth & Helicity Config & Couplings & Structure & Removal \\
\hline\hline
 & $++,++$ & $\frac{h}{\Lambda^2}t$ & fact & suppress\\
\cline{2-5}
$E^2$ & $++,--$ & $\frac{c^2}{\Lambda^2}$ & fact & suppress\\
\cline{2-5}
 & $L,L$ & $\frac{1}{m^2}t\{t,f\}$ & contact & Lie algebra reps \\
 \hline
$E$ & $++,L$ & $\frac{c}{\Lambda} t$ & fact & suppress
\end{tabular}
\end{center}
\caption{Summary of results for the two vector and two scalar amplitude.}\label{tab:2V2S}
\end{table}

Table \ref{tab:2V2S} summarises the results of this subsection.

\subsection{Three vectors and one scalar}\label{3Vec1Scalar}

The unitarisation of Higgs-vector scattering above establishes another block of the Lie algebra commutation relations involving the generators describing the Higgs-vector couplings. The final components of this matrix relation must be derived from unitarising the three vector, one scalar amplitude.

\subsubsection{General amplitude from factorisation}

All three factorisation channels for this amplitude are related by vector boson exchange symmetries. 
\begin{feynfig}
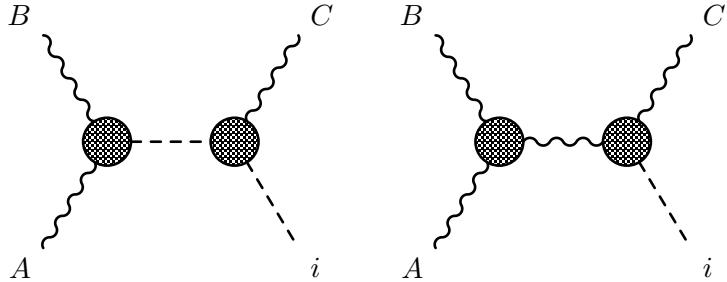
\begin{figure}[h]
\begin{fmffile}{3VecScalar}
 \begin{center}
 \begin{tabular}{c c c}
 & & \\
\begin{fmfgraph*}(120,80)
   \fmfleft{i1,i2}
   \fmfright{o1,o2}
   \fmf{boson}{i2,v1}
   \fmf{boson}{v2,o2}
   \fmf{boson}{i1,v1}
   \fmf{dashes}{v2,o1}
   \fmf{dashes}{v1,v2}
   \fmfv{decor.shape=circle,decor.filled=gray50,decor.size=0.15w}{v1,v2}
   \fmflabel{$A$}{i1}
   \fmflabel{$B$}{i2}
   \fmflabel{$C$}{o2}
   \fmflabel{$i$}{o1}
 \end{fmfgraph*} 
 &\,\,& \begin{fmfgraph*}(120,80)
   \fmfleft{i1,i2}
   \fmfright{o1,o2}
   \fmf{boson}{i2,v1}
   \fmf{boson}{v2,o2}
   \fmf{boson}{i1,v1}
   \fmf{dashes}{v2,o1}
   \fmf{boson}{v1,v2}
   \fmfv{decor.shape=circle,decor.filled=gray50,decor.size=0.15w}{v1,v2}
   \fmflabel{$A$}{i1}
   \fmflabel{$B$}{i2}
   \fmflabel{$C$}{o2}
   \fmflabel{$i$}{o1}
 \end{fmfgraph*}\nonumber\\
 & & 
 \end{tabular}
 \end{center}
\end{fmffile}
\caption{$s$-channel contributions for scattering of three vectors and a scalar.}\label{3V1SOSDi}
\end{figure}
\end{feynfig}
The $s$-channel term arising from scalar exchange (see Fig \ref{3V1SOSDi}) is 
\begin{align}\label{WWWScaGenSSca}
A_{s\varphi}(W_A,W_B,W_C,\varphi_i)&=\frac{1}{\sqrt{2}m_3}\lambda^m_{AB}(t_C)_{mi}\frac{1}{s-m_{s_m}^2}\ds{\bf{12}}\da{\bf{12}}\la{\bf{3}}p_4\rs{\bf{3}},
\end{align}
where the exchanged scalar must also be a Higgs boson. The analogous contribution arising from vector boson exchange is 
\begin{align}\label{WWWScaGenS}
&A_{sW}(W_A,W_B,W_C,\varphi_i)\nonumber\\
&\qquad=\frac{i}{4\sqrt{2}}\frac{1}{s-m_{s_M}^2}\frac{m_{s_M}}{m_1m_2}\lambda^i_{CM}\nonumber\\
&\qquad\qquad\times\Bigg(\Re{f}_{AB}^{\,\,\,\,\,\,\,\,\,M}(\da{\bf{12}}\left(\ds{\bf{13}}\ls{\bf{2}}p_4\ra{\bf{3}}+\ds{\bf{23}}\ls{\bf{1}}p_4\ra{\bf{3}}-2\ds{\bf{13}}\ds{\bf{23}}\right)\nonumber\\
&\qquad\qquad\qquad\qquad\qquad+\ds{\bf{12}}\left(\da{\bf{13}}\la{\bf{2}}p_4\rs{\bf{3}}+\da{\bf{23}}\la{\bf{1}}p_4\rs{\bf{3}}\right)-2\da{\bf{13}}\da{\bf{23}}))\nonumber\\
&\qquad\qquad\qquad\qquad+2m_1\Big(\left(i\Im{f}_{AB}^{\,\,\,\,\,\,\,\,\,M}-{f}_{BM}^{\,\,\,\,\,\,\,\,\,A}\right)\ds{\bf{12}}\ds{\bf{13}}\da{\bf{23}}\nonumber\\
&\qquad\qquad\qquad\qquad\qquad\qquad-\left(i\Im{f}_{AB}^{\,\,\,\,\,\,\,\,\,M}-({f}_{BM}^{\,\,\,\,\,\,\,\,\,A})^*\right)\da{\bf{12}}\da{\bf{13}}\ds{\bf{23}}\Big)\nonumber\\
&\qquad\qquad\qquad\qquad+2m_2\Big(\left(i\Im{f}_{AB}^{\,\,\,\,\,\,\,\,\,M}-{f}_{MA}^{\,\,\,\,\,\,\,\,\,B}\right)\ds{\bf{12}}\da{\bf{13}}\ds{\bf{23}}\nonumber\\
&\qquad\qquad\qquad\qquad\qquad\qquad-\left(i\Im{f}_{AB}^{\,\,\,\,\,\,\,\,\,M}-({f}_{MA}^{\,\,\,\,\,\,\,\,\,B})^*\right)\da{\bf{12}}\ds{\bf{13}}\da{\bf{23}}\Big)\nonumber\\
&\qquad\qquad\qquad\quad\quad+\left(m_1{f}_{BM}^{\,\,\,\,\,\,\,\,\,A}\ds{\bf{12}}+m_2({f}_{MA}^{\,\,\,\,\,\,\,\,\,B})^*\da{\bf{12}}\right)\nonumber\\
&\qquad\qquad\qquad\qquad\quad\qquad\quad\times\left(-2\da{\bf{23}}\ds{\bf{13}}+\frac{1}{m_{s_M}^2}\left(m_1\da{\bf{12}}-m_2\ds{\bf{12}}\right)\la{\bf{3}}p_4\rs{\bf{3}}\right)\nonumber\\
&\qquad\qquad\qquad\quad\quad+\left(m_1({f}_{BM}^{\,\,\,\,\,\,\,\,\,A})^*\da{\bf{12}}+m_2{f}_{MA}^{\,\,\,\,\,\,\,\,\,B}\ds{\bf{12}}\right)\nonumber\\
&\qquad\qquad\qquad\qquad\quad\quad\qquad\times\left(-2\ds{\bf{23}}\da{\bf{13}}+\frac{1}{m_{s_M}^2}\left(m_1\ds{\bf{12}}-m_2\da{\bf{12}}\right)\la{\bf{3}}p_4\rs{\bf{3}}\right)\Bigg).
\end{align}
The $t$-channel terms are obtained by exchanging $2\leftrightarrow 3$, while the $u$-channel is given by $1\leftrightarrow 3$.

Insertions of the $c$-type terms in (\ref{2vec1scalar}) give contributions to the $4$-leg amplitude that I separately present as
\begin{align}\label{WWWScaGenSHDO}
&A_{\varphi F^2}(W_A,W_B,W_C,\varphi_i)\nonumber\\
&=\frac{-1}{s-m_{s_m}^2}\frac{1}{m_3}(t_C)_{im}\la{\bf{3}}p_4\rs{\bf{3}}\left(\frac{c^m_{AB}}{\Lambda}\ds{\bf{12}}^2+\frac{(c^m_{AB})^*}{\Lambda}\da{\bf{12}}^2\right)\nonumber\\
&\quad-\frac{1}{s-m_{s_M}^2}\frac{i}{2m_1m_2}\nonumber\\
&\qquad\times\Bigg( m_{s_M}^2\left(f_{AB}^{\,\,\,\,\,\,\,\,\,M}+(f_{AB}^{\,\,\,\,\,\,\,\,\,M})^*\right)\left(\frac{c_{CM}^i}{\Lambda}\da{\bf{12}}\ds{\bf{13}}\ds{\bf{23}}-\frac{(c_{CM}^i)^*}{\Lambda}\ds{\bf{12}}\da{\bf{13}}\da{\bf{23}}\right)\nonumber\\
&\qquad-\frac{c_{CM}^i}{\Lambda}\bigg(\left(f_{BM}^{\,\,\,\,\,\,\,\,\,A}+(f_{AB}^{\,\,\,\,\,\,\,\,\,M})^*\right)m_1\left(\ds{\bf{13}}\ds{\bf{23}}\left(m_1\da{\bf{12}}-m_2\ds{\bf{12}}\right)-\ds{\bf{13}}^2\la{\bf{2}}p_1\rs{\bf{2}}\right)\nonumber\\
&\qquad\qquad\qquad+\left(f_{MA}^{\,\,\,\,\,\,\,\,\,B}+(f_{AB}^{\,\,\,\,\,\,\,\,\,M})^*\right)m_2\left(\ds{\bf{13}}\ds{\bf{23}}\left(m_2\da{\bf{12}}-m_1\ds{\bf{12}}\right)+\ds{\bf{23}}^2\la{\bf{1}}p_2\rs{\bf{1}}\right)\nonumber\\
&\qquad\qquad\qquad+\left((f_{BM}^{\,\,\,\,\,\,\,\,\,A})^*-(f_{AB}^{\,\,\,\,\,\,\,\,\,M})^*\right)m_1\da{\bf{12}}\ds{\bf{23}}\la{\bf{1}}P_s\rs{\bf{3}}\nonumber\\
&\qquad\qquad\qquad+\left((f_{MA}^{\,\,\,\,\,\,\,\,\,B})^*-(f_{AB}^{\,\,\,\,\,\,\,\,\,M})^*\right)m_2\da{\bf{12}}\ds{\bf{13}}\la{\bf{2}}P_s\rs{\bf{3}}\bigg)\nonumber\\
&\qquad+\frac{(c_{CM}^i)^*}{\Lambda}\bigg(\left((f_{BM}^{\,\,\,\,\,\,\,\,\,A})^*+f_{AB}^{\,\,\,\,\,\,\,\,\,M}\right)m_1\left(\da{\bf{13}}\da{\bf{23}}\left(m_1\ds{\bf{12}}-m_2\da{\bf{12}}\right)-\da{\bf{13}}^2\ls{\bf{2}}p_1\ra{\bf{2}}\right)\nonumber\\
&\qquad\qquad\qquad\quad+\left((f_{MA}^{\,\,\,\,\,\,\,\,\,B})^*+f_{AB}^{\,\,\,\,\,\,\,\,\,M}\right)m_2\left(\da{\bf{13}}\da{\bf{23}}\left(m_2\ds{\bf{12}}-m_1\da{\bf{12}}\right)+\da{\bf{23}}^2\ls{\bf{1}}p_2\ra{\bf{1}}\right)\nonumber\\
&\qquad\qquad\qquad\quad+\left(f_{BM}^{\,\,\,\,\,\,\,\,\,A}-f_{AB}^{\,\,\,\,\,\,\,\,\,M}\right)m_1\ds{\bf{12}}\da{\bf{23}}\ls{\bf{1}}P_s\ra{\bf{3}}\nonumber\\
&\qquad\qquad\qquad\quad+\left(f_{MA}^{\,\,\,\,\,\,\,\,\,B}-f_{AB}^{\,\,\,\,\,\,\,\,\,M}\right)m_2\ds{\bf{12}}\da{\bf{13}}\ls{\bf{2}}P_s\ra{\bf{3}}\bigg)\Bigg).
\end{align}
Contributions from insertions of the $F^3$ interactions are stated in Appendix \ref{app:3V1S} and are less interesting.

\subsubsection{Transverse divergences and covariance of higher dimensional insertions}

There are three classes of helicity configurations for which the amplitude has $\sim E^2$ HEL divergences. The first class has one longitudinal and two opposite transverse helicities. The terms in this class are given by the last two lines of (\ref{WWWScaGenSHDO}) and their parity conjugates. In these cases, there are contributions from two channels, but both still contain their corresponding Mandelstam poles and remain linearly independent. These are the expected massless amplitudes generated from double insertions of $\varphi F^2$ interactions, one of which is the emergent coupling contained in (\ref{3legVecHEL}). These divergences can only be removed by eliminating at least one of these types of couplings. 

The second class of divergent terms have two same-sign transverse polarisations and one longitudinal polarisation. In this case, the amplitude converges to contact terms in the HEL that can interfere across channels. Both (\ref{WWWScaGenS}) and (\ref{WWWScaGenSHDO}) contribute terms. Taking the $(++,++,L)$ configuration, the amplitude converges to contact terms of the form $\ds{12}^2$. Demanding that this cancels gives the constraint
\begin{align}\label{CovSF2}
&(t_C)_{im}\frac{c^m_{AB}}{\Lambda}-\frac{im_3}{2}\lambda^i_{CM}\frac{1}{2\sqrt{2}m_{s_M}}\left(f_{MA}^{\,\,\,\,\,\,\,\,\,B}+f_{MB}^{\,\,\,\,\,\,\,\,\,A}\right)\nonumber\\
&\qquad\qquad\qquad+\frac{i}{2}\left(f_{CM}^{\,\,\,\,\,\,\,\,\,A}+(f_{AC}^{\,\,\,\,\,\,\,\,\,M})^*\right)\frac{c^i_{MB}}{\Lambda}-\frac{i}{2}\left(f_{MC}^{\,\,\,\,\,\,\,\,\,B}+(f_{CB}^{\,\,\,\,\,\,\,\,\,M})^*\right)\frac{c^i_{MA}}{\Lambda}=0.
\end{align}
If the structure constants $f_{CM}^{\,\,\,\,\,\,\,\,\,A}$ and $f_{MC}^{\,\,\,\,\,\,\,\,\,B}$ in the second line of (\ref{CovSF2}) were replaced by $f_{AC}^{\,\,\,\,\,\,\,\,\,M}$ and $f_{CB}^{\,\,\,\,\,\,\,\,\,M}$ respectively, then this equation would be the covariance of the $c$-type couplings under the Lie algebra. The second term extends the internal scalar index on the $\varphi F^2$ coupling $c^i_{AB}$ to include the emergent scalars identified with the longitudinal vectors indexed by $M$, which is given by (\ref{ctof}), and where the generator acting on the tensor is (\ref{HiggsGen}). Where this term exists, it is required to covariantise the couplings $c^i_{AB}$ when $i$ is a Higgs boson. Full covariance is then established by pairing this relation with (\ref{CovSFEm}), in which the external scalar index is identified with a longitudinal mode of a vector. 

Double insertions of $c$-type couplings, explicit or emergent, are responsible for contaminating the interpretation of (\ref{CovSF2}) as coupling covariance. As explained above for the counterpart relation (\ref{CovSFEm}), covariance is recovered in the limit that $c/\Lambda\ll \{f,t\}/m$ (that is, the $\varphi F^2$ are small deformations of theories with softer high energy dependence, in particular spontaneously broken YM). If the $c$-type couplings are permitted to exist, then the inclusion of contact terms $\sim E^2/\Lambda^2$ in the $4$-particle amplitude is consistent with accepted order of truncation in an EFT expansion. Violation of (\ref{CovSF2}) by amounts $\sim m/\Lambda^2$ would seem to be consistent with the existence of these couplings in the first place. 

When one of the vectors is massless, the relation (\ref{CovSF2}) necessarily follows from consistent factorisation through a similar argument to that sketched out around (\ref{MasslessGluonConsFact}) in Section \ref{sec:PartialU} above. Taking $C$ as the massless gluon, then $f_{CM}^{\,\,\,\,\,\,\,\,\,A}=f_{AC}^{\,\,\,\,\,\,\,\,\,M}$ and $f_{MC}^{\,\,\,\,\,\,\,\,\,B}=f_{CB}^{\,\,\,\,\,\,\,\,\,M}$ because of (\ref{PartAntiSym}), while the Higgs coupling is forbidden. The covariance constraint is therefore exact and unavoidable, although this never applies to the case in which the scalars involved are Higgs bosons. 

The next step is to explicitly cancel the divergences between factorisation channels ensured by  (\ref{CovSF2}). These divergent terms can be arranged into the form of (\ref{SymmRed}):
\begin{align}\label{Cred}
\frac{-\ds{\bf{12}}^2}{3m_3}\bigg(\left(C_s-C_t\right)\left(\frac{\ls{\bf{3}}p_4\ra{\bf{3}}}{s-m_{s}^2}-\frac{\ls{\bf{3}}p_1\ra{\bf{3}}}{t-m_{t}^2}\right)&+\left(C_t-C_u\right)\left(\frac{\ls{\bf{3}}p_1\ra{\bf{3}}}{t-m_{t}^2}-\frac{\ls{\bf{3}}p_2\ra{\bf{3}}}{u-m_{u}^2}\right)\nonumber\\
&\qquad+\left(C_u-C_s\right)\left(\frac{\ls{\bf{3}}p_2\ra{\bf{3}}}{u-m_{u}^2}-\frac{\ls{\bf{3}}p_4\ra{\bf{3}}}{s-m_{s}^2}\right)\bigg),
\end{align}
where $C_s$, $C_t$ and $C_u$ are the accompanying collection of coupling constants arising from each channel in (\ref{CovSF2}). The kinematic parts of this expression can be reduced using the syzygy (\ref{SimpSyzFact}) and (\ref{squeeze2int}) to give
\begin{align}\label{SF2red}
&\frac{-\ds{\bf{12}}^2}{m_3}\left(\frac{\ls{\bf{3}}p_4\ra{\bf{3}}}{s-m_{s}^2}-\frac{\ls{\bf{3}}p_1\ra{\bf{3}}}{t-m_{t}^2}\right)\nonumber\\
&=\frac{-\ds{\bf{12}}}{(s-m_{s}^2)(t-m_{t}^2)}\big(\ds{\bf{12}}\la{\bf{3}}p_1p_4\ra{\bf{3}}-\left(s-m_s^2\right)\ds{\bf{13}}\ds{\bf{23}}-\left(m_s^2-m_3^2\right)\ds{\bf{13}}\ds{\bf{23}}\nonumber\\
&\qquad\qquad\qquad\qquad\qquad-m_1\ds{\bf{23}}\ls{\bf{3}}p_4\ra{\bf{1}}-m_2\ds{\bf{13}}\ls{\bf{3}}p_4\ra{\bf{2}}-m_3\ds{\bf{13}}\ls{\bf{2}}p_4\ra{\bf{3}}\nonumber\\
&\qquad\qquad\qquad\qquad\quad+\frac{1}{m_3}\ds{\bf{12}}\left(\left(m_1^2+m_3^2-m_t^2\right)\ls{\bf{3}}p_4\ra{\bf{3}}-\left(m_3^2+m_4^2-m_s^2\right)\ls{\bf{3}}p_1\ra{\bf{3}}\right)\big)
\end{align}
for the $s$ and $t$-channel combinations, while for the $t$ and $u$-channel combination,
\begin{align}\label{SF2redtu}
&\frac{-\ds{\bf{12}}^2}{m_3}\left(\frac{\ls{\bf{3}}p_1\ra{\bf{3}}}{t-m_{t}^2}-\frac{\ls{\bf{3}}p_2\ra{\bf{3}}}{u-m_{u}^2}\right)\nonumber\\
&=\frac{-\ds{\bf{12}}}{(u-m_{u}^2)(t-m_{t}^2)}\big(-\ds{\bf{12}}\la{\bf{3}}p_1p_2\ra{\bf{3}}+\left(s-m_s^2\right)\ds{\bf{13}}\ds{\bf{23}}+m_s^2\ds{\bf{13}}\ds{\bf{23}}\nonumber\\
&\qquad\qquad\qquad\qquad\qquad+m_1\ds{\bf{23}}\ls{\bf{3}}p_4\ra{\bf{1}}+m_2\ds{\bf{13}}\ls{\bf{3}}p_4\ra{\bf{2}}\nonumber\\
&\qquad\qquad\qquad\qquad\qquad-m_3\left(m_1\da{\bf{13}}\ds{\bf{23}}+m_2\ds{\bf{13}}\da{\bf{23}}\right)\nonumber\\
&\qquad\qquad\qquad\qquad\quad-\frac{1}{m_3}\ds{\bf{12}}\left(\left(m_1^2+m_3^2-m_t^2\right)\ls{\bf{3}}p_2\ra{\bf{3}}-\left(m_3^2+m_2^2-m_u^2\right)\ls{\bf{3}}p_1\ra{\bf{3}}\right)\big).
\end{align}
The $s$ and $u$ combination can be obtained by applying particle exchanges to the $s$ and $t$ expression. Parity conjugate terms must also be added to complete this set of terms. Altogether, the expressions still have $\sim E$ divergences for the helicity configurations $(++,++,--)$ and $(++,++,++)$. The latter divergences occur for the single pole terms in (\ref{SF2red}) and (\ref{SF2redtu}). However, these terms directly cancel against those in (\ref{WWWScaGenSHDO}), as expected from the fact that the massless amplitude $A(g_A^+,g_B^+,g_C^+,\varphi_i)$, upon which this structure should match onto in the HEL, is $0$. The $(++,++,--)$ terms in (\ref{Cred}), (\ref{SF2red}) and (\ref{SF2redtu}) directly match onto the expected massless amplitudes involving single $c$-type coupling insertions.

\subsubsection{Longitudinal divergences and the conclusion of the Higgs mechanism}

The remaining $\sim E^2$ HE divergence arises in both scalar (\ref{WWWScaGenSSca}) and vector (\ref{WWWScaGenS}) exchange expressions above for purely longitudinal vector helicities. Contact terms have HEL scaling $\mathcal{O}(E^3)$, so cannot contribute to further cancellation. All three factorisation channels contribute to the $\sim E^2$ HEL with longitudinally polarised vectors. In this limit, the amplitude converges to a linear polynomial in Mandelstam variables. Requiring that this vanishes imposes the condition
\begin{align}\label{BosonCommRel3}
&-m_2\lambda^m_{AB}(t_C)_{mi}+m_3\lambda^i_{CM}\left(\frac{-i}{2m_1m_{s_M}}\left(m_{s_M}^2\Re f_{AB}^{\,\,\,\,\,\,\,\,\,M}+m_1^2\Re f_{BM}^{\,\,\,\,\,\,\,\,\,A}-m_2^2\Re f_{MA}^{\,\,\,\,\,\,\,\,\,B}\right)\right)\nonumber\\
&-\left(-m_3\lambda^m_{AC}(t_B)_{mi}+m_2\lambda^i_{BM}\left(\frac{-i}{2m_1m_{t_M}}\left(m_{t_M}^2\Re f_{AC}^{\,\,\,\,\,\,\,\,\,M}+m_1^2\Re f_{CM}^{\,\,\,\,\,\,\,\,\,A}-m_3^2\Re f_{MA}^{\,\,\,\,\,\,\,\,\,C}\right)\right)\right)\nonumber\\
&\qquad=i\Re f_{CB}^{\,\,\,\,\,\,\,\,\,M}m_{u_M}\lambda^i_{MA}
\end{align}
as well as its counterparts with any of the three vector legs exchanged. This is simply the remaining ``broken'' index components of the Lie algebra commutator involving expressions for generator components given in (\ref{BrokenGenGeneral}) and (\ref{HiggsGen}): 
\begin{align}\label{CommRel3}
(t_B)_{Am}(t_C)_{mi}+(t_B)_{AM}(t_C)_{Mi}-(t_C)_{Am}(t_B)_{mi}-(t_C)_{AM}(t_B)_{Mi}=i\Re f_{CB}^{\,\,\,\,\,\,\,\,\,M}(t_M)_{Ai}.
\end{align}
The couplings in (\ref{BosonCommRel3}) are coefficients of emergent $4$-point contact interactions between four massless scalars, with $\sim E^2$ energy dependence. Notably, there is no such interaction that scales as $\sim E$. As a result, the condition (\ref{BosonCommRel3}) leads directly to full unitarisation of the longitudinal helicity configurations. However, the complete massive amplitude still contains divergent terms $\sim E$ for other helicity configurations. Unsurprisingly, these correspond to single insertions of the $c$-type couplings, both those explicit in (\ref{WWWScaGenSHDO}) and (\ref{SF2red}) and implicit in departures from (\ref{StandardLA}).

With the relation (\ref{CommRel3}), it has been finally established that, if the theory can be extrapolated to high energies, then the scalars, Higgs bosons and longitudinally polarised vector bosons arrange into complete representation spaces of a Lie algebra describing the vector boson self-interactions. In practice, (\ref{BosonCommRel3}) and (\ref{LAbrokengen}) in the previous Section constrain the possible boson masses given the identity of the Lie algebra and the representation of the scalars.

\subsubsection{Final cancellations and unitarisation}

It remains to reduce the three vector, one scalar amplitude to a form without terms with spurious high energy scaling. Firstly, the masses in the propagators should be homogenised as described around (\ref{HomoDen}). Next, the terms diverging as $\sim E^2$ in the HEL occur for fully longitudinally polarised spin configurations and converge to Mandelstam invariants, of which there are only two independent combinations. Once a basis of Lorentz structures converging to these two Mandelstam invariants has been chosen, each of the two sets of terms can be arranged to the form (\ref{SymmRed}). For example, choosing $s$ and $t$ as the independent structures, then the $s$-residue terms become
\begin{align}\label{3vecE2}
&A_{E^2}^{\{st\}s}(W_A,W_B,W_C,\varphi_i)\nonumber\\
&=\frac{1}{6}\bigg(\frac{d_s-d_t}{(s-m_s^2)(t-m_t^2)}\mu^2\ds{\bf{12}}\da{\bf{13}}\la{\bf{2}}p_4\rs{\bf{3}}+\frac{d_u-d_s}{(u-m_u^2)(s-m_s^2)}\mu^2\ds{\bf{12}}\da{\bf{23}}\la{\bf{1}}p_4\rs{\bf{3}}\nonumber\\
&\qquad\qquad\qquad\quad-\left((u-m_u^2)\da{\bf{13}}\la{\bf{2}}p_4\rs{\bf{3}}+(t-m_t^2)\da{\bf{23}}\la{\bf{1}}p_4\rs{\bf{3}}\right)\ds{\bf{12}}\nonumber\\
&\qquad\qquad\qquad\qquad\qquad\times\left(\frac{d_s-d_t}{(s-m_s^2)(t-m_t^2)}+\frac{d_t-d_u}{(t-m_t^2)(u-m_u^2)}+\frac{d_u-d_s}{(u-m_u^2)(s-m_s^2)}\right)\bigg)
\end{align}
plus parity conjugate terms. Here,
\begin{align}
d_s&=\frac{1}{\sqrt{2}m_3}\lambda^m_{AB}(t_C)_{mi}-\frac{i}{2\sqrt{2}m_1m_2m_{s_M}}\lambda^i_{CM}\left(m_{s_M}^2\Re f_{AB}^{\,\,\,\,\,\,\,\,\,M}+m_2^2\Re f_{MA}^{\,\,\,\,\,\,\,\,\,B}-m_1^2\Re f_{BM}^{\,\,\,\,\,\,\,\,\,A}\right)\nonumber\\
d_t&=\frac{i}{\sqrt{2}}\frac{m_{t_M}}{m_1m_3}\lambda^i_{BM}\Re f_{CA}^{\,\,\,\,\,\,\,\,\,M}\nonumber\\
d_u&=\frac{-1}{\sqrt{2}m_3}\lambda^m_{BC}(t_A)_{mi}-\frac{i}{2\sqrt{2}m_2m_3m_{u_M}}\lambda^i_{AM}\left(m_{u_M}^2\Re f_{BC}^{\,\,\,\,\,\,\,\,\,M}+m_2^2\Re f_{MA}^{\,\,\,\,\,\,\,\,\,B}-m_3^2\Re f_{BM}^{\,\,\,\,\,\,\,\,\,A}\right),
\end{align}
so $d_s+d_t+d_u=0$. The mass parameter $\mu^2$ is as defined in (\ref{WWWScaGenSSca}). The remaining $t$-residue terms are obtained by exchanging $2\leftrightarrow3$. The complete amplitude can be made manifestly exchange symmetric again by averaging these expressions over the distinct expressions given by particle exchanges, although I will not bother to do that here. Note that I have implicitly split-up the $s$-pole $s$-residue Lorentz structure into separate terms that converge to $t+u$ in the HEL, which is the reason that (\ref{3vecE2}) appears to contain single $s$-pole terms with $t$ and $u$ residues.

The divergent kinematic structure in the middle line of the right-hand side of (\ref{3vecE2}) can be decomposed using (\ref{3PolSyz}) and (\ref{Squeeze++}). I will present the full result further below, but first focus specifically upon the single pole terms with $\sim E$ scaling for two longitudinal and one transverse vector (which are the last remaining class of divergences). Collecting them all together with those remaining in (\ref{WWWScaGenS}) and (\ref{WWWScaGenSHDO}), the $s$-channel terms are
\begin{align}
&A_{E}(W_A,W_B,W_C,\varphi_i)\nonumber\\
&=\frac{-1}{s-m_{s_M}^2}\frac{c^i_{MC}}{\Lambda}\da{\bf{12}}\ds{\bf{13}}\ds{\bf{23}}\nonumber\\
&\qquad\quad\times\bigg((t_M)_{AB}-\frac{i}{2m_1m_2}\left(m_1^2\left(({f}_{AB}^{\,\,\,\,\,\,\,\,\,M})^*-({f}_{BM}^{\,\,\,\,\,\,\,\,\,A})^*\right)+m_2^2\left(({f}_{AB}^{\,\,\,\,\,\,\,\,\,M})^*-({f}_{MA}^{\,\,\,\,\,\,\,\,\,B})^*\right)\right)\bigg)\nonumber\\
&\qquad-\frac{i}{2\sqrt{2}}\frac{1}{s-m_{s_M}^2}\frac{m_{s_M}}{m_1m_2}\lambda^i_{MC}\nonumber\\
&\qquad\qquad\quad\left(m_1\left({f}_{BM}^{\,\,\,\,\,\,\,\,\,A}-{f}_{AB}^{\,\,\,\,\,\,\,\,\,M}\right)\ds{\bf{12}}\ds{\bf{13}}\da{\bf{23}}+m_2\left({f}_{MA}^{\,\,\,\,\,\,\,\,\,B}-{f}_{AB}^{\,\,\,\,\,\,\,\,\,M}\right)\ds{\bf{12}}\da{\bf{13}}\ds{\bf{23}}\right).
\end{align}
They match onto the expected massless amplitudes involving a single insertion of the $c$-type interaction with a vector exchange, assuming that the double insertion terms are small corrections that can be dropped. Again, these HEL divergent amplitudes are removed by eliminating the $c$-type couplings and demanding (\ref{StandardLA}).

However, regardless of whether these single pole residuals cancel or not, the remaining amplitude, consisting of terms containing pairs of Mandelstam poles, is manifestly unitarised. No $\sim E$ divergences involving $4$-point contact interactions between emerging scalar particles occur, which is consistent with the fact that no such interactions can be Lorentz invariant. Before presenting the result for the unitarised amplitude, it is first useful to establish a basis for the dimension-$4$ Lorentz structures containing equal numbers of left and right-handed spinors for each external leg. Such a basis is given by the structures 
\begin{align}
\mathbf{T}_s&=\da{\bf{12}}\ds{\bf{12}}\la{\bf{3}}p_4\rs{\bf{3}}\nonumber\\
\mathbf{T}_t&=\da{\bf{31}}\ds{\bf{31}}\la{\bf{2}}p_4\rs{\bf{2}}\nonumber\\
\mathbf{T}_u&=\da{\bf{23}}\ds{\bf{23}}\la{\bf{1}}p_4\rs{\bf{1}},
\end{align}
along with the permutation (and parity) antisymmetric structure
\begin{align}
{\bf{A}}_3&=\da{\bf{12}}\ds{\bf{13}}\la{\bf{3}}p_4\rs{\bf{2}}+\da{\bf{12}}\ds{\bf{23}}\la{\bf{3}}p_4\rs{\bf{1}}-\da{\bf{23}}\ds{\bf{13}}\la{\bf{1}}p_4\rs{\bf{2}}\nonumber\\
&\qquad-\da{\bf{23}}\ds{\bf{12}}\la{\bf{1}}p_4\rs{\bf{3}}-\da{\bf{13}}\ds{\bf{12}}\la{\bf{2}}p_4\rs{\bf{3}}+\da{\bf{13}}\ds{\bf{23}}\la{\bf{2}}p_4\rs{\bf{1}}.
\end{align}
This is just the anomalous generalised Chern-Simons quartic (\ref{vectorE3CI}) with the polarisation of particle $4$ replaced with its momentum. Like the aGCS quartic, this structure vanishes in the HEL, so has weaker high-energy scaling than expected from dimensional analysis. However, unlike the aGCS quartic, it can be directly decomposed into dim-$3$ Lorentz structures:
\begin{align}
{\bf{A}}_3&=m_1\da{\bf{12}}\ds{\bf{23}}\da{\bf{31}}+m_2\da{\bf{23}}\ds{\bf{31}}\da{\bf{12}}+m_3\da{\bf{31}}\ds{\bf{12}}\da{\bf{23}}\nonumber\\
&\qquad-m_1\ds{\bf{12}}\da{\bf{23}}\ds{\bf{31}}-m_2\ds{\bf{23}}\da{\bf{31}}\ds{\bf{12}}-m_3\ds{\bf{31}}\da{\bf{12}}\ds{\bf{23}}.
\end{align}
All other non-chiral dim-$4$ structures can be decomposed into this basis e.g.
\begin{align}
\da{\bf{12}}\ds{\bf{13}}\la{\bf{3}}p_4\rs{\bf{2}}=\frac{1}{2}\left({\bf{A}}_3+\mathbf{T}_s+\mathbf{T}_t-\mathbf{T}_u\right).
\end{align}

Neglecting now the potential divergent single-pole terms generated by HDOs, the unitarised amplitude is then
\begin{align}\label{3vecred}
&A\left(W_A,W_B,W_C,\varphi_i\right)\nonumber\\
&=\frac{1}{6}\Bigg(\frac{d_s-d_t}{(s-m_s^2)(t-m_t^2)}\mu^2\left(\mathbf{T}_s+\mathbf{T}_t-\mathbf{T}_u\right)-\frac{d_u-d_s}{(u-m_u^2)(s-m_s^2)}\mu^2\left(\mathbf{T}_s+\mathbf{T}_u-\mathbf{T}_t\right)\nonumber\\
&\qquad\quad-\frac{d_t-d_u}{(t-m_t^2)(u-m_u^2)}\mu^2(m_1\ds{\bf{12}}\da{\bf{23}}\ds{\bf{31}}+m_2\ds{\bf{12}}\ds{\bf{23}}\da{\bf{31}}+m_3\da{\bf{12}}\ds{\bf{23}}\ds{\bf{31}}\nonumber\\
&\qquad\qquad\qquad\qquad\qquad\quad\qquad+m_1\da{\bf{12}}\ds{\bf{23}}\da{\bf{31}}+m_2\da{\bf{12}}\da{\bf{23}}\ds{\bf{31}}+m_3\ds{\bf{12}}\da{\bf{23}}\da{\bf{31}})\nonumber\\
&\qquad\quad+\left(\frac{d_s-d_t}{(s-m_s^2)(t-m_t^2)}+\frac{d_t-d_u}{(t-m_t^2)(u-m_u^2)}+\frac{d_u-d_s}{(u-m_u^2)(s-m_s^2)}\right)\nonumber\\
&\qquad\qquad\times((m_u^2-m_2^2-m_3^2)\left(\mathbf{T}_s+\mathbf{T}_t-\mathbf{T}_u\right)-(m_t^2-m_1^2-m_3^2)\left(\mathbf{T}_s-\mathbf{T}_t+\mathbf{T}_u\right)\nonumber\\
&\qquad\qquad\quad-2m_2m_3\left(\da{\bf{13}}\la{\bf{3}}p_4\rs{\bf{2}}\ds{\bf{12}}+\ds{\bf{13}}\ls{\bf{3}}p_4\ra{\bf{2}}\da{\bf{12}}\right)\nonumber\\
&\qquad\qquad\quad-2m_1m_2\left(\da{\bf{23}}\la{\bf{3}}p_4\rs{\bf{1}}\ds{\bf{12}}+\ds{\bf{23}}\ls{\bf{3}}p_4\ra{\bf{1}}\da{\bf{12}}\right)\nonumber\\
&\qquad\qquad\quad+2m_3m_4^2\left(\ds{\bf{12}}\da{\bf{23}}\da{\bf{31}}+\da{\bf{12}}\ds{\bf{23}}\ds{\bf{31}}\right)\nonumber\\
&\qquad\qquad\quad+\left(m_s^2-m_3^2-m_4^2\right)(m_1\ds{\bf{12}}\da{\bf{23}}\ds{\bf{31}}+m_2\ds{\bf{12}}\ds{\bf{23}}\da{\bf{31}}+m_3\da{\bf{12}}\ds{\bf{23}}\ds{\bf{31}}\nonumber\\
&\qquad\qquad\qquad\qquad\qquad\quad\,\,+m_1\da{\bf{12}}\ds{\bf{23}}\da{\bf{31}}+m_2\da{\bf{12}}\da{\bf{23}}\ds{\bf{31}}+m_3\ds{\bf{12}}\da{\bf{23}}\da{\bf{31}}))\Bigg).
\end{align}
As above, this is to be completed by adding terms obtained by the exchange $2\leftrightarrow 3$ (I have explicitly added the parity conjugate terms in the expression above). 

As expected in the HEL, the only $\sim E^0$ terms correspond to opposite sign Compton scattering or scattering of fully longitudinal polarisations. These correctly match onto the expected massless scalar QCD amplitudes in the HEL with the emergent generators given by (\ref{BrokenGen}) and (\ref{HiggsGen}). The emergent scalar quartic contact interaction in the HEL is 
\begin{align}
&A_c(W_A^L,W_B^L,W_C^L,\varphi_i)\nonumber\\
&\rightarrow\frac{-1}{4}\Bigg(\left(d_u+d_u|_{2\leftrightarrow 3}\right)\left(m_u^2-m_2^2-m_3^2\right)+\left(d_t-2d_s|_{2\leftrightarrow 3}\right)\left(m_t^2-m_1^2-m_3^2\right)\nonumber\\
&\qquad\qquad\qquad+\left(d_t|_{2\leftrightarrow 3}-2d_s\right)\left(m_s^2-m_1^2-m_2^2\right)-2\left(d_s+d_s|_{2\leftrightarrow 3}\right)\left(\sum_i m_i^2-2m_4^2\right)\Bigg).
\end{align}
While not manifest, the combination $d_s+d_s|_{2\leftrightarrow 3}$ is symmetric under exchanges of the vector legs using (\ref{BosonCommRel3}), as are the combination of the other three terms in this expression. 

Since the amplitude (\ref{3vecred}) is entirely proportional to Higgs couplings, its appearance in the $\mathcal{N}=4$ superamplitude (\ref{N=4SYM}) is only non-trivial if the central charges of the external legs are misaligned. The resulting (colour-ordered and stripped) amplitude is then 
\begin{align}
&A[W,W,W,\phi_{(12)}]=\nonumber\\
&\Big(\sin\left(\varphi_1+\varphi_2-\varphi_3-\varphi_4\right)\mathbf{T}_s+\sin\left(\varphi_1+\varphi_3-\varphi_2-\varphi_4\right)\mathbf{T}_t\nonumber\\
&\qquad+\sin\left(\varphi_1+\varphi_3-\varphi_2-\varphi_4\right)\mathbf{T}_u\nonumber\\
&\qquad-\sin\left(\varphi_1-\varphi_4\right)\left(\da{\bf{13}}\ds{\bf{12}}\la{\bf{3}}p_4\rs{\bf{2}}+\ds{\bf{13}}\da{\bf{12}}\ls{\bf{3}}p_4\ra{\bf{2}}\right)\nonumber\\
&\qquad+\sin\left(\varphi_2-\varphi_4\right)\left(\da{\bf{23}}\ds{\bf{12}}\la{\bf{3}}p_4\rs{\bf{1}}+\ds{\bf{23}}\da{\bf{12}}\ls{\bf{3}}p_4\ra{\bf{1}}\right)\nonumber\\
&\qquad-\sin\left(\varphi_3-\varphi_4\right)\left(\da{\bf{13}}\ds{\bf{23}}\la{\bf{1}}p_4\rs{\bf{2}}+\ds{\bf{13}}\da{\bf{23}}\ls{\bf{1}}p_4\ra{\bf{2}}\right)\nonumber\\
&\qquad -m_4\Big(\sin\left(\varphi_1-\varphi_3\right)\left(\ds{\bf{13}}\da{\bf{12}}\da{\bf{23}}+\da{\bf{13}}\ds{\bf{12}}\ds{\bf{23}}\right)\nonumber\\
&\qquad\qquad+\sin\left(\varphi_3-\varphi_2\right)\left(\ds{\bf{13}}\ds{\bf{12}}\da{\bf{23}}+\da{\bf{13}}\da{\bf{12}}\ds{\bf{23}}\right)\nonumber\\
&\qquad\qquad+\sin\left(\varphi_2-\varphi_1\right)\left(\ds{\bf{13}}\ds{\bf{23}}\da{\bf{12}}+\da{\bf{13}}\da{\bf{23}}\ds{\bf{12}}\right)\Big)\Big)\times\frac{2i}{(s-m_s^2)(u-m_u^2)}.
\end{align}
This can be verified to agree with (\ref{3vecred}) once the BPS conditions on the masses are applied and the general couplings are fixed to their $\mathcal{N}=4$ colour factors (including the phase dependence described in the paragraph above (\ref{N=4ScaVec})). Note that, as usual, the numerator accompanying the pair of Mandelstam poles is the same for all channels in $\mathcal{N}=4$ SYM.

In Appendix \ref{app:3V1S}, I include the contributions to this amplitude from the $F^3$-type couplings, which have been omitted in this analysis of this Section for clarity. They are responsible for harder HEL divergences but do not interfere with the results derived here. Table \ref{tab:3V1S} below summarises the qualitative features of the amplitude from this subsection, including the Appendix. 

\begin{table}[h!]
\begin{center}
\begin{tabular}{ c || c c c c c}
Growth & Helicity Config & Couplings & Structure & Removal \\
\hline\hline
$E^3$ & $++,++,--$ & $\frac{h}{\Lambda^2}\frac{c}{\Lambda}$ & fact & suppress\\
\hline
 & $++,++,L$ & $\frac{h}{\Lambda^2}t$ & fact & suppress\\
\cline{2-5}
$E^2$ & $++,--,L$ & $\frac{c^2}{\Lambda^2}$ & fact & suppress\\
\cline{2-5}
 & $++,++,L$ & $\frac{c}{\Lambda}\{\frac{1}{m}t,\frac{c}{\Lambda}\}$ & contact & $\varphi F^2$ cov\\
 \cline{2-5}
 & $L,L,L$ & $\frac{1}{m^2}t\{f,t\}$ & contact & Lie algebra reps\\
 \hline
$E$ & $\begin{array}{l}
++,++,-- \\
++,L,L 
\end{array}$
 & $\frac{c}{\Lambda} \{f,t\}$ & fact & suppress
\end{tabular}
\end{center}
\caption{Summary of results for the three vector and one scalar amplitude.}\label{tab:3V1S}
\end{table}

\section{Vector-Fermion Scattering}\label{VecFer}

It remains to introduce fermions into the theory of massive vector bosons and examine the constraints from tree-unitarity on their interactions. The relevant $3$-particle amplitudes involving fermions were given in Section \ref{sec:AmpList}. 

\subsection{One vector, two fermions and one scalar}\label{sec:Mixed}

\subsubsection{General amplitude and covariant Yukawa couplings}

To begin with, I will construct the amplitude for fermion scattering off a single vector and scalar. 
\begin{feynfig}
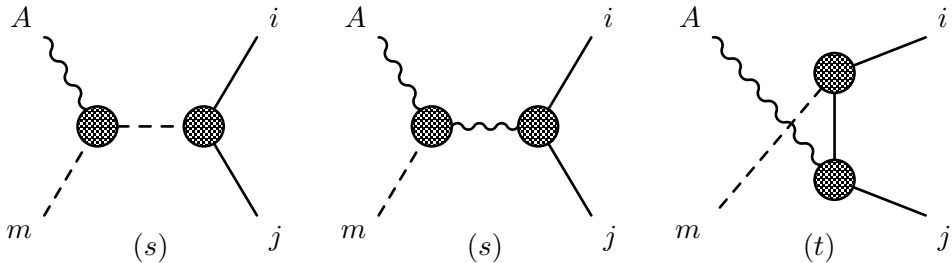
\begin{figure}[h]
\begin{fmffile}{MixedAll}
 \begin{center}
 \begin{tabular}{c c c c c}
 & & & & \\
   \begin{fmfgraph*}(100,67)
   \fmfleft{i1,i2}
   \fmfright{o1,o2}
   \fmf{boson}{i2,v1}
   \fmf{plain}{v2,o2}
   \fmf{dashes}{i1,v1}
   \fmf{plain}{v2,o1}
   \fmf{dashes}{v1,v2}
   \fmfv{decor.shape=circle,decor.filled=gray50,decor.size=0.15w}{v1,v2}
   \fmflabel{$m$}{i1}
   \fmflabel{$A$}{i2}
   \fmflabel{$i$}{o2}
   \fmflabel{$j$}{o1}
 \end{fmfgraph*} 
 & \, & \begin{fmfgraph*}(100,67)
   \fmfleft{i1,i2}
   \fmfright{o1,o2}
   \fmf{boson}{i2,v1}
   \fmf{plain}{v2,o2}
   \fmf{dashes}{i1,v1}
   \fmf{plain}{v2,o1}
   \fmf{boson}{v1,v2}
   \fmfv{decor.shape=circle,decor.filled=gray50,decor.size=0.15w}{v1,v2}
   \fmflabel{$m$}{i1}
   \fmflabel{$A$}{i2}
   \fmflabel{$i$}{o2}
   \fmflabel{$j$}{o1}
 \end{fmfgraph*} 
 & \, & \begin{fmfgraph*}(100,67)
   \fmfleft{i1,i2}
   \fmfright{o1,o2}
   \fmf{phantom}{i2,v1}
   \fmf{plain}{v1,o2}
   \fmf{phantom}{i1,v2}
   \fmf{plain}{v2,o1}
   \fmf{plain}{v1,v2}
   \fmf{dashes,tension=-0.25}{v1,i1}
   \fmf{boson,tension=-0.25}{v2,i2}
   \fmfv{decor.shape=circle,decor.filled=gray50,decor.size=0.15w}{v1,v2}
   \fmflabel{$m$}{i1}
   \fmflabel{$A$}{i2}
   \fmflabel{$i$}{o2}
   \fmflabel{$j$}{o1}
 \end{fmfgraph*}\nonumber\\
 $(s)$ & \, & $(s)$ & \, & $(t)$
 \end{tabular}
 \end{center}
\end{fmffile}
\caption{Mixed scalar-vector-fermion scattering (omitting the $u$-channel).}\label{1V2F1SOSDi}
\end{figure}
\end{feynfig}
Each factorisation channel may be constructed as (see Fig \ref{1V2F1SOSDi}):
\begin{align}\label{FactAllAmp}
A_t\left(\varphi_m,W_A,\psi_i,\psi_j\right)&=\frac{1}{t-m_{t_k}^2}\frac{1}{m_2}\nonumber\\
&\qquad\times\Big(
(y_{ik}^m)^*(t_A)_{kj}\ds{\bf{24}}\la{\bf{2}}P_t\rs{\bf{3}}-y_{ik}^m((t_A)_{kj})^*\da{\bf{24}}\la{\bf{3}}P_t\rs{\bf{2}}\nonumber\\
&\qquad\qquad+y_{ik}^m(t_A)_{kj}m_{t_k}\da{\bf{23}}\ds{\bf{24}}-(y_{ik}^m)^*((t_A)_{kj})^*m_{t_k}\ds{\bf{23}}\da{\bf{24}}\Big)\nonumber\\
A_u\left(\varphi_m,W_A,\psi_i,\psi_j\right)&=\frac{1}{u-m_{u_k}^2}\frac{1}{m_2}\nonumber\\
&\qquad\times\Big(
((t_A)_{ik})^*(y_{kj}^m)^*\ds{\bf{23}}\la{\bf{2}}P_u\rs{\bf{4}}-(t_A)_{ik}y_{kj}^m\da{\bf{23}}\la{\bf{4}}P_u\rs{\bf{2}}\nonumber\\
&\qquad\qquad+(t_A)_{ik}(y_{kj}^m)^*m_{u_k}\da{\bf{23}}\ds{\bf{24}}-((t_A)_{ik})^*y_{jk}^mm_{u_k}\ds{\bf{23}}\da{\bf{24}}\Big)\nonumber\\
A_{s\varphi}\left(\varphi_m,W_A,\psi_i,\psi_j\right)&=\frac{-1}{s-m_{s_n}^2}\frac{1}{m_2}(t_A)_{mn}\la{\bf{2}}p_1\rs{\bf{2}}\left((y_{ij}^n)^*\ds{\bf{34}}+y_{ij}^n\da{\bf{34}}\right)\nonumber\\
A_{sW}\left(\varphi_m,W_A,\psi_i,\psi_j\right)&=\frac{-1}{s-m_{s_M}^2}\frac{\lambda_{AM}^m}{2\sqrt{2}m_{s_M}}\nonumber\\
&\quad\,\times\Big((t_M)_{ij}\left(\la{\bf{2}}p_1\rs{\bf{2}}\left(m_3\ds{\bf{34}}-m_4\da{\bf{34}}\right)-2m_{s_M}^2\da{\bf{23}}\ds{\bf{24}}\right)\nonumber\\
&\quad\quad\,\,+((t_M)_{ij})^*\left(\la{\bf{2}}p_1\rs{\bf{2}}\left(m_4\ds{\bf{34}}-m_3\da{\bf{34}}\right)+2m_{s_M}^2\da{\bf{24}}\ds{\bf{23}}\right)\Big).
\end{align}
The $s$-channel amplitude includes separate contributions from possible scalar and vector exchanges, listed respectively. Insertions of fermion dipoles and the holomorphic interactions in (\ref{2vec1scalar}) are deferred to Appendix \ref{app:1V2P1S}. Both of these couplings contribute terms to the $4$-particle amplitude that scale as $\sim E$ in the HEL for the transversely polarised vector, the elimination of which is only possible by eliminating the interactions themselves.

The candidate terms in the amplitude (\ref{FactAllAmp}) scale as $\sim E$ when the vector is longitudinally polarised and the fermions have the same helicity. All possible contact terms clearly have $\mathcal{O}(E^2)$ scaling, so cannot unitarise the amplitude and need not even be included for consistent power counting in energy. Requiring the divergence cancel imposes the following relation upon the $3$-particle coupling constants:
\begin{align}\label{CovYukawas1}
(t_A)_{ik}y^m_{kj}+(t_A)_{jk}y^m_{ik}+(t_A)_{mn}y^n_{ij}-\frac{m_2}{2\sqrt{2}m_{s_M}}\lambda^m_{AM}\left(m_4(t_M)_{ij}+m_3((t_M)_{ij})^*\right)=0.
\end{align}
This is simply the constraint that the Yukawa couplings be a covariant tensor of the emergent Lie algebra (assuming for the moment that $(t_A)_{ij}$ are, in fact, generators, as will be shown shortly below). Since the scalars are required to combine with the longitudinal modes of the vectors into complete linear representation spaces (as shown in the previous Section), the scalar index of the Yukawa couplings must be covariantly extended to include the emergent Yukawa couplings implicit in (\ref{LongModeLim}):
\begin{align}\label{YukEmerge}
y^A_{ij}=\frac{m_j-m_i}{\sqrt{2}m_A}\Im(t_A)_{ij}-\frac{m_j+m_i}{\sqrt{2}m_A}i\Re(t_A)_{ij}.
\end{align}
Then with the Higgs couplings interpreted as Lie algebra generators (\ref{HiggsGen}), (\ref{CovYukawas1}) reduces to the expected statement of covariance when the free scalar index is identified with a scalar particle. Because of the form of (\ref{YukEmerge}), the covariance relation (\ref{CovYukawas1}) is effectively a constraint relating the fermion masses to the Yukawa couplings, given the Lie algebra representation. 


\subsubsection{Leading high energy cancellations}

An analogous decomposition to the structure of (\ref{AWWPPE2}) can be performed for the HEL divergent terms of this amplitude:
\begin{align}\label{LOmixedAmp}
A_{E}\left(\varphi_m,W_A,\psi_i,\psi_j\right)&=\frac{-1}{2m_2}\left(y_{ik}^m(t_A)_{jk}-y_{kj}^m(t_A)_{ik}\right)\left(\frac{\da{\bf{24}}\la{\bf{3}}P_t\rs{\bf{2}}}{t-m_{t_M}^2}-\frac{\da{\bf{23}}\la{\bf{4}}P_u\rs{\bf{2}}}{u-m_{u_M}^2}\right)\nonumber\\
&\,\quad-\frac{1}{2m_2}\left((t_A)_{mn}y^n_{ij}-\frac{1}{2\sqrt{2}m_{s_{M}}}\lambda^m_{AM}\left(m_4(t_M)_{ij}+m_3((t_M)_{ij})^*\right)\right)\nonumber\\
&\qquad\qquad\quad\,\times\left(-\frac{\da{\bf{24}}\la{\bf{3}}P_t\rs{\bf{2}}}{t-m_{t_M}^2}-\frac{\da{\bf{23}}\la{\bf{4}}P_u\rs{\bf{2}}}{u-m_{u_M}^2}+\frac{2\da{\bf{34}}\la{\bf{2}}p_1\rs{\bf{2}}}{s-m_{s_M}^2}\right)
\end{align}
and likewise for parity conjugate terms. The amplitude can be manifestly unitarised by combining divergent terms from pairs of channels. After homogenising the masses in the propagators and grouping the divergent terms in each channel into the form of (\ref{SymmRed}), identity (\ref{3PolSyz}) can be used to combine and reduce the resulting expressions. Choosing the terms with mostly right-handed spinors and combining them in the $t$ and $u$-channels, then the kinematic factor accompanying $\frac{1}{m_2(t-m_t^2)(u-m_u^2)}$ from the first line becomes
\begin{align}\label{MixAmpAbelian}
&(u-m_u^2)\da{\bf{24}}\la{\bf{3}}p_1+p_3\rs{\bf{2}}+(t-m_t^2)\da{\bf{23}}\la{\bf{4}}p_1+p_4\rs{\bf{2}}\nonumber\\
&=m_2m_3\da{\bf{24}}\la{\bf{2}}p_1\rs{\bf{3}}+m_2m_4\ds{\bf{24}}\ls{\bf{2}}p_1\ra{\bf{3}}+m_2m_4\da{\bf{23}}\la{\bf{2}}p_1\rs{\bf{4}}+m_2m_3\ds{\bf{23}}\ls{\bf{2}}p_1\ra{\bf{4}}\nonumber\\
&\qquad+(u-m_u^2)m_4\da{\bf{23}}\ds{\bf{24}}+(t-m_t^2)m_3\da{\bf{24}}\ds{\bf{23}}-m_2\la{\bf{4}}p_1\rs{\bf{2}}\la{\bf{3}}p_1\rs{\bf{2}}\nonumber\\
&\qquad-\frac{1}{2}\left(m_t^2+m_u^2-m_3^2-m_4^2\right)\left(\da{\bf{23}}\la{\bf{4}}p_1\rs{\bf{2}}+\da{\bf{24}}\la{\bf{3}}p_1\rs{\bf{2}}\right)\nonumber\\
&\qquad+\left(m_t^2+m_u^2-m_3^2-m_4^2\right)\left(m_3\da{\bf{24}}\ds{\bf{23}}+m_4\ds{\bf{24}}\da{\bf{23}}\right)\nonumber\\
&\qquad+\frac{1}{2}\left(m_t^2-m_u^2+m_3^2-m_4^2\right)\da{\bf{34}}\la{\bf{2}}p_1\rs{\bf{2}}-m_1^2m_2\da{\bf{23}}\da{\bf{24}}.
\end{align}
The $P$-conjugate terms have complex conjugate coupling constants and a relative negative sign. Along with the remaining single pole $t$ and $u$-channel terms, these represent the ``Abelian'' part of the amplitude. The non-Abelian parts are obtained by combining the $s$-channel terms with the $t$ and $u$-channel terms in the last line of (\ref{LOmixedAmp}). The combination of divergent $s$ and $t$ terms analogous to those computed above are:
\begin{align}\label{MixAmpNonAbelian}
&(t-m_t^2)\da{\bf{34}}\la{\bf{2}}p_1\rs{\bf{2}}+(s-m_s^2)\da{\bf{24}}\la{\bf{3}}p_1+p_3\rs{\bf{2}}\nonumber\\
&=-m_2m_3\da{\bf{24}}\la{\bf{2}}p_1\rs{\bf{3}}-m_2m_4\ds{\bf{24}}\ls{\bf{2}}p_1\ra{\bf{3}}-m_2m_4\da{\bf{23}}\la{\bf{2}}p_1\rs{\bf{4}}-m_2m_3\ds{\bf{23}}\ls{\bf{2}}p_1\ra{\bf{4}}\nonumber\\
&\qquad+(s-m_s^2)m_4\da{\bf{23}}\ds{\bf{24}}+m_2\la{\bf{3}}p_1\rs{\bf{2}}\la{\bf{4}}p_1\rs{\bf{2}}+m_1^2m_2\da{\bf{23}}\da{\bf{24}}\nonumber\\
&\qquad+\left(m_s^2-m_1^2-m_2^2\right)\left(m_3\da{\bf{24}}\ds{\bf{23}}+m_3\da{\bf{23}}\ds{\bf{24}}\right)\nonumber\\
&\qquad-\frac{1}{2}\left(m_s^2-m_1^2-m_2^2\right)\left(\da{\bf{24}}\la{\bf{3}}p_1\rs{\bf{2}}+\da{\bf{23}}\la{\bf{4}}p_1\rs{\bf{2}}\right)\nonumber\\
&\qquad+\left(m_4^2-m_t^2-\frac{1}{2}\left(m_s^2-m_1^2-m_2^2\right)\right)\da{\bf{34}}\la{\bf{2}}p_1\rs{\bf{2}}
\end{align}
(with a factor of $\frac{1}{m_2(s-m_s^2)(t-m_t^2)}$) and similarly for their $P$-conjugate terms. The $s$ and $u$ combined terms are similar but with $3\leftrightarrow 4$. 

The complete set of remaining $\sim E^0$ HEL terms can be checked for verification with the expected massless amplitudes. This is automatic from (\ref{LOmixedAmp}) for the helicity configurations in which the vector is transversely polarised, given that the relevant terms appear identically in both (\ref{MixAmpAbelian}) and (\ref{MixAmpNonAbelian}). The HEL directly matches the leading order terms onto the expected massless amplitudes (which exist for scalar, vector and same-helicity fermions), identifying the expected emergent couplings (\ref{HiggsGen}) and (\ref{YukEmerge}). It can also be checked that the longitudinally polarised vector amplitudes converges to the expected massless amplitudes (scalar scattering off opposite-helicity fermions). This limit is determined from the single pole terms remaining from the cancellation implemented in (\ref{MixAmpAbelian}) and (\ref{MixAmpNonAbelian}) in (\ref{LOmixedAmp}) and have to be collected together with those from (\ref{FactAllAmp}). The $s$-pole terms are produced only from the $s$-channel terms in (\ref{FactAllAmp}) and match onto massless gluon exchange (between scalars and fermions), while the $t$ and $u$-pole terms assemble into the expected massless fermion exchange amplitudes with a Yukawa coupling given by (\ref{YukEmerge}). Finally, there are no remaining $\sim E^0$ terms that can match onto quartic contact terms. This is expected, since there are no unitary contact terms between two scalars and two fermions. 

As usual, this result can be cross-checked against the amplitude contained in (\ref{N=4SYM}) for $\mathcal{N}=4$ SYM. Firstly, some relevant $3$-particle amplitudes involving fermions can be extracted from (\ref{N=43vec}):
\begin{align}
A[\lambda_1,W,\widetilde{\lambda}^1]&=\frac{1}{m_2}\left(\da{\bf{12}}\ds{\bf{23}}-e^{i\left(\varphi_1-\varphi_3\right)}\da{\bf{32}}\ds{\bf{21}}\right)\nonumber\\
A[\lambda_1,\phi_{(12)},\widetilde{\lambda}^1]&=e^{i\left(\varphi_1-\varphi_2\right)}\ds{\bf{13}}+e^{i\left(\varphi_2-\varphi_3\right)}\da{\bf{13}}\nonumber\\
A[\lambda_1,\widetilde{\phi},\lambda_2]&=e^{i\left(\varphi_1-\varphi_3\right)}\ds{\bf{13}}+\da{\bf{13}}.
\end{align}
The phases from the central charges determine the parity structure of the couplings. The amplitudes for $\lambda_2$ are analogous but with conjugated phases. The $4$-particle amplitudes extracted from (\ref{N=4SYM}) are 
\begin{align}\label{FullMixN=4}
&A[\widetilde{\phi},W,\lambda_1,\lambda_2]\nonumber\\
&=\left(m_1\left(\da{\bf{23}}e^{2i\varphi_1}-\ds{\bf{23}}e^{i(\varphi_2+\varphi_3)}\right)+e^{i\varphi_1}\left(e^{i\varphi_3}\la{\bf{2}}p_1\rs{\bf{3}}-e^{i\varphi_2}\la{\bf{3}}p_1\rs{\bf{2}}\right)\right)\nonumber\\
&\qquad\times \left(m_1\left(\da{\bf{24}}e^{-2i\varphi_1}-\ds{\bf{24}}e^{-i(\varphi_2+\varphi_4)}\right)+e^{i\varphi_1}\left(e^{-i\varphi_4}\la{\bf{2}}p_1\rs{\bf{4}}-e^{-i\varphi_2}\la{\bf{4}}p_1\rs{\bf{2}}\right)\right)\nonumber\\
&\qquad\times\frac{-1}{(t-m_t^2)(u-m_u^2)},
\end{align}
which has the simple factorised form of the numerator $N_{\mathcal{N}=2}[\widetilde{\phi},\chi,\chi,\phi]\times N_{\mathcal{N}=2}[\widetilde{\phi},\chi,\phi,\chi]$ (analogous to (\ref{N=4ScaVec})), and
\begin{align}\label{FullMixN=4Higgs}
& A[\phi_{(12)},W,\lambda_1,\widetilde{\lambda}^1]\nonumber\\
&=\Big((u-m_1^2-m_4^2)\left(e^{i(\varphi_1-\varphi_2)}\da{\bf{23}}\ds{\bf{24}}+e^{i(\varphi_2+\varphi_3-\varphi_1-\varphi_4)}\ds{\bf{23}}\da{\bf{24}}\right)\nonumber\\
&\qquad+(t-m_3^2)\left(e^{i(\varphi_2-\varphi_1)}\da{\bf{23}}\ds{\bf{24}}+e^{i(\varphi_1+\varphi_3-\varphi_2-\varphi_4)}\ds{\bf{23}}\da{\bf{24}}\right)\nonumber\\
&\qquad+e^{i(\varphi_1-\varphi_4)}\la{\bf{3}}p_1\rs{\bf{2}}\la{\bf{4}}p_1\rs{\bf{2}}+e^{i(\varphi_3-\varphi_1)}\la{\bf{2}}p_1\rs{\bf{3}}\la{\bf{2}}p_1\rs{\bf{4}}\nonumber\\
&\qquad+\left(im_1\sin\left(\varphi_1-\varphi_2\right)e^{i(\varphi_1-\varphi_4)}-m_4e^{i\left(\varphi_1+\varphi_2-2\varphi_4\right)}\right)\da{\bf{34}}\la{\bf{2}}p_1\rs{\bf{2}}\nonumber\\
&\qquad-\left(im_1\sin\left(\varphi_1-\varphi_2\right)e^{i(\varphi_3-\varphi_1)}+m_4e^{i\left(\varphi_3+\varphi_4-\varphi_1-\varphi_2\right)}\right)\ds{\bf{34}}\la{\bf{2}}p_1\rs{\bf{2}}\nonumber\\
&\qquad-im_1\sin\left(\varphi_1-\varphi_2\right)\big(e^{i(\varphi_1-\varphi_4)}\left(\da{\bf{23}}\la{\bf{4}}p_1\rs{\bf{2}}+\da{\bf{24}}\la{\bf{3}}p_1\rs{\bf{2}}\right)\nonumber\\
&\qquad\qquad\qquad\qquad\qquad\qquad-e^{i(\varphi_3-\varphi_1)}\left(\ds{\bf{23}}\ls{\bf{4}}p_1\ra{\bf{2}}+\ds{\bf{24}}\ls{\bf{3}}p_1\ra{\bf{2}}\right)\big)\nonumber\\
&\qquad-\left(m_3e^{i\left(\varphi_1-\varphi_4\right)}+m_4e^{i\left(\varphi_3-\varphi_1\right)}\right)\left(\da{\bf{24}}\la{\bf{2}}p_1\rs{\bf{3}}+\ds{\bf{23}}\ls{\bf{2}}p_1\ra{\bf{4}}\right)\nonumber\\
&\qquad-\left(m_3e^{i\left(\varphi_3-\varphi_1\right)}+m_4e^{i\left(\varphi_1-\varphi_4\right)}\right)\left(\da{\bf{23}}\la{\bf{2}}p_1\rs{\bf{4}}+\ds{\bf{24}}\ls{\bf{2}}p_1\ra{\bf{3}}\right)\nonumber\\
&\qquad +m_1\Big(\left(4im_4\sin\left(\varphi_2-\varphi_4\right)-m_1e^{-i\varphi_2}\right)\da{\bf{23}}\ds{\bf{24}}\nonumber\\
&\qquad\qquad\qquad+e^{i(\varphi_3-\varphi_4)}\left(4im_4\sin\left(\varphi_4-\varphi_2\right)-m_1e^{i(\varphi_2-\varphi_1)}\right)\ds{\bf{23}}\da{\bf{24}}\Big)\nonumber\\
&\qquad +m_1^2\left(e^{i(\varphi_3-\varphi_1)}\ds{\bf{23}}\ds{\bf{24}}+e^{i(\varphi_1-\varphi_4)}\da{\bf{23}}\da{\bf{24}}\right)\Big)\times\frac{1}{(t-m_t^2)(u-m_u^2)}.
\end{align}
Applying identity (\ref{LowSyz2}) reveals that (\ref{FullMixN=4}) also contains single pole terms and, once the BPS conditions on the masses are applied, agreement with (\ref{MixAmpAbelian}) and (\ref{MixAmpNonAbelian}) (after the manifestly unitary single pole terms from (\ref{FactAllAmp}) are added). The two distinct amplitudes (\ref{FullMixN=4}) and (\ref{FullMixN=4Higgs}) correspond to (\ref{MixAmpAbelian}) and (\ref{MixAmpNonAbelian}) with different dependencies of the couplings on the phases of the central charges. 

\begin{table}[h!]
\begin{center}
\begin{tabular}{ c || c c c c c}
Growth & Helicity Config & Couplings & Structure & Removal \\
\hline\hline
$E^2$ & $++,-,-$ & $\frac{c}{\Lambda}\frac{m}{\Lambda}$ & fact & suppress\\
\hline
 & $++,+,-$ & $\frac{c}{\Lambda}t$,$\frac{m}{\Lambda}y$ & fact & suppress  \\
\cline{2-5}
$E$ & $L,+,+$ & $\frac{m}{\Lambda} t$ & fact & suppress\\
 &  & $\frac{1}{m}ty$ & contact & Yukawa cov
\end{tabular}
\end{center}
\caption{Summary of results for the mixed all spins amplitude.}\label{tab:1V2P1S}
\end{table}

Table \ref{tab:1V2P1S} summarises the conclusions of this subsection. Unfortunately, the symbol ``$m$'' has double use: if it appears in the numerator then it always represents a dipole coupling, while in the denominator it represents mass. The generic Yukawa coupling ``$y$'' includes those implicit in (\ref{YukEmerge}).

\subsection{Two vectors and two fermions}\label{sec:2Vec2Ferm}

The full covariance of the emergent Yukawa couplings still requires the extension of (\ref{CovYukawas1}) to be established when the external boson index is identified with a longitudinal vector mode. This information lies in the vector-fermion amplitude, which is the last remaining $4$-particle amplitude to be calculated in this analysis.  

\subsubsection{General amplitude, Lie algebra and leading cancellations}

\begin{feynfig}
\begin{figure}[h]
\begin{fmffile}{VecFermion}
 \begin{center}
 \begin{tabular}{c c c c c}
 & & & & \\
 \begin{fmfgraph*}(100,67)
   \fmfleft{i1,i2}
   \fmfright{o1,o2}
   \fmf{boson}{i2,v1}
   \fmf{plain}{v2,o2}
   \fmf{boson}{i1,v1}
   \fmf{plain}{v2,o1}
   \fmf{dashes}{v1,v2}
   \fmfv{decor.shape=circle,decor.filled=gray50,decor.size=0.15w}{v1,v2}
   \fmflabel{$A$}{i1}
   \fmflabel{$B$}{i2}
   \fmflabel{$i$}{o2}
   \fmflabel{$j$}{o1}
 \end{fmfgraph*} 
 &\,& \begin{fmfgraph*}(100,67)
   \fmfleft{i1,i2}
   \fmfright{o1,o2}
   \fmf{boson}{i2,v1}
   \fmf{plain}{v2,o2}
   \fmf{boson}{i1,v1}
   \fmf{plain}{v2,o1}
   \fmf{boson}{v1,v2}
   \fmfv{decor.shape=circle,decor.filled=gray50,decor.size=0.15w}{v1,v2}
   \fmflabel{$A$}{i1}
   \fmflabel{$B$}{i2}
   \fmflabel{$i$}{o2}
   \fmflabel{$j$}{o1}
 \end{fmfgraph*} 
 &\,& \begin{fmfgraph*}(100,67)
   \fmfleft{i1,i2}
   \fmfright{o1,o2}
   \fmf{phantom}{i2,v1}
   \fmf{plain}{v1,o2}
   \fmf{phantom}{i1,v2}
   \fmf{plain}{v2,o1}
   \fmf{plain}{v1,v2}
   \fmf{boson,tension=-0.25}{v1,i1}
   \fmf{boson,tension=-0.25}{v2,i2}
   \fmfv{decor.shape=circle,decor.filled=gray50,decor.size=0.15w}{v1,v2}
   \fmflabel{$A$}{i1}
   \fmflabel{$B$}{i2}
   \fmflabel{$i$}{o2}
   \fmflabel{$j$}{o1}
 \end{fmfgraph*}\nonumber\\
$(s)$ &\,& $(s)$ &\,& $(t)$
 \end{tabular}
 \end{center}
\end{fmffile}
\caption{Vector-fermion scattering (omitting the $u$-channel).}\label{2V2FOSDi}
\end{figure}
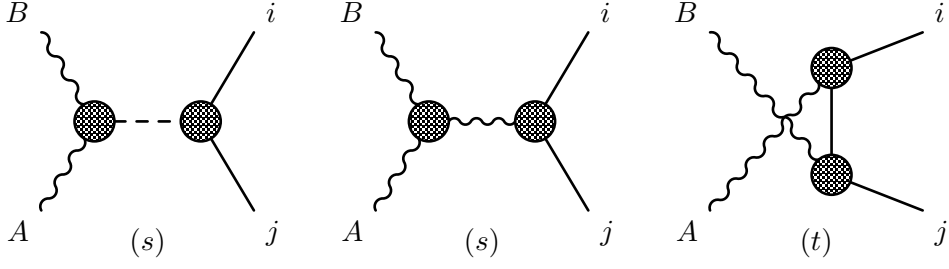
\end{feynfig}

The massive fermion and vector amplitude can be constructed from $t$ and $u$-channel fermion exchange: 
\begin{align}
&A_t\left(W_A,W_B,\psi_i,\psi_j\right)\nonumber\\
&\qquad=\frac{-1}{t-m_{t_k}^2}\frac{1}{m_1m_2}\nonumber\\
&\qquad\qquad\times\Big(
(t_A)_{ik}(t_B)_{kj}\da{\bf{13}}\ds{\bf{24}}\la{\bf{2}}P_t\rs{\bf{1}}+((t_A)_{ik})^*((t_B)_{kj})^*\ds{\bf{13}}\da{\bf{24}}\ls{\bf{2}}P_t\ra{\bf{1}}\nonumber\\
&\qquad\qquad\qquad\qquad+((t_A)_{ik})^*(t_B)_{kj}m_{t_k}\ds{\bf{13}}\ds{\bf{24}}\da{\bf{12}}+(t_A)_{ik}((t_B)_{kj})^*m_{t_k}\da{\bf{13}}\da{\bf{24}}\ds{\bf{12}}\Big),
\end{align}
and the $u$-channel exchange expression is given by swapping the vector bosons above. The $s$-channel can be mediated by a scalar or vector exchange: 
\begin{align}\label{WWFFs}
&A_{s\varphi}\left(W_A,W_B,\psi_i,\psi_j\right)=\frac{-1}{s-m_{s_m}^2}\left(\frac{1}{\sqrt{2}}\lambda_{AB}^m\da{\bf{12}}\ds{\bf{12}}+\frac{c^m_{AB}}{\Lambda}\ds{\bf{12}}^2+\frac{(c^m_{AB})^*}{\Lambda}\da{\bf{12}}^2\right)\nonumber\\
&\qquad\qquad\qquad\qquad\qquad\qquad\qquad\qquad\qquad\qquad\qquad\qquad\qquad\times\left(y_{ij}^m\da{\bf{34}}+(y_{ij}^m)^*\ds{\bf{34}}\right)\nonumber\\
&A_{sW}\left(W_A,W_B,\psi_i,\psi_j\right)\nonumber\\
&=\frac{-1}{s-m_{s_M}^2}\frac{i}{4m_1m_2}\nonumber\\
&\quad\times\Bigg((t_M)_{ij}\bigg(\Re{f}_{AB}^{\,\,\,\,\,\,\,\,\,M}(\da{\bf{12}}\left(\ds{\bf{14}}\ls{\bf{2}}P_s\ra{\bf{3}}+\ds{\bf{24}}\ls{\bf{1}}P_s\ra{\bf{3}}\right)\nonumber\\
&\quad\quad\quad\quad\quad\quad\quad\quad\quad\quad\quad\quad+\ds{\bf{12}}\left(\da{\bf{13}}\la{\bf{2}}P_s\rs{\bf{4}}+\da{\bf{23}}\la{\bf{1}}P_s\rs{\bf{4}}\right))\nonumber\\
&\quad\quad\quad\quad\quad\quad\quad+2\Im{f}_{AB}^{\,\,\,\,\,\,\,\,\,M}(\ds{\bf{14}}\da{\bf{23}}\left(m_1\ds{\bf{12}}-m_2\da{\bf{12}}\right)\nonumber\\
&\quad\quad\quad\quad\quad\quad\quad\quad\quad\quad\quad\quad\quad\quad+\ds{\bf{24}}\da{\bf{13}}\left(m_2\ds{\bf{12}}-m_1\da{\bf{12}}\right))\nonumber\\
&\quad\quad\quad\quad\quad\quad\quad-\left(m_1{f}_{BM}^{\,\,\,\,\,\,\,\,\,A}\ds{\bf{12}}+m_2({f}_{MA}^{\,\,\,\,\,\,\,\,\,B})^*\da{\bf{12}}\right)\nonumber\\
&\quad\quad\quad\quad\quad\quad\quad\quad\times\left(2\da{\bf{23}}\ds{\bf{14}}+\frac{1}{m_{s_M}^2}\left(m_1\da{\bf{12}}-m_2\ds{\bf{12}}\right)\left(m_4\da{\bf{34}}-m_3\ds{\bf{34}}\right)\right)\nonumber\\
&\quad\quad\quad\quad\quad\quad\quad-\left(m_1({f}_{BM}^{\,\,\,\,\,\,\,\,\,A})^*\da{\bf{12}}+m_2{f}_{MA}^{\,\,\,\,\,\,\,\,\,B}\ds{\bf{12}}\right)\nonumber\\
&\quad\quad\quad\quad\quad\quad\quad\quad\times\left(2\da{\bf{13}}\ds{\bf{24}}+\frac{1}{m_{s_M}^2}\left(m_1\ds{\bf{12}}-m_2\da{\bf{12}}\right)\left(m_4\da{\bf{34}}-m_3\ds{\bf{34}}\right)\right)\bigg)\nonumber\\
&\quad\quad-((t_M)_{ij})^*\bigg(\Re{f}_{AB}^{\,\,\,\,\,\,\,\,\,M}(\da{\bf{12}}\left(\ds{\bf{13}}\ls{\bf{2}}P_s\ra{\bf{4}}+\ds{\bf{23}}\ls{\bf{1}}P_s\ra{\bf{4}}\right)\nonumber\\
&\quad\quad\quad\quad\quad\quad\quad\quad\quad\quad\quad\quad\quad+\ds{\bf{12}}\left(\da{\bf{14}}\la{\bf{2}}P_s\rs{\bf{3}}+\da{\bf{24}}\la{\bf{1}}P_s\rs{\bf{3}}\right))\nonumber\\
&\quad\quad\quad\quad\quad\quad\quad\quad+2\Im{f}_{AB}^{\,\,\,\,\,\,\,\,\,M}(\ds{\bf{13}}\da{\bf{24}}\left(m_1\ds{\bf{12}}-m_2\da{\bf{12}}\right)\nonumber\\
&\quad\quad\quad\quad\quad\quad\quad\quad\quad\quad\quad\quad\quad\quad\quad+\ds{\bf{23}}\da{\bf{14}}\left(m_2\ds{\bf{12}}-m_1\da{\bf{12}}\right))\nonumber\\
&\quad\quad\quad\quad\quad\quad\quad\quad-\left(m_1{f}_{BM}^{\,\,\,\,\,\,\,\,\,A}\ds{\bf{12}}+m_2({f}_{MA}^{\,\,\,\,\,\,\,\,\,B})^*\da{\bf{12}}\right)\nonumber\\
&\quad\quad\quad\quad\quad\quad\quad\quad\quad\times\left(2\da{\bf{24}}\ds{\bf{13}}+\frac{1}{m_{s_M}^2}\left(m_1\da{\bf{12}}-m_2\ds{\bf{12}}\right)\left(m_4\ds{\bf{34}}-m_3\da{\bf{34}}\right)\right)\nonumber\\
&\quad\quad\quad\quad\quad\quad\quad\quad-\left(m_1({f}_{BM}^{\,\,\,\,\,\,\,\,\,A})^*\da{\bf{12}}+m_2{f}_{MA}^{\,\,\,\,\,\,\,\,\,B}\ds{\bf{12}}\right)\nonumber\\
&\quad\quad\quad\quad\quad\quad\quad\quad\quad\times\left(2\da{\bf{14}}\ds{\bf{23}}+\frac{1}{m_{s_M}^2}\left(m_1\ds{\bf{12}}-m_2\da{\bf{12}}\right)\left(m_4\ds{\bf{34}}-m_3\da{\bf{34}}\right)\right)\bigg)\Bigg).
\end{align}
The relevant on-shell diagrams are shown in Fig \ref{2V2FOSDi}.

I omit the possible contributions arising from the $h$-type $F^3$ type couplings and the $m$-type dipole couplings in (\ref{FermionMat3leg}). For clarity of exposition, these terms are deferred to Appendix \ref{app:2V2P}. Their inclusion does not modify any of the analysis presented in this Section. 

Now neglecting the dipole interactions, the leading order HEL divergences scale as $\sim E^2$ for helicity configurations in which both vectors are longitudinal and the fermions are opposite. Demanding that the $\sim E^2$ terms cancel forces the couplings to obey the Lie algebra commutator (\ref{LAscalar}). The fermion-vector couplings are therefore identified as Lie algebra generators in some representation (thereby justifying the interpretation of (\ref{CovYukawas1}) as covariance relations). 

The leading HEL terms can then be combined and reduced to leave an amplitude now scaling as $\sim E$ instead. Homogenising the propagator masses with the usual procedure (\ref{HomoDen}), then just as for the scalar case, the leading divergent terms in the amplitude can be grouped into ``Abelian'' and ``non-Abelian'' components:
\begin{align}\label{VecFermE2}
&A_{E^2}(W_A,W_B,\psi_i,\psi_j)=\nonumber\\
&\qquad\frac{-1}{2m_1m_2}\left((t_A)_{ik}(t_B)_{kj}+(t_B)_{ik}(t_A)_{kj}\right)\left(\frac{\da{\bf{13}}\ds{\bf{24}}\ls{\bf{1}}P_t\ra{\bf{2}}}{t-m_{t}^2}+\frac{\da{\bf{23}}\ds{\bf{14}}\ls{\bf{2}}P_u\ra{\bf{1}}}{u-m_{u}^2}\right)\nonumber\\
&\qquad+\frac{1}{2m_1m_2}i\Re {f}_{AB}^{\,\,\,\,\,\,\,\,\,M}(t_M)_{ij}\bigg(-\frac{\da{\bf{13}}\ds{\bf{24}}\ls{\bf{1}}P_t\ra{\bf{2}}}{t-m_{t}^2}+\frac{\da{\bf{23}}\ds{\bf{14}}\ls{\bf{2}}P_u\ra{\bf{1}}}{u-m_{u}^2}\nonumber\\
&\qquad\qquad-\frac{\da{\bf{12}}\left(\ds{\bf{14}}\ls{\bf{2}}P_s\ra{\bf{3}}+\ds{\bf{24}}\ls{\bf{1}}P_s\ra{\bf{3}}\right)+\ds{\bf{12}}\left(\da{\bf{13}}\la{\bf{2}}P_s\rs{\bf{4}}+\da{\bf{23}}\la{\bf{1}}P_s\rs{\bf{4}}\right)}{2(s-m_s^2)}\bigg)
\end{align}
and likewise for the parity conjugate terms, which have been omitted for brevity, but which may be inferred by switching the shape of the spinor brackets and complex conjugating the coefficients. Using (\ref{3PolSyz}), (\ref{LowSyz2}) and preserving the manifest exchange symmetries, the Abelian part can be reduced to 
\begin{align}\label{VecFermMerge1}
&(u-m_u^2)\da{\bf{13}}\ds{\bf{24}}\ls{\bf{1}}P_t\ra{\bf{2}}+(t-m_t^2)\da{\bf{23}}\ds{\bf{14}}\ls{\bf{2}}P_u\ra{\bf{1}}\nonumber\\
&=\frac{1}{2}\Big((u-m_u^2)\left(m_4\ds{\bf{12}}\da{\bf{13}}\da{\bf{24}}+m_3\da{\bf{12}}\ds{\bf{13}}\ds{\bf{24}}\right)\nonumber\\
&\quad\quad\quad-(t-m_t^2)\left(m_4\ds{\bf{12}}\da{\bf{14}}\da{\bf{23}}+m_3\da{\bf{12}}\ds{\bf{14}}\ds{\bf{23}}\Big)\right)\nonumber\\
&\quad+\frac{1}{2}\Big(\la{\bf{2}}p_4\rs{\bf{1}}\Big(m_2m_3\da{\bf{12}}\ds{\bf{34}}+m_1m_4\ds{\bf{12}}\da{\bf{34}}-2m_1m_2\da{\bf{23}}\ds{\bf{14}}\nonumber\\
&\qquad\qquad\qquad\qquad\qquad\qquad\qquad\qquad\qquad\qquad\qquad+\left(m_3^2+m_4^2-2m_u^2\right)\da{\bf{13}}\ds{\bf{24}}\Big)\nonumber\\
&\quad\quad-\la{\bf{1}}p_4\rs{\bf{2}}\Big(m_1m_3\da{\bf{12}}\ds{\bf{34}}+m_2m_4\ds{\bf{12}}\da{\bf{34}}+2m_1m_2\da{\bf{13}}\ds{\bf{24}}\nonumber\\
&\quad\qquad\qquad\qquad\qquad\qquad\qquad\qquad\qquad\qquad\qquad\qquad-\left(m_3^2+m_4^2-2m_t^2\right)\da{\bf{23}}\ds{\bf{14}}\Big)\nonumber\\
&\quad\quad +m_4\da{\bf{12}}\left(m_1m_3\da{\bf{14}}\ds{\bf{23}}-m_2m_3\da{\bf{24}}\ds{\bf{13}}+m_1m_4\da{\bf{13}}\ds{\bf{24}}-m_2m_4\da{\bf{23}}\ds{\bf{14}}\right)\nonumber\\
&\quad\quad +m_2\left(2m_u^2-2m_1^2-m_4^2\right)\da{\bf{13}}\ds{\bf{12}}\ds{\bf{24}}-m_1\left(2m_t^2-2m_2^2-m_4^2\right)\da{\bf{23}}\ds{\bf{12}}\ds{\bf{14}}\nonumber\\
&\quad\quad+m_3\da{\bf{12}}\left(\left(m_u^2-m_1^2-m_4^2\right)\ds{\bf{13}}\ds{\bf{24}}-\left(m_t^2-m_2^2-m_4^2\right)\ds{\bf{14}}\ds{\bf{23}}\right)\nonumber\\
&\quad\quad +m_4\ds{\bf{12}}\left((m_u^2-m_3^2)\da{\bf{13}}\da{\bf{24}}-(m_t^2-m_3^2)\da{\bf{23}}\da{\bf{14}}\right)-2m_1m_2m_4\ds{\bf{12}}^2\da{\bf{34}}\Big).
\end{align}
The $s$-pole term may be separately combined with the $t$ and $u$-pole terms in the non-Abelian part of (\ref{VecFermE2}):
\begin{align}\label{VecFermMergeS}
&-(s-m_s^2)\da{\bf{13}}\ds{\bf{24}}\ls{\bf{1}}P_t\ra{\bf{2}}-\frac{1}{4}(t-m_t^2)(\da{\bf{12}}\left(\ds{\bf{14}}\ls{\bf{2}}P_s\ra{\bf{3}}+\ds{\bf{24}}\ls{\bf{1}}P_s\ra{\bf{3}}\right)\nonumber\\
&\qquad\qquad\qquad\qquad\qquad\qquad\qquad\qquad\qquad\qquad+\ds{\bf{12}}\left(\da{\bf{13}}\la{\bf{2}}P_s\rs{\bf{4}}+\da{\bf{23}}\la{\bf{1}}P_s\rs{\bf{4}}\right))\nonumber\\
&=-\frac{1}{2}(t-m_t^2)\Big(\da{\bf{13}}\ds{\bf{24}}\left(m_1\da{\bf{12}}+m_2\ds{\bf{12}}\right)+\da{\bf{23}}\ds{\bf{14}}\left(m_1\ds{\bf{12}}+m_2\da{\bf{12}}\right)\nonumber\\
&\qquad\qquad\qquad\qquad\qquad\qquad\qquad\qquad\qquad\qquad\qquad+\da{\bf{12}}\ds{\bf{12}}\left(m_3\ds{\bf{34}}-m_4\da{\bf{34}}\right)\Big)\nonumber\\
&\quad-(s-m_s^2)m_3\ds{\bf{13}}\ds{\bf{24}}\da{\bf{12}}\nonumber\\
&\quad+\la{\bf{2}}p_4\rs{\bf{1}}\left(m_1m_4\ds{\bf{12}}\da{\bf{34}}-m_1m_2\ds{\bf{14}}\da{\bf{23}}+\left(m_s^2+m_t^2-m_1^2-m_2^2-m_3^2\right)\ds{\bf{24}}\da{\bf{13}}\right)\nonumber\\
&\quad-\la{\bf{1}}p_4\rs{\bf{2}}\left(m_1m_3\da{\bf{12}}\ds{\bf{34}}+m_1m_2\da{\bf{13}}\ds{\bf{24}}-\left(m_3^2-m_t^2\right)\da{\bf{23}}\ds{\bf{14}}\right)\nonumber\\
&\quad+m_1(m_2^2-m_t^2)\ds{\bf{12}}\ds{\bf{14}}\da{\bf{23}}+m_4(m_3^2-m_t^2)\ds{\bf{12}}\da{\bf{14}}\da{\bf{23}}\nonumber\\
&\quad-m_1m_3m_4\ds{\bf{12}}\ds{\bf{13}}\da{\bf{24}}+m_1m_3m_4\da{\bf{12}}\da{\bf{14}}\ds{\bf{23}}\nonumber\\
&\quad+(m_2^2+m_3^2-m_s^2-m_t^2)\ds{\bf{24}}\left(m_3\ds{\bf{13}}\da{\bf{12}}+m_2\ds{\bf{12}}\da{\bf{13}}\right)\nonumber\\
&\quad+m_1^2m_4\ds{\bf{12}}\da{\bf{13}}\da{\bf{24}}+m_1m_4^2\ds{\bf{24}}\da{\bf{12}}\da{\bf{13}}-m_1m_2m_4\ds{\bf{12}}^2\da{\bf{34}}
\end{align}
for the $st$ combination, while the $su$ combination is obtained by exchanging $1\leftrightarrow 2$ and multiplying by $(-1)$. However, the relation (\ref{VecFermMergeS}) applied to (\ref{VecFermE2}) in this way leads to the breaking of manifest identical fermion exchange (anti)symmetry, even among the terms with leading energy dependence. In order to restore it, I average over this expression and its image obtained by simultaneously parity conjugating and exchanging $3\leftrightarrow 4$ (multiplied by $(-1)$).

\subsubsection{Full unitarisation and Yukawa covariance}

Demanding full unitarisation places new constraints on the couplings. Unitarisation of the amplitudes with a single transversely polarised vector leads to the standard YM properties of the Lie algebra (\ref{StandardLA}), at least for the subset of vectors that couple to the fermions through Lie algebra generators. The terms divergent in these limits then directly cancel once this condition is obeyed. There are also terms that scale as $\sim E$ for fully transverse vectors. These are given by the $s$-channel scalar exchange in (\ref{WWFFs}) mediated by both the $c_{AB}^m$ and Yukawa couplings, but also appear in the vector exchange terms in (\ref{WWFFs}) with the same Lorentz structure. These latter terms have the same form as the scalar exchange terms once the the presence of the emergent Yukawa couplings (\ref{YukEmerge}) and the emergent $c$-type couplings (\ref{3legVecHEL}) are identified. They therefore represent the exchange of the emergent scalar particle identifiable with the longitudinal state of the vector in the HEL. 

An obvious sufficient condition for the elimination of these divergences is the suppression of the $c$-type couplings, both those in (\ref{2vec1scalar}) and (\ref{3legVecHEL}) (the latter being equivalent to (\ref{StandardLA})). It would also seem possible that these terms could cancel their own divergences, potentially leaving behind only mass splitting terms. This would reflect a situation in which, in the UV, no scalar is actually exchanged between the vectors and the fermions through the effective $\varphi F^2$ interaction (that is, none simultaneously participate in Yukawa couplings to the fermions and the $\varphi F^2$ interactions, despite the presence of the latter) but where a vector exchange does still happen (this is analogous to the scenario described in Section \ref{sec:PartialU} above regarding the four fermion amplitude when the mass splitting terms are introduced). Then, in order to accommodate the exchanged vector having mass in the IR, one of the UV scalars must necessarily be drawn into the process to produce the relevant terms in the vector exchange amplitude in (\ref{WWFFs}), only for it to be canceled by a residual scalar exchange (up to mass splitting terms). In either case, unitarity requires that these terms be eliminated in such a way that they are set to zero or directly cancel.

The remaining divergences occur when the vector bosons are both longitudinally polarised and the fermions have the same helicity. Requiring that these divergences cancel imposes the constraint:
\begin{align}\label{CovYukawas2}
&(t_A)_{ik}\left(m_4(t_B)_{kj}+m_{t_k}(t_B)_{jk}\right)+(t_A)_{jk}\left(m_{u_k}(t_B)_{ik}+m_{3}(t_B)_{ki}\right)-\frac{1}{\sqrt{2}}m_1m_2\lambda_{AB}^m y_{ij}^m\nonumber\\
&\qquad-\frac{i}{2}\left(\Re {f}_{AB}^{\,\,\,\,\,\,\,\,\,M}+\frac{1}{m_{s_M}^2}\left(m_2^2\Re {f}_{MA}^{\,\,\,\,\,\,\,\,\,B}-m_1^2\Re {f}_{BM}^{\,\,\,\,\,\,\,\,\,A}\right)\right)\left(m_4(t_M)_{ij}+m_3(t_M)_{ji}\right)=0.
\end{align}
Recognising (\ref{HiggsGen}) and (\ref{BrokenGenGeneral}) in the above expression identifies it as the covariance of the emergent Yukawa couplings (\ref{YukEmerge}) under the emergent Lie algebra: 
\begin{align}
(t_A)_{ik}y_{kj}^B+(t_A)_{jk}y_{ik}^B+(t_A)_{BM}y_{ik}^M+(t_A)_{Bm}y_{ik}^m=0,
\end{align}
which extends and completes the counterpart relation (\ref{CovYukawas1}) to include external boson indices identified as longitudinal modes of the massive vectors. While not manifest, (\ref{CovYukawas2}) is symmetric under the exchange of $1\leftrightarrow 2$. The covariance condition on the emergent couplings, such as the Yukawa couplings here, has implications for the particle spectrum of the theory since, given the Lie algebra, their existence is only possible provided that the states assemble into consistent representations.

Retaining the $c$-type couplings implicit in the three vector amplitude or, equivalently, not assuming (\ref{StandardLA}), then these interactions can induce comparatively small terms in the amplitude that scale $\sim E/\Lambda$ in the HEL if they are present but suppressed. The left hand side of (\ref{CovYukawas2}) is the coupling coefficient of an emergent dim-$5$ $4$-particle coupling between two fermions and two scalars (accompanying the kinematic structure $\ds{34}$). This term scales as $\sim E/m$ in the HEL (omitting the potentially $\sim 1$ accompanying coupling constants). However, accepting the presence of other $\sim E/\Lambda$ effective $3$-particle interactions, it would seem to be consistent to allow for the existence of these $4$-particle terms as well, provided that they are also of order $\sim E/\Lambda$. This will be satisfied as long as (\ref{CovYukawas2}) is obeyed up to $\mathcal{O}\left(m/\Lambda\right)$ corrections, although it is remarkable that, unlike the other dim-$5$ effective interactions discussed in this Section (like the fermion dipole couplings and $\varphi F^2$), the covariance of the Yukawa couplings remains exact even when the vector self-couplings merely obey the partial unitarisation conditions but depart from standard YM. 

Accepting the conditions required to eliminate all remaining HEL divergences, the remaining $\sim E$ divergent terms in the amplitude can be again grouped into a $tu$ ``Abelian'' part and a remaining ``non-Abelian'' part as 
\begin{align}\label{VecFermE1}
&A_{E}(W_A,W_B,\psi_i,\psi_j)=\nonumber\\
&\qquad\frac{-\ds{\bf{12}}}{2m_1m_2}\left(m_{t_k}(t_A)_{ik}(t_B)_{jk}-m_{u_k}(t_B)_{ik}(t_A)_{jk}+\frac{i}{2}f_{ABM}\left(m_4(t_M)_{ij}-m_3(t_M)_{ji}\right)\right)\nonumber\\
&\qquad\qquad\qquad\times\left(\frac{\da{\bf{13}}\da{\bf{24}}}{t-m_{t}^2}+\frac{\da{\bf{14}}\da{\bf{23}}}{u-m_{u}^2}\right)\nonumber\\
&\qquad-\frac{\ds{\bf{12}}}{2m_1m_2}\left(\frac{1}{\sqrt{2}}m_1m_2\lambda_{AB}^my_{ij}^m-\frac{i}{2}f_{ABM}\frac{m_1^2-m_2^2}{m_{s_M}^2}\left(m_4(t_M)_{ij}+m_3(t_M)_{ji}\right)\right)\nonumber\\
&\qquad\qquad\qquad\times\left(\frac{2\da{\bf{12}}\da{\bf{34}}}{s-m_{s}^2}+\frac{\da{\bf{13}}\da{\bf{24}}}{t-m_{t}^2}-\frac{\da{\bf{14}}\da{\bf{23}}}{u-m_{u}^2}\right)
\end{align}
and similarly for parity conjugate terms given by switching the shape of spinor brackets and complex conjugating the coefficients. Obviously my use of the words ``Abelian'' and ``non-Abelian'' parts here have been made to parallel the structure of (\ref{VecFermE2}) and are not intended as precise descriptions of the coupling algebra. Pairs of terms can be combined by applying (\ref{OldSyzFact}) to manifest the cancellation of the HEL. This results in e.g. the $t$ and $u$-pole terms combining in the ``Abelian'' part as
\begin{align}\label{VecFermMerge2}
&(u-m_u^2)\ds{\bf{12}}\da{\bf{13}}\da{\bf{24}}+(t-m_t^2)\ds{\bf{12}}\da{\bf{14}}\da{\bf{23}}\nonumber\\
&=\frac{1}{2}\Big(\ls{\bf{1}}p_4-p_3\ra{\bf{2}}\left(m_1\ds{\bf{12}}\da{\bf{34}}-m_3\ds{\bf{23}}\da{\bf{14}}+m_4\ds{\bf{24}}\da{\bf{13}}\right)\nonumber\\
&\quad+\ls{\bf{2}}p_4-p_3\ra{\bf{1}}\left(m_2\ds{\bf{12}}\da{\bf{34}}+m_3\ds{\bf{13}}\da{\bf{24}}-m_4\ds{\bf{14}}\da{\bf{23}}\right)\nonumber\\
&\quad-2m_3m_4\da{\bf{12}}\left(\ds{\bf{13}}\ds{\bf{24}}+\ds{\bf{14}}\ds{\bf{23}}\right)\nonumber\\
&\quad+\left(m_1^2+m_2^2-2m_t^2\right)\ds{\bf{12}}\da{\bf{14}}\da{\bf{23}}+\left(m_1^2+m_2^2-2m_u^2\right)\ds{\bf{12}}\da{\bf{13}}\da{\bf{24}}\nonumber\\
&\quad-\ds{\bf{12}}\left(m_1m_4\ds{\bf{14}}\da{\bf{23}}+m_2m_4\ds{\bf{24}}\da{\bf{13}}+m_1m_3\ds{\bf{13}}\da{\bf{24}}+m_2m_3\ds{\bf{23}}\da{\bf{14}}\right)\nonumber\\
&\quad+\da{\bf{12}}\left(m_1m_4\ds{\bf{24}}\da{\bf{13}}+m_2m_4\ds{\bf{14}}\da{\bf{23}}+m_1m_3\ds{\bf{23}}\da{\bf{14}}+m_2m_3\ds{\bf{13}}\da{\bf{24}}\right)\Big).
\end{align}
The kinematic parts of the other channels can be combined to give an identical expression by (\ref{OldSyzFact}) up to some subleading terms dependent upon the different internal masses. I leave the expression in the much less symmetric form above for the purposes of reconciling it with the $\mathcal{N}=4$ SYM expression presented further below, rather than preserving the simplicity of (\ref{OldSyzFact}).

It remains to check that the unitarised amplitude matches correctly onto the expected massless amplitudes in the HEL. Firstly, it is immediately clear that the terms corresponding to both transversely polarised vectors correctly match onto the expected massless gluon Compton amplitude. These terms are given by the third terms in the third and fifth lines of (\ref{VecFermMerge1}) and the second terms in the third and fourth lines of the right-hand side of (\ref{VecFermMergeS}), when substituted into (\ref{VecFermE2}). There is no contribution from the reduced $\sim E$ terms in (\ref{VecFermE1}) above. Secondly are the terms corresponding to a single transverse vector. While much less obvious, these can be compiled by collecting the relevant contributing terms in (\ref{VecFermMerge1}), (\ref{VecFermMergeS}), (\ref{VecFermE2}), (\ref{VecFermMerge2}), and (\ref{VecFermE1}) and their parity conjugates, while identifying the emergent scalar couplings in (\ref{BrokenGen}) and (\ref{YukEmerge}). Only a single Lorentz structure contributes to each helicity configuration, so is common to all such terms (e.g. $\ls{\bf{2}}p_4\ra{\bf{1}}\ds{\bf{12}}\da{\bf{34}}$ for the configuration $(L,++,-,-)$). The accompanying combinations of coupling constants then agree with their expected expressions on the residues of each channel. 

This leaves the terms with both vectors longitudinally polarised. In this case, there are two linearly independent Lorentz structures for each possible fermion chirality configuration. For example, for the $(L,L,-,+)$ configuration, these are $\la{\bf{1}}p_4\rs{\bf{2}}\da{\bf{23}}\ds{\bf{14}}$ and $\la{\bf{2}}p_4\rs{\bf{1}}\da{\bf{13}}\ds{\bf{24}}$, which match onto $\frac{1}{4}t\ds{14}\da{13}$ and $-\frac{1}{4}u\ds{14}\da{13}$ respectively in the HEL. Because these contain Mandelstam variables, each of the reduced terms in the massive amplitude match onto terms containing only a single massless pole in the HEL. Additionally, the leading order mass splitting terms also contribute. As required, these partially cancel against some of the terms that contain internal masses in (\ref{VecFermMerge1}) and (\ref{VecFermMergeS}) when substituted into (\ref{VecFermE2}), effectively replacing them with their physical values to ensure that the emergent couplings are correct. Some of the internal masses remain and cancel directly among themselves, which could be made manifest by making some different basis choices in the kinematic numerators of each pair of channels. When the cancellations are not direct, they can be made manifest by applying variations of the identities (\ref{LowSyz2})  and (\ref{3PolSyz}). However, as with the mixed amplitudes with scalars above, I will not bother to explicitly demonstrate these cancellations in the expression already provided for the full massive amplitude. I will simply note that the unitarised massive amplitude does correctly match onto the expected massless amplitudes in the HEL. The $t$ and $u$-channels match onto scalar-fermion scattering amplitude mediated by the emergent scalar-fermion Yukawa  couplings, while the $s$-channel matches onto scalar-fermion scattering mediated by a gluon. Finally, as for the previous mixed case, there is no emergent quartic contact coupling, as expected from the fact that there is no corresponding renormalisable operator. 

As usual, the fermion-vector amplitude can be compared with the special case of $\mathcal{N}=4$ SYM. The fermion-vector colour-stripped amplitude extracted from (\ref{N=4SYM}) is 
\begin{align}
&A[W,W,\lambda_1,\widetilde{\lambda}^1]\nonumber\\
&=\frac{-1}{(s-m_s^2)(u-m_u^2)}\epsilon_\varphi(1,2,3,4)\nonumber\\
&\qquad\times\left(m_4\left(\da{\bf{12}}e^{-2i\varphi_4}-\ds{\bf{12}}e^{-i(\varphi_1+\varphi_2)}\right)+e^{-i\varphi_4}\left(e^{-i\varphi_2}\la{\bf{1}}p_4\rs{\bf{2}}-e^{-i\varphi_1}\la{\bf{2}}p_4\rs{\bf{1}}\right)\right)
\end{align}
which has numerator with the factorised form $N_{\mathcal{N}=2}[\chi,\chi,\chi,\chi] \times N_{\mathcal{N}=2}[\chi,\chi,\phi,\widetilde{\phi}]$. This can be checked for agreement with the combined terms in (\ref{VecFermMerge1}), (\ref{VecFermMergeS}) and (\ref{VecFermMerge2}) once the BPS condition is applied to the masses (I admit that I only did this myself for the Abelian part). 

\begin{table}[h!]
\begin{center}
\begin{tabular}{ c || c c c c c}
Growth & Helicity Config & Couplings & Structure & Removal \\
\hline\hline
$E^3$ & $++,++,-,-$ & $\frac{h}{\Lambda^2}\frac{m}{\Lambda}$ & fact & suppress\\
\hline
 & $++,++,+,-$ & $\frac{h}{\Lambda^2}t$ & fact & suppress  \\
\cline{2-5}
 & $++,--,+,-$ & $\frac{m^2}{\Lambda^2}$ & fact & suppress \\
\cline{2-5}
$E^2$ & $++,L,-,-$ & $\frac{c}{\Lambda}\frac{m}{\Lambda}$ & fact & suppress\\
\cline{2-5}
& $++,L,+,+$ & $\frac{m}{\Lambda}\{\frac{1}{m}\{t,f\},\frac{c}{\Lambda}\}$ & contact & dipole cov\\
\cline{2-5}
& $L,L,+,-$ & $\frac{1}{m^2}t\{t,f\}$ & contact & Lie algebra reps\\
\hline
 & $++,--,+,+$ & $\frac{m}{\Lambda}\{t,f\}$ & fact & suppress\\
 \cline{2-5}
$E$ & $\begin{array}{l}
++,++,+,+ \\
++,++,-,- \\
++,L,+,-
\end{array}$ & $\frac{c}{\Lambda}\{t,y\}$ & fact & suppress\\
 \cline{2-5}
 & $L,L,+,+$ & $\frac{1}{m}t\{t,f,y\}$ & contact & Yukawa cov\\
\end{tabular}
\end{center}
\caption{Summary of results for the vector-fermion amplitude.}\label{tab:2V2P}
\end{table}

Table \ref{tab:2V2P} summarises the qualitative conclusions of this analysis. Again, $m$ in the numerator represents a dipole coupling, $m$ in the denominator represents a particle mass.

\section{Conclusion}

In quantum mechanical descriptions of interacting particles, the combined principles of Lorentz invariance, unitarity and the specific analytic structure of scattering amplitudes are sufficient to bootstrap much of fundamental physics. Assuming that two-to-two scattering amplitudes contain only simple poles as singularities (when viewed as a function of helicity spinors), then the residues of these poles must describe on-shell propagation of an intermediary particle. For massless particles however, these residues can themselves contain poles that must admit a consistent interpretation of a cross-channel exchange. The poles must also be consistent with multiple possible complex approaches or coherent contributions to residues from multiple channels. Consistent factorisation places powerful constraints on permissible structures of the interactions. In this study, I analyse the implications of the all-channel pole for the consistency of gluonic and photonic self-couplings and couplings to matter. This completes the story of the derivation of perturbative YM and electromagnetism from the fundamental principles of the $S$-matrix, encompassing the results of \cite{McGady:2013sga}, \cite{Caron-Huot:2018ape} and \cite{Fonseca:2025mzj} and a few final details.

With massive particles, Lorentz invariance allows for a broader range of independent structures appearing in on-shell amplitudes and consistent factorisation is less demanding. Instead, a successive hierarchy of constraints emerges from demanding degrees of suppression of high energy growth. A partial unitarisation of the amplitudes to the standards of gravity enforces the Lie algebra structure of the $P$ and $T$-symmetric parts of the massive vector boson self-couplings and also allows for $P$ and $T$-violating Generalised Chern-Simons terms. When some of the vector bosons are massless gluons, these results are also required for self-consistency. Demanding full unitarisation leads to the further specifications on the Lie algebra expected for Yang-Mills theory, while the longitudinally polarised vectors are required to combine with the scalar particles into a valid representation space. Minimal matter couplings to vectors are required to be Lie algebra generators, while all others, including those involving longitudinally polarised vectors, must be (approximately) covariant tensors respecting the representations. 
More broadly, I examine the individual conditions required for the unitarisation of each helicity sector in the massive amplitudes and the space of EFTs that emerge from select subsets of conditions.

Finally, I calculate, from first principles, all two-to-two tree-level scattering amplitudes involving massive particles with spin $s\leq 1$ and present them in a form in which their high-energy dependence in manifest. The unitarised expressions for the amplitudes are reconciled with the especially elegant results for $\mathcal{N}=2$ super-QED and $\mathcal{N}=4$ super-Yang-Mills, which are entirely determined by supersymmetry and the BPS condition. 

In a companion work \cite{Trott:2026ozo}, I extend these results to include contributions from graviton exchanges and also investigate connections with supersymmetric theories. Consistent factorisation is used to complete the first-principles derivation of supergravity from the presence of a massless helicity-$3/2$ state. It is also shown that supersymmetric theories of massive BPS particles, with or without gravity, are further constrained by consistent factorisation of superamplitudes, which mandates the emergence of further order on the permissible couplings. In particular, with $\mathcal{N}=2$ supersymmetry, the Jacobi and GCS conditions are forced by self-consistency, affirming the above picture emerging from the weaker requirement of suppressed high-energy scaling. I also elaborate more on the details of the elegant expressions for the $4$-particle amplitudes of BPS particles in extended SUSY theories and identify the special properties of the two equal mass, one massless $3$-particle amplitudes as a special case of general BPS kinematics.

However, other avenues of further work remain. The crossing relations between amplitudes used here and in \cite{Trott:2026ozo} are established either by direct comparison with the Feynman rules or by intelligent guessing. Missing is a first principles derivation (in QFT or otherwise). To my knowledge, this was never done adequately for spinning particles. In any case, the details of the analytic continuation of the (spinning) amplitudes were able to be sidestepped to obtain the results presented here, but an overall sign was still required. 

The bootstrap methodology deduces necessity but not sufficiency. The conditions on the couplings and masses derived here are necessary for suppressed high-energy dependence of the general $4$-leg massive amplitudes that were explicitly calculated. These conditions agree with the expected results established in field theory. The existence of examples validates the conclusions. Extending the analysis here to include loops is an obvious further direction. It has been assumed here that loops are suppressed, but a power counting rule justifying this and guiding a hierarchical ordering of terms in a loop expansion remains a vital further ingredient for a consistent EFT (see \cite{Gavela:2016bzc} for examples). This should exist for a gauged non-linear sigma model, but is unclear for a theory of less structured trivalent couplings. Perturbative unitarity may continue to provide a guide toward this. Establishing the precise relation between the space of consistent EFTs and the space of possible Lorentz structures is an interesting question. The question of convergence is closely related, see, for example, \cite{Alonso:2023upf,Alonso:2025ksv} for recent commentary in the context of HEFT. The off-shell picture of potentials and field geometry is also something that warrants further investigation \cite{Cornwall:1974km,Cheung:2021yog,Derda:2024jvo,Alonso:2023upf,DAuria:1998emc}. In this study, I have identified the relation between the emergent contact interactions and masses and couplings, but not reconstructed the vacuum structure, simply asserting agreement with the expected theories that the HEL matches onto. 

Implicit in the ``high-energy limit'' has been the assumption of a mass gap involving the intermediate exchanged states for each tested amplitude. The emergent Lie algebra representations require the existence of a Higgs boson for a gapped theory to be unitary. Examples of Higgsless Kaluza-Klein Yang-Mills theories exist \cite{Chivukula:2001esy,Csaki:2003dt} in which each amplitude between mass eigenstates has unitary UV behaviour by the standards of this analysis. These theories have properties consistent with the results of this study in spite of not being gapped and not being unitary at high energies. However, the role of the mass gap and the consequences of an unbounded, infinite tower of tree-level exchanges remains to be explored with the level of generality maintained here. See e.g. \cite{Bonifacio:2019ioc} for some discussion of this issue for Kaluza-Klein theories. Exchanges of massive higher spin states also have not been considered here. Obviously this has been a subject of much study in the past \cite{Veneziano:1968yb,tHooft:1973alw} that continue today (see eg. \cite{Basile:2026gnd,Albert:2024yap,Berman:2024wyt} for a handful of recent studies - obviously this subject is vast).

The massive $4$-particle amplitudes calculated in this study were determined entirely from first principles through unitarity methods. They were inferred from demanding that they reproduce the required factorisation channels, which are simply calculated by combining two $3$-particle amplitudes together for kinematic configurations with intermediate on-shell states. Much of the subsequent effort was required to then convert the expression into a form in which the high-energy dependence was manifest. Calculating higher leg amplitudes in this way would be much more complicated. On-shell recursion relations have been successfully used to efficiently calculate multi-leg tree amplitudes for numerous massless theories \cite{Britto:2004ap,Britto:2005fq,Cachazo:2004kj,Risager:2005vk}. Developing these for massive theories remains an open issue. Some recent advances have been made in \cite{Ema:2024rss,Ema:2024vww,Gherghetta:2024tob,Jeong:2026dzt} which adapted the all-line shift of \cite{Cohen:2010mi} to massive theories. These studies demonstrated close connection between the terms in the expansion and a Feynman rule representation for the amplitude under an intelligent gauge choice. The imprints of the Ward identities were also demonstrated as underpinning theories with suppressed high-energy dependence that were amenable to the technique. Alternatively, BCFW recursion \cite{Britto:2004ap} is valid for spontaneously broken $\mathcal{N}=4$ SYM \cite{Herderschee:2019dmc} (since it can be embedded in massless YM in higher dimensions, where BCFW is known to be valid \cite{Arkani-Hamed:2008bsc,Arkani-Hamed:2008owk,Drummond:2008cr,Brandhuber:2008pf}). For massless theories, BCFW is much more efficient than the all-line shift and, for $\mathcal{N}=4$ broken SYM, leads directly to the manifestly high energy unitary representation (\ref{N=4SYM}) containing (\ref{N=44VecComp}). However, this method (or any other related idea egs. \cite{Cachazo:2018hqa,Albonico:2022pmd}) has never been used to calculate massive superamplitudes beyond five legs (where the SUSY structure begins to become non-trivial) in an insightful way, although it should be possible in theory. The significance and usefulness of the different representations of the massive amplitudes (with or without ``spurious'' high energy dependence and with or without residues with other ``spurious'' poles) and the implications for on-shell methods remains open for exploration. 



Performing an analogous analysis for other spacetime backgrounds (AdS, dS, inflation, FRW) is another obvious direction \cite{Baumann:2022jpr}.

\acknowledgments 

I would like to thank Tony Gherghetta and Callum Jones for comments on a draft of this manuscript.


\appendix

\section{Horrible Products of Bilinears}\label{sec:HPoB}

Here is a list of useful algebraic relations and ``syzygies'' for $4$-particle on-shell amplitudes. These relate Lorentz structures of a particular mass dimension to combinations of Lorentz structures of smaller mass dimension (``syzygies'' being such relations that only exist when certain combinations of Mandelstam variables are multiplying the otherwise linearly independent spinor structures). They are useful for manifesting cancellations between combinations of contact and factorisation terms, but may possibly find use in the on-shell operator basis construction of EFTs \cite{DeAngelis:2022qco,Dong:2022mcv,Durieux:2019eor,Durieux:2020gip}. 

The identity
\begin{align}\label{OldSyz}
&(4t-\sum_im_i^2)\ds{\bf{14}}\ds{\bf{32}}+(4u-\sum_im_i^2)\ds{\bf{13}}\ds{\bf{42}}\nonumber\\
&\qquad=m_1\left(\ds{\bf{23}}\la{\bf{1}}p_2-p_3\rs{\bf{4}}+\ds{\bf{24}}\la{\bf{1}}p_2-p_4\rs{\bf{3}}\right)\nonumber\\&\quad\qquad+m_2\left(\ds{\bf{13}}\la{\bf{2}}p_1-p_3\rs{\bf{4}}+\ds{\bf{14}}\la{\bf{2}}p_1-p_4\rs{\bf{3}}\right)\nonumber\\
&\quad\qquad+m_3\left(\ds{\bf{14}}\la{\bf{3}}p_1-p_4\rs{\bf{2}}+\ds{\bf{24}}\la{\bf{3}}p_2-p_4\rs{\bf{1}}\right)\nonumber\\
&\quad\qquad+m_4\left(\ds{\bf{23}}\la{\bf{4}}p_2-p_3\rs{\bf{1}}+\ds{\bf{13}}\la{\bf{4}}p_1-p_3\rs{\bf{2}}\right)\nonumber\\
&\qquad\quad-2\left(m_1m_3\da{\bf{13}}\ds{\bf{24}}+m_1m_4\da{\bf{14}}\ds{\bf{23}}+m_2m_3\da{\bf{23}}\ds{\bf{14}}+m_2m_4\da{\bf{24}}\ds{\bf{13}}\right)\nonumber\\
\end{align}
was found in \cite{Durieux:2020gip} (in a much less manifestly symmetric form). In the $\{\mathbf{L}_i\}$ basis (\ref{L}), (\ref{OldSyz}) becomes
\begin{align}\label{OldSyz2}
&(t-\frac{1}{3}\sum_im_i^2)\ds{\bf{14}}\ds{\bf{32}}+(u-\frac{1}{3}\sum_im_i^2)\ds{\bf{13}}\ds{\bf{42}}=\frac{1}{12}\sum_i\mathbf{L}_i.
\end{align}
This relation encapsulates the deformation of the massless all-channel pole (\ref{AllChanPole}) to massive kinematics, as explained in Section \ref{sec:4vpreamble}.

The following identities have application to the four massive vector amplitude:
\begin{align}\label{LLhatSame}
&\frac{1}{m_1^2}\mathbf{L}_1\hat{\mathbf{L}}_1\nonumber\\
&\quad=24\big((s-m_s^2)\left(\mathbf{P}_t+\mathbf{P}_u-\mathbf{P}_s\right)+(t-m_t^2)\left(\mathbf{P}_u+\mathbf{P}_s-\mathbf{P}_t\right)+(u-m_u^2)\left(\mathbf{P}_s+\mathbf{P}_t-\mathbf{P}_u\right)\big)\nonumber\\
&\qquad+2\Big(\left(\da{\bf{14}}\da{\bf{23}}+\da{\bf{13}}\da{\bf{24}}\right)\mathbf{L}_2+\left(\ds{\bf{14}}\ds{\bf{23}}+\ds{\bf{13}}\ds{\bf{24}}\right)\hat{\mathbf{L}}_2\nonumber\\
&\qquad\qquad-\left(\da{\bf{14}}\da{\bf{23}}-\da{\bf{12}}\da{\bf{34}}\right)\mathbf{L}_3-\left(\ds{\bf{14}}\ds{\bf{23}}-\ds{\bf{12}}\ds{\bf{34}}\right)\hat{\mathbf{L}}_3\nonumber\\
&\qquad\qquad-\left(\da{\bf{12}}\da{\bf{34}}+\da{\bf{13}}\da{\bf{24}}\right)\mathbf{L}_4-\left(\ds{\bf{12}}\ds{\bf{34}}+\ds{\bf{13}}\ds{\bf{24}}\right)\hat{\mathbf{L}}_4\Big)\nonumber\\
&\qquad-36\Big(m_1m_2\ds{\bf{34}}\da{\bf{34}}\left(\ds{\bf{12}}^2+\da{\bf{12}}^2\right)+m_3m_4\ds{\bf{12}}\da{\bf{12}}\left(\ds{\bf{34}}^2+\da{\bf{34}}^2\right)\nonumber\\
&\qquad\qquad\quad+m_1m_3\ds{\bf{24}}\da{\bf{24}}\left(\ds{\bf{13}}^2+\da{\bf{13}}^2\right)+m_2m_4\ds{\bf{13}}\da{\bf{13}}\left(\ds{\bf{24}}^2+\da{\bf{24}}^2\right)\nonumber\\
&\qquad\qquad\quad+m_2m_3\ds{\bf{14}}\da{\bf{14}}\left(\ds{\bf{23}}^2+\da{\bf{23}}^2\right)+m_1m_4\ds{\bf{23}}\da{\bf{23}}\left(\ds{\bf{14}}^2+\da{\bf{14}}^2\right)\Big)\nonumber\\
&\qquad-36\Big(m_3m_4\left(\ds{\bf{13}}\ds{\bf{23}}\da{\bf{14}}\da{\bf{24}}+\da{\bf{13}}\da{\bf{23}}\ds{\bf{14}}\ds{\bf{24}}\right)\nonumber\\
&\qquad\qquad\quad+m_2m_3\left(\ds{\bf{12}}\ds{\bf{24}}\da{\bf{34}}\da{\bf{13}}+\da{\bf{12}}\da{\bf{24}}\ds{\bf{34}}\ds{\bf{13}}\right)\nonumber\\
&\qquad\qquad\quad-m_2m_4\left(\ds{\bf{12}}\ds{\bf{23}}\da{\bf{34}}\da{\bf{14}}+\da{\bf{12}}\da{\bf{23}}\ds{\bf{34}}\ds{\bf{14}}\right)\Big)\nonumber\\
&\qquad+4\left(m_2^2\left(\mathbf{P}_s+4\left(\mathbf{P}_t+\mathbf{P}_u\right)\right)+m_3^2\left(\mathbf{P}_t+4\left(\mathbf{P}_s+\mathbf{P}_u\right)\right)+m_4^2\left(\mathbf{P}_u+4\left(\mathbf{P}_s+\mathbf{P}_t\right)\right)\right)\nonumber\\
&\qquad+24\left(m_s^2\left(\mathbf{P}_t+\mathbf{P}_u-\mathbf{P}_s\right)+m_t^2\left(\mathbf{P}_u+\mathbf{P}_s-\mathbf{P}_t\right)+m_u^2\left(\mathbf{P}_s+\mathbf{P}_t-\mathbf{P}_u\right)\right)\nonumber\\
&\qquad+8\left(\mathbf{P}_s+\mathbf{P}_t+\mathbf{P}_u\right)
\end{align}
and
\begin{align}\label{LLhatDiff}
&\frac{1}{m_1m_2}\mathbf{L}_2\hat{\mathbf{L}}_1\nonumber\\
&\quad=144\left(s-m_s^2\right)\ds{\bf{14}}\ds{\bf{13}}\da{\bf{24}}\da{\bf{23}}\nonumber\\
&\qquad-2\frac{m_2}{m_1}\left(\ds{\bf{14}}\ds{\bf{23}}+\ds{\bf{24}}\ds{\bf{13}}\right)\hat{\mathbf{L}}_1-2\frac{m_1}{m_2}\left(\da{\bf{14}}\da{\bf{23}}+\da{\bf{24}}\da{\bf{13}}\right)\mathbf{L}_2\nonumber\\
&\qquad-36m_3m_4\left(\da{\bf{24}}^2\ds{\bf{13}}^2+\da{\bf{23}}^2\ds{\bf{14}}^2\right)\nonumber\\
&\qquad+36\ds{\bf{34}}\da{\bf{34}}\left(m_1^2\da{\bf{12}}^2+m_2^2\ds{\bf{12}}^2\right)\nonumber\\
&\qquad+3m_1m_4\ds{\bf{13}}\da{\bf{24}}\left(6\da{\bf{24}}\da{\bf{13}}-14\left(\da{\bf{14}}\da{\bf{23}}-\da{\bf{12}}\da{\bf{34}}\right)\right)\nonumber\\
&\qquad+3m_1m_3\ds{\bf{14}}\da{\bf{23}}\left(6\da{\bf{23}}\da{\bf{14}}-14\left(\da{\bf{24}}\da{\bf{13}}+\da{\bf{12}}\da{\bf{34}}\right)\right)\nonumber\\
&\qquad+3m_2m_4\ds{\bf{14}}\da{\bf{23}}\left(6\ds{\bf{23}}\ds{\bf{14}}-14\left(\ds{\bf{24}}\ds{\bf{13}}+\ds{\bf{12}}\ds{\bf{34}}\right)\right)\nonumber\\
&\qquad+3m_2m_3\ds{\bf{13}}\da{\bf{24}}\left(6\ds{\bf{24}}\ds{\bf{13}}-14\left(\ds{\bf{14}}\ds{\bf{23}}-\ds{\bf{12}}\ds{\bf{34}}\right)\right)\nonumber\\
&\qquad+72\Big(m_2m_3\ds{\bf{12}}\ds{\bf{14}}\da{\bf{23}}\da{\bf{34}}-m_2m_4\ds{\bf{12}}\ds{\bf{13}}\da{\bf{24}}\da{\bf{34}}\nonumber\\
&\qquad\qquad\quad+m_1m_4\ds{\bf{14}}\ds{\bf{34}}\da{\bf{12}}\da{\bf{23}}-m_1m_3\ds{\bf{13}}\ds{\bf{34}}\da{\bf{12}}\da{\bf{24}}\nonumber\\
&\qquad\qquad\qquad+\frac{1}{2}\left(4m_s^2-m_3^2-m_4^2\right)\ds{\bf{14}}\ds{\bf{13}}\da{\bf{23}}\da{\bf{24}}\Big)\nonumber\\
&\qquad-4m_1m_2\left(17\mathbf{P}_s+2\left(\mathbf{P}_t+\mathbf{P}_u\right)\right).
\end{align}
These can be combined together to give 
\begin{align}\label{LLSum}
&\sum_j\mathbf{L}_j\sum_k\hat{\mathbf{L}}_k\nonumber\\
&\quad=144m_1m_2(s-m_s^2)\ds{\bf{14}}\ds{\bf{13}}\da{\bf{23}}\da{\bf{24}}+\text{ex.}\nonumber\\
&\qquad+24\sum_im_i^2\big((s-m_s^2)\left(\mathbf{P}_t+\mathbf{P}_u-\mathbf{P}_s\right)\nonumber\\
&\qquad\qquad\qquad\qquad\qquad+(t-m_t^2)\left(\mathbf{P}_u+\mathbf{P}_s-\mathbf{P}_t\right)+(u-m_u^2)\left(\mathbf{P}_s+\mathbf{P}_t-\mathbf{P}_u\right)\big)\nonumber\\
&\qquad-144\left(\prod_im_i\right)\da{\bf{12}}^2\ds{\bf{34}}^2+\text{ex.}\nonumber\\
&\qquad+144m_1m_2\left(m_s^2-m_3^2-m_4^2\right)\ds{\bf{13}}\ds{\bf{14}}\da{\bf{23}}\da{\bf{24}}+\text{ex.}\nonumber\\
&\qquad-16(m_1m_2)^2\left(8\mathbf{P}_s-\left(\mathbf{P}_t+\mathbf{P}_u\right)\right)+\text{ex.}\nonumber\\
&\qquad+24\sum_im_i^2\left(m_s^2\left(\mathbf{P}_t+\mathbf{P}_u-\mathbf{P}_s\right)+m_t^2\left(\mathbf{P}_u+\mathbf{P}_s-\mathbf{P}_t\right)+m_u^2\left(\mathbf{P}_s+\mathbf{P}_t-\mathbf{P}_u\right)\right)\nonumber\\
&\qquad+8\sum_im_i^4\left(\mathbf{P}_s+\mathbf{P}_t+\mathbf{P}_u\right),
\end{align}
where ``ex.'' indicates all distinct terms obtained by exchanging particle labels in the term listed in the same line. 

Other useful identities are
\begin{align}\label{LtoLhat}
&\frac{1}{6}\da{\bf{12}}\frac{1}{m_3}\mathbf{L}_3-\frac{1}{6}\ds{\bf{12}}\frac{1}{m_4}\hat{\mathbf{L}}_4\nonumber\\
&=\frac{1}{3}m_4\ds{\bf{12}}\left(\da{\bf{13}}\da{\bf{24}}+\da{\bf{14}}\da{\bf{23}}\right)-\frac{1}{3}m_3\da{\bf{12}}\left(\ds{\bf{13}}\ds{\bf{24}}+\ds{\bf{14}}\ds{\bf{23}}\right)\nonumber\\
&\qquad+\left(m_1\da{\bf{12}}-m_2\ds{\bf{12}}\right)\da{\bf{13}}\ds{\bf{24}}+\left(m_2\da{\bf{12}}-m_1\ds{\bf{12}}\right)\da{\bf{23}}\ds{\bf{14}}
\end{align}
and
\begin{align}\label{LowSyz}
&\la{\bf{1}}p_3\rs{\bf{2}}\la{\bf{3}}p_1\rs{\bf{4}}-\left(u-\frac{1}{2}\sum_im_i^2\right)\ds{\bf{24}}\da{\bf{13}}\nonumber\\
&\qquad\qquad\qquad\qquad\qquad\qquad\qquad\qquad=-\frac{1}{2}\da{\bf{13}}\left(m_4\la{\bf{4}}p_3-p_1\rs{\bf{2}}+m_2\la{\bf{2}}p_3-p_1\rs{\bf{4}}\right)\nonumber\\
&\qquad\qquad\qquad\qquad\qquad\qquad\qquad\qquad\qquad\qquad-m_1m_3\ds{\bf{14}}\ds{\bf{23}}+\frac{1}{2}\left(m_1^2+m_3^2\right)\ds{\bf{24}}\da{\bf{13}}
\end{align}
and
\begin{align}\label{LowSyz2}
\la{\bf{2}}p_4\rs{\bf{1}}&\la{\bf{1}}p_4\rs{\bf{3}}+\left(m_1^2+m_4^2-u\right)\ds{\bf{13}}\da{\bf{12}}\nonumber\\
&\qquad\qquad\qquad\qquad\qquad=m_4^2\ds{\bf{13}}\da{\bf{12}}-m_1\da{\bf{12}}\la{\bf{1}}p_4\rs{\bf{3}}-m_1\ds{\bf{13}}\ls{\bf{1}}p_4\ra{\bf{2}}\nonumber\\
&\qquad\qquad\qquad\qquad\qquad\qquad\qquad+m_3\da{\bf{23}}\la{\bf{1}}p_4\rs{\bf{1}}-m_2\ds{\bf{23}}\la{\bf{1}}p_4\rs{\bf{1}}
\end{align}
and
\begin{align}
&\la{\bf{1}}p_3\rs{\bf{1}}\la{\bf{2}}p_4\rs{\bf{2}}-\la{\bf{1}}p_4\rs{\bf{1}}\la{\bf{2}}p_3\rs{\bf{2}}\nonumber\\
&\qquad\qquad=\left(t-u\right)\ds{\bf{12}}\da{\bf{12}}\nonumber\\
&\qquad\qquad\qquad-\frac{1}{2}\left(m_1\da{\bf{12}}+m_2\ds{\bf{12}}\right)\la{\bf{1}}p_3-p_4\rs{\bf{2}}
-\frac{1}{2}\left(m_2\da{\bf{12}}+m_1\ds{\bf{12}}\right)\la{\bf{2}}p_3-p_4\rs{\bf{1}}
\end{align}
and
\begin{align}\label{squeeze2int}
&\ds{\bf{12}}\ls{\bf{3}}p_1p_4\rs{\bf{3}}+\left(s-m_1^2-m_2^2\right)\ds{\bf{13}}\ds{\bf{23}}\nonumber\\
&\qquad=\left(m_1^2-m_2^2+m_3^2\right)\ds{\bf{13}}\ds{\bf{23}}-m_1\ds{\bf{23}}\la{\bf{1}}p_4\rs{\bf{3}}-m_3\ds{\bf{13}}\la{\bf{3}}p_4\rs{\bf{2}}-m_2\ds{\bf{13}}\la{\bf{2}}p_4\rs{\bf{3}}
\end{align}
and
\begin{align}\label{Squeeze++}
&\ds{\bf{23}}\la{\bf{2}}p_4\rs{\bf{1}}\la{\bf{3}}p_4\rs{\bf{1}}
-(u-m_1^2-m_4^2)\ds{\bf{12}}\ds{\bf{13}}\da{\bf{23}}\nonumber\\
&=m_4^2\ds{\bf{12}}\ds{\bf{13}}\da{\bf{23}}-\left(m_2\da{\bf{23}}-m_3\ds{\bf{23}}\right)\ds{\bf{13}}\la{\bf{2}}p_4\rs{\bf{1}}-\left(m_3\da{\bf{23}}-m_2\ds{\bf{23}}\right)\ds{\bf{12}}\la{\bf{3}}p_4\rs{\bf{1}}\nonumber\\
&\qquad-\frac{1}{2}m_1\left(\ds{\bf{23}}\left(\da{\bf{12}}\la{\bf{3}}p_4\rs{\bf{1}}+\da{\bf{13}}\la{\bf{2}}p_4\rs{\bf{1}}\right)+\da{\bf{23}}\left(\ds{\bf{12}}\la{\bf{1}}p_4\rs{\bf{3}}+\ds{\bf{13}}\la{\bf{1}}p_4\rs{\bf{2}}\right)\right).
\end{align}
and the syzygies
\begin{align}\label{SimpSyz}
\left(t-m_1^2-m_3^2\right)\la{\bf{3}}p_4\rs{\bf{3}}-\left(s-m_3^2-m_4^2\right)\la{\bf{3}}p_1\rs{\bf{3}}=m_3\left(\ls{\bf{3}}p_1p_4\rs{\bf{3}}+\la{\bf{3}}p_1p_4\ra{\bf{3}}\right)
\end{align}
(which was mentioned already in (\ref{SimpSyzFact})) and
\begin{align}\label{3PolSyz}
&-(u-m_1^2-m_4^2)\da{\bf{21}}\la{\bf{3}}p_4\rs{\bf{2}}-(s-m_3^2-m_4^2)\da{\bf{23}}\la{\bf{1}}p_4\rs{\bf{2}}\nonumber\\
&=-(t-m_2^2-m_4^2)\left(m_3\da{\bf{12}}\ds{\bf{23}}+m_1\ds{\bf{12}}\da{\bf{23}}\right)+m_2\la{\bf{1}}p_4\rs{\bf{2}}\la{\bf{3}}p_4\rs{\bf{2}}\nonumber\\
&\qquad+m_2\left(m_3\da{\bf{12}}\la{\bf{2}}p_4\rs{\bf{3}}-m_1\da{\bf{23}}\la{\bf{2}}p_4\rs{\bf{1}}+m_1\ds{\bf{12}}\la{\bf{3}}p_4\rs{\bf{2}}-m_3\ds{\bf{23}}\la{\bf{1}}p_4\rs{\bf{2}}\right)\nonumber\\
&\qquad+m_4^2\left(\da{\bf{23}}\la{\bf{1}}p_4\rs{\bf{2}}-\da{\bf{12}}\la{\bf{3}}p_4\rs{\bf{2}}\right)+\left(m_3^2-m_1^2\right)\da{\bf{13}}\la{\bf{2}}p_4\rs{\bf{2}}-m_2m_4^2\da{\bf{12}}\da{\bf{23}}
\end{align}
and
\begin{align}\label{BigSyz}
&\left(s-m_1^2-m_2^2\right)\la{\bf{1}}p_4\rs{\bf{2}}\la{\bf{2}}p_4\rs{\bf{1}}+\left(t-m_2^2-m_4^2\right)\left(u-m_1^2-m_4^2\right)\ds{\bf{12}}\da{\bf{12}}\nonumber\\
&\qquad=m_1\left(t-m_2^2-m_4^2\right)\left(\ds{\bf{12}}\la{\bf{2}}p_4\rs{\bf{1}}+\da{\bf{12}}\la{\bf{1}}p_4\rs{\bf{2}}\right)\nonumber\\
&\qquad\qquad-m_2\left(u-m_1^2-m_4^2\right)\left(\ds{\bf{12}}\la{\bf{1}}p_4\rs{\bf{2}}+\da{\bf{12}}\la{\bf{2}}p_4\rs{\bf{1}}\right)\nonumber\\
&\qquad\qquad+m_1m_2\left(\la{\bf{2}}p_4\rs{\bf{1}}^2+\la{\bf{1}}p_4\rs{\bf{2}}^2\right)+m_4^2\left(s-m_1^2-m_2^2\right)\da{\bf{12}}\ds{\bf{12}}\nonumber\\
&\qquad\qquad+m_1m_2m_4^2\left(\da{\bf{12}}^2+\ds{\bf{12}}^2\right).
\end{align}
Reversing the spinor chiralities gives analogous identities. The latter two syzygies can both be derived by contracting various bilinears against (\ref{OldSyz2}).

\section{Special 3-Particle Massless Kinematics}\label{sec:S3PMK}

Here I just give some simple complex momentum configurations that realise the on-shell factorisation channels for the four leg amplitudes $A(1,2\rightarrow 3,4)$ of massless particles. 

Denote by $p_i$ the real-valued external particle momenta chosen so that the amplitude factorises on the $s$-channel. Define $\hat{p}_i$ as the corresponding on-shell momenta under a complex deformation made so that on the amplitude remains factorised on the $s$-channel. Then within the $3$-particle factor amplitude $A(\hat{1},\hat{2},\hat{P}_s)$, the momenta are 
\begin{align}
\hat{p}_1&=p_1=E_1(1,0,0,1)\nonumber\\
\hat{p}_2&=p_2-r,\qquad p_2=E_2(1,0,0,1)\nonumber\\
P_s&=-(p_1+p_2)\nonumber\\
\hat{P}_s&=-(\hat{p}_1+\hat{p}_2)=-(p_1+p_2)+r=P_s+r,
\end{align}
where $E_{1,2}$ are the particles' energies and $r$ is a complex null vector satisfying $r\cdot r=0$ and $r\cdot p_2=0$. A possible choice is $r=z(0,1,\pm i,0)$ for some constant $z$. The corresponding spinors are then
\begin{align}
\rs{i}=\begin{pmatrix}
0\\
\sqrt{2E_i}
\end{pmatrix}\qquad\qquad
\la{i}=\begin{pmatrix}
0 &
\sqrt{2E_i}
\end{pmatrix}
\end{align}
for either $i=1,2$. The choice of sign in $r$ corresponds to choice of spinors
\begin{align}
\rs{r_+}=\begin{pmatrix}
-z\sqrt{2/e}\\
0
\end{pmatrix}\qquad\qquad
\la{r_+}=\begin{pmatrix}
0 &
\sqrt{2e}\end{pmatrix}\nonumber\\
\rs{r_-}=\begin{pmatrix}
0\\
\sqrt{2e}
\end{pmatrix}\qquad\qquad
\la{r_-}=\begin{pmatrix}
z\sqrt{2/e} & 0
\end{pmatrix}.
\end{align}
where the factor of $e$ represents the $GL(1,\mathbb{C})$ rescaling redundancy in the complexified spinors. The choice of $\pm$ clearly corresponds to the choice of alignment of each chirality of spinor. 

On the opposite side of the channel, the outgoing particles can be treated similarly. The complex momentum deformation can be absorbed into a shift of leg $4$ in exactly the same way that it deforms $p_2$. This describes the limit $\da{\hat{1}\hat{2}},\da{\hat{3}\hat{4}}\rightarrow 0$ or $\ds{\hat{1}\hat{2}},\ds{\hat{3}\hat{4}}\rightarrow 0$, which, of course, activates all $s,t,u$-channels simultaneously and describes the all-channel pole. The channels with vanishing spinor bilinears of opposite chirality on each side requires $p_2$ and $p_4$ to be instead deformed by $r$ and $r^*$ respectively, rather than both by $r$, as is enforced by momentum conservation above. This alternative can be arranged by also deforming $p_1$ and $p_3$ by $-r$ and $-r^*$ respectively as well, which ensures that either $\da{\hat{1}\hat{2}},\ds{\hat{3}\hat{4}}\rightarrow 0$ or $\ds{\hat{1}\hat{2}},\da{\hat{3}\hat{4}}\rightarrow 0$.

\section{Multipoles and the Geometry of Gauge Theories}\label{sec:ExoticGT}

This Appendix links vector boson cubic couplings to Lie algebra geometry for a simple theory consisting of a massless photon and a charge conjugate pair of $W$-bosons. 
The massless photon immediately demands conserved electric charge. There are therefore two possible $T$-symmetric couplings parameterising a Lie algebra and one independent $T$-violating GCS coupling. The Lie algebra describing the $T$-symmetric couplings is, in general, defined by the commutation relations
\begin{align}\label{LA3}
[T_{W_1},T_{W_2}]=is\alpha T_\gamma\qquad\qquad[T_{\gamma},T_{W_1}]=i\beta T_{W_2}\qquad\qquad[T_{\gamma},T_{W_2}]=-i\beta T_{W_1}.
\end{align}
Here $T_\gamma$ is the generator associated with the photon and $T_{W_1}$ and $T_{W_2}$ are the generators of the $W$-bosons in a self-conjugate representation. The signature of the Killing form is determined by $s\in\{0,\pm 1\}$. The parameters $\alpha,\beta>0$ are continuous and describe the normalisations of the photon and $W$-boson generators. On-shell, $\beta$ corresponds to the gauge coupling, while the ratio $s\alpha/\beta$ parameterises the magnetic dipole moments of the $W$-bosons (as well as the electric quadrupole moment). The $3$-particle amplitude is then 
\begin{align}\label{WEM3leg}
A\left(W_1,W_2,\gamma^+\right)&=i\beta\left(\frac{1}{mx}\da{\bf{12}}^2-\frac{1-s\alpha/\beta}{2m^2}\ds{3\bf{2}}\ds{3\bf{1}}\da{\bf{12}}\right).
\end{align}
The electric dipole moment (and magnetic quadrupole moment) corresponds to the possible GCS term, which may be accounted for by complexifying $\alpha$ in this expression.

In general, the Lie algebra is identified as $\mathfrak{su}_2$ if $s=1$ and, furthermore, the case of homogeneous generator normalisation required for spontaneously broken YM is given by $\alpha=\beta$. In this case, the coupling is clearly minimal. If $s=-1$, then the Lie algebra is the non-compact $\mathfrak{su}_{1,1}$, while the degenerate case with $s=0$ (or equivalently, $\alpha\rightarrow 0$) corresponds to the non-semisimple $\mathfrak{e}_2$ (the Euclidean algebra of the $2$d plane). The relative generator normalisations match onto the departure of the $W$-boson's multipoles from minimal coupling. In this sense, the dipole moments encode the geometric ``phase'' of the theory. Incidentally, if the possible $F^3$ interactions are neglected, then any departure of the amplitude (\ref{WEM3leg}) from minimal coupling has a Lie algebra interpretation. The partial unitarisation of the four vector boson amplitudes is therefore automatic. 

The weak boson amplitudes match onto terms in an effective Lagrangian of anomalous cubic couplings. The multipole expansion applied here is (somewhat) traditionally parameterised as
\begin{align}
A\left(W,\overline{W},\gamma^+\right)&=-\frac{e}{mx}\ds{\bf{12}}^2-\frac{2(\mu+id)}{m}\ds{\bf{12}}\ds{3\bf{1}}\ds{3\bf{2}}+\frac{(Q+i\widetilde{Q})}{m}x\ds{3\bf{1}}^2\ds{3\bf{2}}^2,
\end{align}
where $e$ is the electric charge, $\mu$ and $d$ are the magnetic and electric dipole moments respectively and $Q$ and $\widetilde{Q}$ are electric and magnetic quadrupole moments. 
Both $d$ and $\widetilde{Q}$ are $P$ and $T$-violating interactions. 

Off-shell, each of these couplings can be matched onto effective operators. A suitable basis of terms in the effective Lagrangian is \cite{Hagiwara:1986vm}
\begin{align}\label{WelectroLag}
\mathcal{L}_{W\text{multi}}&=ie\kappa W^+_\mu W^-_\nu F^{\mu\nu}+ie\tilde{\kappa}W^+_\mu W^-_\nu \widetilde{F}^{\mu\nu}+i\frac{e\lambda}{m^2} W^{+\rho}_\mu W^{-\mu}_\nu F^{\,\,\nu}_\rho+i\frac{e\tilde{\lambda}}{m^2}W^{+\rho}_\mu W^{-\mu}_\nu \widetilde{F}^{\,\,\nu}_\rho
\end{align}
defining $W^+_{\mu\nu}=D_\mu W^+_\nu-D_\nu W^+_\mu$, $W^-_{\mu\nu}=D^\dagger_\mu W^-_\nu-D^\dagger_\nu W^-_\mu$ and otherwise following the conventions and definitions of \cite{Srednicki:2007qs}. For brevity, I omit the standard kinetic terms that, by themselves, would otherwise describe regular $\mathfrak{su}_2\rightarrow \mathfrak{u}_1$ spontaneously broken YM under which the $W$-bosons have electric charge $e$. Of course, this effective Lagrangian should be supplemented with appropriate quartic terms, but these do not affect the matching onto the $3$-particle amplitude, which is the meager goal here. 

The operators in (\ref{WelectroLag}) can be evaluated for on-shell particles. The YM field and its Hodge dual become
\begin{align}
F&=\frac{1}{\sqrt{2}}\ls{\bf{p}}S_L\rs{\bf{p}}-\frac{1}{\sqrt{2}}\la{\bf{p}}S_R\ra{\bf{p}}\nonumber\\
*F&=\frac{i}{\sqrt{2}}\ls{\bf{p}}S_L\rs{\bf{p}}+\frac{i}{\sqrt{2}}\la{\bf{p}}S_R\ra{\bf{p}}.
\end{align}
The first term is the self-dual component and the second is the anti-self-dual component:
\begin{align}
G=\frac{1}{\sqrt{2}}\ls{\bf{p}}S_L\rs{\bf{p}}\qquad\qquad 
G^\dagger=-\frac{1}{\sqrt{2}}\la{\bf{p}}S_R\ra{\bf{p}}.
\end{align}
The gauge connections $A$ map directly onto polarisations:
\begin{align}
\varepsilon=\frac{1}{\sqrt{2}m}\ls{\bf{p}}\sigma\ra{\bf{p}}.
\end{align}
An extra factor of $\sqrt{2}$ must be introduced for the longitudinally polarised mode to be correctly normalised, besides evaluating the suppressed spin indices at opposite values and symmetrising them. 

Adding to these effective $3$-particle interactions the minimal coupling $-\frac{e}{mx}\da{\bf{12}}^2$ term, the multipoles are matched onto as
\begin{align}
\begin{split}
\mu&=\frac{e}{2m}\left(2+\frac{1}{2}(\lambda-\kappa)\right)\\
Q&=\frac{e}{m^2}\left(-1+\frac{1}{2}\kappa\right)
\end{split}
\begin{split}
d&=\frac{e}{4m}\left(\tilde{\lambda}-\tilde{\kappa}\right)\\
\widetilde{Q}&=\frac{e}{2m^2}\tilde{\kappa}
\end{split}
\end{align}
I will use the word ``anomalous'' to mean departure from minimal coupling, which is represented by the constants $\kappa$, $\tilde{\kappa}$, $\lambda$ and $\tilde{\lambda}$ above. The significance of ``minimal coupling'' as defined in \cite{Arkani-Hamed:2017jhn} is that it ultimately corresponds to the amplitude required for tree-level unitarity in the HEL. These definitions disagree with conventions established in previous studies of electroweak bosons \cite{Hagiwara:1986vm}. Matching onto the specific gauged NL$\Sigma$M expression (\ref{WEM3leg}) corresponds to $e=\beta$, $\kappa=-(1-s\alpha/\beta)$ and $\lambda=0$.

With $\mathcal{N}=1$ supersymmetry, assuming that the massive vectors belong to massive vector multiplets, operators of the form $F^3$ are prohibited, so $\lambda=\tilde{\lambda}=0$ and $\kappa$ and $\tilde{\kappa}$ represent the only independent combination of anomalous multipoles (the quadrupole is effectively determined by the dipole). 

In theories of supergravity with BPS vectors, the graviphoton commonly has a coupling of the form
\begin{align}
A\left(W,\overline{W},\gamma^+\right)=\frac{m}{M_{Pl}}\frac{1}{mx}\ds{\bf{12}}\da{\bf{12}}.
\end{align}
This corresponds to a large ``anomalous'' MDM with $g=1$ and zero quadrupole ($g$ being the gyromagnetic ratio). Matching onto (\ref{WEM3leg}), this theory of a single BPS vector (and its conjugate) in extended SUGRA is described by a non-compact $\mathfrak{su}_{1,1}$ Lie algebra \cite{deWit:1983xe}. Because the coefficient, or electric charge, is $m/M_{Pl}$, the minimal coupling term vanishes in the massless limit, while the anomalous dipole term converges to a gravitational-strength effective interaction between same-helicity vectors and a scalar (arising from the longitudinal mode). This is the typical interaction involving the graviphoton that is expected for massless matter in extended SUGRA. See \cite{Trott:2026ozo} for review and further details. This is an example of a theory that ``ungauges'' in the massless limit (which coincides here to switching off gravity), rather than converging to massless non-Abelian Yang-Mills. The cut-off of the underlying gauged NL$\Sigma$M is pushed to the Planck length through the very weak gauge coupling.



\section{Extra Amplitudes}\label{sec:HDOInsertions}

In this Appendix, I compile contributions to the massive $4$-particle amplitudes in Sections \ref{Sec:LowSpinAmp} and \ref{VectorHiggs} that are predominantly tangential to the main narrative. These mostly arise from insertions of $3$-particle amplitudes that, off-shell, correspond to effective operators of the form $F^3$, $\varphi F^2$ and $\chi F\chi$ (or their parity-violating counterparts, which I will include implicitly in reference to these labels). On-shell, these $3$-particle amplitudes are respectively given by the $h$-terms in (\ref{3legVec}), the $c$-terms in (\ref{2vec1scalar}) and the $m$-terms in (\ref{FermionMat3leg}). 

The contributions presented here will mostly be the uninteresting ones from the viewpoint of seeking constraints from high-energy unitarity. More precisely, these are the terms with high-energy dependence that can only be softened by the outright elimination or suppression of their couplings. There are nevertheless a few exceptional cases in which these insertions do participate non-trivially in constraints, invariably related to their covariantisation. For clarity of exposition, the constraints associated to both the $F^3$ and $\chi F\chi$ are derived in this Appendix, while the $\varphi F^2$ case is discussed in the main body of the text in Sections \ref{VectorHiggs} and \ref{3Vec1Scalar} since it is inextricably tied to the structure of the $f$-type couplings in the three vector amplitude. 

Additionally, the calculation of the single vector and three scalar amplitude will also be presented here. This case is less interesting for the purposes of deriving constraints from high-energy unitarity because, in the absence of $\varphi F^2$ insertions, it is manifestly unitary upon construction. It may nevertheless have relevance to the reconstruction of the emergent scalar potential in the HEL and can have implications for perturbative unitarity.

\subsection{Low spin amplitudes}\label{app:LowSpinGen}

In this Appendix, I collect the contributions to the amplitudes in Sections \ref{sec:QED} and \ref{sec:4fermion} mediated by the fermion dipole interactions in (\ref{FermionMat3leg}). Additionally, I also give the general expressions for the low spin amplitudes mediated by massive vector exchange. 

The general massive scalar amplitude induced by massive vector exchange is
\begin{align}\label{sQCD}
A\left(\varphi_i,\varphi_j,\varphi_k,\varphi_l\right)=\frac{1}{s-m_{s_M}^2}(t_M)_{ij}(t_M)_{kl}\frac{1}{2m_{s_M}^2}\left(m_{s_M}^2\left(t-u\right)+\left(m_1^2-m_2^2\right)\left(m_3^2-m_4^2\right)\right).
\end{align}
The scalar QED amplitude (\ref{sQED}) is (equivalent to) a special case of this (modulo the quartic contact term). 

The general mixed fermion-scalar amplitude mediated by a massive vector is 
\begin{align}
&A\left(\psi_i,\psi_j,\varphi_k,\varphi_l\right)\nonumber\\
&=\frac{1}{s-m_{s_M}^2}\frac{1}{2}(t_M)_{kl}\Bigg((t_M)_{ij}\left(\la{\bf{1}}p_3-p_4\rs{\bf{2}}+\frac{m_3^2-m_4^2}{m_{s_M}^2}\left(m_1\ds{\bf{12}}-m_2\da{\bf{12}}\right)\right)\nonumber\\
&\qquad\qquad\qquad\qquad\qquad-(t_M)_{ji}\left(\la{\bf{2}}p_3-p_4\rs{\bf{1}}-\frac{m_3^2-m_4^2}{m_{s_M}^2}\left(m_2\ds{\bf{12}}-m_1\da{\bf{12}}\right)\right)\nonumber\\
&\qquad\qquad\qquad\qquad\qquad+\frac{2((m_M)_{ij})^*}{\Lambda}\ls{\bf{1}}p_4p_3-p_3p_4\rs{\bf{2}}-\frac{2(m_M)_{ij}}{\Lambda}\la{\bf{1}}p_4p_3-p_3p_4\ra{\bf{2}}\Bigg).
\end{align}
Again, this clearly agrees with the QED amplitude (\ref{sfQED}).

The general $4$-fermion amplitude is given in (\ref{GenMoller}) without inclusion of the dipole $m$-type couplings. The remaining terms are given by
\begin{align}
&A_{\chi F\chi}(\psi_i,\psi_j,\psi_k,\psi_l)\nonumber\\
&=\frac{-1}{s-{m_{s_M}^2}}\bigg(-\frac{2((m_M)_{ij})^*(m_M)_{kl}}{\Lambda^2}\left(\ls{\bf{1}}P_s\ra{\bf{3}}\ls{\bf{2}}P_s\ra{\bf{4}}+\ls{\bf{1}}P_s\ra{\bf{4}}\ls{\bf{2}}P_s\ra{\bf{3}}\right)\nonumber\\
&\qquad\qquad\qquad-\frac{2(m_M)_{ij}((m_M)_{kl})^*}{\Lambda^2}\left(\la{\bf{1}}P_s\rs{\bf{3}}\la{\bf{2}}P_s\rs{\bf{4}}+\la{\bf{1}}P_s\rs{\bf{4}}\la{\bf{2}}P_s\rs{\bf{3}}\right)\nonumber\\
&\qquad\qquad\qquad+\frac{2(m_M)_{ij}(m_M)_{kl}}{\Lambda^2}m_{s_M}^2\left(\da{\bf{13}}\da{\bf{24}}+\da{\bf{14}}\da{\bf{23}}\right)\nonumber\\
&\qquad\qquad\qquad+\frac{2((m_M)_{ij}(m_M)_{kl})^*}{\Lambda^2}m_{s_M}^2\left(\ds{\bf{13}}\ds{\bf{24}}+\ds{\bf{14}}\ds{\bf{23}}\right)\nonumber\\
&\qquad\qquad\qquad\qquad+(t_M)_{ij}\frac{((m_M)_{kl})^*}{\Lambda}\left(\ds{\bf{23}}\ls{\bf{4}}P_s\ra{\bf{1}}+\ds{\bf{24}}\ls{\bf{3}}P_s\ra{\bf{1}}\right)\nonumber\\
&\qquad\qquad\qquad\qquad+(t_M)_{ji}\frac{(m_M)_{kl}}{\Lambda}\left(\da{\bf{23}}\la{\bf{4}}P_s\rs{\bf{1}}+\da{\bf{24}}\la{\bf{3}}P_s\rs{\bf{1}}\right)\nonumber\\
&\qquad\qquad\qquad\qquad-(t_M)_{ij}\frac{(m_M)_{kl}}{\Lambda}\left(\da{\bf{13}}\la{\bf{4}}P_s\rs{\bf{2}}+\da{\bf{14}}\la{\bf{3}}P_s\rs{\bf{2}}\right)\nonumber\\
&\qquad\qquad\qquad\qquad-(t_M)_{ji}\frac{((m_M)_{kl})^*}{\Lambda}\left(\ds{\bf{13}}\ls{\bf{4}}P_s\ra{\bf{2}}+\ds{\bf{14}}\ls{\bf{3}}P_s\ra{\bf{2}}\right)\nonumber\\
&\qquad\qquad\qquad\qquad+(t_M)_{kl}\frac{((m_M)_{ij})^*}{\Lambda}\left(\ds{\bf{14}}\ls{\bf{2}}P_s\ra{\bf{3}}+\ds{\bf{24}}\ls{\bf{1}}P_s\ra{\bf{3}}\right)\nonumber\\
&\qquad\qquad\qquad\qquad+(t_M)_{lk}\frac{(m_M)_{ij}}{\Lambda}\left(\da{\bf{14}}\la{\bf{2}}P_s\rs{\bf{3}}+\da{\bf{24}}\la{\bf{1}}P_s\rs{\bf{3}}\right)\nonumber\\
&\qquad\qquad\qquad\qquad-(t_M)_{kl}\frac{(m_M)_{ij}}{\Lambda}\left(\da{\bf{13}}\la{\bf{2}}P_s\rs{\bf{4}}+\da{\bf{23}}\la{\bf{1}}P_s\rs{\bf{3}}\right)\nonumber\\
&\qquad\qquad\qquad\qquad-(t_M)_{lk}\frac{((m_M)_{ij})^*}{\Lambda}\left(\ds{\bf{13}}\ls{\bf{2}}P_s\ra{\bf{4}}+\ds{\bf{23}}\ls{\bf{1}}P_s\ra{\bf{3}}\right)\bigg).
\end{align}
I explicitly state parity conjugate structures here simply because I do in (\ref{GenMoller}). As in the main text, $P_s=p_3+p_4$.

Clearly, these diverge as $\sim E^2$ for the $(+,+,-,-)$ helicity configuration (and its parity conjugate). For this amplitude, the lowest dimension contact terms also share this degree of high energy dependence, so should generally be included at this order. However, they cannot cancel the leading HEL divergence from the factorisation terms constructed above. This is clear from converting to the $\{\mathbf{L}_i\}$ basis using (\ref{E4toL}). The leading $\sim E^2$ terms are then of the form $\frac{(u-t)}{s-m_{s_M}^2}\ds{\bf{12}}\da{\bf{34}}\rightarrow \frac{(u-t)}{s}\ds{12}\da{34}$, which clearly retains the pole in the HEL and therefore cannot be canceled by contact terms. It is also clear that these divergent terms cannot be canceled by other terms in different channels, since the divergent helicity configurations in the other channels are distinct (they correspond to different pairings of the same-helicity fermions). The only way for them to be removed is by eliminating the dipole interactions.

\subsection{Four vectors}\label{app:4V}
The $F^3$-type $3$-particle couplings induce contributions to the four vector amplitude. The double insertion terms are given by
\begin{align}\label{F3double}
&A_{sF^3F^3}\left(W_A,W_B,W_C,W_D\right)\nonumber\\
&=\frac{-1}{s-m_{s_M^2}}\frac{1}{2\Lambda^4}\bigg(h_{ABM}(h_{MCD})^*\ds{\bf{12}}\da{\bf{34}}\nonumber\\
&\qquad\qquad\qquad\times\bigg(\left(u-t\right)\ds{\bf{12}}\da{\bf{34}}+\frac{1}{6}\left(\frac{m_2}{m_1}\hat{\mathbf{L}}_1+\frac{m_1}{m_2}\hat{\mathbf{L}}_2+\frac{m_4}{m_3}\mathbf{L}_3+\frac{m_3}{m_4}\mathbf{L}_4\right)\nonumber\\
&\qquad\qquad\qquad\qquad+\frac{2}{3}m_1m_2\left(\da{\bf{13}}\da{\bf{24}}+\da{\bf{23}}\da{\bf{14}}\right)+\frac{2}{3}m_3m_4\left(\ds{\bf{13}}\ds{\bf{24}}+\ds{\bf{23}}\ds{\bf{14}}\right)\bigg)\nonumber\\
&\qquad\qquad\qquad\qquad\qquad\qquad\qquad+h_{ABM}h_{MCD}m_{s_M}^2\ds{\bf{12}}\ds{\bf{34}}\left(\ds{\bf{13}}\ds{\bf{24}}+\ds{\bf{14}}\ds{\bf{23}}\right)\bigg)\nonumber\\
&\qquad\qquad\qquad\qquad\qquad+\text{Parity conj.}
\end{align}
The first Lorentz structure on the right-hand side has been converted into a form with transparent HEL scaling through application of (\ref{E4toL}). The leading HEL divergence clearly scales as $\sim E^4$ for the MHV helicity configuration and is given by the term proportional to $u-t$. This can't be canceled by contact terms because the Lorentz structure contains a pole (and agrees with the expected massless amplitude containing a pole), nor can it be canceled by terms from other channels. As long as the $h$-type couplings are present, these terms are present. 

The single insertion terms are given by 
\begin{align}\label{SingleF3Res}
&A_{sF^3}\left(W_A,W_B,W_C,W_D\right)\nonumber\\
&=\frac{-1}{s-m_{s_M^2}}\frac{1}{4\Lambda^2}\frac{1}{m_3m_4}h_{ABM}\ds{\bf{12}}\nonumber\\
&\qquad\times\Bigg((f_{CD}^{\,\,\,\,\,\,\,\,\,M})^*\bigg(\ds{\bf{34}}\bigg(\frac{1}{6}\left(\frac{m_4}{m_3}\mathbf{L}_3+\frac{m_3}{m_4}\mathbf{L}_4\right)+\frac{4}{3}m_1m_2\left(\da{\bf{13}}\da{\bf{24}}+\da{\bf{14}}\da{\bf{23}}\right)\nonumber\\
&\qquad\qquad\qquad\qquad\qquad\qquad\qquad\qquad\qquad+\frac{2}{3}m_3m_4\left(\ds{\bf{13}}\ds{\bf{24}}+\ds{\bf{14}}\ds{\bf{23}}\right)\bigg)\nonumber\\
&\qquad\qquad\qquad\qquad-\frac{1}{6}\da{\bf{34}}\left(\mathbf{L}_3+\mathbf{L}_4\right)-\left(m_3\ds{\bf{34}}-m_4\da{\bf{34}}\right)\left(m_1\da{\bf{14}}\ds{\bf{23}}+m_2\da{\bf{24}}\ds{\bf{13}}\right)\nonumber\\
&\qquad\qquad\qquad\qquad\qquad-\left(m_4\ds{\bf{34}}-m_3\da{\bf{34}}\right)\left(m_1\da{\bf{13}}\ds{\bf{24}}+m_2\da{\bf{23}}\ds{\bf{14}}\right)\nonumber\\
&\qquad\qquad\qquad\qquad+\left(m_{s_M}^2-\frac{1}{3}\left(m_3^2+m_4^2+2m_1^2+2m_2^2\right)\right)\da{\bf{34}}\left(\ds{\bf{13}}\ds{\bf{24}}+\ds{\bf{14}}\ds{\bf{23}}\right)\bigg)\nonumber\\
&\qquad\qquad\qquad+f_{CD}^{\,\,\,\,\,\,\,\,\,M}m_{s_M}^2\da{\bf{34}}\left(\ds{\bf{13}}\ds{\bf{24}}+\ds{\bf{14}}\ds{\bf{23}}\right)\nonumber\\
&\qquad\qquad\qquad+\left(f_{DM}^{\,\,\,\,\,\,\,\,\,
C}m_3\ds{\bf{34}}+(f_{MC}^{\,\,\,\,\,\,\,\,\,
D})^*m_4\da{\bf{34}}\right)\nonumber\\
&\qquad\qquad\times\left(\frac{1}{6m_4}\mathbf{L}_4+m_1\ds{\bf{23}}\da{\bf{14}}+m_2\ds{\bf{13}}\da{\bf{24}}+\frac{1}{3}m_4\left(\ds{\bf{13}}\ds{\bf{24}}+\ds{\bf{14}}\ds{\bf{23}}\right)\right)\nonumber\\
&\qquad\qquad\qquad+\left(f_{MC}^{\,\,\,\,\,\,\,\,\,
D}m_4\ds{\bf{34}}+(f_{DM}^{\,\,\,\,\,\,\,\,\,
C})^*m_3\da{\bf{34}}\right)\nonumber\\
&\qquad\qquad\times\left(\frac{1}{6m_3}\mathbf{L}_3+m_1\ds{\bf{24}}\da{\bf{13}}+m_2\ds{\bf{14}}\da{\bf{23}}+\frac{1}{3}m_3\left(\ds{\bf{13}}\ds{\bf{24}}+\ds{\bf{14}}\ds{\bf{23}}\right)\right)\Bigg)\nonumber\\
&\qquad\qquad\qquad\qquad\qquad\qquad+\text{Parity conj.}.
\end{align}
To this expression, terms generated by the simultaneous exchanges $1\leftrightarrow 3$ and $2\leftrightarrow 4$ should be added. This amplitude has been converted to the $\{\mathbf{L}_i\}$ basis using (\ref{E3toL}), as well as identity (\ref{E4toL}) again. The $t$ and $u$-channel expressions are inferable from exchanges. This time, contact interactions do have to be considered and tuned against the leading HEL divergent terms. This has been implicitly performed in the expression above so that the leading HEL scaling arises at $\sim E^3$. Finally, the identity (\ref{LtoLhat}) has also been applied to convert all terms containing factors of $\mathbf{L}_i$ or $\hat{\mathbf{L}}_i$ into a linearly independent basis. I choose this basis to consist of Lorentz structures in which the spinor bilinear prefactors accompanying the factors of $\mathbf{L}_i$ are as holomorphic as possible. 

The amplitude's terms (\ref{SingleF3Res}) presented above that are generated by $F^3$ insertions have leading order HEL scaling $\sim E^3$ after tuning by contact terms. This occurs for two classes of helicity configurations. The first consists of one longitudinal and three transverse helicities, of which one is opposite the other two. This combination receives contributions from the double insertion terms in (\ref{F3double}), both as the leading order limit of the terms proportional to $\mathbf{L}_i$ and $\hat{\mathbf{L}}_i$, and as subleading limits of the $\sim E^4$ terms. Because there are contributions from subleading terms, these limits contain residual reference spinors (``gauge artifacts'').

The single $F^3$ insertion terms that scale with this configuration are expected to match onto massless amplitudes generated by massless $F^3$ couplings and the emergent $\varphi F^2$ couplings contained within (\ref{3legVecHEL}). This can be verified for e.g. the $(++,++,--,L)$ configuration, where the $s$-channel terms converge to 
\begin{align}\label{handc}
\frac{1}{2}\frac{1}{m_4}h_{ABM}\left((f_{DM}^{\,\,\,\,\,\,\,\,\,
C})^*-(f_{CD}^{\,\,\,\,\,\,\,\,\,
M})^*\right)\frac{\ds{12}^2\ds{14}\da{13}\da{34}}{s}
\end{align}
This still contains the $s$-pole, so is linearly independent of the other channels. Identifying (\ref{ctof}), these terms have the expected structure of the massless amplitude mediated by both dim-$6$ and dim-$5$ insertions. They cannot be reduced without the elimination of (at least one of) these couplings. 

The second class of leading helicity configurations consists of three transverse helicities of the same sign and one longitudinal helicity. In these cases, the amplitude converges to contact terms which can interfere between channels. For example, the $(++,++,++,L)$ configuration terms converge to the Lorentz structure $\ds{12}\ds{13}\ds{23}$. Requiring that this cancels gives the condition
\begin{align}\label{CovF3}
\left(f_{MD}^{\,\,\,\,\,\,\,\,\,
C}+(f_{DC}^{\,\,\,\,\,\,\,\,\,
M})^*\right)h_{ABM}+\left(f_{MD}^{\,\,\,\,\,\,\,\,\,
B}+(f_{DB}^{\,\,\,\,\,\,\,\,\,
M})^*\right)h_{AMC}+\left(f_{MD}^{\,\,\,\,\,\,\,\,\,
A}+(f_{DA}^{\,\,\,\,\,\,\,\,\,
M})^*\right)h_{MBC}=0.
\end{align}
This is close to the expected covariance condition on the $h_{ABC}$ couplings, but differs by contaminant terms proportional to $c_{CM}^Dh_{ABM}$ (and others related by index permutations). This is similar to the other covariance conditions on the $c$-type and $m$-type couplings derived in Section \ref{3Vec1Scalar} and Appendix \ref{app:2V2P} below. When both the $c_{CM}^D$ and $h_{ABC}$ couplings are suppressed compared to the terms in the full amplitude proportional to the $f$-couplings, the contaminant terms are a small correction. 

When one of the gluons is massless (choose $D$ in (\ref{CovF3})), then (\ref{PartAntiSym}) ensures that the relation becomes exactly the covariance of the $h$-couplings. This is enforced by consistent factorisation and is not an optional property that could emerge from high energy unitarity. 

It remains to explicitly perform the cancellations facilitated by (\ref{CovF3}). The terms contributing to the limit are given by
\begin{align}
\frac{-1}{24}\frac{1}{m_4^2\Lambda^2}\mathbf{L}_4\Bigg(&\left(f_{DM}^{\,\,\,\,\,\,\,\,\,
C}+(f_{CD}^{\,\,\,\,\,\,\,\,\,
M})^*\right)h_{MAB}\frac{\ds{\bf{12}}\ds{\bf{34}}}{s-m_{s_M}^2}+\left(f_{DM}^{\,\,\,\,\,\,\,\,\,
B}+(f_{BD}^{\,\,\,\,\,\,\,\,\,
M})^*\right)h_{MAC}\frac{\ds{\bf{13}}\ds{\bf{24}}}{t-m_{t_M}^2}\nonumber\\
&\qquad\qquad\qquad\qquad\qquad\qquad\qquad\quad+\left(f_{DM}^{\,\,\,\,\,\,\,\,\,
A}+(f_{AD}^{\,\,\,\,\,\,\,\,\,
M})^*\right)h_{MCB}\frac{\ds{\bf{32}}\ds{\bf{14}}}{u-m_{u_M}^2}\Bigg).
\end{align}
The cancellation clearly proceeds through another application of (\ref{OldSyzFact}), although I will not bother to explicitly present this. 

Notably, none of the divergent $\sim E^3$ terms in the amplitudes presented here arise for the three longitudinal, one transverse helicity configuration for which the amplitude constructed from $f$-type couplings in Section \ref{sec:PartialU} diverges. The presence of the $h$-type couplings therefore does not modify any of the conclusions presented there. 

In summary, the $F^3$ couplings are responsible for the strongest high energy dependence of the four vector amplitude in the absence of $4$-particle contact terms. The amplitude scales like $A\sim (hE^2/\Lambda^2)^2$ as a result. The only way in which they can viably admit an extended energy range far beyond the particle masses ($E\gg m$) is if $hE^2/\Lambda\ll \sqrt{4\pi}$. 
Having suppressed the $\sim E^4$ terms, the $\sim E^3$ terms become leading. There are three such classes of terms: the two mentioned above with single $F^3$ insertions and the class discussed in Section \ref{sec:PartialU} involving only the $f$-type couplings. As explained above, suppression of the terms with only $f$-type couplings can be arranged non-trivially by demanding that they obey the Jacobi and GCS constraints. The other two classes of terms contain single insertions of $h/\Lambda^2$, so can be made comparatively small. However, as is clear in (\ref{SingleF3Res}), there are single insertion terms of the form $\frac{h}{\Lambda^2}\frac{f}{m}E^3$ and $\frac{h}{\Lambda^2}fE^2$ and terms of the latter form cannot be canceled in the HEL. The relation (\ref{CovF3}) must be satisfied in order for the $E^3$ terms not to dominate over the weaker $E^2$ at high enough energies. Covariance emerges in the further limit that $\frac{cE}{\Lambda}$ are also small deformations, which is also required for the associated suppression of the $\sim E^3$ terms described around (\ref{handc}).

\subsection{Two vectors and two scalars}\label{app:2V2S}

This Appendix compiles the contributions to the two vector, two scalar amplitude from Section \ref{sec:2vec2sca} induced by insertions of $F^3$ and $\varphi F^2$. The $s$-channel scalar exchange is modified to include
\begin{align}
A_{s\varphi F^2}(W_A,W_B,\varphi_i,\varphi_j)=\frac{-1}{s-m_{s_m}^2}C_{ijm}\left(\frac{c^m_{AB}}{\Lambda}\ds{\bf{12}}^2+\frac{(c^m_{AB})^*}{\Lambda}\da{\bf{12}}^2\right),
\end{align}
which is manifestly unitary in the HEL, while the vector exchange must also include
\begin{align}\label{2V2SF3insert}
A_{sF^3}(W_A,W_B,\varphi_i,\varphi_j)&=\frac{-1}{s-m_{s_M}^2}\frac{i}{2}(t_M)_{ij}\bigg(\frac{h_{ABM}}{\Lambda^2}\ds{\bf{12}}\ls{\bf{1}}p_3p_4-p_4p_3\rs{\bf{2}}\nonumber\\
&\qquad\qquad\qquad\qquad\qquad\qquad+\frac{(h_{ABM})^*}{\Lambda^2}\da{\bf{12}}\la{\bf{1}}p_3p_4-p_4p_3\ra{\bf{2}}\bigg).
\end{align}
The $t$-channel vector exchange amplitude is extended to include
\begin{align}\label{2V2SSF2insert}
&A_{t\varphi F^2}(W_A,W_B,\varphi_i,\varphi_j)\nonumber\\
&=\frac{-1}{t-m_{t_M}^2}\Bigg(
\frac{c^i_{AM}(c^j_{BM})^*}{\Lambda^2}\ls{\bf{1}}P_t\ra{\bf{2}}^2+\frac{(c^i_{AM})^*c^j_{BM}}{\Lambda^2}\la{\bf{1}}P_t\rs{\bf{2}}^2\nonumber\\
&\qquad\qquad\qquad\qquad+\frac{c^i_{AM}c^j_{BM}}{\Lambda^2}m_{t_M}^2\ds{\bf{12}}^2+\frac{(c^i_{AM}c^j_{BM})^*}{\Lambda^2}m_{t_M}^2\da{\bf{12}}^2\nonumber\\
&\qquad\qquad\qquad-\frac{1}{2\sqrt{2}}m_{t_M}\Bigg(\lambda^i_{AM}\left(\frac{c^j_{BM}}{\Lambda}\ds{\bf{12}}\ls{\bf{2}}P_t\ra{\bf{1}}+\frac{(c^j_{BM})^*}{\Lambda}\da{\bf{12}}\la{\bf{2}}P_t\rs{\bf{1}}\right)\nonumber\\
&\qquad\qquad\qquad\qquad\qquad\qquad+\lambda^j_{BM}\left(\frac{c^i_{AM}}{\Lambda}\ds{\bf{12}}\ls{\bf{1}}P_t\ra{\bf{2}}+\frac{(c^i_{AM})^*}{\Lambda}\da{\bf{12}}\la{\bf{1}}P_t\rs{\bf{2}}\right)\Bigg)\Bigg)
\end{align}
(I omit the contribution induced purely from the Higgs interactions described already in Section \ref{sec:2vec2sca}). The $u$-channel expressions can be inferred by exchanging $1\leftrightarrow 2$. 

Clearly these terms have $\sim E^2$ HEL scaling. This occurs for same sign transverse helicities in (\ref{2V2SF3insert}) and opposite sign transverse helicities in (\ref{2V2SSF2insert}). In the case of the former, the divergence cannot be canceled by a contact term because the massless Lorentz structure to which the amplitude converges to still contains a pole. It can only be removed if $h_{ABM}(t_M)_{ij}=0$, meaning that, in some basis, the vectors coupling to the scalars (at least through (\ref{ScalarMat3leg})) cannot participate in $F^3$ interactions. In the latter case, the contributing terms in the $t$ and $u$-channels cannot cancel in the massless limit because the presence of uncanceled $t$ and $u$ poles ensures that they remain linearly independent. These terms can only be removed by eliminating the $c$-type couplings. None of these terms interfere with the limits analysed in Section \ref{sec:2vec2sca}.

\subsection{Three vectors and one scalar}\label{app:3V1S}

I give here the expressions for the amplitudes induced by insertion of the $F^3$ terms. These are
\begin{align}
&A_{F^3}\left(W_A,W_B,W_C,\varphi_i\right)\nonumber\\
&=\frac{1}{s-m_{s_M}^2}\frac{i}{\Lambda^2}h_{ABM}\ds{\bf{12}}\nonumber\\
&\qquad\times\bigg(\frac{c^i_{MC}}{\Lambda}m_{s_M}^2\ds{\bf{13}}\ds{\bf{23}}+\frac{(c^i_{MC})^*}{\Lambda}\left(\ls{\bf{1}}p_4\ra{\bf{3}}-m_3\ds{\bf{13}}\right)\left(\ls{\bf{2}}p_4\ra{\bf{3}}-m_3\ds{\bf{23}}\right)\nonumber\\
&\qquad\qquad\qquad\qquad\qquad-\frac{1}{2\sqrt{2}}m_{s_M}\lambda^i_{MC}\left(\ds{\bf{13}}\ls{\bf{2}}p_4\ra{\bf{3}}+\ds{\bf{23}}\ls{\bf{1}}p_4\ra{\bf{3}}-2m_3\ds{\bf{13}}\ds{\bf{23}}\right)\bigg)\nonumber\\
&\qquad\qquad\qquad\qquad\qquad+\text{Parity conj.},
\end{align}
where the $P$-conjugate terms are given by switching bracket shapes and complex conjugating the accompanying couplings. As usual, the $t$-channel terms are obtained by exchanging $2\leftrightarrow 3$ and the $u$-channel from exchanging $1\leftrightarrow 3$. The single insertion terms of $\varphi F^2$ without $F^3$ are covered in Section \ref{3Vec1Scalar}. 

The leading order high energy dependence is given by helicity configurations of the form $(++,++,--)$. This only receives contributions from the $h_{ABM}(c_{MC}^i)^*$ term, is $\sim E^3$ and cannot be canceled by a contact term (because of the presence of a pole). Switching off the $\varphi F^2$ couplings, the remaining leading divergence is $\sim E^2$ and arises for configurations like $(++,++,L)$. While this is the same configuration for which leading transverse divergences appeared in (\ref{WWWScaGenS}), the Lorentz structures are linearly independent. The conditions obtained by demanding the softening of these terms therefore do not interfere with each other. This leads to the requirement that $h_{ABM}(t_M)_{Ci}=0$ and therefore that the vectors participating in $F^3$ interactions cannot couple to scalars.

\subsection{One vector, two fermions and one scalar}\label{app:1V2P1S}

Here is the collection of contributions to the mixed amplitude in Section \ref{sec:Mixed} involving insertions of $\varphi F^2$ and $\chi F\chi$. The $t$-channel is modified to include
\begin{align}
A_{t\chi F\chi}\left(\varphi_m,W_A,\psi_i,\psi_j\right)&=\frac{-2}{t-m_{t_k}^2}\nonumber\\
&\quad\times\bigg(y^m_{ik}\frac{((m_A)_{kj})^*}{\Lambda}\ds{\bf{24}}\la{\bf{3}}P_t\rs{\bf{2}}-(y^m_{ik})^*\frac{(m_A)_{kj}}{\Lambda}\da{\bf{24}}\la{\bf{2}}P_t\rs{\bf{3}}\nonumber\\
&\qquad\quad+(y^m_{ik})^*\frac{((m_A)_{kj})^*}{\Lambda}m_{t_k}\ds{\bf{23}}\ds{\bf{24}}-y^m_{ik}\frac{(m_A)_{kj}}{\Lambda}m_{t_k}\da{\bf{23}}\da{\bf{24}}\bigg).
\end{align}
The $u$-channel can be inferred from exchanging the fermions. The $s$-channel vector exchange terms are modified to include
\begin{align}\label{sHDOMixed}
&A_{s\{\varphi F^2,\chi F\chi\}}\left(\varphi_m,W_A,\psi_i,\psi_j\right)\nonumber\\
&=\frac{-1}{s-m_{s_M}^2}
\bigg(\frac{c^m_{AM}}{\Lambda}\frac{2(m_M)_{ji}}{\Lambda}\ls{\bf{2}}P_s\ra{\bf{3}}\ls{\bf{2}}P_s\ra{\bf{4}}+\frac{(c^m_{AM})^*}{\Lambda}\frac{2((m_M)_{ij})^*}{\Lambda}\la{\bf{2}}P_s\rs{\bf{3}}\la{\bf{2}}P_s\rs{\bf{4}}\nonumber\\
&\qquad\qquad\qquad\qquad+\frac{c^m_{AM}}{\Lambda}\frac{2((m_M)_{ij})^*}{\Lambda}m_{s_M}^2\ds{\bf{23}}\ds{\bf{24}}+\frac{(c^m_{AM})^*}{\Lambda}\frac{2(m_M)_{ji}}{\Lambda}m_{s_M}^2\da{\bf{23}}\da{\bf{24}}\nonumber\\
&\qquad\qquad\qquad\qquad+\frac{c_{AM}^m}{\Lambda}(t_M)_{ij}\ds{\bf{24}}\ls{\bf{2}}P_s\ra{\bf{3}}-\frac{c_{AM}^m}{\Lambda}((t_M)_{ij})^*\ds{\bf{23}}\ls{\bf{2}}P_s\ra{\bf{4}}\nonumber\\
&\qquad\qquad\qquad\qquad+\frac{(c_{AM}^m)^*}{\Lambda}(t_M)_{ij}\da{\bf{23}}\la{\bf{2}}P_s\rs{\bf{4}}-\frac{(c_{AM}^m)^*}{\Lambda}((t_M)_{ij})^*\da{\bf{24}}\la{\bf{2}}P_s\rs{\bf{3}}\nonumber\\
&\qquad\qquad\qquad\qquad+\frac{m_{s_M}}{\sqrt{2}}\lambda_{AM}^m\bigg(\frac{((m_M)_{ji})^*}{\Lambda}\left(\ds{\bf{23}}\la{\bf{2}}P_s\rs{\bf{4}}+\ds{\bf{24}}\la{\bf{2}}P_s\rs{\bf{3}}\right)\nonumber\\
&\qquad\qquad\qquad\qquad\qquad\qquad\qquad\qquad+\frac{(m_M)_{ij}}{\Lambda}\left(\da{\bf{23}}\ls{\bf{2}}P_s\ra{\bf{4}}+\da{\bf{24}}\ls{\bf{2}}P_s\ra{\bf{3}}\right)\bigg)\bigg).
\end{align}
This includes single and double insertions of ``higher-dimensional operators''. 

The $s$-channel double insertion terms (those with both $\varphi F^2$ and $\chi F\chi$) scale as $\sim E^2$ in the HEL for the $(++,-,-)$ configuration and its parity conjugate. This can only be eliminated by eliminating the associated combination of couplings (it cannot be canceled by a contact term because the Lorentz structure still contains a pole in the HEL). At $\sim E$, the $s$-channel single insertions of fermion dipoles contribute for the same helicity configurations ($(L,+,+)$ and conjugate) as the leading divergences of the amplitude presented in (\ref{FactAllAmp}). However, the terms in (\ref{sHDOMixed}) converge to linearly independent Lorentz structures containing a pole, whereas the terms in (\ref{FactAllAmp}) converge to contact amplitudes. The limits of these terms therefore do not interfere with each other. Finally, the expressions above also include $\sim E$ divergences for the helicity configurations $(++,-,+)$ and $(++,+,-)$, as well as their parity conjugates. These are generated by the $\varphi F^2$ couplings in the $s$-channel and the dipole couplings in the $t$ and $u$-channels. However, the terms emerging in the HEL in each channel retain their poles and remain linearly independent, so cannot cancel each other. Again, the only way to eliminate these divergences is by precluding the accompanying combination of couplings. This all amounts to a forbidding of vectors or fermions involved in the $\varphi F^2$ or $\chi F\chi$ interactions (in some basis, possibly not the mass eigenbasis) from coupling either to each other or through the other matter couplings ($y_{ij}^m$, $(t_M)_{ij}$ and $\lambda_{AM}^m$) involved in this amplitude.

\subsection{Two vectors and two fermions}\label{app:2V2P}

This Appendix begins with the collection of terms in the massive vector and fermion amplitude from Section \ref{sec:2Vec2Ferm} that are induced by $F^3$ interactions. The contributions purely from the dipole moments of the fermions are given further below, while the $\varphi F^2$ interactions are covered in Section \ref{sec:2Vec2Ferm}. With the inclusion of $F^3$, only the $s$-channel is modified:
\begin{align}
&A_{sF^3}\left(W_A,W_B,\psi_i,\psi_j\right)\\\nonumber
&\qquad\qquad=\frac{-1}{s-m_{s_M}^2}\frac{i}{2\Lambda^2}h_{ABM}\ds{\bf{12}}\nonumber\\
&\qquad\qquad\qquad\times\bigg(\frac{2(m_M)_{ij}}{\Lambda}\left(\ls{\bf{2}}P_s\ra{\bf{3}}\ls{\bf{1}}P_s\ra{\bf{4}}+\ls{\bf{2}}P_s\ra{\bf{4}}\ls{\bf{1}}P_s\ra{\bf{3}}\right)\nonumber\\
&\qquad\qquad\qquad\qquad\qquad\qquad\qquad+\frac{2((m_M)_{ji})^*}{\Lambda}m_{s_M}^2\left(\ds{\bf{13}}\ds{\bf{24}}+\ds{\bf{23}}\ds{\bf{14}}\right)\nonumber\\
&\qquad\qquad\qquad\qquad\qquad\qquad\qquad-(t_M)_{ij}\left(\ds{\bf{24}}\ls{\bf{1}}P_s\ra{\bf{3}}+\ds{\bf{14}}\ls{\bf{2}}P_s\ra{\bf{3}}\right)\nonumber\\
&\qquad\qquad\qquad\qquad\qquad\qquad\qquad+((t_M)_{ij})^*\left(\ds{\bf{23}}\ls{\bf{1}}P_s\ra{\bf{4}}+\ds{\bf{13}}\ls{\bf{2}}P_s\ra{\bf{4}}\right)\bigg)\nonumber\\
&\qquad\qquad\qquad\qquad\qquad\qquad\qquad+\text{Parity conj.}
\end{align}

The leading HEL scaling is $\sim E^3$ for the $(++,++,-,-)$ helicity configuration (and conjugate). This is proportional to $h_{ABM}(m_M)_{ij}$ and can only be removed by direct elimination of this factor (again, the massless Lorentz structure that these terms match onto contains a pole and cannot be canceled by a contact term). Doing so leaves behind $\sim E^2$ divergent terms for $(++,++,-,+)$ and $(++,++,+,-)$ configurations, as well as their conjugates, which are respectively proportional to $h_{ABM}(t_M)_{ij}$, $h_{ABM}((t_M)_{ij})^*$ and their conjugates. These configurations are distinct from those that yielded the Lie algebra commutation relations for the $(t_M)_{ij}$ couplings, as well as those induced by the dipole moment insertions, so don't interfere with the arguments in Section \ref{sec:2Vec2Ferm} and below. They can only be eliminated by eliminating the associated coupling constants, necessitating that, in an appropriate basis (possibly not the mass eigenbasis), the vector bosons with $F^3$ interactions cannot couple to the fermions. 

I next present separately the terms induced from insertions of the fermion dipole interactions (the $m$-type couplings in (\ref{FermionMat3leg})). They are given by
\begin{align}\label{tMDM}
&A_{t\chi F\chi}\left(W_A,W_B,\psi_i,\psi_j\right)\nonumber\\
&=\frac{-1}{t-m_{t_k}^2}\Bigg(\frac{4((m_A)_{ik})^*(m_B)_{jk}}{\Lambda^2}\ds{\bf{13}}\da{\bf{24}}\ls{\bf{1}}P_t\ra{\bf{2}}-\frac{4((m_A)_{ik}(m_B)_{kj})^*}{\Lambda^2}m_{t_k}\ds{\bf{12}}\ds{\bf{13}}\ds{\bf{24}}\nonumber\\
&\qquad-\frac{2((m_A)_{ik})^*}{\Lambda}\frac{1}{m_2}(t_B)_{kj}\ds{\bf{13}}\ds{\bf{24}}\ls{\bf{1}}P_t\ra{\bf{2}}+\frac{2((m_B)_{kj})^*}{\Lambda}\frac{1}{m_1}((t_A)_{ik})^*\ds{\bf{13}}\ds{\bf{24}}\la{\bf{1}}P_t\rs{\bf{2}}\nonumber\\
&\qquad\qquad-\frac{2((m_A)_{ik})^*}{\Lambda}((t_B)_{kj})^*\frac{m_{t_k}}{m_2}\da{\bf{24}}\ds{\bf{13}}\ds{\bf{12}}+\frac{2((m_B)_{kj})^*}{\Lambda}(t_A)_{ik}\frac{m_{t_k}}{m_1}\ds{\bf{24}}\da{\bf{13}}\ds{\bf{12}}\Bigg)\nonumber\\
&\qquad\qquad\qquad\qquad\qquad+\text{Parity conj.}
\end{align}
(with $u$-channel obtained by swapping the vectors) and 
\begin{align}\label{sMDM}
&A_{s\chi F\chi}\left(W_A,W_B,\psi_i,\psi_j\right)\nonumber\\
&\qquad=\frac{-1}{s-m_{s_M}^2}\frac{i}{2m_1m_2}\frac{((m_M)_{ij})^*}{\Lambda}\nonumber\\
&\qquad\qquad\times\Bigg(\left(\left({f}_{BM}^{\,\,\,\,\,\,\,\,\,A}+({f}_{AB}^{\,\,\,\,\,\,\,\,\,M})^*\right)m_1\ds{\bf{12}}+\left(({f}_{MA}^{\,\,\,\,\,\,\,\,\,B})^*-({f}_{AB}^{\,\,\,\,\,\,\,\,\,M})^*\right)m_2\da{\bf{12}}\right)\nonumber\\
&\qquad\qquad\qquad\qquad\qquad\qquad\qquad\qquad\times\left(\ds{\bf{13}}\la{\bf{2}}P_s\rs{\bf{4}}+\ds{\bf{14}}\la{\bf{2}}P_s\rs{\bf{3}}\right)\nonumber\\
&\qquad\qquad\qquad+\left(\left({f}_{MA}^{\,\,\,\,\,\,\,\,\,B}+({f}_{AB}^{\,\,\,\,\,\,\,\,\,M})^*\right)m_2\ds{\bf{12}}+\left(({f}_{BM}^{\,\,\,\,\,\,\,\,\,A})^*-({f}_{AB}^{\,\,\,\,\,\,\,\,\,M})^*\right)m_1\da{\bf{12}}\right)\nonumber\\
&\qquad\qquad\qquad\qquad\qquad\qquad\qquad\qquad\times\left(\ds{\bf{23}}\la{\bf{1}}P_s\rs{\bf{4}}+\ds{\bf{24}}\la{\bf{1}}P_s\rs{\bf{3}}\right)\nonumber\\
&\qquad\qquad\qquad\qquad-\left({f}_{AB}^{\,\,\,\,\,\,\,\,\,M}+({f}_{AB}^{\,\,\,\,\,\,\,\,\,M})^*\right)m_{s_M}^2\da{\bf{12}}\left(\ds{\bf{13}}\ds{\bf{24}}+\ds{\bf{23}}\ds{\bf{14}}\right)\Bigg)\nonumber\\
&\qquad\qquad\qquad\qquad\qquad\qquad+\text{Parity conj.}
\end{align}
The parity conjugate terms are given, as usual, by swapping the bracket shapes and complex conjugating the coefficients (I omit writing them explicitly for brevity). 

The double insertion terms of the dipoles scale as $\sim E^2$ for the $(++,--,+,-)$ helicity configuration (and its conjugate). All contact terms have scaling $\mathcal{O}(E^3)$, so cannot be involved in reducing the HEL. The divergent terms can only be removed by eliminating the dipole couplings, which is consistent with off-shell expectations. The terms with single insertions also scale as $\sim E^2$, this time for helicity configurations with one longitudinal and one transverse vector. The configuration $(L,++,-,-)$ only receives contributions from $s$-channel terms proportional to $\frac{1}{m_1}\left({f}_{MA}^{\,\,\,\,\,\,\,\,\,B}-{f}_{AB}^{\,\,\,\,\,\,\,\,\,M}\right)(m_M)_{ij}\propto c_{BM}^A(m_M)_{ij}$, (the configuration $(--,L,+,+)$ and both of their parity conjugates give similar results). This is the expected dependence arising from insertions of both a fermion dipole and an emergent $\varphi F^2$ operator in the HEL in which the scalar is identified with the longitudinal vector. The divergence for these configurations can only be reduced by removing $c_{BM}^A$ (equivalently (\ref{StandardLA})) or the dipole couplings. 

The divergences for the configurations $(--,L,-,-)$ (and those related by parity and exchanges) are more intricate and receive contributions from all three factorisation channels. This cancellation requires
\begin{align}\label{DipoleCov}
(t_B)_{ik}(m_A)_{kj}+(t_B)_{jk}(m_A)_{ik}+\frac{i}{2}\left(({f}_{MA}^{\,\,\,\,\,\,\,\,\,B})^*+{f}_{AB}^{\,\,\,\,\,\,\,\,\,M}\right)(m_M)_{ij}=0,
\end{align}
which is almost the expected expression for covariance of the dipole interactions under the Lie algebra. The covariance relation is contaminated by the presence of $({f}_{MA}^{\,\,\,\,\,\,\,\,\,B}-{f}_{AB}^{\,\,\,\,\,\,\,\,\,M})^*(m_M)_{ij}\propto m_1(c^A_{BM})^*(m_M)_{ij}$, in spite of the fact that the emergent $\varphi F^2$ interaction is not expected to mediate the massless amplitude upon which this helicity configuration should match onto. This is similar to the previous cases of the near covariance of the $F^3$ and $\varphi F^2$ interactions explained in Sections \ref{sec:FullUni}, \ref{3Vec1Scalar} and \ref{app:4V}. Just as for those cases, the relation (\ref{DipoleCov}) is only significant when the emergent contact term that these terms converge to otherwise carries the leading order energy dependence. The cancellation of this term provided by (\ref{DipoleCov}) ensures that the cut-off of the theory is extended as far as possible while allowing for these effective interactions to exist. If all of the conditions required for high energy unitarity are otherwise met and small dim-$5$ deformations are additionally introduced, then (\ref{DipoleCov}) is required to remove what would otherwise be the single leading class of terms in the HEL obstructing perturbative unitarity (with a cut-off $\sim \sqrt{m\Lambda}$ instead of $\Lambda$). 

Note that, if vector $A$ is massless, then (\ref{PartAntiSym}) ensures that (\ref{DipoleCov}) becomes the expected uncontaminated covariance relation. Furthermore, this constraint becomes necessitated by consistent factorisation, which can be demonstrated by a similar argument to that sketched-out above for demonstrating covariance of other higher-dim operator insertions described in previous Sections under analogous conditions. 

It remains to convert the residues in (\ref{tMDM}) and (\ref{sMDM}) into a form in which the cancellation for the $(--,L,-,-)$ configuration (and its partners) is made manifest. This simply involves another application of (\ref{OldSyzFact}) and I will not bother to present its implementation here. I will, however, mention that the residual terms have leading $\sim E$ HEL divergences for helicity configurations in which both vectors are transverse and both fermion helicities are the same. These do not interfere with any of the $\sim E$ divergent terms encountered in the analysis in Section \ref{sec:2Vec2Ferm}, which all arise for different helicity configurations.

\subsection{One vector and three scalars}

The final amplitude to complete the catalogue of tree-level $4$-particle massive amplitudes of particles of spin $\leq 1$ is $A(\varphi_i,\varphi_j,\varphi_k,W_A)$. All three factorisation channels for this amplitude are again related by exchange symmetries, so only the $s$-channel need be explicitly calculated. This is shown in Fig \ref{1V3SOSDi}.
\begin{feynfig}
\begin{figure}[h]
\begin{fmffile}{1Vec3Scalar}
 \begin{center}
 \begin{tabular}{c c c}
 & & \\
\begin{fmfgraph*}(120,80)
   \fmfleft{i1,i2}
   \fmfright{o1,o2}
   \fmf{dashes}{i2,v1}
   \fmf{dashes}{v2,o2}
   \fmf{dashes}{i1,v1}
   \fmf{boson}{v2,o1}
   \fmf{dashes}{v1,v2}
   \fmfv{decor.shape=circle,decor.filled=gray50,decor.size=0.15w}{v1,v2}
   \fmflabel{$i$}{i1}
   \fmflabel{$j$}{i2}
   \fmflabel{$k$}{o2}
   \fmflabel{$A$}{o1}
 \end{fmfgraph*} 
 &\,\,& \begin{fmfgraph*}(120,80)
   \fmfleft{i1,i2}
   \fmfright{o1,o2}
   \fmf{dashes}{i2,v1}
   \fmf{dashes}{v2,o2}
   \fmf{dashes}{i1,v1}
   \fmf{boson}{v2,o1}
   \fmf{boson}{v1,v2}
   \fmfv{decor.shape=circle,decor.filled=gray50,decor.size=0.15w}{v1,v2}
   \fmflabel{$i$}{i1}
   \fmflabel{$j$}{i2}
   \fmflabel{$k$}{o2}
   \fmflabel{$A$}{o1}
 \end{fmfgraph*}\nonumber\\
 & & 
 \end{tabular}
 \end{center}
\end{fmffile}
\caption{$s$-channel contributions for scattering of a vector and three scalars.}\label{1V3SOSDi}
\end{figure}
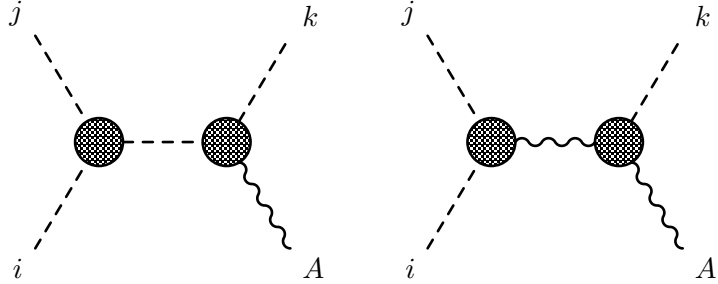
\end{feynfig}

The $s$-channel scalar exchange term is
\begin{align}
A_{s\varphi}\left(\varphi_i,\varphi_j,\varphi_k,W_A\right)=\frac{-(t_A)_{km}C_{ijm}}{s-m_{s_m}^2}\frac{1}{m_4}\la{\bf{4}}p_3\rs{\bf{4}}.
\end{align}
Here $C_{ijm}$ is a scalar trilinear coupling introduced in Section \ref{sec:2vec2sca}. Notably, this amplitude does not grow with energy in the HEL, so is automatically consistent with perturbative unitarity provided that the scalar trilinear is not too much larger than the mass of the vector boson. The vector exchange terms are generated from insertions of (\ref{2vec1scalar}) and are given by
\begin{align}\label{vecEx3scalar}
&A_{sW}\left(\varphi_i,\varphi_j,\varphi_k,W_A\right)\nonumber\\
&=\frac{-1}{2}\frac{1}{s-m_{s_M}^2}(t_M)_{ij} \bigg(-\frac{1}{\sqrt{2}}\frac{m_1^2-m_2^2}{m_{s_M}}\lambda_{MA}^k\la{\bf{4}}p_3\rs{\bf{4}}\nonumber\\
&\qquad\qquad\qquad\qquad\qquad\qquad+\frac{c_{MA}^k}{\Lambda}\ls{\bf{4}}p_1p_2-p_2p_1\rs{\bf{4}}+\frac{(c_{MA}^k)^*}{\Lambda}\la{\bf{4}}p_1p_2-p_2p_1\ra{\bf{4}}\bigg).
\end{align}
As usual, the $t$-channel terms are obtained by exchanging $2\leftrightarrow 3$ and the $u$-channel from exchanging $1\leftrightarrow 3$. 

The terms in the second line in (\ref{vecEx3scalar}) scale as $\sim E$ in the HEL when the vector is transversely polarised. Contact terms for this amplitude scale as $\sim E^2$ at mildest (in agreement with dimensional analysis), so cannot cancel these divergences, while each channel's terms in this limit remain linearly independent. The only possibility for eliminating this divergence is to exclude the $c_{MA}^k$ couplings. The remaining amplitude does not grow with energy in the HEL and is manifestly unitary. 

When the external vector is a massless gluon, consistent factorisation demands that the scalar trilinear be covariant:
\begin{align}\label{CovTri}
(t_A)_{im}C_{mjk}+(t_A)_{jm}C_{imk}+(t_A)_{km}C_{ijm}=0.
\end{align}
This can be concluded through arguments paralleling those sketched out around (\ref{MasslessGluonConsFact}) in Section \ref{sec:PartialU} and is an unavoidable consistency requirement. When the vector is not massless however, no such constraint on the trilinear scalar coupling appears necessary since the amplitude does not grow with energy. Nevertheless, perturbative unitarity is an implicitly assumed condition on the scalar trilinear coupling, which is dimensionful. This bounds from above ratios of the form $C_{ijk}/m$, where $m$ is a relevant particle's mass. So in the limit that $m_4\rightarrow 0$, (\ref{CovTri}) becomes a necessary condition for the longitudinal amplitude to remain perturbatively unitary. 

In the HEL however, this covariance condition is not essential. The emergent scalar quartic is
\begin{align}
&A_{c}\left(\varphi_i,\varphi_j,\varphi_k,W_A^L\right)\nonumber\\
&\rightarrow\frac{i}{\sqrt{2}}\Bigg(\frac{1}{m_4}\left((t_A)_{im}C_{mjk}+(t_A)_{jm}C_{imk}+(t_A)_{km}C_{ijm}\right)\nonumber\\
&\qquad\qquad-\frac{1}{2\sqrt{2}}\left((t_M)_{ij}\lambda_{MA}^k\frac{m_1^2-m_2^2}{m_{s_M}}+(t_M)_{ki}\lambda_{MA}^j\frac{m_3^2-m_1^2}{m_{t_M}}+(t_M)_{jk}\lambda_{MA}^i\frac{m_2^2-m_3^2}{m_{u_M}}\right)\Bigg).
\end{align}
If $C_{ijk}\gg m_4$, then terms in the first line could dominate again and perturbative unitarity would require (\ref{CovTri}) so that the scalar quartic is not too large. However, this need not generally be the case.

I will not bother to reconcile the general amplitude constructed above with the specific case contained in the $\mathcal{N}=4$ superamplitude (\ref{N=4SYM}). The general expression is simple enough. However, I will note that this comparison is not trivial and requires application of the syzygy (\ref{SimpSyzFact}), since the $\mathcal{N}=4$ amplitude contains factors with two Mandelstam poles, meaning that terms from two channels in this non-SUSY calculation would need to be amalgamated together. 

\begin{table}[h!]
\begin{center}
\begin{tabular}{ c || c c c c c}
Growth & Helicity Config & Couplings & Structure & Removal \\
\hline\hline
$E$ & $++$ & $\frac{c}{\Lambda}t$ & fact & suppress\\
\hline
$E^0$ & $L$ & $\frac{C}{m}t$, $tt$ & contact & $C/m\ll 4\pi$  \\
\end{tabular}
\end{center}
\caption{Summary of results for single vector and three scalar amplitude.}\label{tab:1V3S}
\end{table}

Table \ref{tab:1V3S} summarises the (brief) conclusions of this calculation. In this one case, I list in the bottom row the leading energy dependence of the unitarised amplitude. The $C/m\ll 4\pi$ condition is necessary for perturbative unitarity at any energy and is not a condition for further cancellation (I generally won't otherwise state perturbative unitarity bounds on dimensionless couplings in the unitarised theories).

\bibliography{superspace}{}
\bibliographystyle{JHEP}
\end{document}